\documentclass[a4paper]{aa}

\usepackage{graphics,epsfig,txfonts}
\usepackage[utf8]{inputenc}
\usepackage[section]{placeins}
\usepackage{amsmath}
\usepackage{color}
\usepackage{xcolor}
\usepackage{natbib}
\bibpunct{(}{)}{;}{a}{}{,}

\newcommand{\kms}{km s$^{-1}$}

\def\vmeas {\ifmmode{\mathrm{v}} \else {$\mathrm{v}$}\fi} 

\def\vsys {\ifmmode{V_\mathrm{sys}} \else {$V_\mathrm{sys}$}\fi} 

\def\PAkin {\ifmmode{PA_\mathrm{{kin}}} \else {$PA_\mathrm{{kin}}$}\fi}  

\def\PAphot {\ifmmode{PA_\mathrm{{phot}}} \else {$PA_\mathrm{{phot}}$}\fi}  

\def\re {\ifmmode{R_\mathrm{{e}}} \else {$R_\mathrm{{e}}$}\fi}

\usepackage{color}

\DeclareMathOperator{\arctanh}{arctanh}
\begin{document}


\title{The extended Planetary Nebula Spectrograph (ePN.S) early-type galaxy survey: the kinematic diversity of stellar halos and the relation between halo transition scale and stellar mass}

\author{C. Pulsoni\inst{1,2}
     \and{O. Gerhard\inst{1}}
     \and{M. Arnaboldi\inst{3}}
     \and{L. Coccato\inst{3}}
     \and{A. Longobardi\inst{4}}
     \and{N. R. Napolitano\inst{5}}
     \and{E. Moylan\inst{6}}
     \and{C. Narayan\inst{1,8}}
     \and{V. Gupta\inst{1,7}}
     \and{A. Burkert\inst{9}}
     \and{M. Capaccioli\inst{10}}
     \and{A. L. Chies-Santos\inst{11}} 
     \and{A. Cortesi\inst{12}} 
     \and{K. C. Freeman\inst{13}} 
     \and{K. Kuijken\inst{14}} 
     \and{M. R. Merrifield\inst{15}} 
     \and{A. J. Romanowsky\inst{16,17}} 
     \and{C. Tortora\inst{18}}  }

\titlerunning{The extended PN.S ETG survey - velocity fields at large radii}
\authorrunning{C. Pulsoni et al.}

\institute{Max-Planck-Institut f\"ur extraterrestrische Physik, Giessenbachstra{\ss}e, 85748 Garching, Germany
	   \and Excellence Cluster Universe, Boltzmannstra{\ss}e 2, 85748, Garching, Germany 
           \and European Southern Observatory, Karl-Schwarzschild-Stra{\ss}e 2, 85748 Garching, Germany
           \and Kavli Institute for Astronomy and Astrophysics, Peking University, 5 Yiheyuan Road, Haidian District, Beijing 100871, PR China
           \and INAF - Astronomical Observatory of Capodimonte, Salita Moiariello, 16, 80131, Naples, Italy
           \and School of Civil Engineering The University of Sydney NSW 2006, Australia
           \and Department of Physics, Cornell University, Ithaca, New York 14853, USA
           \and Current address: 71, Akashganga IUCAA Post Bag 4, Ganeshkhind Pune University Campus, Pune - 411007, India
           \and University Observatory Munich, Scheinerstra{\ss}e 1, 81679 Munich, Germany
	   \and University of Naples "Federico II", Department of physics "Ettore Pancini", C.U. Monte Sant'Angelo, via Cinthia, I-80126, Naples, Italy 
           \and Departamento de Astronomia, Instituto de Fisica, Universidade Federal do Rio Grande do Sul, Porto Alegre, R.S 90040-060, Brazil
           \and Departamento de Astronomia, Instituto de Astronomia, Geofisica e Ciencias Atmosfericas da USP, Cidade Universitaria, CEP:05508900 Sao Paulo, SP, Brazil
           \and Research School of Astronomy and Astrophysics, Mount Stromlo Observatory, Cotter Road, Weston Creek, ACT 2611, Australia
	   \and Leiden Observatory, Leiden University, PO Box 9513, NL-2300 RA Leiden, the Netherlands
	   \and School of Physics and Astronomy, The University of Nottingham, University Park, Nottingham, NG7 2RD, UK
	   \and Department of Physics and Astronomy, San Jose State University, One Washington Square, San Jose, CA 95192, USA
	   \and University of California Observatories, 1156 High Street, Santa Cruz, CA 95064, USA
	   \and Kapteyn Astronomical Institute, University of Groningen, Postbus 800, NL-9700 AV Groningen, the Netherlands}

\abstract
{In the hierarchical two-phase formation scenario, the 
  halos of early type galaxies (ETGs) are expected to have different
  physical properties from the galaxies' central regions.}
{The ePN.S survey characterizes the kinematic properties of ETG
  halos using planetary nebulae (PNe) as tracers, overcoming the 
  limitations of absorption line spectroscopy at 
  low surface brightness.}
{ The survey is based on data from the custom built Planetary Nebula
  Spectrograph (PN.S), supplemented with PN kinematics from
  counter-dispersed imaging and from high-resolution PN
  spectroscopy.
  We present two-dimensional velocity and velocity
  dispersion fields for 33 ETGs, including both fast (FRs) and slow
  rotators (SRs), making this the largest kinematic survey to-date of
  extragalactic PNe. 
  The velocity fields are reconstructed from the measured PN
  velocities using an adaptive kernel procedure validated with
  simulations, and extend to a median of 5.6 effective radii
  ($\re$), with a range $[3 \re - 13 \re]$.
  We complemented the PN kinematics with absorption line data from the
  literature, for a complete description of the kinematics from the
  center to the outskirts.  }
{ We find that ETGs typically show a kinematic transition between
  inner regions and halo. Estimated transition radii in units of $\re$
  anti-correlate with stellar mass. SRs have increased but still modest
  rotational support at large radii, while most of the FRs show a
  decrease in rotation, due to the fading of the inner disk in the
  outer, more slowly rotating spheroid. $30\%$ of the FRs are dominated
  by rotation also at large radii.  Most ETGs have flat or slightly
  falling halo velocity dispersion profiles, but $15\%$ of the sample
  have steeply falling profiles.  All of the SRs and 40\% of the FRs
  show signatures of triaxial halos such as kinematic twists,
  misalignments, or rotation along two axes. We show with
    illustrative photometric models that this is consistent with the
    distribution of isophote twists from  extended photometry.}
{ETGs have more diverse kinematic properties in their halos than in
  the central regions.  FRs do contain inner disk components but these frequently
  fade in outer spheroids which are often triaxial.  The observed
  kinematic transition to the halo and its dependence on stellar mass
  is consistent with $\Lambda$CDM simulations and supports a two-phase
  formation scenario. }


\maketitle  

\section{Introduction}
 \label{sec:introduction}
 
 Observations \citep[e.g.][]{2006ApJ...650...18T,
     2010ApJ...709.1018V} as well as simulations
   \citep[e.g.][]{2010ApJ...725.2312O,2016MNRAS.458.2371R,2017MNRAS.464.1659Q}
   suggest a two phase scenario for the formation of early type
   galaxies (ETGs). An initial fast assembly stage, in which the
 ETGs grow through rapid star formation fueled by the infall of cold
 gas ($z\gtrsim1.5$) or through major merger events
 \citep{2010ApJ...722.1666W,2010ApJ...721.1755S,2011ApJ...730....4B,
   2016MNRAS.456.1030W}, is followed by a series of merger episodes
 which enrich the galaxy halos of stars and make them grow efficiently
 in size \citep{2010ApJ...725.2312O,2012MNRAS.427.1816G,
   2012MNRAS.425..641L,2017MNRAS.466.4888B}.  The hierarchical
 accretion scenario finds its best evidence in the observations of a
 rapid growth of stellar halos at redshift $\lesssim 2$ with little or
 no star formation
 \citep[e.g.][]{2005ApJ...626..680D,2007MNRAS.382..109T,2010ApJ...709.1018V,2011ApJ...739L..44D}.
 In this context ETGs are layered structures in which the central
 regions are the remnant of the primordial stars formed in-situ, while
 the external halos are principally made of accreted material. The
 consequence is that the halos are expected to show significant
 variation with radius of properties such as light profiles
 \citep[][]{2013ApJ...768L..28H,2014MNRAS.443.1433D,2016ApJ...820...42I,2017A&A...603A..38S},
 and kinematics
 \citep[][]{2009MNRAS.394.1249C,2012ApJS..203...17R,2014ApJ...791...80A,2016MNRAS.457..147F}.
 
 Long slit spectroscopic observations of ETGs
 \citep[e.g.][]{1983ApJ...266...41D, 1989ApJ...344..613F,
   1994MNRAS.269..785B} revealed that this apparently homogeneous
 class of objects actually displays a kinematic diversity which also
 correlates with the isophote shape
 \citep{1988A&A...193L...7B,1996ApJ...464L.119K}. Disky ellipticals
 generally rotate fast, while slowly rotating ellipticals have a
 rather boxy shape.  A remarkable step forward in the comprehension of
 the nature of ETGs was attained by the $\mathrm{ATLAS^{3D}}$ project
 \citep[][]{2011MNRAS.413..813C}, which for the first time applied
 integral-field spectroscopy (IFS) over a statistically-significant
 sample, mapping kinematics, dynamics, and stellar population
 properties within one effective radius ($\re$). A new paradigm for
 ETGs was proposed, which distinguishes between FRs and SRs
 according to the central projected specific angular momentum,
 $\lambda_{R}$ \citep[][]{2007MNRAS.379..401E}.  FRs include
 also S0 galaxies and represent the great majority ($86\%$) of
 ETGs. These are apparently oblate systems with regular disk-like
 kinematics along the photometric major axis. SRs, on the
 other hand, often display kinematic features such as counter-rotating
 cores or twist of the kinematic position angle. They are relatively
 rounder systems, mildly triaxial, and tend to be massive
 \citep[][]{2013MNRAS.432.1862C}.
 
 The two classes have been interpreted as the result of the variety of
 processes shaping galaxies, leading to a sequence of baryonic angular
 momentum \citep[][]{2011MNRAS.414..888E,
   2014MNRAS.444.3357N,2014MNRAS.438.2701W}. On-going
 surveys like MANGA \citep{2015ApJ...798....7B}, CALIFA
 \citep{2012A&A...538A...8S}, SAMI
 \citep{2012MNRAS.421..872C,2015MNRAS.447.2857B}, and MASSIVE
 \citep{2014ApJ...795..158M} are currently working on increasing the
 size of the sample of IFS mapped objects, and extending the study to
 a wider range of environment and mass.
 
 However, a classification scheme based on the characteristics of the
 galaxies in the central regions (inside $\sim 1\re$) may not be fully
 representative of the nature of these objects
 \citep[e.g.][]{2017MNRAS.470.1321B}, raising the question of how
 complete our understanding is without a full knowledge of their
 properties on larger scales. The outer regions beyond $R_e$
 in fact contain half of the galaxies' stars and most of their
 dynamical mass. Dark matter is known to dominate there
 \citep[e.g][]{2006MNRAS.368..715M,2006ApJ...646..899H,2009ApJ...703L..51K,
   2010MNRAS.404.1165C} and dynamical modeling of the outskirts is
 essential to constrain its distribution at intermediate radii
 \citep[e.g.][]{2001AJ....121.1936G,
   2003Sci...301.1696R,2009MNRAS.393..641T,
   2011MNRAS.411.2035N,2013MNRAS.431.3570M}.  Stellar halos are
 predicted to host mostly accreted star material as shown by particle
 tagging simulations \citep{2013MNRAS.434.3348C} and hydro-dynamical
 simulations \citep{2016MNRAS.458.2371R}. In addition, these regions
 provide insight into the most recent dynamical phase of the
 galaxy. In the halos the settling times are of order 1\,Gyr and so
 signatures of the most recent assembly events may still be apparent,
 providing a mine of information about their formation and evolution
 mechanisms \citep[e.g.][]{2005ApJ...635..931B, 2009AJ....138.1417T,
   2012ApJ...748...29R,2013MNRAS.436.1322C,2015MNRAS.446..120D,2015A&A...579L...3L}.
 Thus extending investigations to the outer halos is crucial for
 having a complete picture of ETGs.

 However kinematic measurements are not easily obtained for ETG halos,
 which generally lack cold gas (and so the 21 cm HI emission) used to
 probe the outer parts of spiral galaxies. Since the continuum light
 from the stars quickly drops with radius, absorption line
 spectroscopy is challenging beyond $1-2$ $\re$. This limits the
 assessment of the complicated dynamics of ETGs which, because
 dominated by dispersion, necessitates a good knowledge of the higher
 moments of the line of sight velocity distribution (LOSVD) in order
 to alleviate the anisotropy-potential-degeneracy
 \citep[e.g.][]{1993MNRAS.265..213G,1997ApJ...488..702R,2009MNRAS.393..641T,2009MNRAS.395...76D,2009MNRAS.393..329N}.
 
 Kinematic studies of ETGs from integrated-light spectra out to large
 radii have been performed by
 \citet{2002ApJ...576..720K,2009MNRAS.398..561W,
   2010A&A...519A..95C,2011ApJ...729..129M, 2018A&A...609A..78B} using
 long slit spectroscopy or IFS on individual objects. More recently
 the SLUGGS survey \citep{2014ApJ...791...80A,
   2016MNRAS.457..147F,2017MNRAS.467.4540B}, the MASSIVE survey
 \citep{2014ApJ...786...23R,2017MNRAS.464..356V}, and
 \citet{2017MNRAS.471.4005B} generated kinematic data from integral
 field spectrographs (IFSs) for larger samples of ETGs, but never
 reaching beyond $3-4\re$\footnote{The values of $\re$ used by most
   kinematic studies are measured in the bright central regions of
   galaxies, and may underestimate the half light radii (see
   discussion in section
   \ref{subsec:DISCUSSION_kinematic_transition_radius}).}.
 
 The only possibility to probe the kinematics of a large sample of
 galaxies out to the very outskirts is through kinematic tracers that
 overcome the limit of the decreasing surface brightness, like
 globular clusters
 \citep[e.g.][]{2010A&A...513A..52S,2011ApJS..197...33S} or planetary
 nebulae.
 
 Planetary nebulae (PNe) are established probes of the stellar
 population in ETG halos \citep[e.g.][]{ 2013A&A...558A..42L,
   2017A&A...603A.104H}.  Their bright [OIII] line stands out against
 the faint galaxy background, making them relatively easy to
 detect. Since they are drawn from the main stellar population, their
 kinematics traces the bulk of the host-galaxy stars, and are directly
 comparable to integrated light measurements
 \citep{1995ApJ...449..592H,1996ApJ...472..145A,2001ApJ...563..135M,
   2009MNRAS.394.1249C,2013A&A...549A.115C}. This makes PNe the ideal
 kinematic probes for the halos of ETGs. Globular clusters do
   not generally follow the surface brightness distribution of the
   stars and do not trace the stellar kinematics
 \citep[e.g.][]{2006ARA&A..44..193B, 2013MNRAS.436.1322C,
   2014MNRAS.442.2929V}, and their color bimodality, suggesting two
 distinct formation mechanisms \citep[][and references
 therein]{2017MNRAS.465.3622R}, complicates the interpretation of
 their use as kinematic tracers. The pioneering work of
 \citet{2009MNRAS.394.1249C} studied the kinematics of 16 ETGs traced
 with PNe out to $8\re$, finding evidence for kinematic transitions at
 large radii from the trends observed in the central regions
 \citep[see also][]{2014ApJ...791...80A}. 
 
 The extended Planetary Nebula Spectrograph (ePN.S) survey in based on
 observation mostly done with the PN.S, and consists of catalogs of
 PNe for 33 ETGs. This dataset is the largest survey to-date of
 extragalactic PNe identified in the halos of ETGs, complementing the
 absorption line kinematics of the central regions available in the
 literature. The rationale of the survey, the sample definition, and
 the construction of the catalogs are described in detail in Arnaboldi
 et al. (in preparation).  Section \ref{sec:sample} is a brief
 description of the ePN.S sample. Section \ref{sec:method} describes
 the general procedure adopted for extracting the mean velocity fields
 from the measured radial velocities of PNe, and reviews the adaptive
 kernel smoothing technique introduced by
 \citet{2009MNRAS.394.1249C}. In section \ref{sec:systemic_velocity}
 we evaluate the systemic velocity of the galaxies. The point-symmetry
 of the smoothed velocity fields is studied in section
 \ref{sec:Point_symmetry_analysis}, while the trends of the kinematic
 parameters, such as rotational velocity, kinematic position angle,
 and velocity dispersion, are derived in section
 \ref{sec:the_halo_kinem_of_ETGs}. The results are described in
 section \ref{sec:results_per_family} with a detailed analysis of SRs and FRs.  A discussion of the results is presented
 in section \ref{sec:discussion}. In section
 \ref{sec:summary_and_conclusions} we give a summary of the work and
 draw our conclusions.

\section{Description of the Sample - Observations - Data reduction}
\label{sec:sample}

This work is based mostly on data collected with the Planetary Nebula
Spectrograph (PN.S) at the William Herschel Telescope in La Palma. The
PN.S is a custom-built instrument designed for counter-dispersed
imaging \citep{2002PASP..114.1234D}. Arnaboldi et al. (in
  prep.) collected catalogs of PNe for 25 galaxies from PN.S, 11 of
  which new, to which they added 6 further catalogs from the
  literature and 2 additional new catalogs, for a total of 33
  ETGs. This ePN.S sample is magnitude limited and covers a wide
range of internal parameters, such as luminosity, central velocity
dispersion, ellipticity, boxy/diskyness (table \ref{tab:galaxies}
summarizes the properties of the sample and the origin of the
catalogs). Our catalogs contain a total of 8636 PNe, with data
covering 4, 6, and 8 effective radii (\re) for, respectively,
  85\%, 41\%, and 17\% of the sample, with median extension of 5.6
  $\re$ (see $R_\mathrm{max}$ values of the last radial bins in table
  \ref{tab:galaxies}). This makes the ePN.S the largest kinematic
survey to-date of extragalactic PNe in the outer halos of ETGs.

Arnaboldi et al. (in prep.) give a full discussion of the extraction
and validation of the catalogs; here we provide a brief description of
the adopted procedures.  All the datasets (the new catalogs, as well
as the PN catalogs from the literature) are uniformly (re)analyzed, in
order to obtain a homogeneous sample of ETG kinematics whose
properties can be consistently compared. The new PN.S
  observations, the 2 additional new catalogs, and the reanalyzed
  catalogs will be described in Arnaboldi et al. (in prep.).

 For each galaxy, after the raw catalog has been obtained, it is
  uniformly cleaned from possible spurious sources and from so-called
  velocity "outliers". The first step for the removal of outliers
  among the PN candidates is the exclusion of all the detections with
signal-to-noise ratio below a given threshold. We adopted $S/N\geq2.5$
as good compromise value between a reasonable signal-to-noise and the
number of detections that satisfy this requirement.

Next we separate PNe belonging to any satellites from those in
  the hosts. For this we use the probability membership method from
\citet{2012A&A...539A..11M}, which uses both kinematic and photometric
information to assign to each star a probability of belonging to the
satellite or host. Membership to the host galaxy is assigned only if
the probability is greater than 90\%.
 
The last step is the removal of outliers in the remaining host PN
  velocity distribution. Such outliers could, e.g., arise because of
contamination from other narrow emission-line sources (i.e. background
star-forming galaxies) that are not resolved and appear point-like in
the counter-dispersed images, similar to the monochromatic [OIII]
$5007\mathring{\mathrm{A}}$ emission from a PN. We identify
  outliers using a robust mean/sigma clipping procedure. The
algorithm derives a robust mean velocity ($\mathrm{v_{mean}}$) and
velocity dispersion $\sigma$, using a running average in a 2D phase
space (coordinate, $\mathrm{v}$) with a window of $N$ data points
($15\lesssim N\lesssim30$, according to the number of tracers in each
galaxy) and a 3 data points step. An iterative procedure clips the PN
candidates whose $|\vmeas-\mathrm{v_{mean}}|> 2\sigma$, and evaluates
$\mathrm{v_{mean}}$ and $\sigma$ until the number of clipped objects
stabilizes. In each iteration $\sigma$ is corrected by a factor 1.14
to account for the $2\sigma$ cut of the LOSVD tails. Finally, to
  the 95\% of a galaxy's PNe thus validated, the remaining 5\% of the
  PNe are added back to the sample, using those clipped objects which
  are closest in $\mathrm{v}$ to the $2\sigma$ contours, so that the
  final PN sample also includes the objects expected in the
  approximately Gaussian tails of the LOSVD.

In the case of disk galaxies, dominated by rotation, a disk/spheroid
decomposition was performed following \citet{2011MNRAS.414..642C}:
using both photometric data and kinematic information we assigned to
each PN the probability of being associated with each photometric
component. The tagging of the outliers from the disk and the spheroid
separately allows us to account for their different kinematics when
using the robust mean/sigma clipping procedure. The disk is processed
first, and its flagged PNe are added to the PNe of the bulge. Eventually
the flagged bulge PNe are considered as
outliers of the entire galaxy.
 
 Finally, we estimated the number of background
  emitting galaxies, whose emission line might fall in
  the range of velocities determined in the procedure above. We employed
  the approach adopted in \citet{2013A&A...558A..42L}, which uses the
  Lyman alpha (Ly$\alpha$) luminosity function by \citet{2007ApJ...667...79G}, and adapted
  it to the ePN.S survey. Given the limiting magnitude, the area
  coverage, and the filter band-passes of the PN.S, the number of
  expected background galaxies is 2. This is an upper limit, as the
  \citet{2007ApJ...667...79G} sample also includes [OII] emitters at
  $z\simeq0.34$. These [OII] emitters are characterized by the oxygen doublet at
  $3726-3729\mathring{\mathrm{A}}$ that is resolved in wavelength at
  the resolution of the PN.S, so the [OII] emitters have already been
  removed from the PN candidate sample because they are not
  monochromatic emission. We discuss the effect of the Ly$\alpha$ background
  contaminants on the kinematics in section
  \ref{subsubsec:Errors_on_the_fitted_parameters}. 
 
The datasets processed in this way are the Bona Fide PNe catalogs used
in the following analysis.

\section{Kernel smoothing method}
\label{sec:method}

The measured line-of-sight (LOS) velocities of the PNe are random
samplings of the galaxy LOSVD function at the position of the
source. Therefore each velocity measurement randomly deviates from the
local mean velocity by an amount that depends on the local LOS
velocity dispersion.

In order to extract the mean LOS velocity and the LOS velocity
dispersion fields from this discrete velocity field, we use an
adaptive kernel smoothing technique, as described in
\citet{2009MNRAS.394.1249C}, that performs a local average of the
measured discrete LOS velocities. In the following section we briefly
review the smoothing technique, while we refer to
\citet{2009MNRAS.394.1249C} for a more detailed discussion. In
appendix \ref{subsec:discussion_about_the_method} we validated the
adopted procedure on simulated data, in order to test the effects of
different statistical realizations of a given sample of tracers,
  the dependency on the number of tracers, and different $V/\sigma$
ratios on the estimated kinematic parameters.

\subsection{Averaging the discrete velocity field with the adaptive kernel smoothing technique}
\label{subsec:smoothing}
The smoothing of the discrete velocity field is carried out by
computing the velocity at each position $(x,y)$ in the sky as a
weighted mean $\tilde{\vmeas}(x, y)$ of all the PN LOS velocities
\vmeas
\begin{equation}
  \tilde{\vmeas}(x, y) = \frac{\sum_i \vmeas_i w_{i,p}}{\sum_i w_{i,p}}\newline\newline
    \label{eq:vel_smoothed}
\end{equation}
while the velocity dispersion $\tilde{\sigma}(x, y)$ is given by the
square root of the variance of \vmeas\ with respect to
$\tilde{\vmeas}$
  \begin{equation}
   \begin{split}
    \tilde{\sigma}(x, y) &= (\langle \vmeas^2\rangle - \langle \vmeas\rangle^2 - \delta \vmeas^2)^{1/2}
\\
    &= \left[\frac{\sum_i \vmeas^2_i w_{i,p}}{\sum_i w_{i,p}}-\tilde{\vmeas}(x, y)^2-\delta\vmeas^2\right]^{1/2}\newline\newline 
    \label{eq:sigma}
   \end{split}
  \end{equation}
The weight $w_{i,p}$ of each PN is defined using a Gaussian Kernel that depends on the distance of the PN from the
position $(x,y)$, normalized by a kernel width $K$
\begin{equation}
 w_{i,p} = \exp \frac{-D_i^2}{2K(x,y)^2} \qquad ; \qquad    K(x,y) = A\sqrt{\frac{M}{\pi\rho}} + B
\label{eq:kernel_2}
\end{equation}
The latter controls the spatial scale of the region over which the
smoothing is performed, and hence the spatial resolution of the
kinematic study. Large values of $K$, in fact, lead to smoother
profiles in the mean LOS velocity fields, highlighting the general
trends, but also suppressing the small scale structures, while smaller
values of $K$ allow a better spatial resolution but may amplify any
noise pattern. Hence the optimal $K$ should be a compromise between
spatial resolution and statistical noise smoothing. 

The width $K$ is therefore defined to be linearly dependent on
the distance between the position $(x,y)$ and the M$^{th}$ closest PN,
so that $K$ is a function of the local density of tracers $\rho(x,y)$.
This allows $K$ to be smaller in the innermost, dense regions of
galaxies, and larger in the outskirts, where their density is usually
lower. The optimal kernel parameters $A$ and $B$ are derived as
described in section \ref{subsec:derive_A_B}.  We chose $M =20$, but
\citet{2009MNRAS.394.1249C} tested the procedure with $10<M<60$
finding no significant differences in the results.

\begin{table*}
\caption[]{Properties of the ETG sample analyzed in this paper, and list of references}
\begin{center}
\begin{tabular}{llllrrlllll}
\hline\hline\noalign{\smallskip}
  \multicolumn{1}{l}{Galaxy} &
  \multicolumn{1}{l}{$M_K$ \tablefootmark{(a)}} &
  \multicolumn{1}{l}{D  \tablefootmark{(b)}} &
  \multicolumn{1}{l}{class  \tablefootmark{(c)}} &
  \multicolumn{1}{l}{\re  \tablefootmark{(d)}} &
  \multicolumn{1}{l}{$R_\mathrm{max}/\re$\tablefootmark{(e)}} &
  \multicolumn{1}{l}{\PAphot \tablefootmark{(f)}} &
  \multicolumn{1}{l}{$\epsilon$   \tablefootmark{(g)}}&
  \multicolumn{1}{l}{$N_\mathrm{PNe}$  \tablefootmark{(h)}}&
  \multicolumn{1}{l}{References \tablefootmark{(i)}}  &
  \multicolumn{1}{l}{References \tablefootmark{(l)}} \\
  \multicolumn{1}{l}{NGC} &
  \multicolumn{1}{l}{[mag]} &
  \multicolumn{1}{l}{[Mpc]} &
  \multicolumn{1}{l}{} &
  \multicolumn{1}{l}{[arcsec]} &
  \multicolumn{1}{l}{} &
  \multicolumn{1}{l}{[degrees]} &
  \multicolumn{1}{l}{} &
  \multicolumn{1}{l}{} &
  \multicolumn{1}{l}{PN data}&
  \multicolumn{1}{l}{abs.line data} \\
\noalign{\smallskip}\hline\noalign{\smallskip}
   0584	&  $-24.23$ & 20.2 & F   & 33	  (1)&  7.4   &  63	    & $ 0.339			 $   &   25	& (7)	     &   (21)		  \\
   0821 &  $-23.99$ & 23.4 & F   & 40	  (2)&  4.8   &  31.2	    & $ 0.35			 $   &   122	& (8)	     &   (18);(40);(41) 	  \\ 
   1023 &  $-23.89$ & 10.5 & F   & 48	  (2)&  6.8   &  83.3	    & $ 0.63			 $   &   181	& (8);(9)   &	 (18);(30)	  \\
   1316 &  $-26.02$ & 21.0 & F   & 109    (3)&  4.7   &  50	    & $ 0.29  \tablefootmark{*}  $   &   737	& (10)       &   (22)		  \\
   1344 &  $-24.21$ & 20.9 & F   & 30	  (4)&  7.8   &  167	    & $ 0.333			 $   &   192	& (12)       &   (23)		  \\
   1399 &  $-25.29$ & 20.9 & S   & 127    (3)&  4.    &  110	    & $ 0.1   \tablefootmark{*}  $   &   145	& (11)       &   (24)		  \\
   2768 &  $-24.77$ & 22.4 & F   & 63	  (2)&  6.2   &  91.6	    & $ 0.57			 $   &   312	& (9)	     &   (18);(31)	  \\ 
   2974 &  $-23.76$ & 22.3 & F   & 38	  (2)&  5.8   &  44.2	    & $ 0.37			 $   &   22	& (7)	     &   (18)		  \\
   3115 &  $-24.02$ & 9.5  & F   & 93	  (6)&  4.7   &  43.5	    & $ 0.607			 $   &   183	& (9)	     &   (18);(25)	  \\
   3377 &  $-22.78$ & 11.0 & F   & 35.5   (2)&  7.7   &  46.3	    & $ 0.33			 $   &   136	& (8)	     &   (18); (33)	  \\
   3379 &  $-23.80$ & 10.3 & F   & 40	  (2)&  5.3   &  68.2	    & $ 0.13  \tablefootmark{*}  $   &   189	& (8)	     &   (19);(32);(40)   \\
   3384 &  $-23.51$ & 11.3 & F   & 32.5   (2)&  6.8   &  50	    & $ 0.5			 $   &   85	& (9)	     &   (19)		  \\
   3489 &  $-23.04$ & 12.0 & F   & 22.5   (2)&  4.8   &  70.5	    & $ 0.45			 $   &   57	& (9)	     &   (19)		  \\
   3608 &  $-23.69$ & 22.8 & S   & 29.5   (2)&  8.2   &  82	    & $ 0.2   \tablefootmark{*}  $   &   92	& (8)	     &   (18)		  \\
   3923 &  $-25.33$ & 23.1 & S   & 86.4   (1)&  4.9   &  48	    & $ 0.271			 $   &   99	& (15)       &   (26)		  \\
   4278 &  $-23.80$ & 15.6 & F   & 31.5   (2)&  7.6   &  39.5	    & $ 0.09  \tablefootmark{*}  $   &   69	& (7)	     &   (18)		  \\
   4339 &  $-22.62$ & 17.0 & F   & 30	  (2)&  3.    &  15.7	    & $ 0.07  \tablefootmark{*}  $   &   44	& (7)	     &   (20);(38)	  \\
   4365 &  $-25.19$ & 23.1 & S   & 52.5   (2)&  5.6   &  40.9	    & $ 0.24  \tablefootmark{*}  $   &   227	& (7)	     &   (18)		  \\
   4374 &  $-25.12$ & 18.5 & S   & 52.5   (2)&  5.9   &  128.8      & $ 0.05  \tablefootmark{*}  $   &   445	& (8)	     &   (18)		  \\
   4472 &  $-25.73$ & 16.7 & S   & 95.5   (2)&  8.4   &  154.7      & $ 0.19  \tablefootmark{*}  $   &   431	& (7)	     &   (20);(37)	  \\
   4473 &  $-23.76$ & 15.2 & F   & 27.    (2)&  5.6   &  92.2	    & $ 0.43			 $   &   153	& (7)	     &   (18)		  \\
   4494 &  $-24.17$ & 17.1 & F   & 49	  (2)&  4.8   &  176.3      & $ 0.14  \tablefootmark{*}  $   &   255	& (8)	     &   (18);(36)	  \\
   4552 &  $-24.32$ & 16.0 & S   & 34.    (2)&  9.2   &  132	    & $ 0.11  \tablefootmark{*}  $   &   227	& (7)	     &   (19);(38)	  \\
   4564 &  $-23.10$ & 15.9 & F   & 20.5   (2)&  6.5   &  47	    & $ 0.53			 $   &   47	& (8)	     &   (18)		  \\
   4594 &  $-24.93$ & 9.5  & F   & 102    (5)&  4.    &  88	    & $ 0.521			 $   &   258	& (16)       &   (27)		  \\
   4636 &  $-24.35$ & 14.3 & S   & 89.    (2)&  3.    &  144.2      & $ 0.23  \tablefootmark{*}  $   &   189	& (7)	     &   (20);(39)	  \\
   4649 &  $-25.35$ & 16.5 & F   & 66	  (2)&  4.5   &  91.3	    & $ 0.16  \tablefootmark{*}  $   &   281	& (13)       &   (18);(34)	  \\
   4697 &  $-24.14$ & 12.5 & F   & 61.5   (1)&  4.5   &  67.2	    & $ 0.32			 $   &   525	& (14)       &   (18);(35)	  \\ 
   4742 &  $-22.60$ & 15.8 & F   & 14.4   (4)&  13.1  &  80	    & $ 0.351			 $   &   64	& (7)	     &   (28)		  \\
   5128 &  $-24.16$ & 4.1  & F   & 162.6  (1)&  11.9  &  30	    & $ 0.069  \tablefootmark{*} $   &   1222	& (17)       &   (29)		  \\ 
   5846 &  $-25.04$ & 24.6 & S   & 59	  (2)&  4.3   &  53.3	    & $ 0.08   \tablefootmark{*} $   &   118	& (8)	     &   (18)		  \\
   5866 &  $-23.99$ & 14.8 & F   & 36	  (2)&  9.4   &  125	    & $ 0.58			 $   &   150	& (7)	     &   (18)		  \\
   7457 &  $-22.38$ & 12.9 & F   & 36	  (2)&  3.2   &  124.8      & $ 0.47			 $   &   108	& (9)	     &   (18)		  \\

\noalign{\smallskip}\hline
\end{tabular}
\tablefoot{\\
\tablefoottext{a}{$M_K$ is the total absolute luminosity in the K band. These values are obtained from the total apparent total magnitudes $K_T$ of the 2MASS atlas \citep{2006AJ....131.1163S} using the distance D, and correcting for foreground galactic extinction $A_B$ \citep[][]{1998ApJ...500..525S}: $M_K=K_T-5\log_{10}D-25-A_B/11.8$. We assume $A_B/A_K=11.8$, consistently with \citet{2011MNRAS.413..813C}. The $K_T$ magnitudes are from integrating the surface brightness profiles ($\propto \exp(-r/r_{\alpha})^{(1/\beta)}$), extrapolated from the 20 mag/arcsec$^2$ isophote to $\sim5 r_{\alpha}$  \citep{2003AJ....125..525J}}\\
 \tablefoottext{b}{Distances of galaxies derived from the surface brightness fluctuation method. Whenever possible we adopt the distance moduli measured by \citet[][B09]{2009ApJ...694..556B}, otherwise we used the values from \citet[][J03]{2003ApJ...583..712J} or from \citet[][T01]{2001ApJ...546..681T}. The distance moduli from J03 were rescaled to the zero-point calibration of B09 by  applying a shift of +0.1 mag, while the distance moduli from T01 were zero-point- and bias-corrected using the formula from \citet{2010ApJ...724..657B} and the data quality factor Q given by T01.}\\
 \tablefoottext{c}{The sample is divided into SRs (S) and FRs (F), according to the definition of \citet{2011MNRAS.414..888E}, from the kinematics within $1\re$. }\\
 \tablefoottext{d}{Adopted effective radius. The index in parenthesis corresponds to the reference: (1) \citet{2012yCat..21970021H}, (2) \citet{2011MNRAS.413..813C}, (3) \citet{1994A&AS..106..199C}, (4) \citet{2001MNRAS.327.1004B}, (5) \citet{1989ApJ...338..752K},   (6) \citet{1987AJ.....94.1519C}.}\\
 \tablefoottext{e}{Mean radius of the last radial bin in units of effective radii.}\\
 \tablefoottext{f}{Average value of the photometric position angle and \\
  \tablefoottext{g} {ellipticity} ($\epsilon$), from \citet{2011MNRAS.414.2923K} (within $2.5-3 \re$) and \citet{2011ApJS..197...21H} (in the outer regions, where they converge to a constant value) For NGC 3384 and NGC 4564 we used the $\PAphot$ from \cite{2007AN....328..562M} and \cite{1994A&AS..104..179G}, respectively. } \tablefoottext{*} {Objects for which we used circular radial bins ($\epsilon=0$)}, see section \ref{subsec:fit_rotation_model}. \\
\tablefoottext{h}{Number of detected PNe.}\\	
 \tablefoottext{i}{References for the PNe datasets: (7) are new PN.S catalogs presented in Arnaboldi et al. (in prep.), (8) \citet{2009MNRAS.394.1249C}, (9) \citet{2013A&A...549A.115C}, (10) \citet{2012A&A...539A..11M}, (11) \citet{2010A&A...518A..44M}, (12) \citet{2005ApJ...635..290T}, (13) \citet{2011ApJ...736...65T}, (14) \citet{2009ApJ...691..228M}, (15) unpublished data from counter dispersed imaging (Arnaboldi et al. in prep.),  (16) unpublished data from narrow band imaging and spectroscopic follow up (Arnaboldi et al. in prep.), (17) \cite{2004ApJ...602..685P} and \cite{2015A&A...574A.109W}}\\
\tablefoottext{l}{References for absorption line data:  kinemetry from (18) \cite{2016MNRAS.457..147F} on SLUGGS+$\mathrm{ATLAS^{3D}}$ data, from (19) \cite{2008MNRAS.390...93K} and from (20) \citet{2011MNRAS.414.2923K}, major axis long slit spectroscopy from (21) \citet{1983ApJ...266..516D}, (22) \citet{2006MNRAS.371.1912B}, (23) \citet{2005ApJ...635..290T}, (24) \citet{2000AJ....119..153S} and \citet{2014MNRAS.441..274S}, (25) \citet{2006MNRAS.367..815N}, (26) \citet{1998MNRAS.294..182C}, (27) \citet{1982ApJ...256..460K}, (28) \citet{1983ApJ...266...41D}, (29) \citet{1983IAUS..100..335M}), (30) \cite{1997A&AS..126..519S}, (31) \cite{1997A&AS..122..521S}, (32) \cite{1999AJ....117..839S}, (33) \cite{2009MNRAS.394.1249C}, (34) \cite{2001ApJ...546..903D}, (35) \cite{2008MNRAS.385.1729D}, (36) \cite{2009MNRAS.393..329N}, (37) \cite{2017MNRAS.464..356V}, (38) \cite{1997A&AS..126...15S}, (39) \cite{2011RAA....11..909P}, (40) \cite{2009MNRAS.398..561W}, (41) \cite{2010ApJ...716..370F}.}
 }
\end{center}
\label{tab:galaxies}
\end{table*}

\subsubsection{Deriving the optimal A and B kernel parameters}
\label{subsec:derive_A_B}
The parameters $A$ and $B$ in equation \eqref{eq:kernel_2} are chosen
so that the best compromise between spatial resolution and noise
smoothing is achieved.  We developed an iterative procedure in order
to derive the optimal kernel parameters that realize this condition.

We first estimated the velocity gradient to be resolved by the
smoothing procedure by performing a preliminary averaging with a fully
adaptive kernel ($A=1$ and $B=0$).  The derived mean velocity field,
$\tilde{\vmeas}_{A=1,B=0}$, is fitted with a cosine function, which,
in general, approximately describes the velocity profiles of early
type galaxies \citep[see][]{1997ApJ...486..230C}:

  \begin{equation}
  \tilde{\vmeas}_{A=1,B=1}(\phi)=V_\mathrm{max}\cos(\phi-\PAkin)+\mathrm{const}
  \end{equation}
  This interpolation function provides a measure of the position angle
  of the kinematic major axis ($\PAkin$), along which the steepest
  velocity gradient $d\tilde{\vmeas}$ is expected to lie. The gradient
  $d\tilde{\vmeas}$ is obtained by fitting a straight line to the
  velocities $\tilde{\vmeas}$ of the PNe lying in a section along the
  $\PAkin$ direction, as a function of the radius.

The best kernel parameters that allow to resolve spatial substructures
with typical velocity gradient $d\tilde{\vmeas}$ are derived by
building simulated sets of PNe \citep[see][]{2009MNRAS.394.1249C}. The
stars are spatially distributed according to a given density
$\overline\rho$, while their velocities are assigned using the derived
velocity gradient and adding a dispersion equal to the standard
deviation of the observed radial velocities.  The artificial sets are
processed using different values of the kernel $K$ until the simulated
input velocity field is recovered: this provides the best $K$ for this
$\overline\rho$. The procedure is repeated for different values of
$\overline\rho$, and the optimal A and B are the best fit values of
equation \eqref{eq:kernel_2} based on on the derived best $K$ as a
function of $\overline\rho$.

\subsubsection{Errors in the derived velocity fields}

  Errors on $\tilde{\vmeas}(x,y)$ and $\tilde{\sigma}(x,y)$ are
  obtained using Monte Carlo simulations, as also discussed in
  \citet{2009MNRAS.394.1249C}. For each galaxy, 100 PN datasets are
  built with simulated radial velocities at the same positions as the
  observed PNe. The radial velocity for each simulated object is
  obtained from the two-dimensional smoothed velocity field, adding a
  random value from a Gaussian distribution centered at 0 and with
  dispersion $\sigma=\sqrt{\tilde{\sigma}^2+\delta \vmeas^2}$ where
  $\delta \vmeas$ is the velocity error. These simulated datasets are
  smoothed with the same kernel $K$ as the real sample, and the
  standard deviations of the simulated velocity and velocity
  dispersion fields give the errors on $\tilde{\vmeas}$ and
  $\tilde{\sigma}$. This procedure is validated in appendix
  \ref{subsubsec:tests_on_simulations} with PN velocity distributions
  generated from a simulated merger remnant galaxy, which are analyzed
  in an identical manner as the real galaxy PN samples. As an
  additional test, we have also extracted and analyzed kinematic
  information from 1000 subsamples of the original PN sample for all
  our galaxies and found kinematic parameters consistent with our
  full-sample results, as described in section
  \ref{subsubsec:Errors_on_the_fitted_parameters}.

\subsection{Fitting a rotation model}
\label{subsec:fit_rotation_model}

The mean velocity fields, derived from smoothing the discrete
velocities, are divided into radial bins with equal numbers of PNe such
that they contain at least 30 stars. If a galaxy contains less than
$60$ tracers, we divide the sample in two bins in order to study
possible radial trends.

The bins are circular for galaxies either with small
flattening, i.e. $10\times\epsilon < 3 $, or whose PN spatial
distribution has a rather square shape. For all the other galaxies we
use elliptical bins oriented along the photometric major axis, with a
flattening equal to a characteristic ellipticity of the isophotes of
the galaxy (the adopted ellipticity, $\epsilon$, and photometric
position angle $\PAphot$ are given in table \ref{tab:galaxies}).

We found that the results of our analysis do not depend on the chosen
flattening but, since the spatial distribution of the PNe follows the
light, and so it may be rather flattened, elliptical bins help to
sample the field homogeneously in an azimuthally-unbiased way.

In each radial bin, we fitted the PN velocities $\tilde{v_i}$
  of position ($R_i$, $\phi_i$) with a rotation model
  $\tilde{v}(\phi,R)$ (see section \ref{subsubsec:model}). Here
  $\phi_i$ is the eccentric anomaly of the PN in the bin of
  coordinates $(x_i,y_i)$, $\phi_i(x_i,y_i)=\arctan[y_i/((1-\epsilon)
  x_i)]$, and $\epsilon$ is the ellipticity. $R_i$ is the major axis
  distance of each PN and $R$ is the mean $R_i$ of the PNe in each bin.

\subsubsection{Point-symmetric rotation model}
\label{subsubsec:model}

A velocity map $\tilde{\vmeas}(R,\phi)$ is a periodic function in
$\phi$, so it can be expanded in a Fourier series and approximated by
a finite number of harmonics:
\begin{equation}
    \tilde{\vmeas}(R,\phi)=a_0(R)+\sum_{n=1}^{N} a_{n} (R) \cos(n\phi) + \sum_{n=1}^{N} b_n (R) \sin(n\phi)
\label{eq:fourier1}
\end{equation}

Elliptical galaxies in dynamical equilibrium are triaxial systems
\citep[e.g.][ and references therein]{1994ApJ...425..500S}, so the
projection on the sky of the mean velocity field should be
\emph{point-symmetric} with respect to the systemic velocity $a_0$
\citep[see][]{2006MNRAS.366..787K,2013MNRAS.436.1322C}, i.e. symmetric
positions have equal velocities with opposite sign
($\tilde{\vmeas}(\phi)=-\tilde{\vmeas}(\phi+\pi)$). Deviations from
this behavior arise from perturbations from equilibrium that may be
due to interaction or merger episodes. If one of these processes,
which plays a role in the formation and evolution of early type
galaxies, occurred relatively recently (a few Gyrs ago), it is likely
that some signatures in the kinematics and orbital structure of the
galaxy are still observable, especially in the halo where the dynamic
time-scales are longer.

The requirement of point-symmetry on $\tilde{\vmeas}(R,\phi)$, namely
$\tilde{\vmeas}(R,\phi)-a_0=-[\tilde{\vmeas}(R,\phi+\pi)-a_0]$, allows
only odd values for $n$ in equation \eqref{eq:fourier1}.

The expansion in equation \eqref{eq:fourier1} can be rewritten in a
more direct way, as a rotation around the kinematic axis plus higher
order modes.  This is achieved through a rotation such that
\begin{equation}
     \begin{split}
     &\tilde{\vmeas}(R,\phi) = a_0(R) +\sum_{n=1,3,...} c_{n}(R) \cos(n\phi-n\alpha) +\\
     &+ \sum_{n=1,3,...} s_n(R) \sin(n\phi-n\alpha)
     \end{split}
    \label{eq:fourier2}
\end{equation}
    with 
\begin{equation}
     \begin{cases}
	a_n = c_n\cos(n\alpha)-s_n\sin(n\alpha) \\
	b_n = c_n\sin(n\alpha)+s_n\cos(n\alpha)
    \end{cases}	  
\end{equation}
The phase  $\alpha$ can be chosen so that the amplitude of the first order sine term is 0: $s_1=0$ if $\alpha =\arctan(b_1/a_1)$. This implies that 
\begin{equation}
     \begin{split}
      c_1 &= \sqrt{a_1^2+b_1^2}\\
      c_n &= \frac{b_n}{\sin(n\alpha)}-\frac{s_n}{\tan(n\alpha)}\\
      s_n &= \left(\frac{b_n/a_n - \tan(n\alpha)}{1+\tan(n\alpha)b_n/a_n}\right)c_n
    \end{split}
\end{equation} 
 and 
\begin{equation}
\begin{split}
      &\tilde{\vmeas}(R,\phi)=a_0(R)+c_1(R) \cos(\phi-\alpha(R))+\\
      &+\sum_{n=3,5,...} c_{n} (R) \cos(n\phi-n\alpha(R)) + \sum_{n=3,5,...} s_n (R) \sin(n\phi-n\alpha(R))
    \label{eq:fourier3}
    \end{split}
\end{equation}
In this notation $\alpha$ coincides with the position angle of the kinematic major axis, $\PAkin$, $a_0$ is the mean
velocity of the PNe in the bin, and $c_1$ is the amplitude of the projected rotation, $V_\mathrm{rot}$. The amplitudes
of the higher order harmonics, $c_k$ and $s_k$, are corrections that account for deviations of the galaxy motion from
the simple cosine rotation.

In practice the series in equation \eqref{eq:fourier3} can be truncated to the third order as the higher order
  harmonics are generally zero within the errors. The resulting function is fitted to the mean velocity estimates at all
  PN positions in each radial bin, with the position angle $\PAkin$, the constant $a_0$, and the amplitudes
  $V_\mathrm{rot}$, $s_3$, and $c_3$, as free parameters. The fit of the parameter $a_0$ gives an estimate of the
  systemic velocity $V_\mathrm{sys}$ in the halo (see section \ref{sec:systemic_velocity}). Once $V_\mathrm{sys}$ is
  subtracted from the velocity fields, their point-symmetry can be studied (section \ref{sec:Point_symmetry_analysis})
  and used to produce the final mean velocity fields for the point symmetric galaxies (see section
  \ref{subsec:velocity_fields}). The final mean velocity fields, after subtracting $V_\mathrm{sys}$, are eventually fitted
  with the function
\begin{equation}
\begin{split}
 \tilde{V}(R,\phi) & = V_\mathrm{rot}(R) \cos(\phi-\PAkin(R))+s_3(R) \sin(3\phi-3\PAkin(R)) \\
                    & +c_3(R) \cos(3\phi-3\PAkin(R))
\end{split}
\label{eq:fourier_definitivo} 
\end{equation}
where the only free parameters are $\PAkin$, $V_\mathrm{rot}$, $s_3$, and $c_3$ (see section \ref{subsec:kinematic_parameters}).

The kinematic quantities $\PAkin$ and $V_\mathrm{rot}$ obtained fitting the model in equation
\eqref{eq:fourier_definitivo} on the smoothed velocity fields are comparable to the results from a kinemetry fit to IFS
data \citep{2006MNRAS.366..787K,2011MNRAS.414.2923K,2016MNRAS.457..147F}. However, we do not apply kinemetry, because
this would mean fitting ellipses to the PN smoothed velocity fields. Since these have been derived from small samples of
discrete tracers which, by nature, have lower spatial resolution and signal-to-noise ratio, a more straightforward
approach, with fewer free parameters, is preferable.

\subsubsection{Errors on the fitted parameters}
\label{subsubsec:Errors_on_the_fitted_parameters}

The errors on the fitted parameters, $a_0(R)$, $\PAkin(R)$, $V_\mathrm{rot}(R)$, $c_3(R)$, and $s_3(R)$, are evaluated via Monte Carlo simulations: the 100 simulated datasets produced for deriving the errors on $\tilde{\vmeas}$ and $\tilde{\sigma}$ (see section \ref{subsec:smoothing}) are divided into radial bins and modeled with equation \eqref{eq:fourier3}. The errors are the standard deviations on the fitted parameters.
 
\begin{figure}[h]
\begin{center}

  \includegraphics[width=\columnwidth]{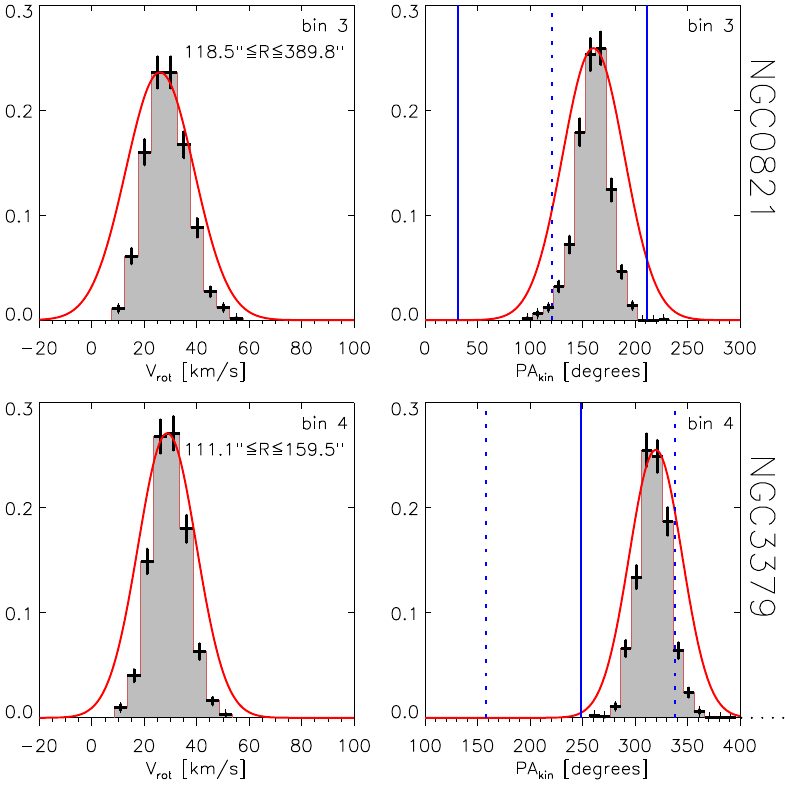}
    
\caption{ Distribution of the fitted $V_\mathrm{rot}$ and $\PAkin$ in one radial bin obtained from 1000 PN subsamples
  extracted for NGC 0821 and NGC 3379 (gray histograms).  The red curves are Gaussians centered on the $V_\mathrm{rot}$
  and $\PAkin$ fitted on the full dataset, and with dispersion given by the Monte Carlo simulations of the
  galaxy under study. The vertical solid lines show the position angle of the photometric major axis for the two
  galaxies; the dotted lines show the photometric minor axis. The values for $\PAphot$ are listed in table
  \ref{tab:galaxies}. }
\label{fig:resampling_distribution}
 
\end{center}
\end{figure}

 We tested whether the mean velocity field extracted through the smoothing procedure is sensitive to the relatively
  large velocities of objects belonging to the tails of the LOSVD. For this we selected subsamples of the observed
  dataset, re-extracted the kinematics, and studied the distribution of the fitted kinematic parameters. For each galaxy
  we used 1000 subsamples built with 80\% of the observed PNe each. (We do not use bootstrap with replacement as it
  would not be consistent with the constraint for the PNe to follow light.) Figure \ref{fig:resampling_distribution}
  shows the distributions of the fitted $V_\mathrm{rot}$ and $\PAkin$ in one radial bin for the PN subsamples extracted for NGC 0821 and
  NGC 3379. For all the ePN.S galaxies these distributions fall well within the statistical uncertainties of the fitted
  parameters from the full datasets. This shows that the values of $V_\mathrm{rot}$ and $\PAkin$ measured are not driven
  by a few high velocity objects, but are properties of the whole PN sample.  
  
   In addition, we simulated the effect of the contamination by
    background Ly$\alpha$ emitters by adding two random
    velocity measurements (see section \ref{sec:sample}), uniformly drawn in velocity from the
    filter band-pass used for each individual galaxy in our sample. We
    then re-extracted all the kinematic observables, and found that
    they are well within the 1 sigma errors of the
    measurements without contaminants.

 \section{Systemic velocity subtraction} 
\label{sec:systemic_velocity}

A measure of the systemic velocity of the galaxies is provided by the
fit of the PN smoothed velocity field in radial bins with the harmonic
expansion in equation \eqref{eq:fourier3}. The bins are built as
described in section \ref{subsec:fit_rotation_model} and the adopted
geometry for each galaxy (i.e. ellipticity, $\epsilon$, and
photometric position angle, \PAphot) is listed in table
\ref{tab:galaxies}. The $a_0 (R)$ parameter, in fact, represents the
mean velocity of the tracers in the radial bin with radius $R$.  When
the galaxy does not display kinematic substructures (bulk motions),
this mean velocity is an estimate of the systemic velocity of the
galaxy which is constant with radius for a gravitationally bound
system.

Since the PNe are not distributed uniformly on the sky, $a_0 (R)$
gives actually a more precise evaluation of the systemic velocity than
a straight average of the measured LOS velocities. The fit of equation
\eqref{eq:fourier3} removes any contribution to the mean from rotation
and is not sensitive to azimuthal completeness.  Hence we can perform
the fit leaving the parameter $a_0(R)$ free to vary in each bin. We
find that $a_0(R)$ is generally constant with radius within the
errors. Therefore we adopt, for each galaxy, a mean systemic velocity,
\vsys, defined as a mean of the $a_0 (R)$ values, weighted with the
errors on the fit:
\begin{equation}
\vsys=\frac{\sum_\mathrm{bins} a_0(R_\mathrm{bin})/\Delta a_0^2(R_\mathrm{bin})}{\sum_\mathrm{bins} 1/\Delta a_0^2(R_\mathrm{bin})}
\end{equation}
We conservatively consider as error on \vsys, $\Delta \vsys$, the mean
of the errors $\Delta a_0(R)$, since the single measures of $a_0(R)$,
coming from a smoothed velocity field, are not independent quantities.

We find that $a_0(R)$ does sometimes display a trend with radius
within the errors. This is due to the interplay between spatial
inhomogeneities and smoothing, which may result in a slight asymmetry
of one side of the galaxy with respect to the other. This effect
naturally disappears as soon as the catalogs are folded by
point-symmetry transformation (see section
\ref{sec:Point_symmetry_analysis}), but we keep track of it in the
uncertainties, by adding in quadrature the scatter of the $a_0(R)$
values to the error $\Delta \vsys$.

NGC 1316 and NGC 5128 are treated separately with respect to the other
galaxies. Their fitted $a_0(R)$ are constant in most radial bins, but
they deviate in localized bins from this constant by more than twice
the errors. At these radii the galaxies display important features in
their velocity fields which cause an offset of the average velocity
from the systemic value. We masked the most irregulars bins and use
the fitted $a_0(R)$ on the others to compute the mean \vsys.

The measured values of $\vsys$ for all the galaxies are reported in
table \ref{tab:smoothing_parameters}. We do not observe any systematic
bias in the measured values, and they all agree within twice the error
on $\vsys$ with the literature. Hereafter we will refer to the
barycentric velocities using $\mathrm{V}$, and to the smoothed
barycentric velocities using $\tilde{\mathrm{V}}$.

\section{Point symmetry analysis of the sample}
\label{sec:Point_symmetry_analysis}

In this section we investigate whether the galaxies in the ePN.S
sample show any deviation from point-symmetry.

\begin{figure*}[h]
   \centering
   \begin{minipage}[b]{0.320\linewidth}
     \includegraphics[width=\columnwidth]{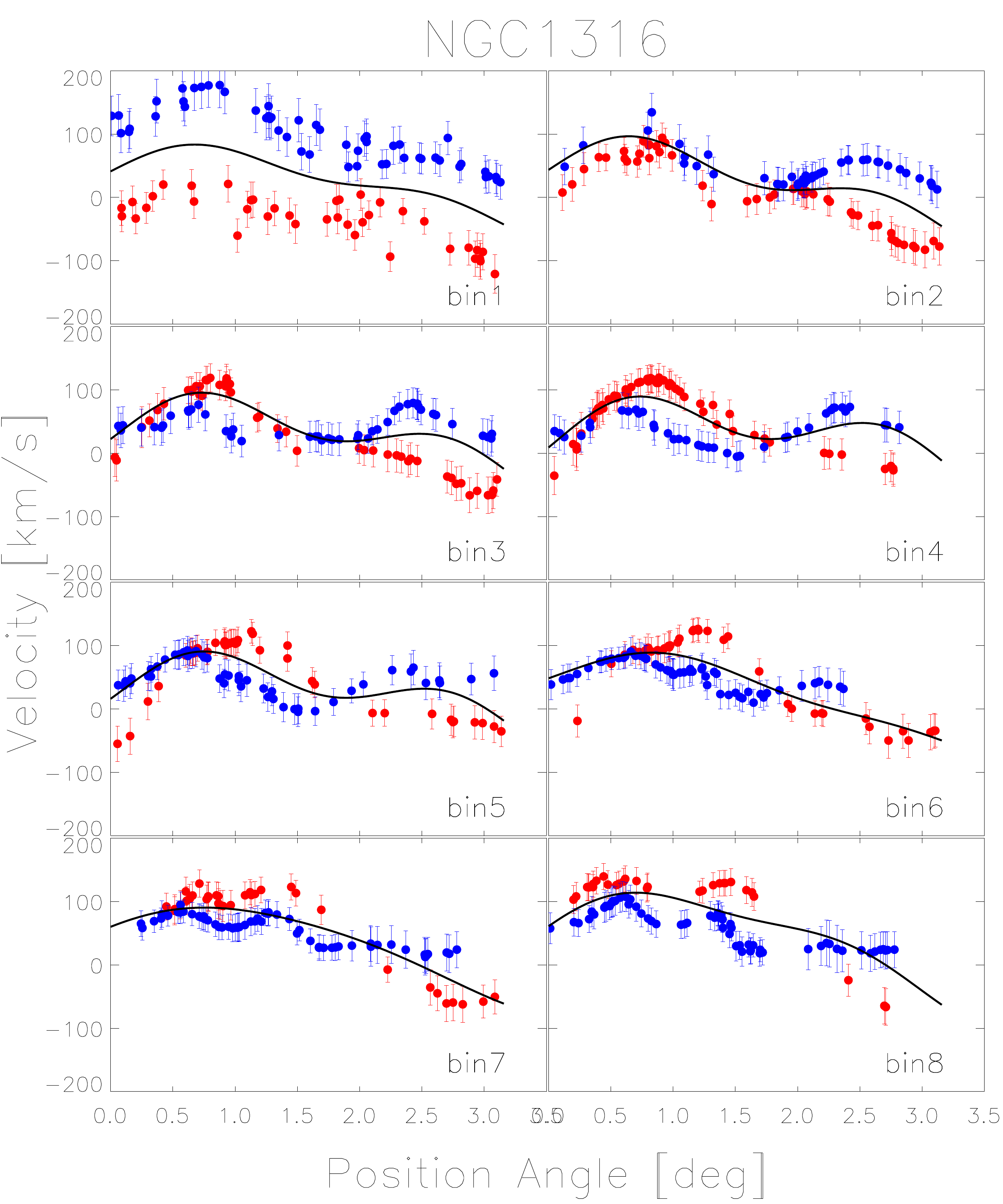}
     \includegraphics[width=\columnwidth]{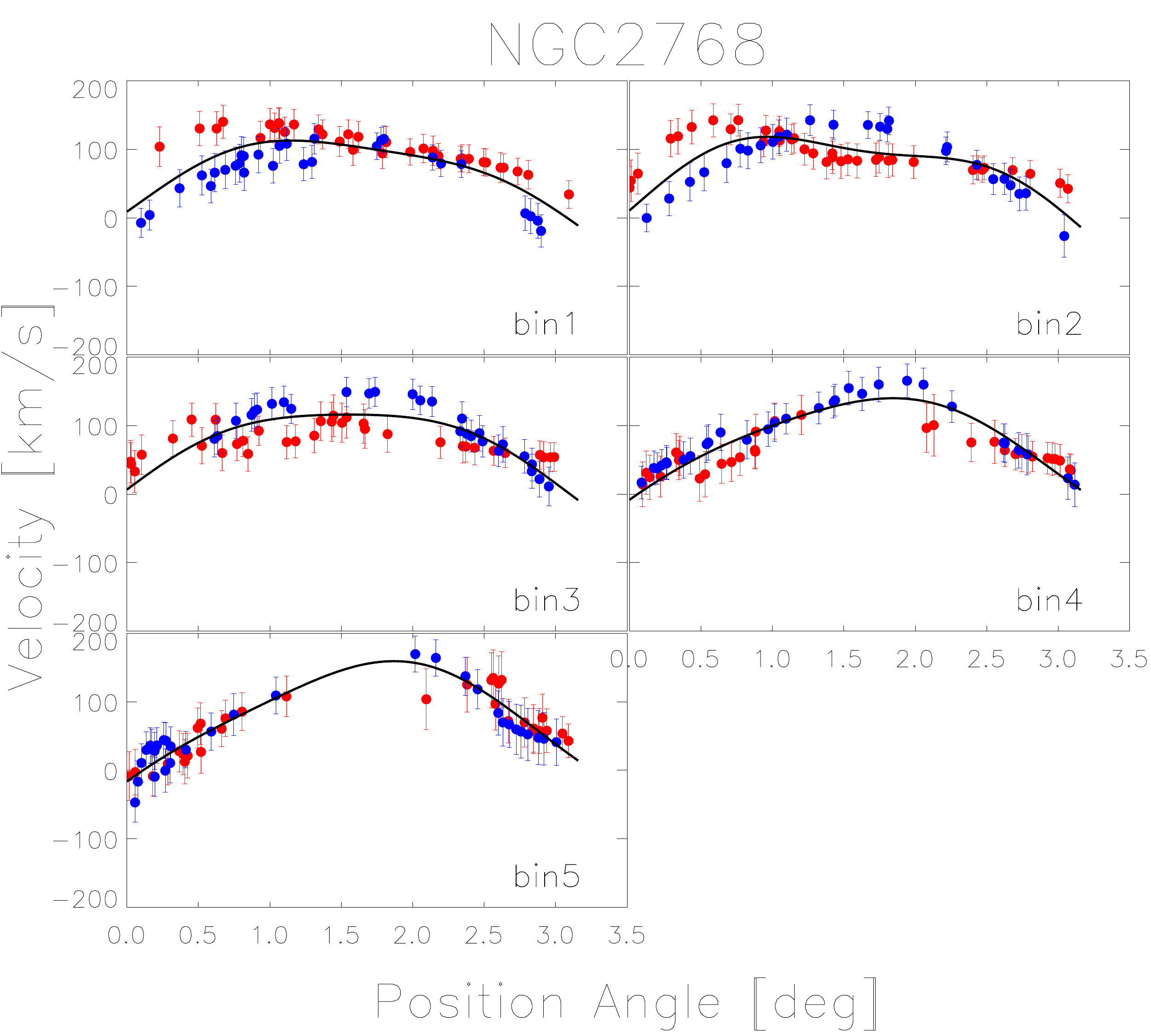}
   \end{minipage}
     \hspace{0.2cm}
   \begin {minipage}[b]{0.320\linewidth}
     \includegraphics[width=\columnwidth]{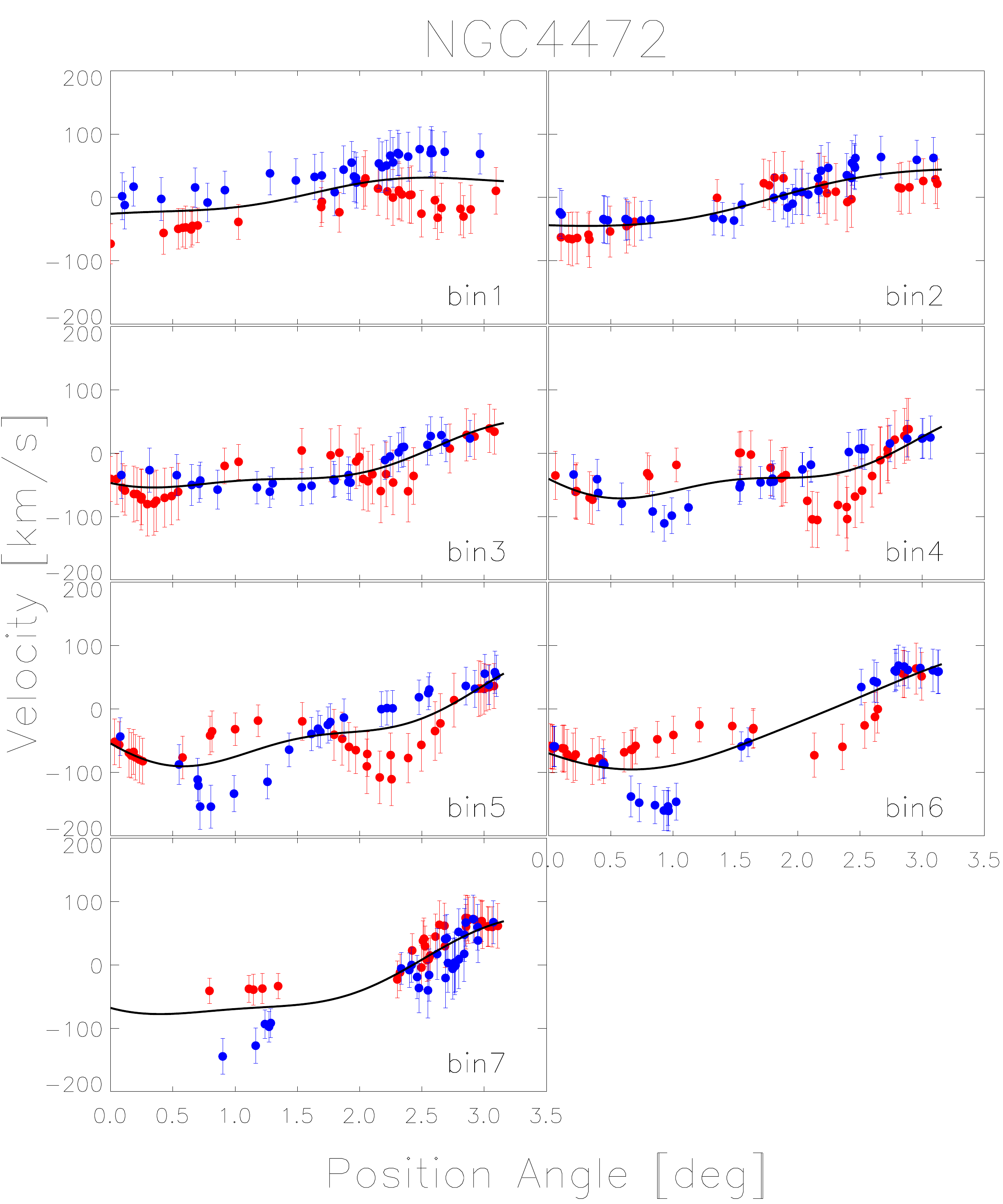}
     \includegraphics[width=\columnwidth]{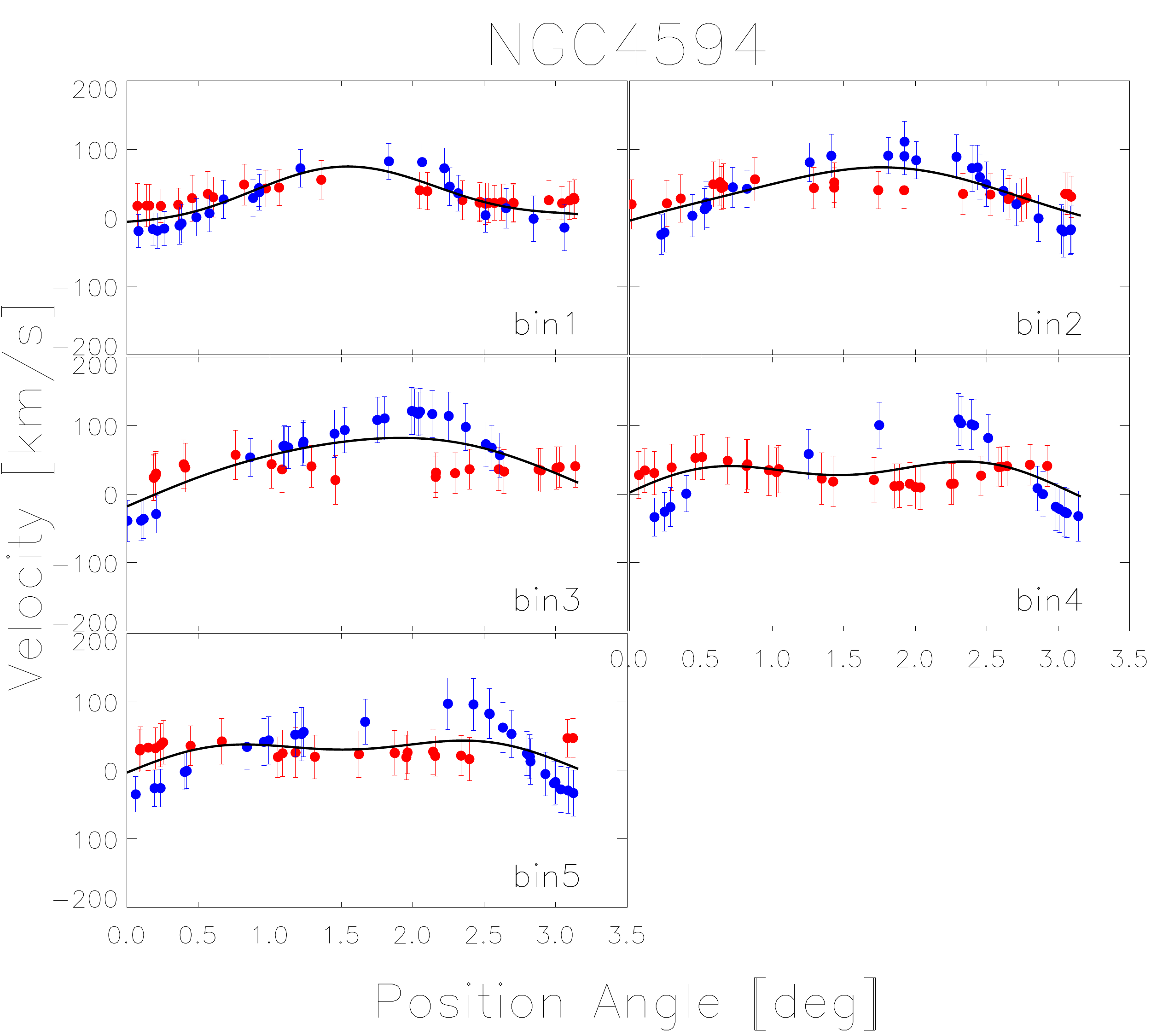}
   \end{minipage}
   \begin {minipage}[b]{0.320\linewidth}
     \includegraphics[width=\columnwidth]{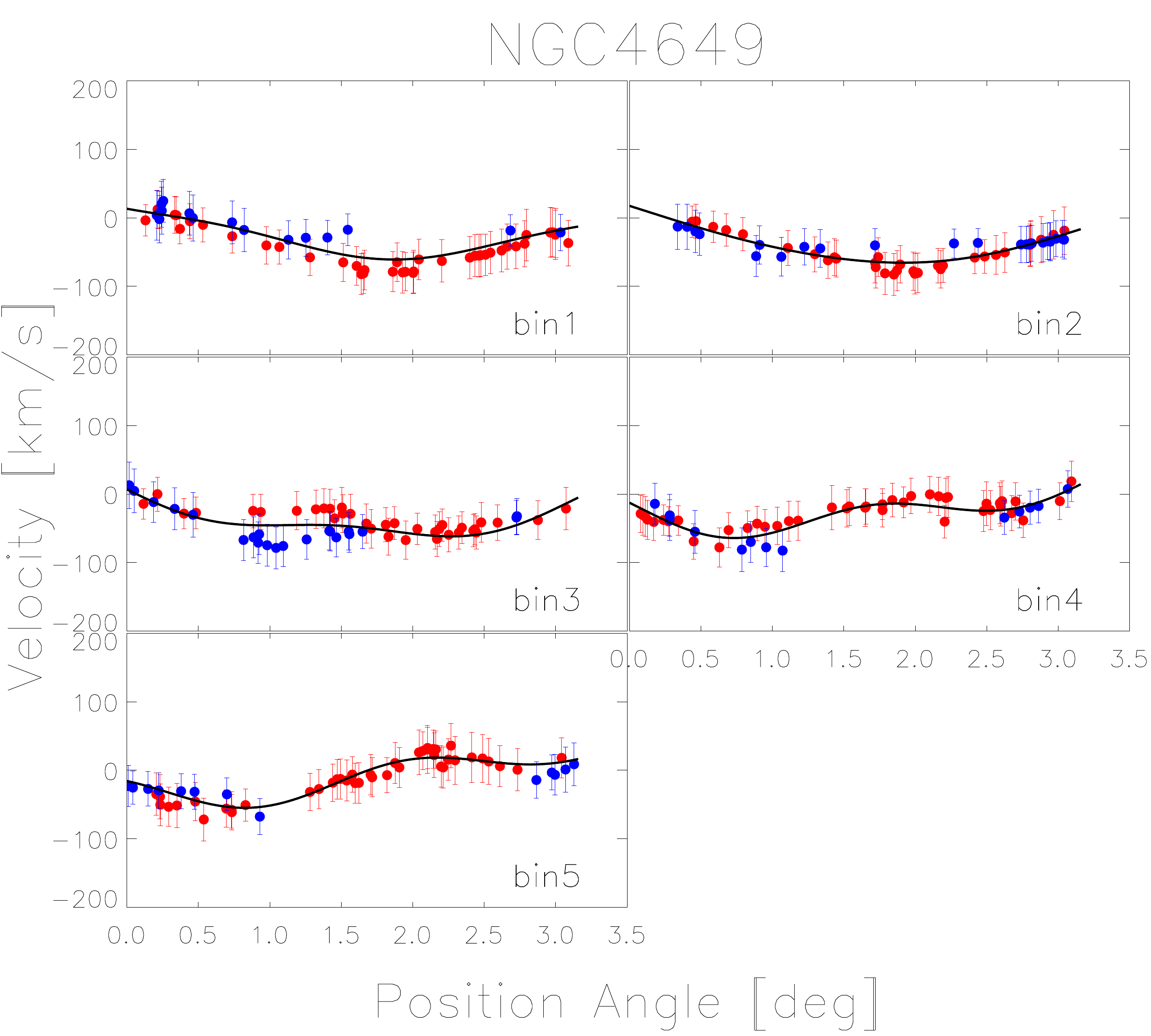}
     \includegraphics[width=\columnwidth]{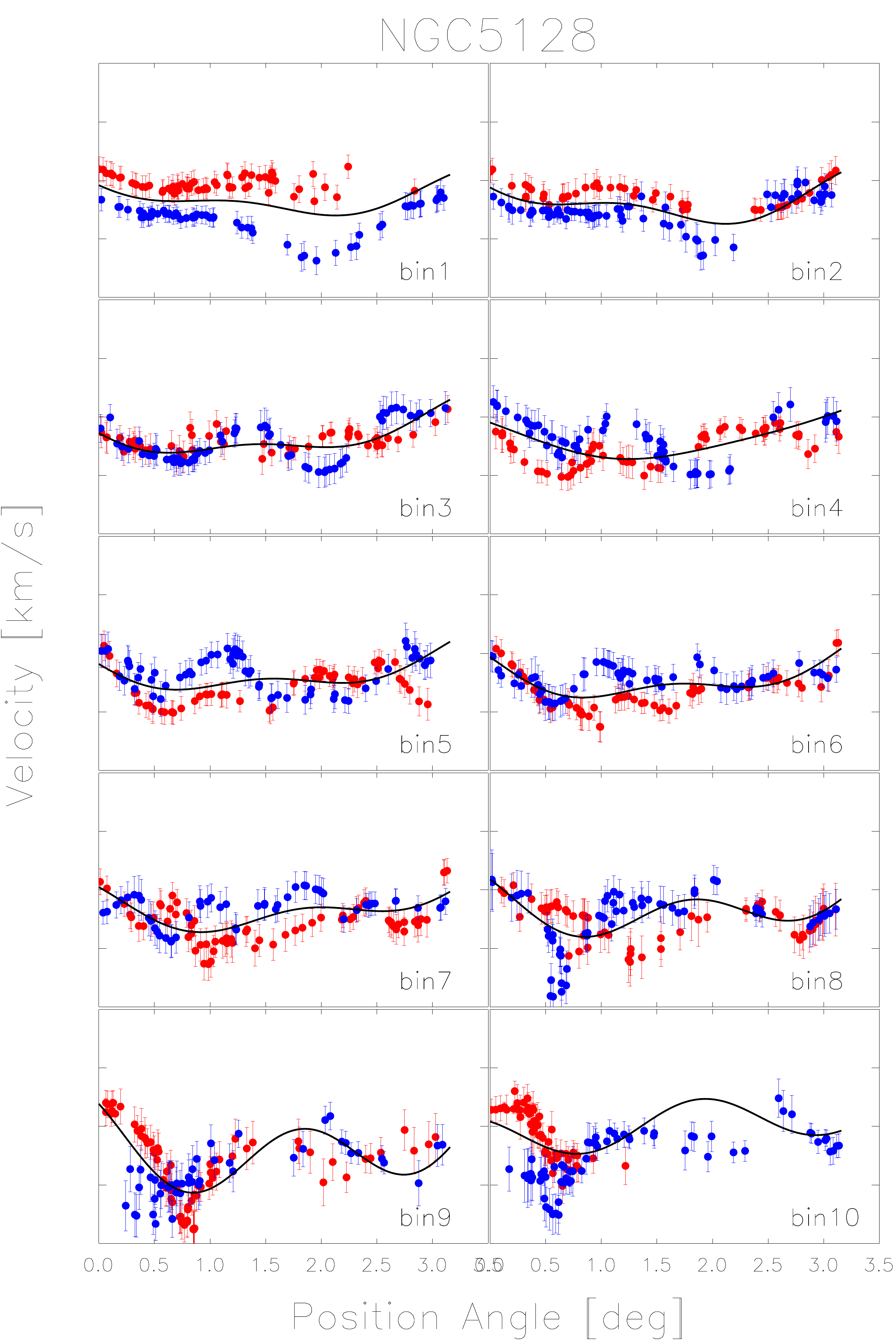}
   \end{minipage}
   
   \caption{ Galaxies with deviations from point symmetry. Mean
     velocity $\tilde{V}(R,\phi)$ as a function of the PN
     eccentric anomaly $\phi$, folded around $\phi=\pi$, for each
     radial bin of NGC 1316, NGC 2768, NGC 4472, NGC 4594, NGC 4649,
     and NGC 5128. The black solid line is the fitted point-symmetric
     rotation model. The red points are the PN  positions with
     $0\leq\phi_i<\pi$; the blue points are those located at
     $\pi\leq\phi_i<2\pi$, with their velocity changed in sign and
     coordinate $\phi$ shifted by $\pi$, i.e. $V(\phi)$ are
     compared with $-V(\phi+\pi)$. The overlap of PNe of opposite
     sides shows possible asymmetries in the velocity fields. NGC 2768 and NGC 4594 show localized small scale
     deviations from point symmetry, that do not influence the
     kinematic analysis. NGC 1316, NGC 4472, and NGC 5128 have
     non-point-symmetric velocity fields.  By comparison NGC 4649
       is point-symmetric.}
    \label{fig:point_sym}
   \end{figure*}

We studied the point-symmetry of the velocity fields of the galaxies
by comparing the velocities $\tilde{V}(R,\phi)$ with $0\leq\phi<\pi$
with those with $\pi\leq\phi<2\pi$, changed in sign, in each radial
bin. Asymmetries in the velocity fields are visible where these
quantities significantly differ from each other. Figure
\ref{fig:point_sym} shows a few examples of this analysis. NGC 4649 is
point-symmetric, while the others show significant deviations. In the
sample of 33 galaxies, 5 are found to be non-point-symmetric: NGC
1316, NGC 2768, NGC 4472, NGC 4594 and NGC 5128. The galaxies for which we find evidence for asymmetries are those with the
richest PN catalogs. For these objects the kinematic details are best
recovered. 

The others galaxies are
consistent with point-symmetry, so for these systems we used the
folded catalogs to reconstruct the final velocity and velocity
dispersion fields, as described in section
\ref{subsec:velocity_fields}. 
Since the mean velocity fields are the result of a
smoothing procedure, their point symmetry does not rule out kinematic
asymmetries on smaller scales.

\subsection{Testing the significance of the deviations from point-symmetry}
\label{subsec:test_point-symmetry}

 As first step, the PN smoothed velocities $\tilde{V_i}$ are
  modeled with the harmonic expansion in equation
  \eqref{eq:fourier3}, which also provides a good description of the
  galaxy velocity field where the spatial distribution of the tracers
  is not azimuthally complete. We consider as possible deviations from
  point-symmetry any groups of at least three tracers whose velocity
  $\tilde{V_i}$ deviates more than twice the errors from the fitted
  point symmetric model. 

 We evaluate the significance of the observed deviations from
  point symmetry by using 100 models of the galaxies, constructed as
  described in appendix \ref{subsubsec:statistical_tests}. These are
  built using the positions $(x_i,y_i)$ of the PNe from the real
  dataset, and, by construction, have a point symmetric mean velocity
  field. If similar local deviations from point symmetry that we
  observe in the velocity field of the galaxy appear also in the
  smoothed velocity fields of the models, then we know that they are
  artifacts of the smoothing over that particular spatial
  distribution, and not properties of the intrinsic galaxy velocity
  field. Hence, for each feature in the galaxy, we select in the
  models the groups of PNe having the same coordinates as the feature,
  and compute the distribution of deviations of their velocities
  from the harmonic expansion fitted to each
  model. This distribution will give the probability of occurrence of
  the feature due to statistical fluctuations.

We found that the features in NGC 2768 and NGC 4594 have a probability
$<1\%$ to happen in the symmetric models so they are likely
real. Those in NGC 2768 might be related to asymmetries in light distribution clearly visible in 
deep optical images
\citep[e.g. the g and r maps of][]{2015MNRAS.446..120D}\footnote{http://www-astro.physics.ox.ac.uk/atlas3d/}. The
features in NGC 4594 are more likely due to extinction effects from
its dusty disk, which hampers the detection of a complete sample of
PNe in that area. In both cases the deviations of the velocity
  fields from point symmetry are localized and do not influence the
  kinematic analysis.

For NGC 1316, NGC 4472, and NGC 5128 the velocity offsets and the
phase-angle shifts of the $\tilde{V}(R,\phi)$ in ($0\leq\phi<\pi$, red
in figure \ref{fig:point_sym}) with respect to ($\pi\leq\phi<2\pi$,
blue in figure \ref{fig:point_sym}) cannot be reproduced by the point
symmetric models.  These galaxies are well known recent mergers. Their
halos are dominated by the recently accreted component which is not
yet in a phase-mixed equilibrium with the surroundings and hence it
still maintains peculiar kinematics (see also appendix
\ref{sec:notes_on_single_galaxies}).

One may be tempted to identify the groups of PNe whose velocity
significantly deviate from the model as those associated to the
structure, but we need to keep in mind that their velocities are the
result of an averaging procedure, and that the different kinematic
components can only be separated by analyzing the full phase space
\citep[see e.g. the GMM modeling of][]{2015A&A...579L...3L}; such a
study is beyond the scope of this paper.

\section{The halo kinematics of ETGs} \label{sec:the_halo_kinem_of_ETGs}

\subsection{Velocity fields}
\label{subsec:velocity_fields}

A point symmetric system is, by definition, such that each point of
the phase space $(x,y,\mathrm{V})$ has a point-reflected counterpart
$(-x,-y,-\mathrm{V})$. For the galaxies that do not show any
significant deviation from point symmetry (section
\ref{sec:Point_symmetry_analysis}), we assume that point symmetry
holds. In these cases we can double the number of data-points by
adding to the observed dataset its mirror dataset, and creating in
this way a \emph{folded catalog} \citep[e.g][]{1998ApJ...507..759A,
  2001A&A...377..784N,2004ApJ...602..685P,2009MNRAS.394.1249C}. This
helps in reducing the fluctuations in the recovered velocity
fields. The results obtained using the folded catalogs are consistent
with those from the unfolded datasets within the errors. Therefore
for the galaxies consistent with point symmetry we will use the folded
catalogs to produce the final mean velocity fields; for the others
(i.e. NGC 1316, NGC 2768, NGC 4472, NGC 4594 and NGC 5128) the
original catalogs are used.

\begin{figure}[!h]
  \centering
  \begin{minipage}{0.485\linewidth}
  \includegraphics[width=\columnwidth]{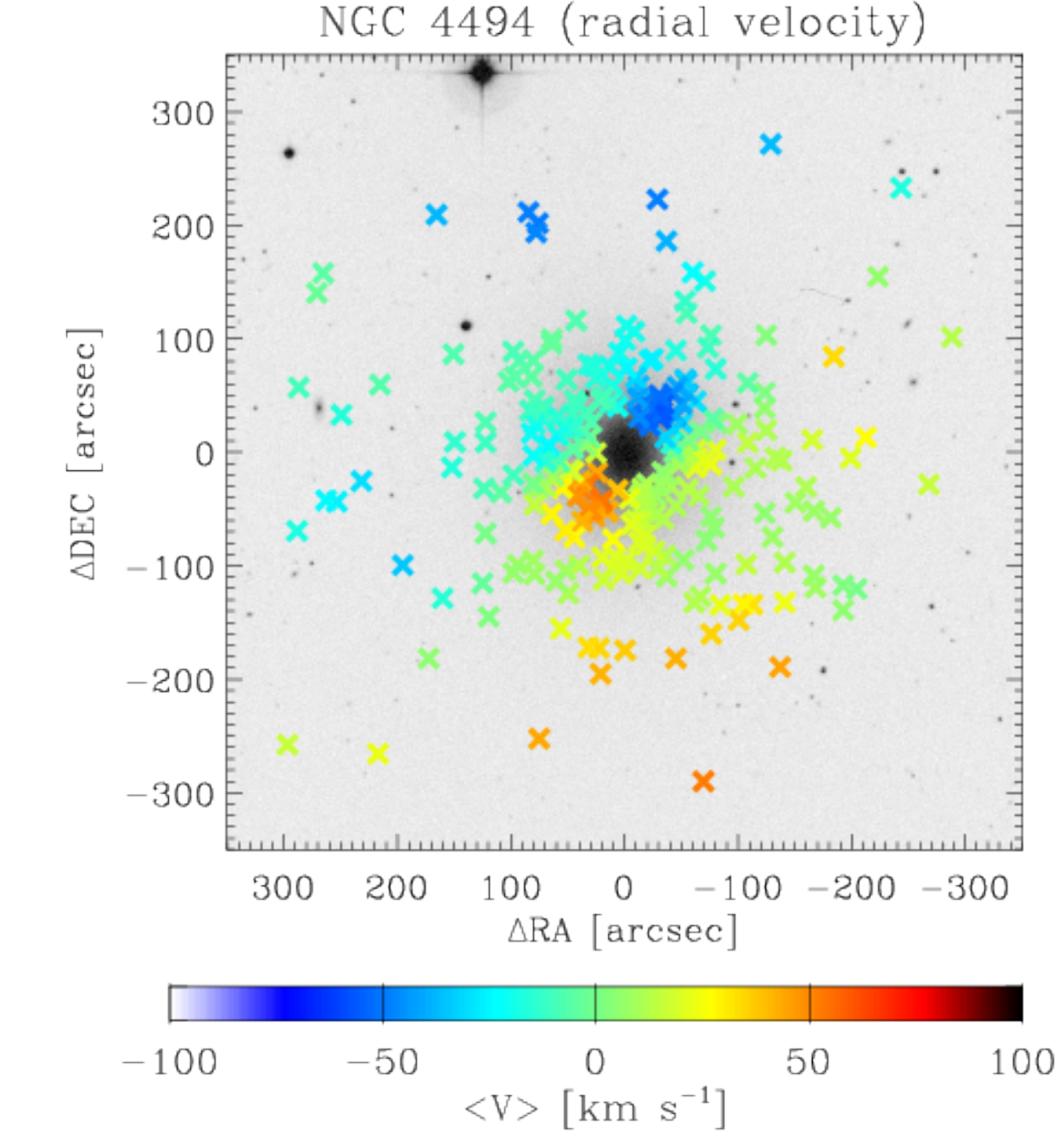}
  \vspace{0.1cm}
   \includegraphics[width=\columnwidth]{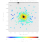}
  \end{minipage}
  \hspace{0.05cm}
   \begin{minipage}{0.485\linewidth}
  \includegraphics[width=\columnwidth]{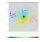}
  \vspace{0.1cm}
   \includegraphics[width=\columnwidth]{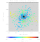}
  \end{minipage}

  \caption{Top row: smoothed velocity fields of NGC 4494 and NGC 4552;
    bottom row: velocity dispersion fields. The fields are built using
    the \emph{folded catalogs}, but only the positions of the actual
    PN data points are shown. The images in the background are from
    the Digitized Sky Survey (DSS); north is up, east is left.}
\label{fig:esempio_4494_4552}
\end{figure}

Figure \ref{fig:esempio_4494_4552} shows the result for two galaxies
with a similar number of tracers, NGC 4494 and NGC 4552. Both are
point symmetric, so the velocity fields in figure
\ref{fig:esempio_4494_4552} are built using the folded catalogs. NGC
4494 is a FR showing some rotation also in the halo. Its
velocity dispersion field reveals that $\sigma$ decreases with
radius. The SR NGC 4552, by contrast, displays increasing
rotation velocity about two perpendicular axes, and increasing
velocity dispersion with radius.

The smoothed velocity fields for all the galaxies of the ePN.S sample
are shown in appendix \ref{sec:figures_vel_fields}.  For a more
immediate visualization we present interpolations of the velocity
fields, based on computing $\tilde{V}$ and $\tilde{\sigma}$ on a
regular grid.  The kinematics typically extends a median of 5.6 $\re$,
covering a minimum of $3\re$ to a maximum of $13$ \re. The adopted
$\re$ values are listed in table \ref{tab:galaxies}. Table
\ref{tab:galaxies} also shows the mean radius of the last radial bins,
in which we can statistically determine $\tilde{\mathrm{V}}$ and
$\tilde{\sigma}$.

The typical errors on the mean velocities and on the velocity
dispersions, evaluated with Monte Carlo simulations, range from 10 to
40 \kms, being smaller for galaxies with a larger number of tracers
and higher $\tilde{V}/\tilde{\sigma}$.

These errors on the mean velocity fields, the mean errors on the
radial velocity measurement, the kernel parameters $A$ and $B$ used in
the smoothing procedure, and the systemic velocities subtracted are
reported in table \ref{tab:smoothing_parameters}.

A visual comparison with the kinematic maps published by the
$\mathrm{ATLAS^{3D}}$ \citep{2011MNRAS.414.2923K} and SLUGGS
\citep{2014ApJ...791...80A,2016MNRAS.457..147F} surveys shows a
general good agreement for all the galaxies in the regions of overlap.
By consistency/good agreement we mean that the values and trends in $V_\mathrm{rot}, \sigma$ 
from either kinemetry \citep[by][]{2008MNRAS.390...93K, 2016MNRAS.457..147F} or slit absorption line kinematics agree within the errors with the kinematics from the PNe in the regions of overlap, or the latter extend such radial trends to the outer regions.
See the appendix \ref{sec:notes_on_single_galaxies} for a detailed
description of the individual objects.

\subsection{Kinematic parameters}
\label{subsec:kinematic_parameters}
 
We quantify the properties of the reconstructed mean velocity
fields by evaluating the amplitude of rotation $V_\mathrm{rot}$, the
variation of the $\PAkin$ with radius, and the possible misalignments
with \PAphot.  Therefore we model the velocities in each elliptical bin as
a function of the angle $\phi$ (positive angles are from North to
East, with the zero at North) with the rotation model in equation
\eqref{eq:fourier_definitivo}, as described in section \ref{subsec:fit_rotation_model}. 

  \begin{figure*}[t]
   \centering    
   \begin{minipage}[b]{0.47\linewidth}
     \includegraphics[width=\columnwidth]{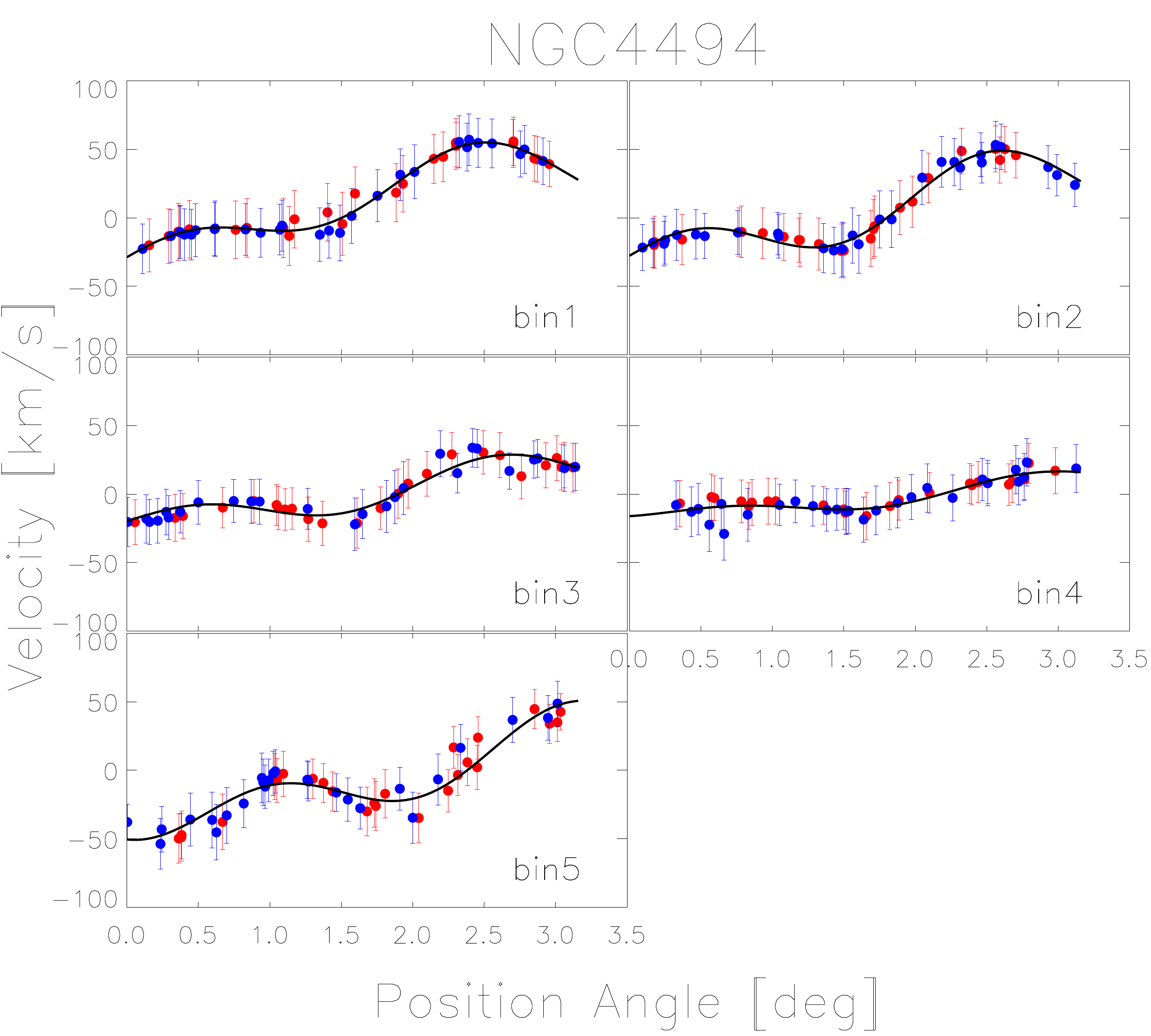}
     \includegraphics[width=\columnwidth]{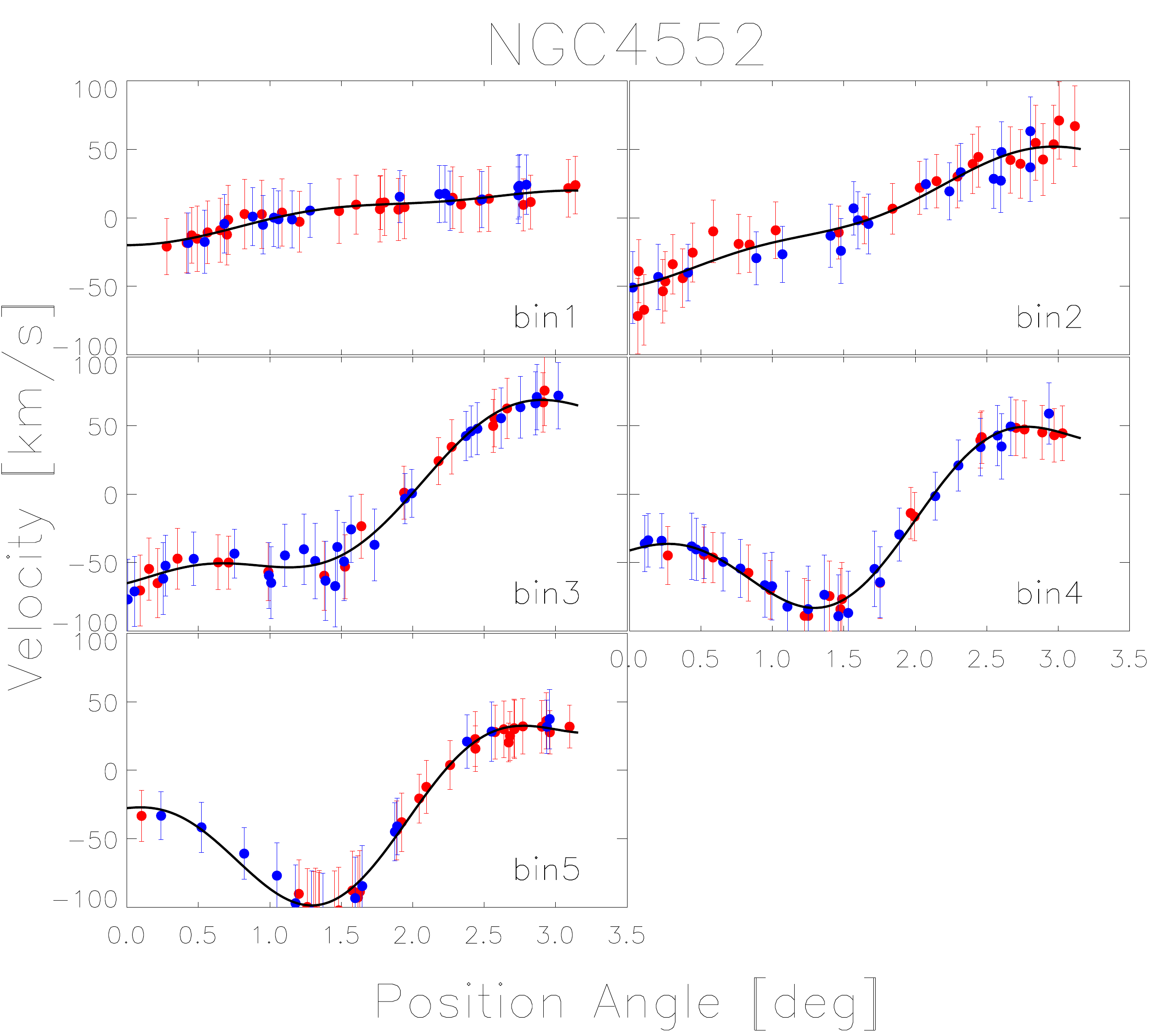}
   \end{minipage}
   \begin{minipage}[b]{0.47\linewidth}
     \includegraphics[width=\columnwidth]{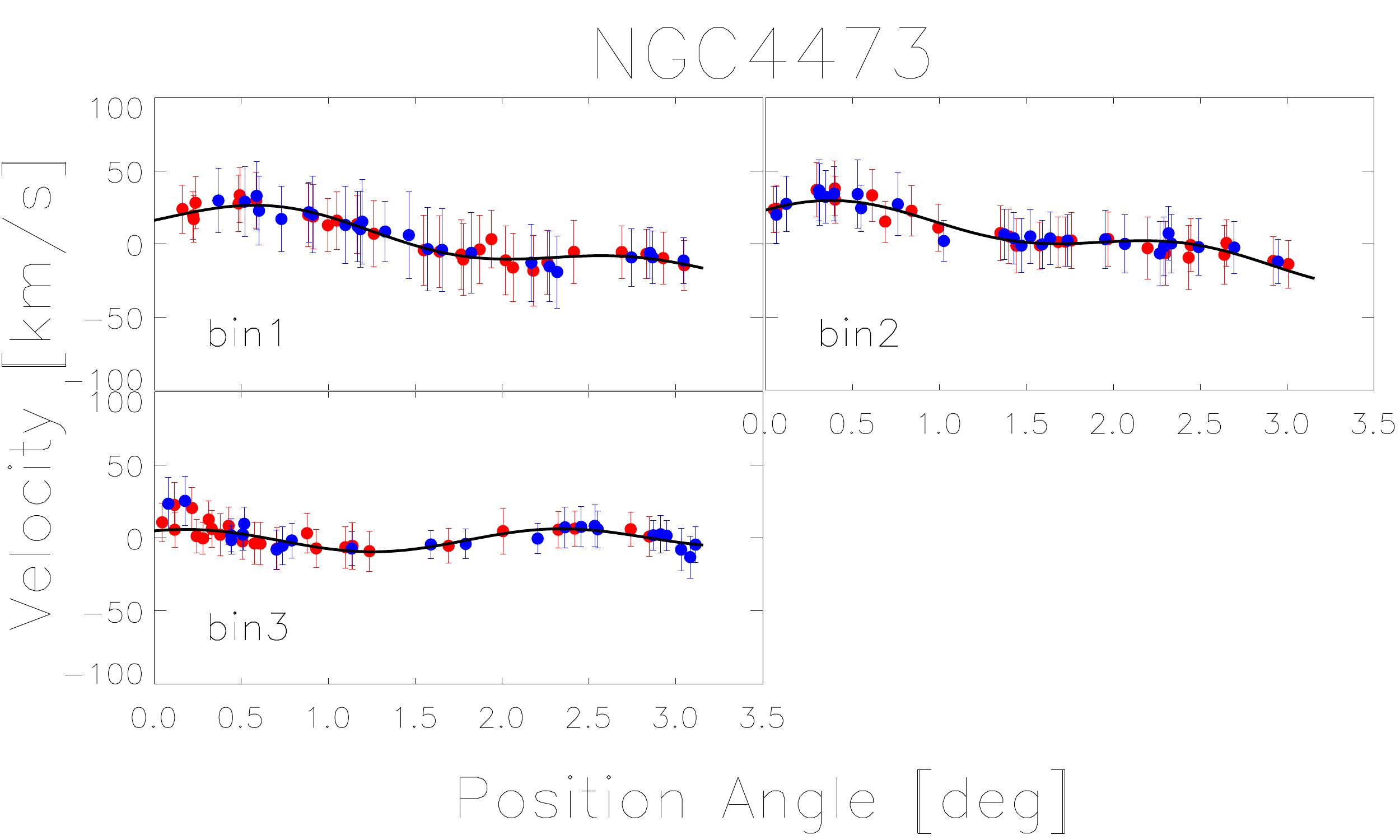}
     \includegraphics[width=\columnwidth]{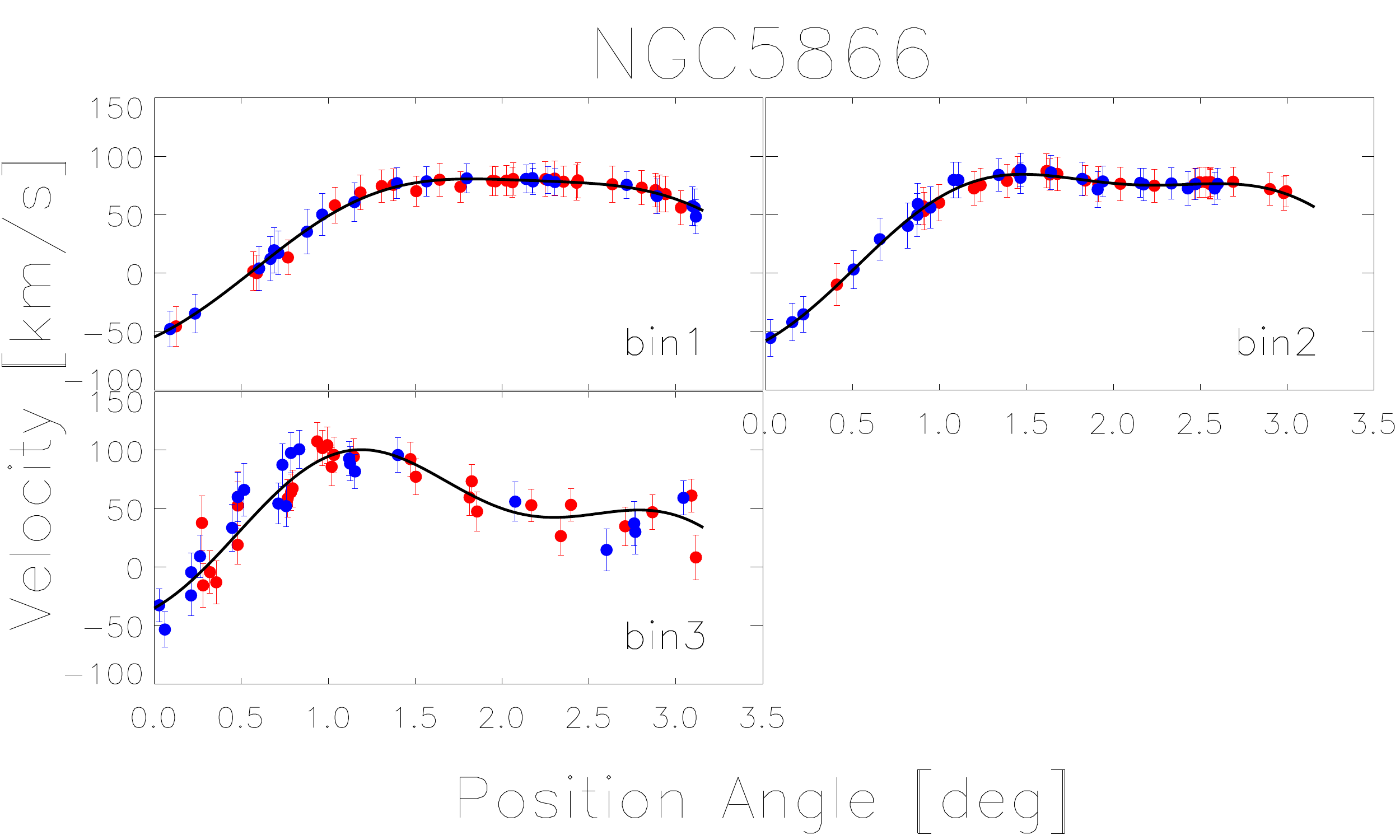}
      \includegraphics[width=\columnwidth]{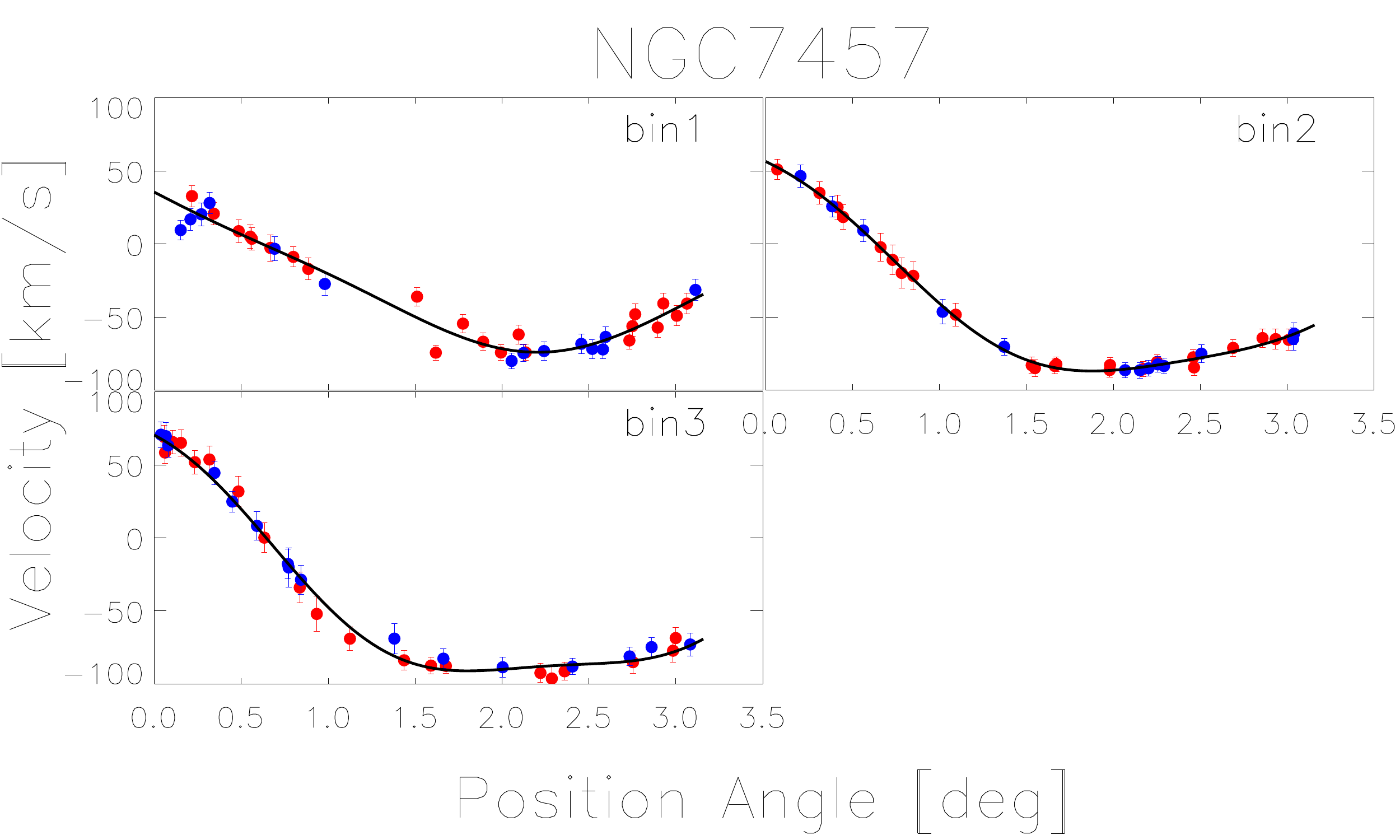}
   \end{minipage}
    
   \caption{ Point symmetric galaxies. Mean velocity field
       $\tilde{V}(R,\phi)$ in elliptical annuli as a function of the
       PN eccentric anomaly $\phi$, folded around $\phi=\pi$ for the
       FRs NGC 4473, NGC 4494, NGC 5866 and NGC 7457, and
       the SR NGC 4552 (colors as in figure
       \ref{fig:point_sym}). $\tilde{V}(R,\phi)$ is reconstructed
       using the folded catalogs, and shown at the position $(R_i,\phi_i)$
       of the actually observed PNe. The solid lines are the best
       rotation model (see equation \ref{eq:fourier_definitivo}) fit
       to the velocity field. }
    \label{fig:azimuthal_plots}
   \end{figure*}
   
    Figure \ref{fig:azimuthal_plots} shows the smoothed velocity
     field $\tilde{V}(R,\phi)$ in each elliptical radial bin for a
     subsample of galaxies: a SR, NGC 4552, and four FRs, NGC 4473, NGC 4494, NGC 5866 and NGC 7457. The solid
     lines are the rotation models that give the best fit to the data,
     and from which we derive the kinematic parameters. 
     The errors on the fitted parameters are
     derived from Monte Carlo simulations as described in section
     \ref{subsubsec:Errors_on_the_fitted_parameters}, and depend on
     the number of tracers and the ratio
     $\tilde{V}/\tilde{\sigma}$. In the case of the galaxies show in
     figure \ref{fig:azimuthal_plots}, they are largest for NGC 4473,
     which has very low rotation in the halo, and smallest for the
     lenticular galaxy NGC 7457, which is dominated by rotation up to
     large radii. 

\begin{figure*}
\begin{center}
   \includegraphics[width=5.5cm,angle=90]{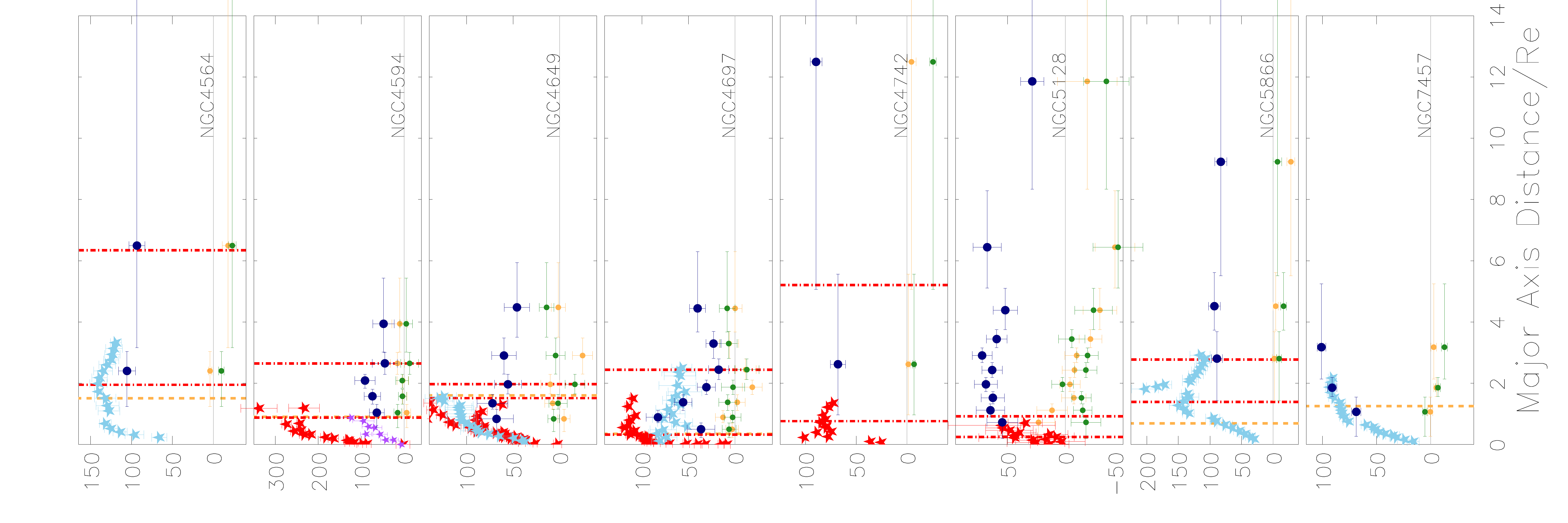}
   \includegraphics[width=5.5cm,angle=90]{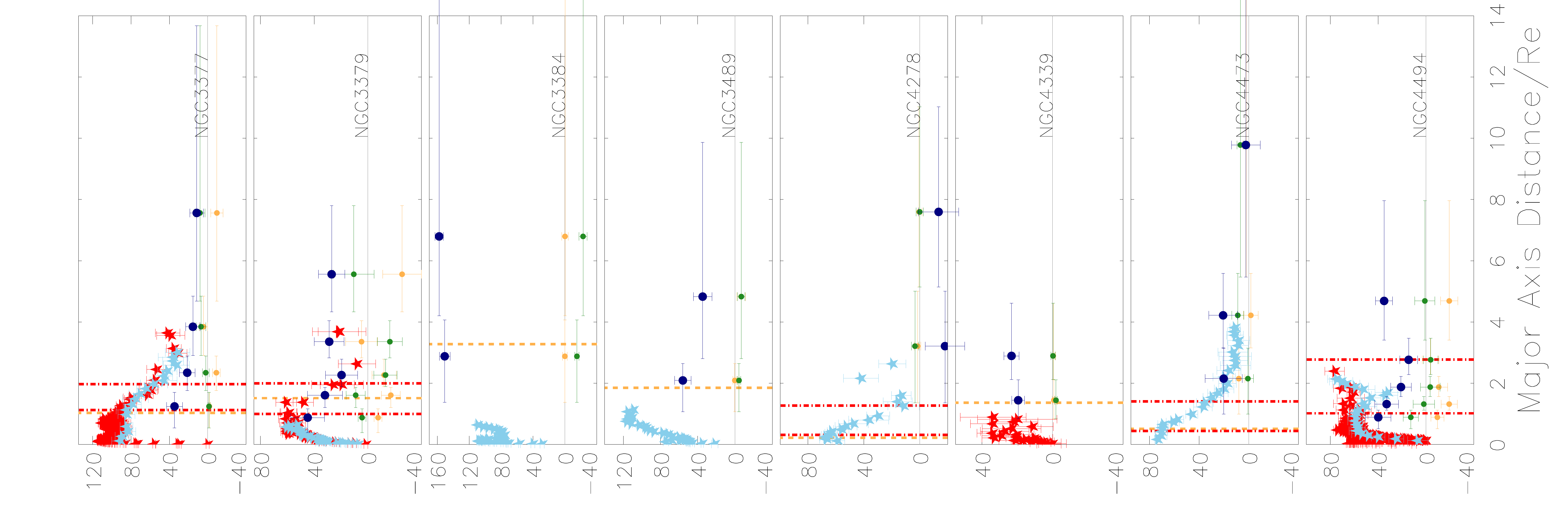}
   \includegraphics[width=5.5cm,angle=90]{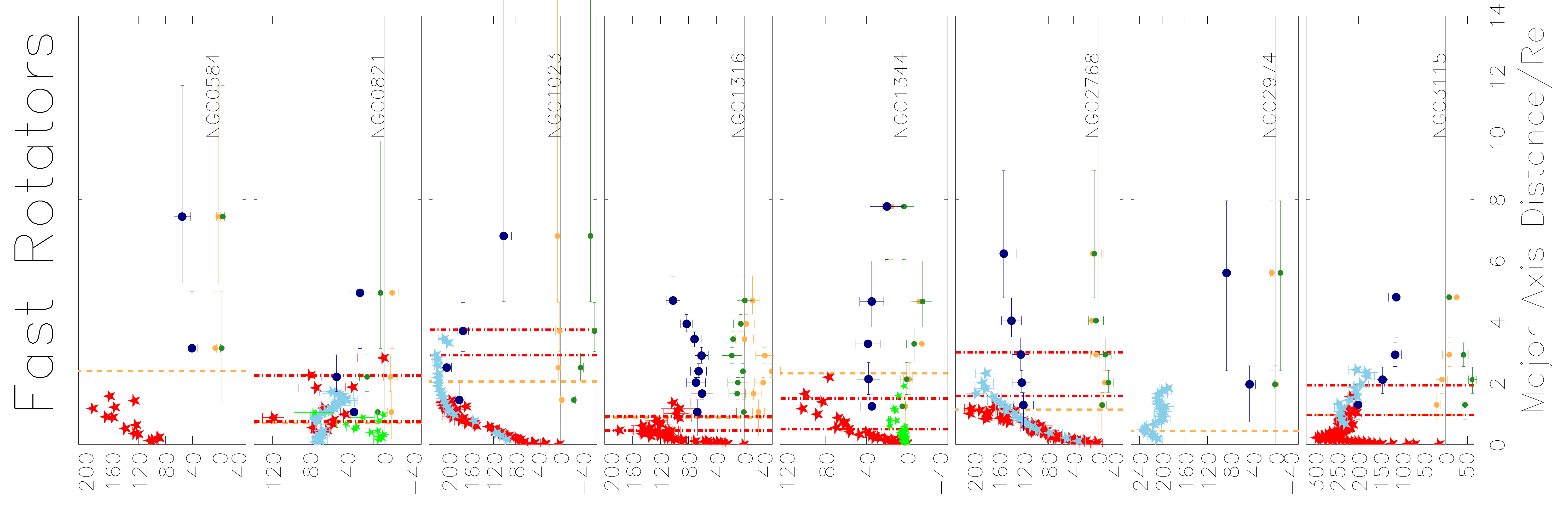}
   \includegraphics[width=6.2cm,angle=90]{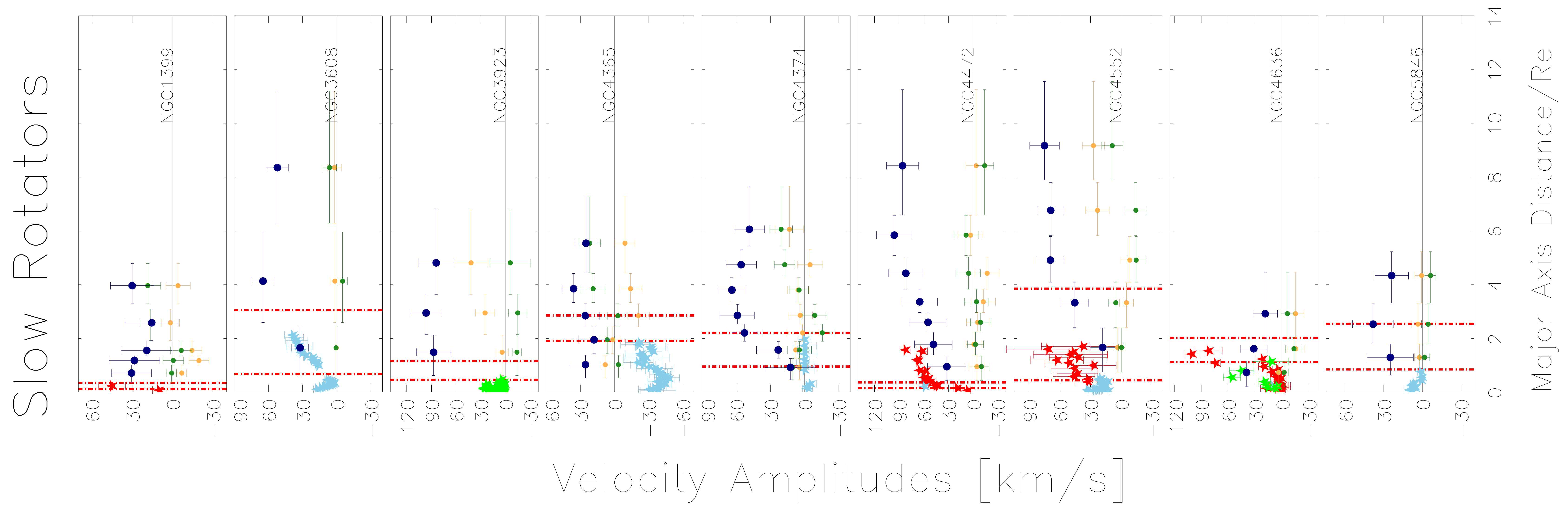}
   \caption{Fitted rotation velocities $V_{\mathrm{rot}} (R)$ (full circles) and third order harmonics amplitudes, $c_3(R)$ in green and $s_3(R)$ in orange, as functions of the major axis distance for SRs and FRs. The comparison values for $V_{\mathrm{rot}} $ from absorption line data from the literature are shown with colored stars. Whenever available we show kinemetric profiles (light blue stars); for the other galaxies we show velocities from long slit spectroscopy along the photometric major axis (red stars) or minor axis (green stars). The references are in table \ref{tab:galaxies}. For NGC 4594 we show in addition the stellar kinematics from a slit along $\PAphot$, offset by 30 arcsec (purple stars). The dashed vertical lines for the FRs show the disk half light radius $R_{1/2}$, see section \ref{subsubsec:embedded_disks}. The red dot-dashed vertical lines report the radial transition range $R_T\pm\Delta R_T$, see section \ref{subsec:DISCUSSION_kinematic_transition_radius}.}
   \label{fig:fitted_velocity_amplitudes}  
\end{center}
\end{figure*}

\begin{figure*}
\begin{center}
   \includegraphics[width=5.5cm,angle=90]{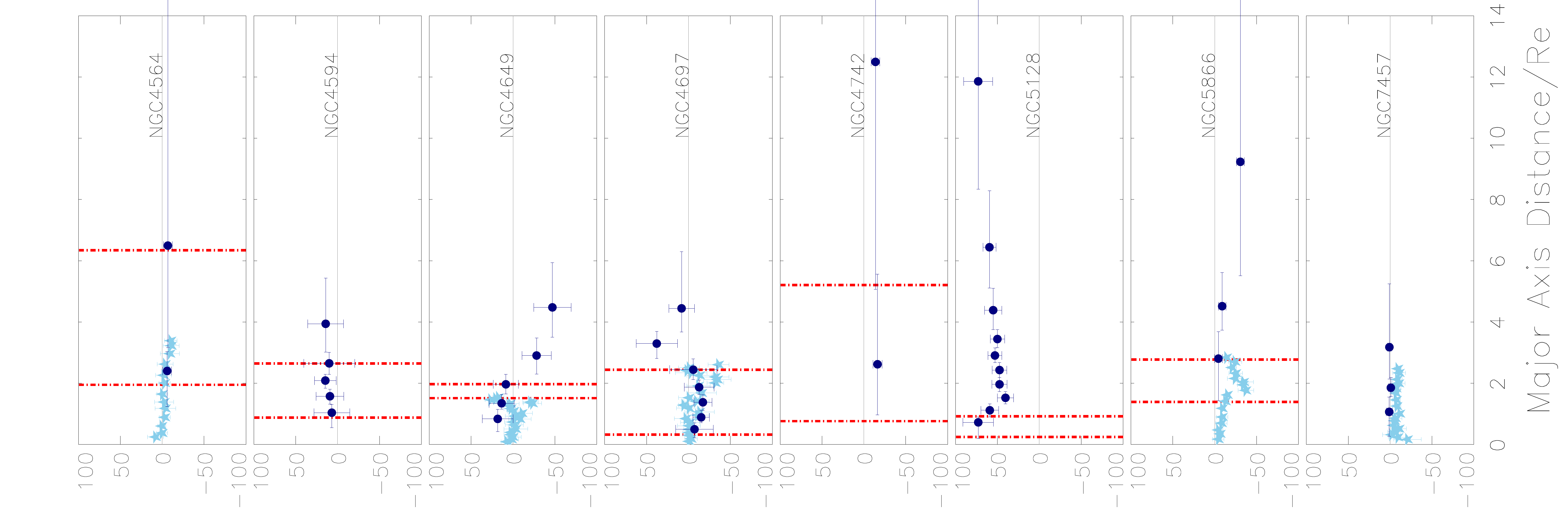}
   \includegraphics[width=5.5cm,angle=90]{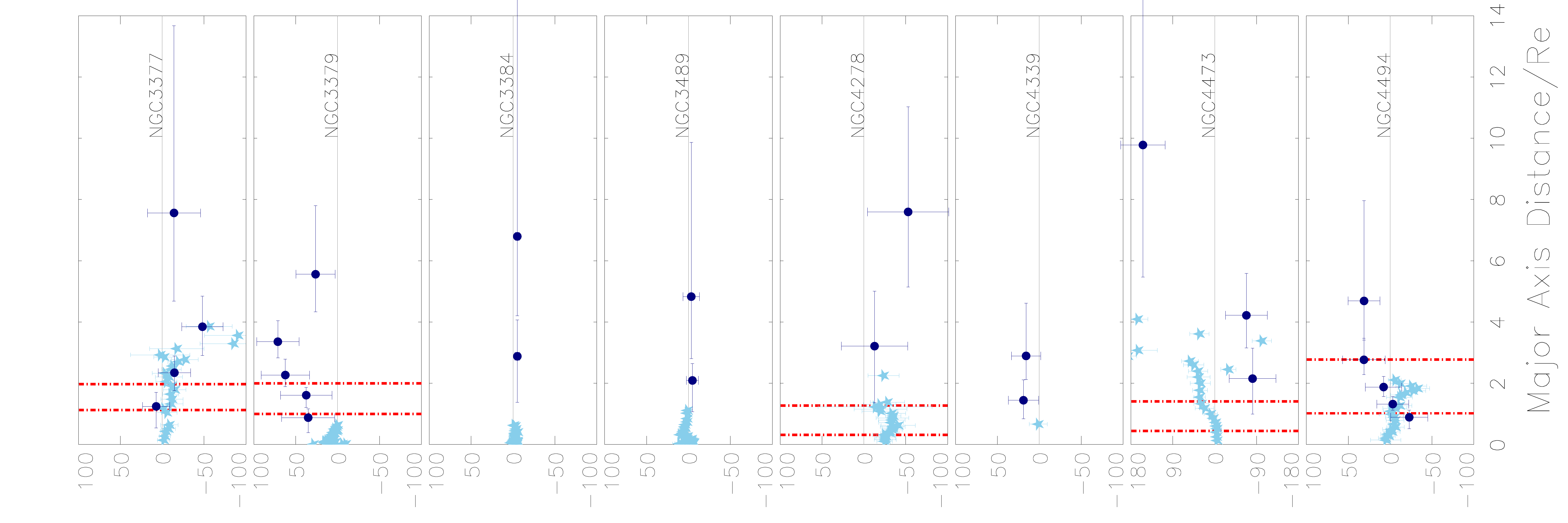}
   \includegraphics[width=5.5cm,angle=90]{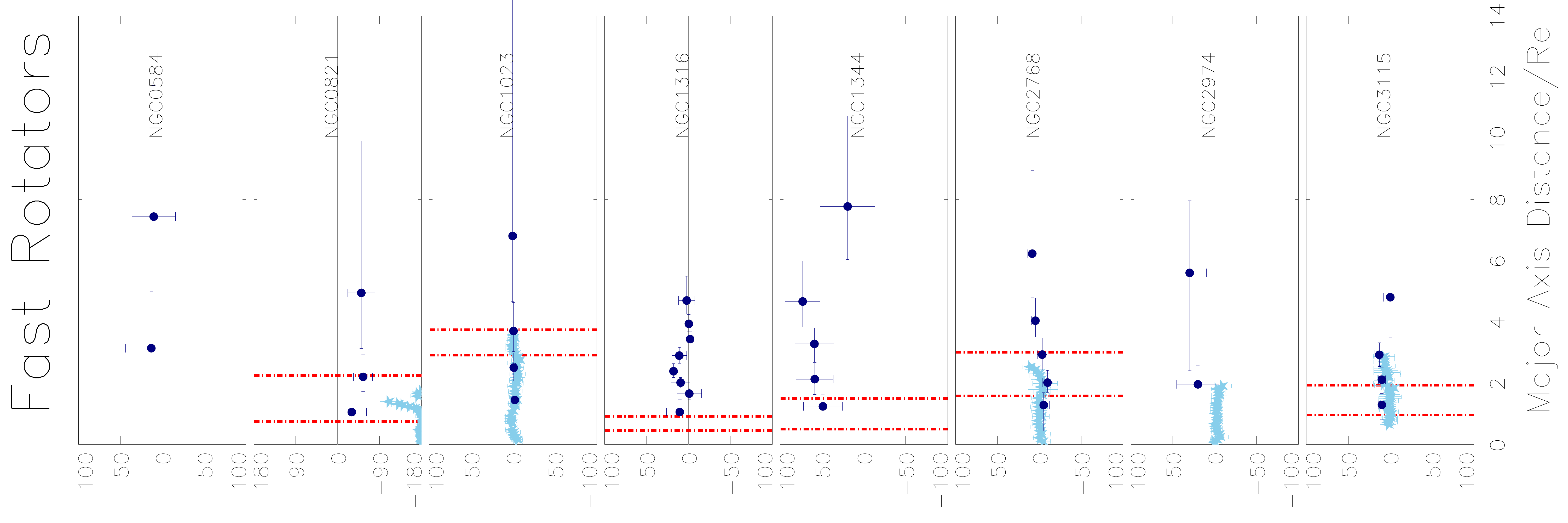}
   \includegraphics[width=6.2cm,angle=90]{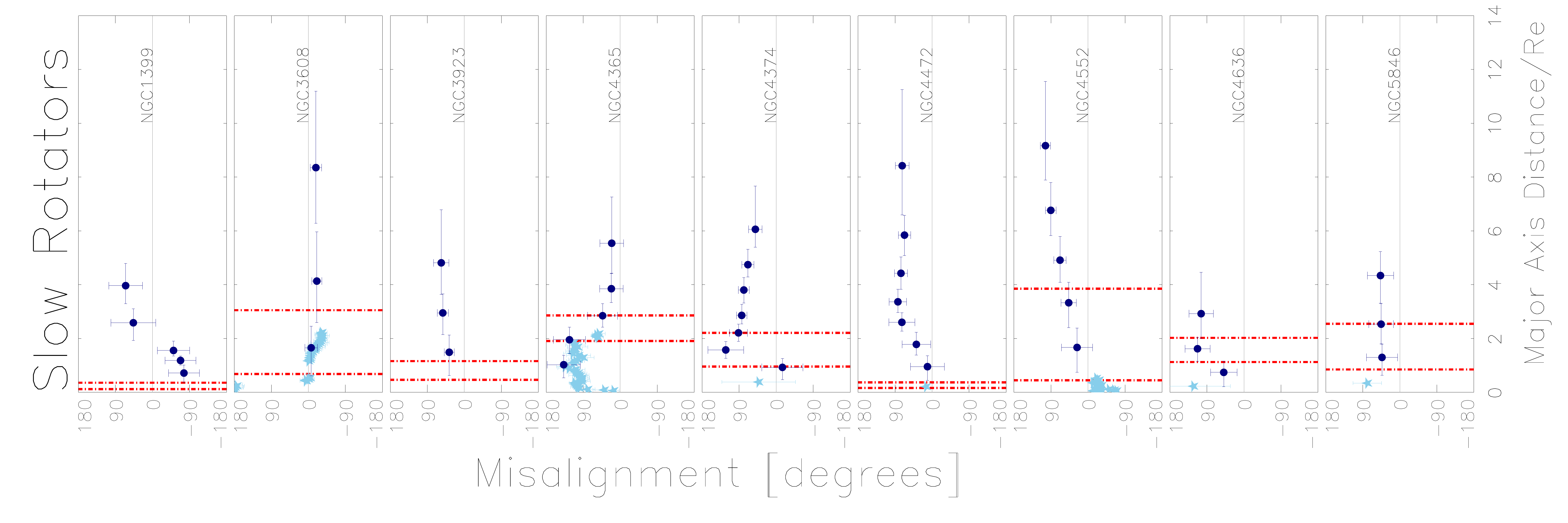}
   \caption{Misalignments $\Psi(R)$ (full circles) as function of the major axis distance for FRs and SRs. The horizontal solid line shows the $\Psi=0$ axis. The light blue stars are the $\Psi$ values calculated on $\PAkin$ from the kinemetry of \citet{2016MNRAS.457..147F}, \citet{2008MNRAS.390...93K}, and \citet{2011MNRAS.414.2923K}; for the other galaxies the $\PAkin$ is not previously available in the literature. The red dot-dashed vertical lines report the radial transition range $R_T\pm\Delta R_T$, see section \ref{subsec:DISCUSSION_kinematic_transition_radius}. Signatures of triaxial halos are seen in NGC 0821, NGC 1344, NGC 1399, NGC 3377, NGC 3379, NGC 3608, NGC 3923, NGC 4365, NGC 4374, NGC 4472, NGC 4473, NGC 4494, NGC 4552, NGC 4636, NGC 4649, NGC 4742, NGC 5128, NGC 5846, and NGC 5866.}
   \label{fig:fitted_misalignments}  
\end{center}
\end{figure*}

   We divided the sample of ETGs into FRs and SRs according
   to the definition of \citet{2011MNRAS.414..888E}, see table
   \ref{tab:galaxies}.  In figure \ref{fig:fitted_velocity_amplitudes}
   we show separately for both families the fitted parameters
   $V_\mathrm{rot}$, $s_3$ and $c_3$, as functions of the major axis
   distance $R$ in units of \re.  This is a reasonable choice in case
   of flattened systems rotating along the photometric major axis. In
   case of misalignment or twist of the $\PAkin$, $R$ does not
   correspond to the position of the peak in $\tilde{V}$ but to the
   major axis of the elliptical bin in which the amplitude $\tilde{V}$
   is calculated.  Figure \ref{fig:fitted_misalignments} shows the
   misalignment $\Psi$ of $\PAkin$ with respect to $\PAphot$,
   $\Psi=\PAkin(R)-\PAphot$. If the difference
   $\PAkin(\mathrm{bin1})-\PAphot$ (where $\PAkin(\mathrm{bin1})$ is
   the value measured in the first radial bin), is greater than 90
   degrees, we define $\Psi$ as $\PAkin(R)-\PAphot-\pi$. Since
   $\PAphot$ is a constant value for each galaxy, a variation of
   $\PAkin$ with radius corresponds to a variation of $\Psi$.  We do
   not use the definition of \citet{1991ApJ...383..112F},
   $\sin\Psi=\sin(\PAkin(R)-\PAphot)$, as it does not allow the
   description of large position angle twists. The values and the
   references for the $\PAphot$ used are in table \ref{tab:galaxies}.

   Both $V_\mathrm{rot}$ and $\Psi$ are compared with literature
   values in figures \ref{fig:fitted_velocity_amplitudes} and
   \ref{fig:fitted_misalignments}. When available, we show the
   profiles from the kinemetric analysis of
   \citet{2016MNRAS.457..147F} on the SLUGGS+$\mathrm{ATLAS^{3D}}$
   data, or the kinemetric profiles from \citet{2008MNRAS.390...93K}.
   In these cases we rescale the radii of the profiles to major axis
   distances using the flattening $q_\mathrm{kin}=q_\mathrm{phot}$
   given by \citet{2016MNRAS.457..147F}, or
   $\left< q_\mathrm{kin}\right>$ given by
   \citet{2008MNRAS.390...93K}.  For the other galaxies we plot the
   corresponding quantities from the kinemetry of \citet[][namely
   $k_1^\mathrm{max}$ and $PA_\mathrm{kin}$ from their table
   D1]{2011MNRAS.414.2923K}, or the kinematic profiles from long slit
   spectroscopy similarly rescaled (references in table
   \ref{tab:galaxies}).  While comparing with the literature, it is
   important to note the following effect. A kinematic measurement
   from a slit along the major axis of an edge-on fast rotating galaxy
   will give high velocities and low dispersions.  On the other hand,
   the PN velocity fields are the results of a smoothing procedure,
   which averages together PNe belonging to the very flat disk with PN
   belonging to the spheroid. This might result in a systematically
   lower rotation and higher velocity dispersion (see equation
   \ref{eq:sigma}) in the PN velocity fields.  A disk/spheroid
   decomposition of the PNe in some ETGs has already been performed by
   \citet{2013MNRAS.432.1010C}, and it is beyond the scope of this
   paper to extend this to all FRs. In addition, if the
   number of tracers or the ratio $V/\sigma$ is low, our kinematic
   analysis provides a lower limit for the rotation velocity and an
   upper limit for the velocity dispersion. This issue is addressed in
   appendix \ref{subsec:discussion_about_the_method}.  In such cases,
   the kinematics traced by the PNe may show systematic differences
   from that in the integrated light as consequence of the discrete
   spatial sampling of the velocity field by the adopted tracers.

   The higher order harmonics amplitudes $s_3$ and $c_3$ differ from
   zero whenever the smoothed velocity field deviates from simple
   disk-like rotation, i.e. if it is cylindrical (see for example the
   case of NGC 3115), or in correspondence to components rotating at
   different position angles (e.g. NGC 4649).

Misalignments and twists of the $\PAkin$ are typically displayed by
triaxial galaxies, see section \ref{subsec:triaxiality}. Figure
\ref{fig:fitted_misalignments} shows that both FRs and SRs
can have radial variations of the $\PAkin$ or a constant
non-negligible $\Psi$. These galaxies may have a triaxial halo. A few
galaxies instead have kinematically decoupled halos with respect to
the regions $\lesssim 2\re$. Section \ref{subsec:triaxiality}
validates these results for each galaxy using models.

The asymmetric galaxies (i.e. NGC 1316, NGC 2768, NGC 4472, NGC 4594,
and NGC 5128) are, by construction, not well represented by the point
symmetric model and increasing the number of harmonics does not
improve the quality of the fit. We can however still use the fitted
parameters to obtain an approximate description of the shape of their
velocity field.

\subsection{Velocity dispersion profiles}
\label{subsec:velocity_dispersion_profiles}

\begin{figure*}
\begin{center}
   \includegraphics[width=5.3cm,angle=90]{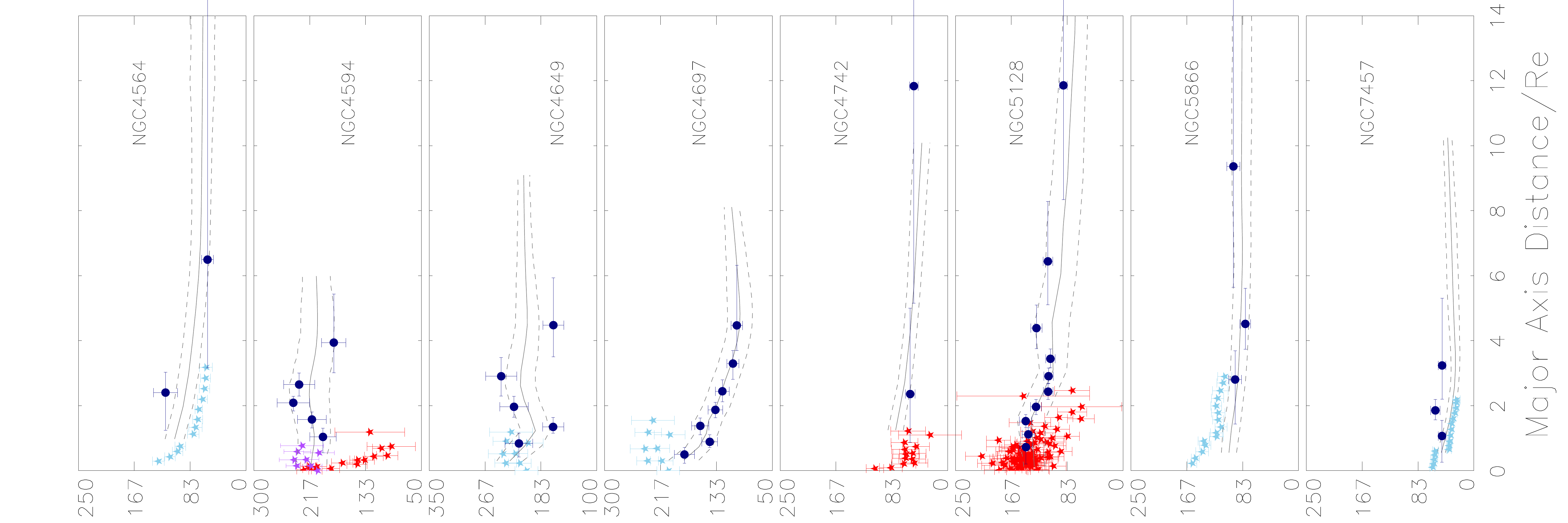}
   \includegraphics[width=5.3cm,angle=90]{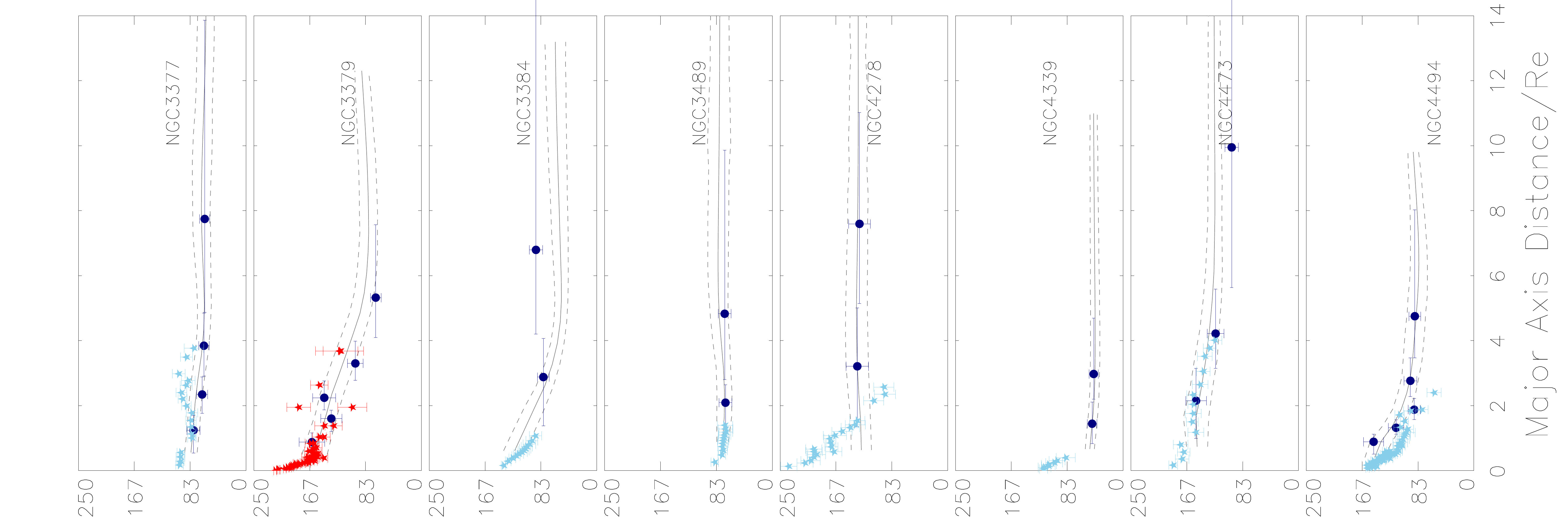}
   \includegraphics[width=5.3cm,angle=90]{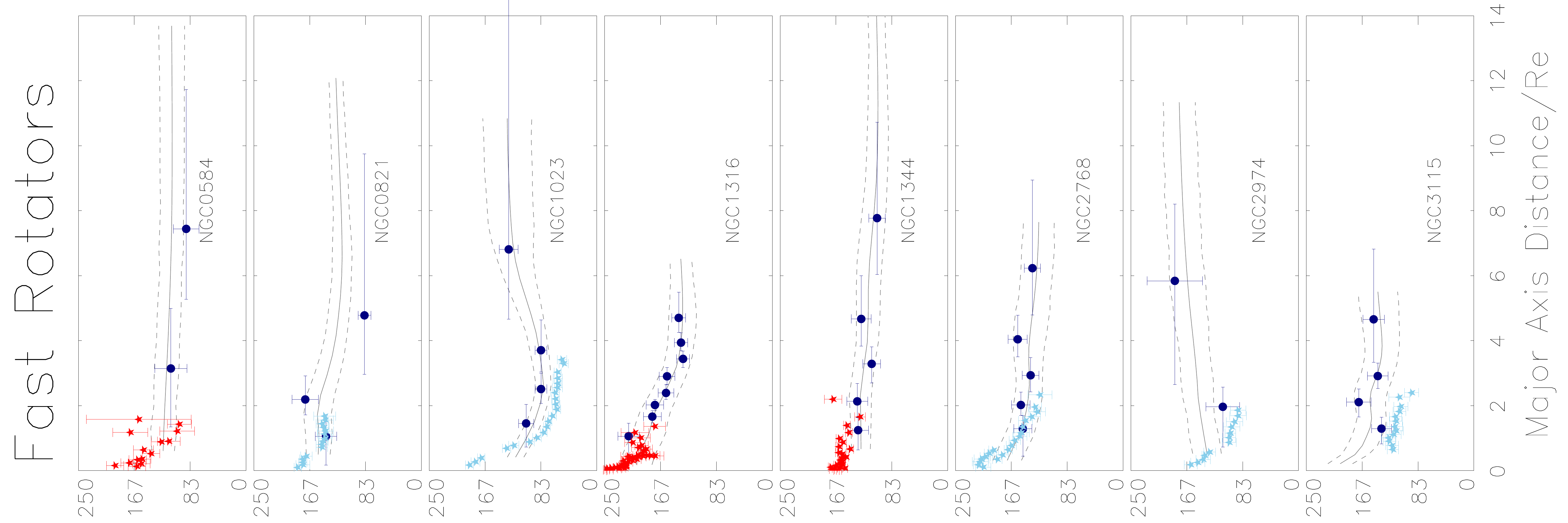}
   \includegraphics[width=6cm,angle=90]{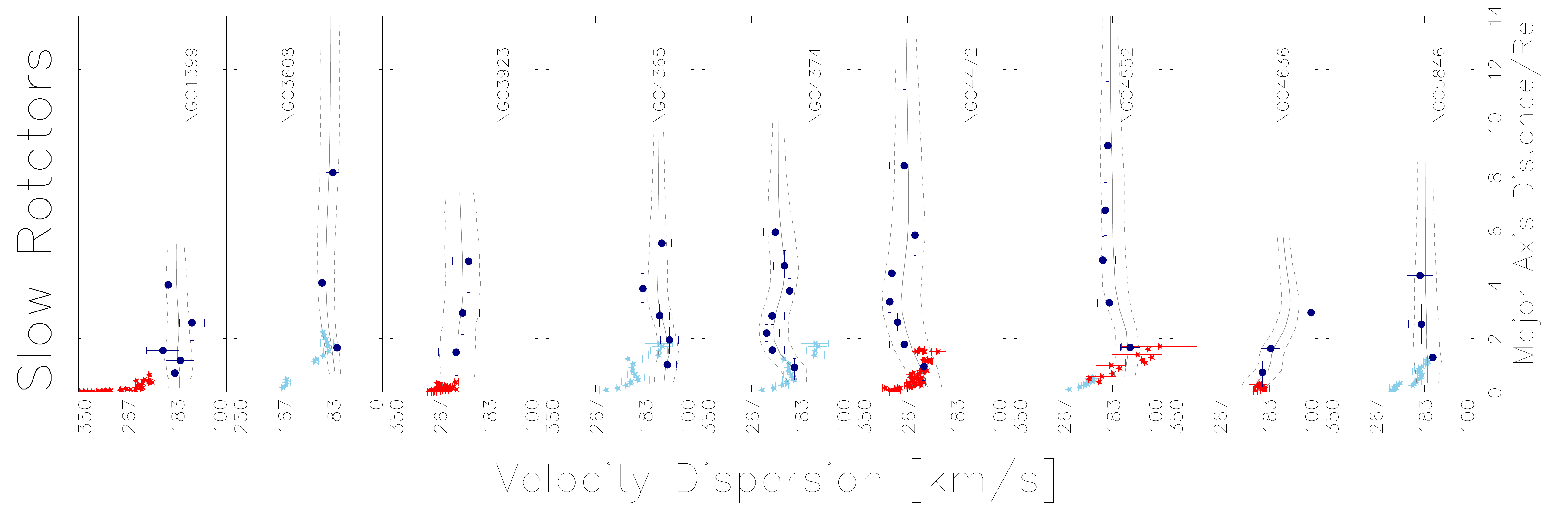}
   \caption{Azimuthally averaged velocity dispersion profiles as
     functions of the major axis distance in units of $\re$ for SRs
     and FRs (full circles). The gray solid lines represent
     the profiles from the interpolated velocity fields with their 
     errors (dashed lines, see text). The stars show
     dispersion from integrated light: when available we plot the
     kinemetric analysis of \citet{2016MNRAS.457..147F} on the
     SLUGGS+$\mathrm{ATLAS^{3D}}$ data in elliptical bins (light blue
     stars); for NGC 3384, NGC 3489, NGC 4339, NGC 4472, NGC 4552, and
     NGC 4636, we show azimuthally averaged profiles from
     $\mathrm{ATLAS^{3D}}$ data
     \citep[][]{2011MNRAS.413..813C,2011MNRAS.414..888E}; for the
     other galaxies we show data from long slit spectroscopy along
     $\PAphot$ (red stars, references in table
     \ref{tab:galaxies}). For NGC 4594 we also plot the stellar
     velocity dispersion profile in a slit parallel to the major axis
     but 30 arcsec offset from the center (purple stars).}
   \label{fig:sigma_profiles}  
\end{center}
\end{figure*}

Figure \ref{fig:sigma_profiles} shows the velocity dispersion
profiles, azimuthally averaged in radial bins. These have been
calculated using two different methods. In the first we used the
interpolated velocity dispersion field $\tilde\sigma(x,y)$ in
elliptical annuli of growing radius, with position angle and
ellipticity as in table \ref{tab:galaxies}. The values shown in
  the plots (solid lines) are averages over each elliptical annulus,
  and the errors (dotted lines) are taken conservatively as the means
  of the Monte Carlo simulated errors in the elliptical annulus
  (section \ref{subsubsec:Errors_on_the_fitted_parameters}).  The
second method is binning the measured radial velocities $\vmeas_i$ of
the PNe in the radial bins built as described in section
\ref{subsec:fit_rotation_model}. The PN catalogs are folded by point
symmetry and the dispersion $\sigma_\mathrm{bin}$ with respect to the
weighted mean velocity is computed in each bin. The weights are
computed from the measurement errors. The errors on the dispersion are
given by the expression: $\Delta \sigma_\mathrm{bin}=
\sigma_\mathrm{bin}/\sqrt{2(N_\mathrm{bin}-1)}$, where
$N_\mathrm{bin}$ is the number of PNe in each bin. The values and the
trends given by the two methods are generally in good agreement.

The profiles obtained are compared with dispersion profiles from
integrated light (red stars). For the galaxies in common with the
SLUGGS survey, we show the profiles from the kinemetric analysis of
\citet{2016MNRAS.457..147F} on the SLUGGS+$\mathrm{ATLAS^{3D}}$ data
in elliptical radial bins. For NGC 3384, NGC 3489, NGC 4339, NGC 4552,
and NGC 4636, we extracted the azimuthally averaged profiles from the
$\mathrm{ATLAS^{3D}}$ data
\citep[][]{2011MNRAS.413..813C,2011MNRAS.414..888E} in elliptical bins
(geometry in table \ref{tab:galaxies}). For the other galaxies we show
the velocity dispersion along the $\PAphot$ from long slit
spectroscopy (references in table \ref{tab:galaxies}).

Our dispersion profiles generally compare well with the literature in
the regions of overlap (typically $R\lesssim2\re$).  The results are
described in section \ref{sec:results_per_family}, separately for FRs
and SRs.

We tested whether it is possible that the large scale trends in the
velocity dispersion profiles are the result of statistical and
smoothing effects, by using 100 models of the galaxies, built as
described in appendix \ref{subsubsec:statistical_tests}, created with
a constant dispersion profile with radius. The velocity dispersion
profiles are recovered with the same procedure as for the measured PN
sample. We find that although artificial local structures may
sometimes appear in the velocity dispersion maps, they are not such as
to influence the trends with radius of the large scale velocity
dispersion fields, and typically the error bars from Monte Carlo
simulations give a good estimate of the uncertainties.

%

\subsection{Triaxiality}
\label{subsec:triaxiality}

Significant twists of the $\PAkin$, as well as its departures from
$\PAphot$ imply an intrinsic triaxial shape for the system. In an
axisymmetric object both the projected photometric minor axis and the
intrinsic angular momentum are aligned with the symmetry axis of the
system, while in a triaxial galaxy the rotation axis can be in any
direction in the plane containing both the short and long axis.  This
is because in a triaxial potential the main families of stable
circulating orbits are tube orbits which loop around the minor (z-tube
orbits) or the major axis (x-tube orbits) \citep[see
  e.g.][]{1985MNRAS.216..273D, 1987ApJ...321..113S,
    1993ApJ...409..563S}.  The relative number of z- and x-tube
orbits determines the direction of the intrinsic angular
momentum. Thus, depending on the variation of this ratio with radius
we can have that
\begin{itemize} 
\item the measured $\PAkin$ shows a smooth radial twist (e.g. NGC 4552
  in figure \ref{fig:fitted_misalignments}, see section
  \ref{subsubsec:triax_twist});
 \item $\PAkin$ has a sudden change in direction (i.e. the galaxy has a kinematically decoupled halo, like NGC 1399, see section \ref{subsubsec:triax_decoupledhalo});
 \item $\PAkin$ has a constant misalignment with respect to the $\PAphot$  (e.g. NGC 3923, see section \ref{subsubsec:triax_constmisal}).
\end{itemize}

Therefore we consider as triaxial galaxies the objects displaying at
least one of these features in their velocity fields with statistical
significance. We do not consider in this analysis galaxies that show significant variation of the kinematic position angle in the last radial bin only, as the geometrical shape of the survey area might prevent an azimuthally complete detection of PNe in the outermost bin. 
The statistical significance is determined by our MC
modeling  as described in appendix
  \ref{subsubsec:statistical_tests}, which gives us the probability
  that a measured property of the smoothed velocity field is obtained
  in sequences of galaxy-specific simulated PN data sets without this
  feature. Table \ref{tab:results} provides a summary of the
results, which are discussed in the following sections.

\subsubsection{Galaxies with radial variation in the kinematic position angle}\label{subsubsec:triax_twist}

 Figure \ref{fig:fitted_misalignments} shows that the fitted $\PAkin$
may show a smooth variation with radius. This happens in NGC 3377, NGC
3379, NGC 4494, NGC 4697, NGC 4649, NGC 5128, and NGC 5866, among the FRs, and in NGC 3608, NGC 4472, and NGC 4552 among the SRs.

We tested whether the variation with radius of the $\PAkin$ for a
galaxy is an artifact of the combination of a small number of tracers
and the smoothing procedure. We looked at the fitted $\PAkin$ in 100
models of each galaxy, built as described in appendix
\ref{subsubsec:statistical_tests}, which have by construction the
$\PAkin$ of the mean velocity field aligned with $\PAphot$.  For each
radial bin we computed the probability of obtaining the observed
misalignment $\Psi$ from the distribution of the misalignments in the
models.  We found that the probability of observing a twist of
  $48\pm24$ degrees in models of NGC 3377 is $\sim4\%$; it is
  $\sim3\%$ for the twist of $71\pm25$ degrees in NGC 3379, $\sim2\%$
  for the twist of $20\pm12$ degrees in NGC 3608, and $\sim4\%$ for
  the twist $32\pm19$ degrees in NGC 4494. For the other galaxies,
none of the 100 models produces the observed trends of $\PAkin$.

For NGC 4472 the determination of $\PAkin$ is influenced by the
kinematics of the in-falling satellite UGC 7636. An inspection of its
smoothed velocity field suggests that the main body of the galaxy has
approximately major axis rotation, once the PNe of the satellite are
excluded. Nevertheless we include this galaxy in the sample of
potentially triaxial galaxies, and we refer to
\citet{2018arXiv180503092H} for a more detailed study.

NGC 5128 shows non point-symmetric kinematics, and rotation along both
the photometric major and minor axes. The high number of tracers
available for this galaxy (1222 PNe) makes this kinematic signature
unambiguous.

NGC 1316 shows a small but significant jump of the $\PAkin$ at
$\sim 200-250$ arcsec, that is related to the perturbed kinematics of
this galaxy.

NGC 4697 has a constant $\PAkin$ profile consistent with the 
$\PAphot$, with a sudden localized variation at $\sim 3 \re$. The 
study of \cite{2006AJ....131..837S} on the same PN dataset showed 
evidence for a secondary PN population in this galaxy that is not 
in dynamical equilibrium with the main population, and which has not 
been excluded in this analysis. The presence of this population 
does not determine any significant deviation from point symmetry 
of its smoothed velocity field. However we refrain 
to include this galaxy in the sample of galaxies with triaxial halo.

Therefore, excluding NGC 1316 and NGC 4697 that have local and irregular variations
of \PAkin, there are 9 galaxies in the sample showing significant
kinematic twist, of which 6 are FRs and 3 SRs.

\subsubsection{Galaxies with kinematically decoupled halos}\label{subsubsec:triax_decoupledhalo}

Galaxies with kinematically decoupled halos are galaxies whose
outskirts rotate about a different direction than their inner regions,
hence $\PAkin$ shows a step function with radius. NGC 1399 and NGC
4365 both show this feature beyond $2\re$.

NGC 1399 is found to be slowly rotating around its $\PAphot$
(i.e. $\Psi\sim90$ degrees) at $\sim30$ \kms inside 1\re, in very good
agreement with the integral field spectroscopic data of
\citet{2014MNRAS.441..274S}.

The halo $\PAkin$, by contrast, is almost aligned with the $\PAphot$
(i.e. $\Psi=0$) degrees. We studied whether such a misalignment is an
artifact of our procedure, using models that mimic the inner
kinematics, i.e. with $\PAkin$ aligned with the photometric minor axis
. The probability of measuring a misalignment of the halo similar to
the observed one is $2\%$.

The $\PAkin$ of NGC 4365 is ill-constrained in the innermost regions
where the kinematics is compatible with no rotation. At the center we
do not recover the rolling about the minor axis visible in velocity
fields from absorption line data
\citep{2011MNRAS.414..888E,2014ApJ...791...80A}, because of smoothing
over the inner velocity gradients. In these regions the bright
background of the galaxy hampers the detection of PNe, and the
resulting low number of tracers combined with the low $V/\sigma$ leads
to heavily smoothed velocities.  We do detect a significant outer
($R\gtrsim3 \re$) rotation of $\sim 50$ \kms along $\PAphot$
($\Psi\sim0$ degrees), misaligned with respect to the inner kinematics
reported in the literature. So we built mock models as described in
appendix \ref{subsubsec:statistical_tests} but with $\PAkin$ given by
IFS data up to 1 \re (references are given in table
\ref{tab:galaxies}). We found that none of the models displays the
observed step function in the $\PAkin$ values.

We therefore conclude that the signature of a kinematically decoupled halo has a high probability to be real in both galaxies.

\subsubsection{Galaxies with constant offset between photometric and kinematic major axis}\label{subsubsec:triax_constmisal}

The galaxies showing an approximately constant misalignment of the
$\PAkin$ with respect to the $\PAphot$ are NGC 0821, NGC 1344, NGC 3115, NGC
4473, and NGC 4742 among the FRs, and NGC 3923, NGC 4374,
NGC 4636, and NGC 5846 among the SRs (see figure
\ref{fig:fitted_misalignments}).

In these cases we can define the $\PAkin$ using all the PNe, without
radial binning. The derived quantities can be compared to the $\PAkin$
measured in mock models built as described in appendix
\ref{subsubsec:statistical_tests}.

For most of the listed galaxies none of the models reproduces the
observed misalignments. NGC 0821 has a misalignment of $50$ degrees
with $\sim4\%$ probability in the Monte Carlo models, while the
misalignment of $22$ degrees of NGC 4742 has $1\%$
probability. Because of the small number of tracers and the low
$V/\sigma$ ratio this probability is higher for NGC 5846 ($9\%$).

To these objects we add NGC 4473 whose $\PAkin$ is not well determined
from the PN kinematics because of the very low $V/\sigma$ ratio in its halo, but \citet{2013MNRAS.435.3587F}
already detected a significant rotation along the minor axis using
absorption line data, showing this object to be triaxial.

The PN velocity field of NGC 3115 shows a constant misalignment of $\sim 10$ degrees with respect to $\PAphot$. This misalignment is probably related to perturbations at the interface between the disk and the spheroid, visible as deviations from axisymmetry in the photometry of the disk component \citep[][see also discussion in appendix \ref{sec:notes_on_single_galaxies}]{1987AJ.....94.1519C}. NGC 3115 is a complicated case, and is not included in the sample of galaxies with triaxial halo. 

Therefore in the ePN.S sample a total of 8 galaxies, 4 FRs and 4 SRs, show a significant constant misalignment of $\PAkin$ with
\PAphot.

\subsubsection{Summary}

We conclude that a total of 19 galaxies ($\sim 60$\%) of the ePN.S
sample show smoothed velocity fields that reveal their non-axisymmetric
nature. 9 objects (6 FRs and 3 SRs) have significant
kinematic twists, and 8 (4 FRs and 4 SRs) show a
significant constant misalignment of $\PAkin$ with \PAphot. In
addition two SRs have a kinematically decoupled halo.

The observed features are more than 2 sigma significant for most of
the cases (1.7 sigma for NGC 5846), and they are not effects of the
folding operation on the catalogs nor of the smoothing procedure.

All in all, we found that all the SRs and 10 out of 24 FRs 
show indications of intrinsic triaxial morphology in the PN
kinematics. We will discuss the signature of triaxiality in the
photometry in section
\ref{subsec:DISCUSSION_signatures_of_triaxiality_in_photometry}.

\section{Results per family}
\label{sec:results_per_family}

\subsection{Slow rotators}
\label{subsec:velocity_fields_SR}

In the sample of 33 galaxies 9 are SRs. Figure
\ref{fig:fitted_velocity_amplitudes} shows that they typically display
some more pronounced rotation at large radii when compared to rotation
in their central regions as measured from absorption line
spectroscopy. The PN velocity fields show gently increasing profiles
for the $V_\mathrm{rot}$ amplitude which, eventually, flatten around
$\sim 50$ \kms.  Twists or misalignments of the $\PAkin$ are commonly
observed, so that all the SRs show signatures of a triaxial
halo (see figure \ref{fig:fitted_misalignments}).  In particular, we
found that the halos of NGC 1399 and NGC 4365 are kinematically
decoupled with respect to the innermost regions as mapped by
\cite{2014MNRAS.441..274S} and \cite{2014ApJ...791...80A}.  NGC 4472
has a non point-symmetric velocity field, as a result of a recent
accretion event.  The complicated kinematics of the SRs is
also reflected in the amplitudes of the third order harmonics, which
describe the presence of additional kinematic components, and twists
of the $\PAkin$.

The velocity dispersion profiles, shown in figure
\ref{fig:sigma_profiles}, are generally flat in the halo. Some
galaxies (e.g. NGC 1399, NGC 3608) reach such a constant value around
$\sim 1 - 2 \re$. Others (e.g. NGC 4374, NGC 4552) flatten only beyond
$\sim 4-5 \re$, after a small increase. NGC 4636 is the only SR showing a falling profile.

\subsection{Fast rotators}
\label{subsec:velocity_fields_FR}

In the ePN.S sample 24 galaxies are classified as FRs (table
\ref{tab:results}).  Figures \ref{fig:fitted_velocity_amplitudes} and
\ref{fig:fitted_misalignments} show that the majority of the objects
have regular rotation along the photometric major axis. The comparison
between inner and outer parts reveals that the rotation amplitudes
$V_{\mathrm{rot}} $ show declining trends with more or less steep
gradients for 14 out of 24 FRs: some galaxies show very
small or no rotation in the outskirts (e.g. NGC 4278 and NGC 4473),
while others reach a minimum, after which their rotation increases
(e.g. NGC 4494 or NGC 4697).  Among the remaining galaxies, 3 have
fairly constant profiles (NGC 2768, NGC 4564, and NGC 5866), and 3
show increasing rotation (NGC 3384, NGC 4742, and NGC 7457). For NGC
0584, NGC 2974, NGC 3489 and NGC 4339, the limited number of tracers
leads to heavy smoothing, allowing only an estimate of a lower
limit for the rotation in the halo.

The $\PAkin$ is well aligned with the photometric major axis in the
majority of cases, but 10 out of 24 galaxies display a kinematic
twist (NGC 3377, NGC 3379, NGC 4494, NGC 4649, NGC 5128,
and NGC 5866) or a constant misalignment with $\PAphot$ (i.e. NGC
0821, NGC 1344, NGC 4473, and NGC 4742).  The smoothed velocity fields
of four galaxies (NGC 3379, NGC 4649, NGC 5128, and NGC
4494, see figures in appendix \ref{sec:figures_vel_fields}) show 
  indications of additional components rotating along the minor axis
of the system, while NGC 1344 has prolate rotation.  All these
features are generally interpreted as evidences of triaxiality of the
systems (see section \ref{subsec:triaxiality}) and $40\%$ of the
fast rotating galaxies of the sample display them in the halo.

The azimuthally averaged velocity dispersion profiles (figure
\ref{fig:sigma_profiles}) are found to be either constant (e.g. NGC
3377), or decreasing with radius. Some profiles decline gently
(e.g. NGC 1344), while others decrease steeply in the halo (e.g. NGC
3379).  This diversity between flat and falling profiles is also
reflected in the variety of the $V_\mathrm{rms}$ profiles, as already
observed by \citet{2009MNRAS.394.1249C}, and is the result of
differences in the mass distributions
\citep{2001AJ....121.1936G,2003Sci...301.1696R,2012ApJ...748....2D,2013MNRAS.431.3570M,2015ApJ...804L..21C,2016MNRAS.460.3838A,2018MNRAS.473.5446V}
or the presence of radial anisotropy in the orbits
\citep[][]{1998MNRAS.295..197G,2009MNRAS.395...76D,2009MNRAS.393..329N},
which may contribute to a lower projected velocity dispersion.  For
some lenticular galaxies (e.g. NGC 1023 or NGC 3115), the
two-dimensional velocity dispersion maps (see figures in appendix
\ref{sec:figures_vel_fields}) reveal the presence of the colder disk
along its major axis, while the dispersion is higher along the minor
axes. For NGC 1023 the presence of the disk is evident also in the
azimuthally averaged velocity dispersion profile, which increases with
radius as the contribution of the disk to the light decreases.

Among the fast rotating galaxies there are two mergers, NGC 1316 and
NGC 5128, whose velocity fields are highly disturbed by the recent
accretion events and are not very well described by point-symmetric
rotation models (see appendix \ref{sec:notes_on_single_galaxies} for
more details).

The comparison of the PN kinematics with integrated light data shows a
general good agreement in the overlapping regions, confirming once
again that the PNe are reliable tracers of the kinematics of the
parent stellar population.  The tension with the stellar kinematics in
cases of high rotation and low dispersion (see e.g. NGC 1023, NGC
2768, NGC 4594) is primarily related to the presence of a known near
edge-on very flat disk that dominates the major axis stellar kinematics but
not that of the PNe. Because our velocity fields are the results of an
averaging operation that does not distinguish the PNe from the disk
from those belonging to the spheroid, the velocity gradients are
underestimated.  Likewise the discrepancy between the integrated light
data of NGC 4494 and its PNe could be explained by a face on
disk fading into the spheroid at radii beyond the coverage of the
stellar kinematics, combined with the lower spatial resolution of the
PN smoothed velocity field. 

\begin{figure*}[!ht]
\begin{center}

  \includegraphics[width=2\columnwidth]{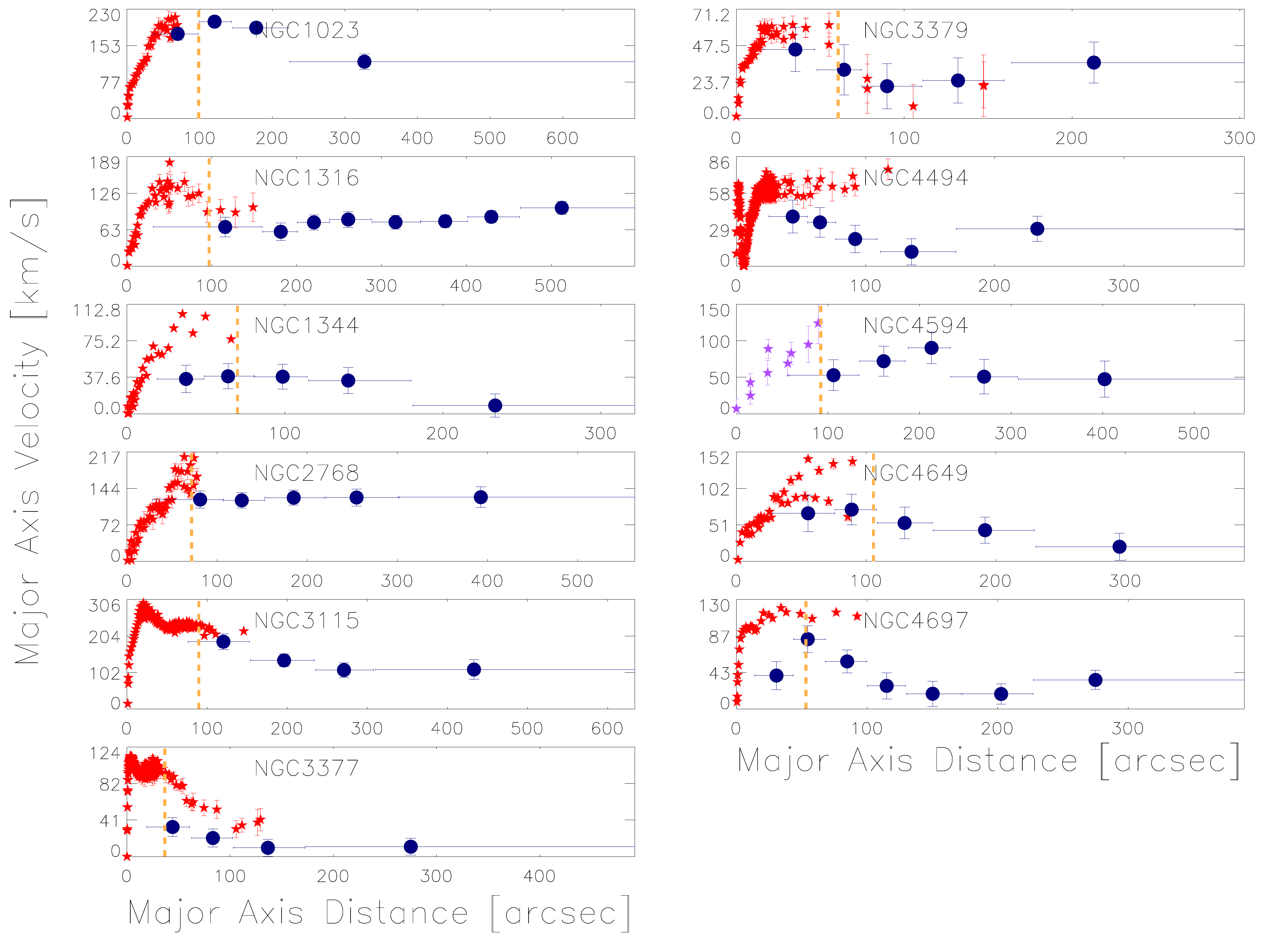}
    
  \caption{Rotation velocity profiles for the FRs with
    highest number of tracers along the major axis.  The full circles
    show the fitted amplitude $V_{\mathrm{maj.ax.}}$ on the PN
    velocity fields (see equation \ref{eq:single_cosine}).  The red stars are stellar kinematics along the
    photometric major axis from the literature (references in table
    \ref{tab:galaxies}). The PN kinematics of NGC 4594 is compared
    with the stellar kinematics from a slit parallel to the major
    axis, offset by 30 arcsec (purple stars). The orange vertical
    lines mark the sizes of the photometric disks (see text). NGC 4494
    shows a decrease in $V_{\mathrm{maj.ax.}}$ but there is no
    photometric evidence for a stellar disk. When the disk fades in
    the slowly rotating spheroid, the rotation velocity decreases. In
    addition, the PN rotation velocity may be lower than the values
    from absorption line spectroscopy at the same radii because of the
    smoothing that averages disk and spheroid PNe for near edge-on
    disks.}
\label{fig:PD+FR_literature+disk}
 
\end{center}
\end{figure*}

  In addition, if the disk is obscured by dust, the
  fraction of observed disk PNe is reduced, causing a drop in the
  measured rotation and a higher dispersion. An example is the case of
  NGC 4594. This galaxy has a dusty disk which affects the
  detections of PNe in that region, leading to heavily absorbed light
  in this component and hence inhibit the PNe detections in that
  region where the rotation is highest. The rotational velocity and
  the dispersion profiles from PNe agree well with slit data along a
  direction slightly offset ($30$ arcsec) from the disk plane (purple
  stars in figures \ref{fig:fitted_velocity_amplitudes},
  \ref{fig:sigma_profiles} and \ref{fig:PD+FR_literature+disk}), but
  are offset with respect to the major axis profiles. With this in
  mind, we shall consider the smoothed velocity fields as giving a
  global description of the halo kinematics, but not of their
  small-scale spatial structures, unless the number of tracers is very
  large (as, for example, in the case of NGC 1316).

\subsubsection{Embedded disks in fast rotators}
\label{subsubsec:embedded_disks}

The observed negative gradients of $V_{\mathrm{rot}} (R)$ observed for
many FRs at large radii has been interpreted as signatures
of a rotating disk component embedded in a dispersion dominated
spheroid \citep{2009MNRAS.394.1249C, 2014ApJ...791...80A}.  Using a
photometric disk-spheroid decomposition and maximum-likelihood fit,
\citet{2013MNRAS.432.1010C} reconstructed the kinematics of the disk
and the spheroid separately in 6 of the ePN.S lenticular galaxies (NGC
1023, NGC 2768, NGC 3115, NGC 3384, NGC 3489 and NGC 7457).
\citet{2014ApJ...791...80A} reproduced the rotation profile of the
E5-6 galaxy NGC 3377 also performing a disk-bulge decomposition.
These works showed that the variation in rotation reflects the
transition between disk and bulge dominance in light, and their
different spatial contribution in each galaxy explains the variety in
the observed $\lambda_R$ profiles (the detailed study of the
$\lambda_R$ profiles for the ePN.S sample is the subject of a separate
paper, Coccato et al. in prep.). In the following we use the term
  "disk component" for a highly flattened, but definitely
  three-dimensional, oblate rotating structure (see also section
  \ref{subsec:DISCUSSION_signatures_of_triaxiality_in_photometry}).

Here we can verify the interpretation for the negative gradients
  of $V_\mathrm{rot}(R)$ by estimating and comparing the "size" of
the disk components given by the photometry, and the radius
at which we observe the decrease in rotation.  Figure
\ref{fig:PD+FR_literature+disk} shows a subsample of the galaxies with
the highest number of tracers, in order to have the best possible
statistics. Most of them show a decreasing rotation amplitude. We
fitted their radially binned smoothed velocity fields with a cosine,
whose position angle is aligned with the photometric major axis
\begin{equation}
     \tilde{V}(R,\phi) = V_\mathrm{maj.ax.}(R) \cos(\phi-\PAphot)
    \label{eq:single_cosine}
\end{equation}
in order to extract the velocity profiles along the major axis. The
fitted amplitudes are shown in figure
\ref{fig:PD+FR_literature+disk}. Overlaid in red is the stellar
kinematics along the photometric major axis from literature slit data
(references in table \ref{tab:galaxies}).  The orange vertical lines
indicate the characteristic scale of the disks. For all the galaxies
shown, the quantity plotted is the disk half light radius:
$R_{1/2}\simeq1.67 R_h$. $R_h$ is the disk scale length from an
exponential fit of the disk component \citep[$R_h$ from][for figures
\ref{fig:fitted_velocity_amplitudes} and
\ref{fig:PD+FR_literature+disk}]{1986AJ.....91..777B,1998A&AS..131..265S,
  1999astro.ph..6378P, 2010MNRAS.405.1089L,2011MNRAS.418L...6B,
  2013MNRAS.432.1768K,2013MNRAS.432.1010C}.

We can see that the radial distance at which the decrease in the
rotation occurs is consistent with the size of the disk. Therefore we
interpret this behavior as the transition between a flat component
that rotates fast and a dispersion dominated rounder spheroid. For NGC
4494 there is no photometric evidence for a disk component. In this
case the stellar disk component might be absent or
very faint if the galaxy is seen nearly face on ($\epsilon\sim0.14$
for NGC 4494).  In the other FRs that are not shown in
figure \ref{fig:PD+FR_literature+disk}, but displaying a drop in the amplitude
of rotation, the comparison between slow rotating PN system and rapid
rotation in the absorption line kinematics, suggests similar
transition at smaller radii.

Figure \ref{fig:fitted_velocity_amplitudes} shows that the three
galaxies with approximately constant $V_\mathrm{rot}$ profiles (NGC
2768, NGC 4564, and NGC 5866) do actually show a small decrease in
rotation in correspondence to $R_{1/2}$, when compared with the values
from absorption line spectroscopy. At larger radii their fast-rotating
spheroidal halo \citep[NGC 2768,][]{2013MNRAS.432.1010C}, or
alternatively an unidentified outer disk component, dominates the
kinematics. Among the remaining three galaxies in figure
\ref{fig:fitted_velocity_amplitudes} with increasing $V_\mathrm{rot}$
profiles, two of them (NGC 3384 and NGC 7457) have an extended disk
component, to which most of the PNe belong
\citep{2013MNRAS.432.1010C}. NGC 4742, by contrast, does not have any
photometric evidence for a disk component, hence the high rotation at
large radii is likely associated with the spheroid.

\subsection{Summary}

\begin{figure}[h]
\begin{center}

  \includegraphics[width=\columnwidth]{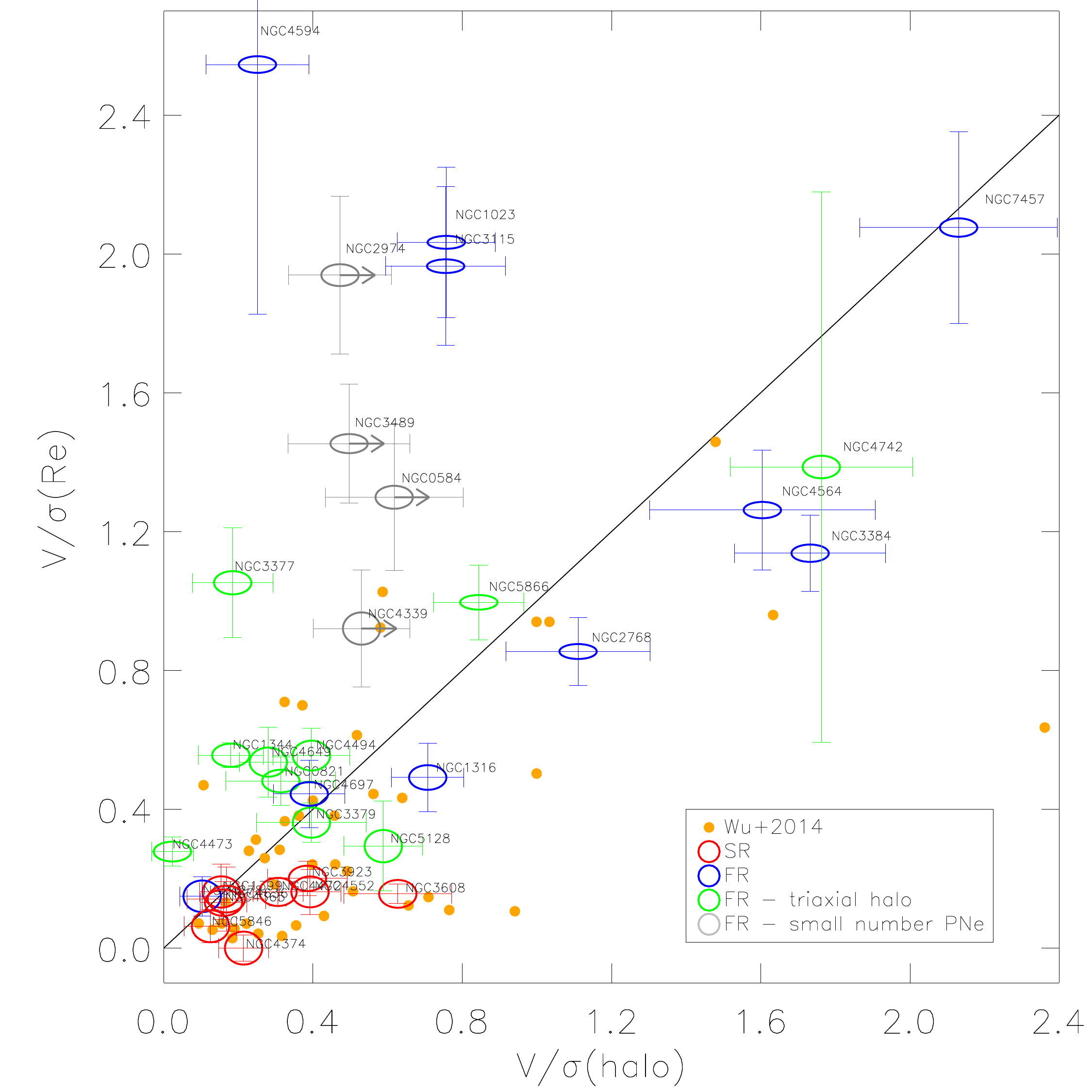}
    
  \caption{$V/\sigma(\re)$ from absorption line data compared with
    $V/\sigma(\mathrm{halo})$ from PN data. References for the
    absorption line data are in table \ref{tab:galaxies}; note that
    for NGC 1399 and NGC 3923 data are available up to $\re/2$ and
    $\re/4$, respectively. The flattening of the ellipses used to plot
    the galaxies correspond to the $\epsilon$ values of table
    \ref{tab:galaxies}. FRs and SRs are shown with
    different colors, as are the FR with triaxial halos. The gray open
    ellipses represent the galaxies (all FRs) with fewer
    tracers, for which our analysis provides a lower limit for
    $V/\sigma(\mathrm{halo})$. The solid line traces equal values of
    $V/\sigma(\re)$ and $V/\sigma(\mathrm{halo})$. The orange circles
    shows the $V/\sigma$ of simulated galaxies from
    \cite{2014MNRAS.438.2701W}, divided by a factor 0.57 to rescale
    their two dimensional flux-weighted measurement to one-dimensional
    quantities \citep{2007MNRAS.379..418C}. While the SRs show increased rotation in their halos, the majority of the FRs show a drop in rotation at large radii.} 
\label{fig:Vsigma_haloVScenter}
 
\end{center}
\end{figure}

Our results show that the kinematics of ETGs at large radii can be substantially different from that in the inner regions. For the SRs we observe a growth in the amplitude of rotation. For the FRs, this variation in the kinematics manifests as a decrease in the amplitude of rotation or a twist in the $\PAkin$. We interpreted this behavior as the transition from the inner disk component into the spheroidal halo, which is dispersion dominated and might deviate from axisymmetry.  

Figure \ref{fig:Vsigma_haloVScenter} illustrates this conclusion by comparing the $V/\sigma$ ratio in the halo and in the inner regions (at 1\re). The $V/\sigma(R_e)$ values are derived by interpolating the $V_\mathrm{rot}$ and $\sigma$ profiles from integrated light (shown in figures \ref{fig:fitted_velocity_amplitudes} and \ref{fig:sigma_profiles}, see references in table \ref{tab:galaxies}) at $R=\re$, while $V/\sigma(\mathrm{halo})$ is the ratio of $V_\mathrm{rot}$ and $\sigma$ estimated in the outermost radial bin of the PN velocity fields (see table \ref{tab:galaxies} for the outermost mean radius ).

All the SRs fall below the 1:1 line, showing higher
rotational support at large radii. The spread of the FRs in
the diagram reflects their different intrinsic structure and
kinematics. The halos of most FRs have $V/\sigma$ ratio
similar to the $V/\sigma(\mathrm{halo})$ of SRs.  Among
these the scatter in $V/\sigma(\re)$ is probably driven by the
presence of a more or less prominent disk component seen at
different inclinations, and embedded into the dispersion-dominated
spheroid. The flattening of the ellipses in figure
\ref{fig:Vsigma_haloVScenter}, in fact, shows that galaxies with
higher $\epsilon$ also display higher $V/\sigma(\re)$.

A second group of FRs with high rotational support in the
halo populate the diagram on the right of the 1:1 line. These galaxies
are either dominated by disk rotation at all radii (NGC 3384 and NGC
7457), or have a rapidly rotating spheroid \citep[NGC
2768,][]{2013MNRAS.432.1010C}, or either of these (NGC 4564, NGC
4742).

The FRs with triaxial halos typically show equal values of
$V/\sigma(\re)$ and $V/\sigma(\mathrm{halo})$, spanning all values in
$V/\sigma(\mathrm{halo})$.  It is possible that the group of galaxies
near the 1:1 line but without signature of triaxiality in the current
data could show such signatures with higher resolution in the
kinematics. Therefore we cannot presently determine whether these last
two subset of FRs are different or not.

The orange full circles in the $V/\sigma$ plot show the inner-halo
kinematics from \cite{2014MNRAS.438.2701W}, who studied the kinematics
of ETGs in cosmological zoom simulations \citep{2010ApJ...725.2312O,
  2014MNRAS.444.3357N} out to $5\re$. Their $V/\sigma(\re)$ and
$V/\sigma(5\re)$ were divided by a factor 0.57 to rescale their two
dimensional flux-weighted measurements to one-dimensional quantities
\citep{2007MNRAS.379..418C}. The comparison with the observations
shows that these simulations do not adequately reproduce the observed
properties of fast rotating galaxies, which span a much wider spectrum
of kinematic properties at large radii.

\section{Discussion}
\label{sec:discussion}

\subsection{Halo rotation versus central rotation}\label{subsec:DISCUSSION_halo_vs_central_rotation}

The PN velocity fields show that ETGs may have very different kinematics at large radii compared to
their inner regions.  We found that SRs typically increase their rotation in the halo, while most of
the FRs display a decrease of $V_\mathrm{rot}$ towards large radii.

The observed variety of radial trends for the rotation amplitudes is consistent with the different
shapes of the $\lambda_R$ parameter profiles observed in the smaller sample of
\citet{2009MNRAS.394.1249C}, and confirmed by \citet{2014ApJ...791...80A} and
\citet{2016MNRAS.457..147F}. The study of the $\lambda_R$ profiles for the current sample of
galaxies is the subject of a future paper (Coccato et al. in prep.).

The existence of these radial trends was recently questioned by \citet{2014ApJ...786...23R} and
\citet{2017MNRAS.471.4005B}, who found no evidence for a change in their stellar kinematics beyond
\re. The divergence from our results may arise from the different radial coverage; their kinematics
does not reach to the distances where the drop in rotation typically occurs, beyond $\sim1\re$. An
example is the lenticular galaxy NGC 1023, in common between the two samples. For this galaxy
\citet{2017MNRAS.471.4005B} report a rising $\lambda_R$ profile to almost 2\re, in agreement with
our observation of a decrease in rotation only beyond $\sim3\re$.

The onset of rotation of the SRs at larger radii, often along directions that do not coincide with
the major axis of the galaxy, may trace the accreted stellar component, which maintains a
memory of the orbital angular momentum of the accreted progenitors.

\subsection{Fast rotators with disks embedded in slowly rotating halos} \label{subsec:DISCUSSION_fast_rotators_with_disks_embedded}

The sharp drop in angular momentum in some FRs has been interpreted as the fading of an embedded
disk structure of in-situ stars in a dispersion dominated spheroidal halo
\citep[][]{2009MNRAS.394.1249C,2014ApJ...791...80A}. We qualitatively verified this scenario in
section \ref{subsubsec:embedded_disks}, where we observed that the radius at which the rotation
along the major axis drops is consistent with the half light radius of the inner disk component.

Our results suggest that FRs contain a more or less extended rotating disk component, embedded in a
more or less prominent halo with its own kinematic signature. This is reflected in the distribution
of these objects in the $V/\sigma$ plane (figure \ref{fig:Vsigma_haloVScenter}). We found that while a smaller group of
FRs shows high rotation also in the halo, the
halos of most FRs have rotational support comparable to that of the halos of SRs. This result is corroborated by the observation that the ePN.S FRs tend to become rounder at large radii (see figure \ref{fig:maxtwist_maxell}).

This variety of kinematic properties is consistent with the variety of physical processes that may
drive the evolution of these objects \citep[minor mergers, major mergers that lead to a spin-up of
  the remnant, gas accretion, interactions with the environment, secular
  evolution;][]{2011MNRAS.417..863D,2014MNRAS.444.3357N, 2017MNRAS.468.3883P,2018MNRAS.473.2679S}
while preserving a rotating disk structure at the center.  The comparison of our observations with
simulated data from \cite{2014MNRAS.438.2701W} in figure \ref{fig:Vsigma_haloVScenter} indicates,
however, that those simulations do not yet reproduce the full diversity in the variety of these
processes.

\subsection{Signatures of triaxial halos in the kinematics} \label{subsec:DISCUSSION_signatures_of_triaxial_halos}

In section \ref{subsec:triaxiality} we studied the velocity fields and their kinematic position
angle profiles $\PAkin (R)$, and linked the observed misalignments and twists with the triaxiality
of the halos.  SRs are known to be mildly triaxial in their central regions, while the FRs are
predominantly oblate \citep{2014MNRAS.444.3340W, 2017MNRAS.472..966F}.  Kinematic twists and
misalignments for the FRs are rare and small in the central regions. By sampling the kinematics at
more than twice the distances probed by previous studies, the PNe line-of-sight velocities show that
these features become more frequent and pronounced in the outskirts. Based on these signatures, we
classified the halos of 40\% of the FR galaxies as triaxial, while the remaining ones are still
consistent with being axisymmetric at the resolution of the ePN.S survey. For these objects the
triaxiality of the halos is not ruled out.

On the other hand, it is in principle possible that axisymmetric galaxies with a recent merger might
display kinematic twists and misalignments because of the contribution of the unrelaxed accreted
component to the PN velocity field. However, if the size of the accreted satellite is such as to
significantly contribute to the PN population of the the host halo, and so as to produce features in
the PN velocity field, then the effect is short lived. This makes the probability of occurrence of
such circumstance low. Deviations from a point-symmetric velocity field were found for 5/33
  cases in section~\ref{sec:Point_symmetry_analysis}, probably due to recent accretion/mergers in
  four cases and due to dust in one case. These galaxies are not included in the triaxial
  classification.

An examination of the smoothed velocity fields in appendix \ref{sec:figures_vel_fields} shows that
for some galaxies the trends in $\PAkin$ with radius reflect a two-component velocity field in which
rotation along one axis dominates at intermediate radii and rotation along the other axis dominates
further out. This is the case for NGC 3379 and NGC 4649 among the FRs, and for NGC 4552 and NGC 4636
among the SRs.  The velocity field of NGC 1344, instead, shows that this galaxy is a prolate
rotator. Up to now there are not many observations of this phenomenon \citep[][and references
  therein]{2017A&A...606A..62T}, which is probably related to a recent merger
\citep{2017ApJ...850..144E}, as the shells of this galaxy attest. Our sample contains two other
galaxies (NGC 3923 and NGC 4365) known to have prolate rotation at their centers, but their halos
are decoupled with respect to the inner kinematics.

A previous study of the intrinsic shapes of ETG halos was performed by \citet{2016MNRAS.457..147F}
who, using the observed distribution of ellipticities and misalignments for the SLUGGS galaxies out
to $\sim 2.5\re$ were unable to rule out triaxiality in the outskirts of FRs. The same authors had
previously found strong evidence for minor axis rotation, hence a triaxial halo, in NGC 4473
\citep{2013MNRAS.435.3587F}.  \citet{2018MNRAS.474L..47V} recently observed a strong kinematic twist
beyond 1.5 \re for the massive FR NGC 5626.  The extended PN kinematics allow us to generalize this
result to a significant fraction of the FRs.

\subsection{Signatures of triaxial halos in extended photometry}
\label{subsec:DISCUSSION_signatures_of_triaxiality_in_photometry}

If the kinematic twists seen in the PN velocity fields for a significant fraction of ETGs indeed
indicate triaxial halos in these systems, one expects additional evidence for triaxiality to come
from observations of isophotal twists \citep{1978MNRAS.182..797C, 1981ApJ...248..485L}. Hence in
this section we compare with photometric data from the literature, We do not expect to see a
one-to-one correspondence between variations of $\PAkin$ and variations of $\PAphot$ \citep[see
  also][]{2016MNRAS.457..147F}, as the former are driven by the total potential of the galaxy,
including dark matter, while the latter trace variations of the distribution of the light alone.

The isophotes of ETGs are well-described by ellipses, with deviations from perfect elliptical shape
typically at the $\sim 0.5\%$ level \citep[e.g.][]{1988A&AS...74..385B}. Hence the constant
luminosity surfaces of ETGs can be approximated by ellipsoids \citep[for a
  review see][]{1992BAAS...24..522M}. The projection on the sky of a coaxial triaxial ellipsoidal
distribution of stars can result in twisting isophotes only if their axial ratios varies with radius
\citep{1977ApJ...213..368S, 1980MNRAS.193..885B}.
The effects of triaxiality on the $\PAphot$ profiles are model dependent
\citep[e.g.][]{1980MNRAS.193..885B, 1988MNRAS.231..285F,1990A&A...234..119M}. In general considerable twists
($>10$ degrees) can be produced by moderately triaxial models, but their relative frequency is low,
$\lesssim10\%$, as the effects on the $\PAphot$ are small whenever the systems are viewed close
enough to one of their principal planes, and the magnitude of the twists depends on how much the
axial ratios change.
  
 Figure \ref{fig:photometry} in appendix \ref{sec:appendix_photometry} shows extended ellipticity
 and $\PAphot$ profiles from the literature for both FRs and SRs, for three quarters of the ePN.S
 sample. We chose to consider only galaxies with photometric data reaching at least 4 $\re$. 
Twists of $\PAphot$
 are observed for almost all the galaxies classified as triaxial in section
 \ref{subsec:triaxiality}, and for several other FRs with no evidence for triaxiality at the spatial
 sampling of the PN kinematics. It is evident, however, that the variation of $\PAphot$ measured for
 the FRs is modest ($<10$ degrees) in $12/19 = 63\%$ of the cases.  Thus we verify whether the
 occurrence of small photometric twists is consistent with the presence of a triaxial halo. We
   do this by constructing illustrative, observationally motivated photometric models for triaxial
   FR galaxies, and comparing the statistical distribution of photometric twists in these models
 with that observed for the ePN.S sample of FR galaxies.
  
 The models are built on the kinematic evidence found in the previous sections suggesting that
   centrally flat FRs are embedded in dispersion dominated spheroids, which can be triaxial. The
 intrinsic shape of the central regions of FRs were investigated by \citet[][$\mathrm{ATLAS^{3D}}$
   survey]{2014MNRAS.444.3340W} and \citet[][SAMI survey]{2017MNRAS.472..966F}, who found that these
 galaxies are close to oblate ($p=1$)\footnote{Here $p$ and $q$ are the intrinsic axis ratios, such
   that $1\geq p \geq q \geq 0$.} with mean flattening $q\sim0.3$. SRs are, instead, less flattened
 ($q\sim 0.6$), and mildly triaxial but, because of the small number of these galaxies in the
 $\mathrm{ATLAS^{3D}}$ and SAMI samples, their intrinsic shape distribution is not well constrained.

  In our models, the central regions of the FRs have intrinsic shape in agreement with the
   previous results. In their halos, a significant fraction of the FRs has kinematics similar to the
   SRs (see figure~\ref{fig:Vsigma_haloVScenter}). This is consistent with the expectation that the
   satellites accreted into the halo should be largely uncorrelated with the dynamics of the central galaxy.  Thus for our first two halo models we assume outer halo axis
   ratios like those for the SRs, in which the accreted stars are expected to reach down to
   small radii \citep{2013MNRAS.434.3348C,2016MNRAS.458.2371R}. This choice is supported by the evidence that the
   ellipticity profiles of FRs generally show decreasing trends at large radii (see figures \ref{fig:maxtwist_maxell} and 
   \ref{fig:photometry} in appendix). However,
   some FRs in figure \ref{fig:Vsigma_haloVScenter} rotate rapidly also in their halos; thus we also
   investigate triaxial halos with flatter isophotes.  We parametrize the transition between inner
 and outer regions in different ways. The models are described in detail in appendix
 \ref{sec:photometric_models}. In summary we use:
  
  \begin{itemize}
  \item Model 1: a single component with a S\'ersic $n = 4$ density
    distribution and variable flattening: $q = 0.3$ and $p = 1$ at the
    center, $q=0.6$ and $p=0.9$ in the outskirts. The transition
    between these regions happens at $R = <R_{T}> = 1.8 \re$, where $R_{T}$ is
    the kinematic transition radius defined in section
    \ref{subsec:DISCUSSION_kinematic_transition_radius}. 
  \item Model 2: a two-component model including a \emph{cored} S\'ersic
    $n = 4$  halo with intrinsic constant flattening $q=0.6$ and
    $p=0.9$ plus an embedded spheroidal-exponential disk with flattening
    $q=0.3$ and $p=1$. The scale length of the disk is taken to be such that 
    $1.67 h = <R_{T}>$, as qualitatively observed in section \ref{subsubsec:embedded_disks}. Within $r = R_T$ the total luminosity of the
    disk in this model is $\sim3.5$ times that of the halo.
  \item Model 3: as Model 2, but with a maximally triaxial halo with
    $q=0.6$, $p=0.8$.
  \item Model 4: as Model 2, but with strongly flattened, slightly
    prolate-triaxial halo with $q=0.4$, $p=0.8$.
  \end{itemize}

  Figure \ref{fig:maxtwist_maxell} shows the maximum photometric twist
  and the mean projected ellipticity measured for the subsample of
  the ePN.S FRs with extended photometry (references in
  table \ref{tab:smoothing_parameters}; the profiles are shown in
  appendix \ref{sec:photometric_profiles}). 
  The distribution of the
  galaxies in this diagram is compared with the same quantities
  from the four triaxial models, each observed at 100 random viewing
  angles.  
  We find that:\\
  \\
  1) the fraction of twists larger than 10 degrees is of order 30\% (45\% for Model 3). Large twists occur for viewing angles $\theta \lesssim 50$ degrees for Models 1 and 2, and for $\theta \lesssim 65$ degrees for Models 3 and 4 ($\theta=0$ for face-on view). \\
  \\
  2) The comparison of the model prediction of twist angles with the data
  points for the ePN.S FRs in figure \ref{fig:maxtwist_maxell} shows that most of the model projections have low twists of the same magnitude as the majority (63\%) of the galaxies. Within the statistical uncertainties, the models represent well the locations of the galaxies in figure \ref{fig:maxtwist_maxell}: an average $\sim 35\%$ of the models are above the 10 degrees threshold. 
  From the models distribution $\sim 1-2$ galaxies with photometric twists larger than $30$ degrees and mean ellipticity $\sim 0.1$ would be expected but are not observed. The missing ePN.S FRs in that region of the diagram could be a result of small number statistics, observational bias for near face-on systems, or model details.
  \\
  \\
  3) Since the photometric twists are thus expected to be small in most
   galaxies, the most obvious signature of a change in the intrinsic
   shape is the variation of the projected ellipticity with radius, as
   shown in figure \ref{fig:photometric_models} in the appendix. The
   change in ellipticity observed in the majority of the FRs
   in figure \ref{fig:photometry} (see also appendix
  \ref{sec:photometric_profiles}) independently suggests the fading of
   the central disk component in a more spheroidal halo, as we
   previously inferred from rotation profiles.\\
   
  This analysis based on illustrative oblate-triaxial photometric
  models shows that the presence of small photometric twists, as
  observed for the majority of the ePN.S FRs, is consistent
  with the presence of a triaxial halo. I.e., it is likely that individual oblate-triaxial models for these galaxies can be constructed that
  are consistent with the measured photometric twist angles.

\begin{figure}[ht]
 \begin{minipage}[b]{0.8\linewidth}
 \begin{center}
 \includegraphics[width=1.2\linewidth]{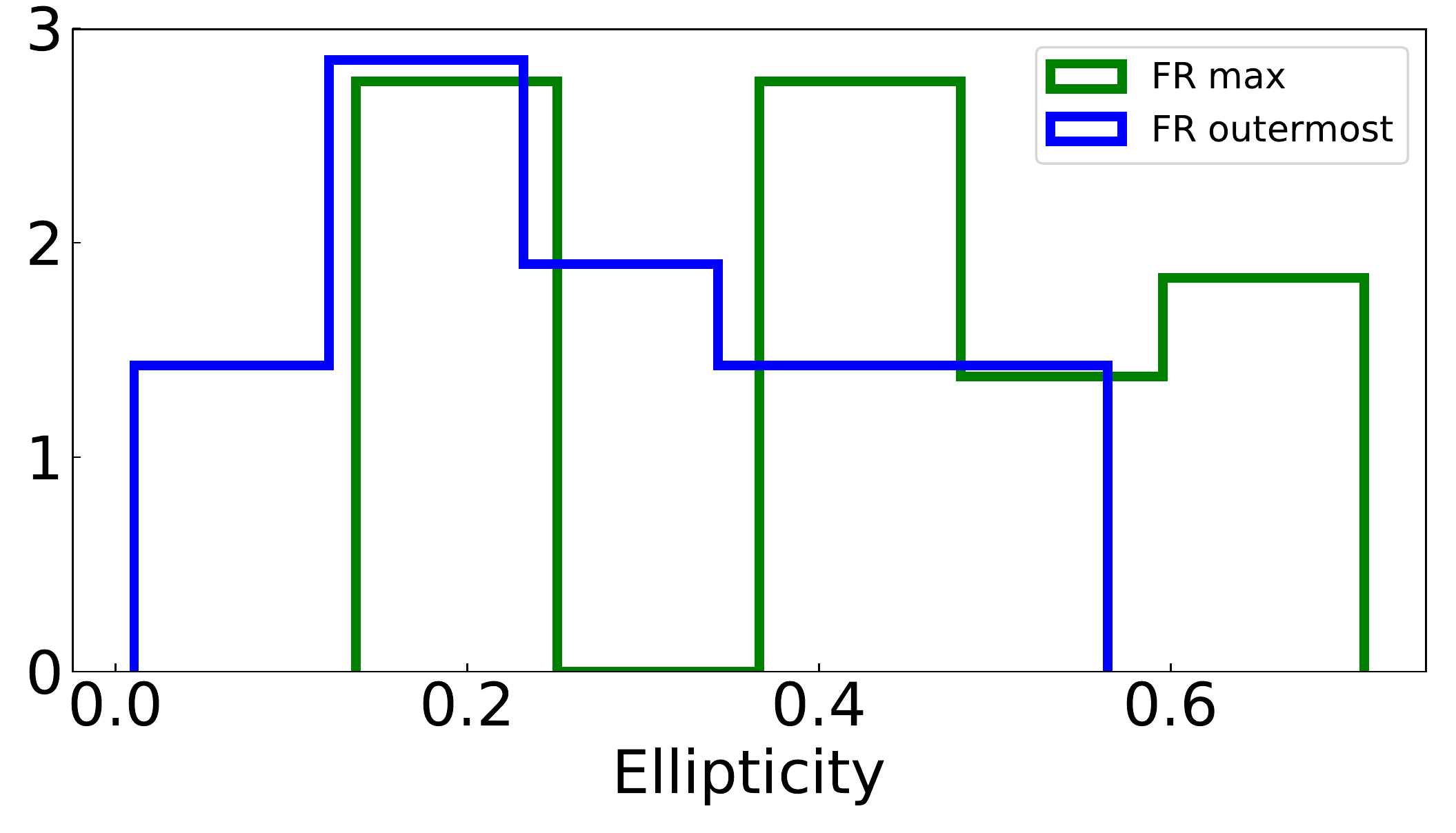}
  \includegraphics[width=1.2\linewidth]{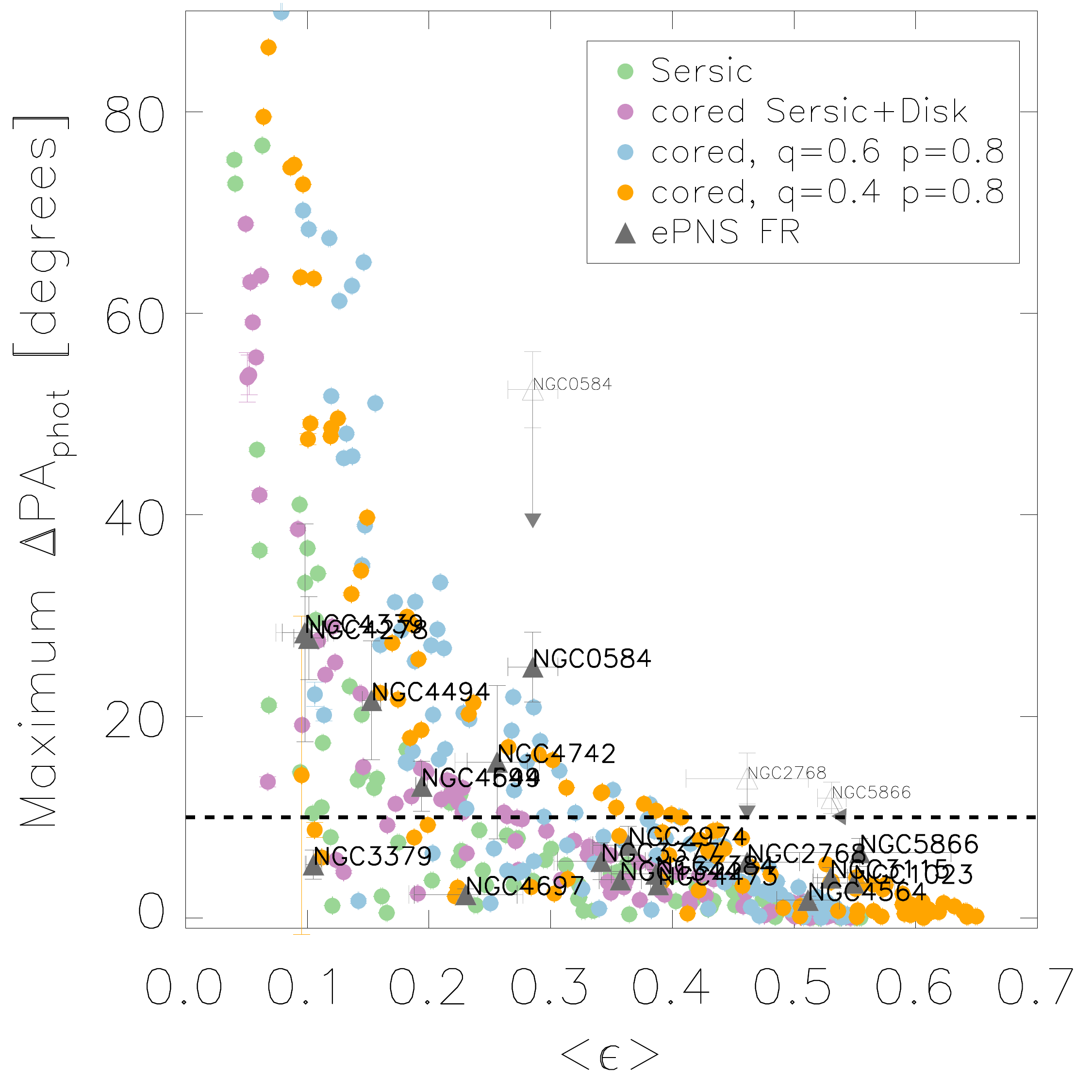}
 \end{center}
  \end{minipage}
 \caption{ Upper panel: Ellipticity distribution for the FRs. The
    histogram of the maximum measured ellipticity at $R>0.5 \re$ is in
   green; the histogram of the outermost measured values is in
   blue. The distributions show that the FRs tend to have
   halos rounder than their central regions. 
 Lower panel: Maximum photometric twist versus mean projected ellipticity for a subsample of the
   ePN.S FRs with extended photometric data (gray triangles; see text and appendix
   \ref{sec:photometric_profiles}, references in table \ref{tab:smoothing_parameters}). The errors reflect the scatter between contiguous data points in the photometric profiles. For the galaxies with indications for ongoing interactions in the outermost photometry (see appendix \ref{sec:photometric_profiles} and \ref{sec:notes_on_single_galaxies}) we  estimated the mean ellipticity and maximum position angle twist both from the complete profiles (open symbols) and considering only the regions with regular photometry (filled symbols). For NGC 1023 and NGC 3384 we show quantities derived excluding the
   regions where the bar dominates. The horizontal dashed line marks $\Delta \PAphot =10$
   degrees. The statistical distribution of points is consistent with simple photometric triaxial
   models, shown each with 100 random projections (solid circles as described in the legend; see
   text and appendix \ref{sec:photometric_models}).}
\label{fig:maxtwist_maxell}
\end{figure}

  We infer that many of the ePN.S FRs show small
  photometric twists because of the highly flattened, oblate shape of
  their central regions and the gradual transition to a triaxial halo. The consequence is a near-alignment within $\sim10$ degrees between the inner
  kinematics and outer photometry of most these galaxies,
  as analogously observed for most of the FRs \citep[][ however, this
  numerical value also depends on sample selection effects, see the
  following section]{2011MNRAS.414.2923K,2015MNRAS.454.2050F}. Hence
  extended kinematic studies that sample the outer regions of these
  FRs are important to unveil the kinematic transition and
  thus the change of the intrinsic shape.

  Both PN kinematics and photometry are therefore consistent with a
  picture in which FRs have a central disk component
  embedded in a spheroidal dispersion dominated component. We found
  this component to be triaxial for at least $40\%$ of the FRs in 
  the ePN.S sample. The presence of photometric twists
  also for other FRs without evidence for a kinematic twist
  in the PN data, suggests that the fraction of galaxies with a
  triaxial halo might be higher. In the following sections we will see
  that the spheroidal component is most prominent in the most massive
  objects, and that the comparison with cosmological simulations
  suggests an ex-situ origin for it.

 \subsection{Comparison with the results of the ATLAS$^{\rm 3D}$ survey} \label{subsec:DISCUSSION_A3D}

 The results discussed in the previous sections challenge a simple
 interpretation of the ATLAS$^{\rm 3D}$ survey results in terms of a
 dichotomy between disk FRs and triaxial SRs
 \citep[see][for a review on the subject]{2016ARA&A..54..597C}.

 The ATLAS$^{\rm 3D}$ (A3D) survey carried out IFS and photometric
 observations of the central regions of a complete sample of 260 ETGs,
 morphologically selected from a volume-limited sample of galaxies
 brighter than $M_{K}<-21.5$ \citep{2011MNRAS.413..813C}. Their sample
 of ETGs spans luminosities $-21.57 \geq M_K \geq -25.78$ and mostly
 contains lenticular galaxies, i.e. galaxies with T type $T>-3.5$:
 only $24\%$ of their ETGs are ellipticals \citep[$T\leq -3.5$,
 ][]{2011MNRAS.413..813C}. The ePN.S sample is, on average, 1 mag more
 luminous in the K band ($-22.38 \geq M_K \geq -26.02$), and the
 majority of the galaxies ($73\%$) are ellipticals, according to their
 T type from HyperLeda (see Arnaboldi et al. in prep. for more details
 on the sample selection). The A3D IFS kinematics covers typically
 radii up to $\sim 1\re$ \citep[][]{2011MNRAS.414..888E}.  The
 fraction of $\re$ covered is actually a decreasing function of galaxy
 luminosity, as $\re$ depends on $M_K$ \citep{2011MNRAS.413..813C},
 hence, for the ePN.S galaxies in common with the A3D sample, the mean
 radial coverage is $0.49 \re$. Finally if one considers that the
 values adopted for $\re$ may underestimate the half light radii for
 the brightest objects (see discussion in section
 \ref{subsec:DISCUSSION_kinematic_transition_radius}), this fraction
 is actually smaller: $0.35R_{\rm halflight}$, where
 $\mathrm{R_{halflight}}$ is defined in section
 \ref{subsec:DISCUSSION_kinematic_transition_radius}.

 \begin{figure}[ht]
 \begin{center}
   \includegraphics[width=\linewidth]{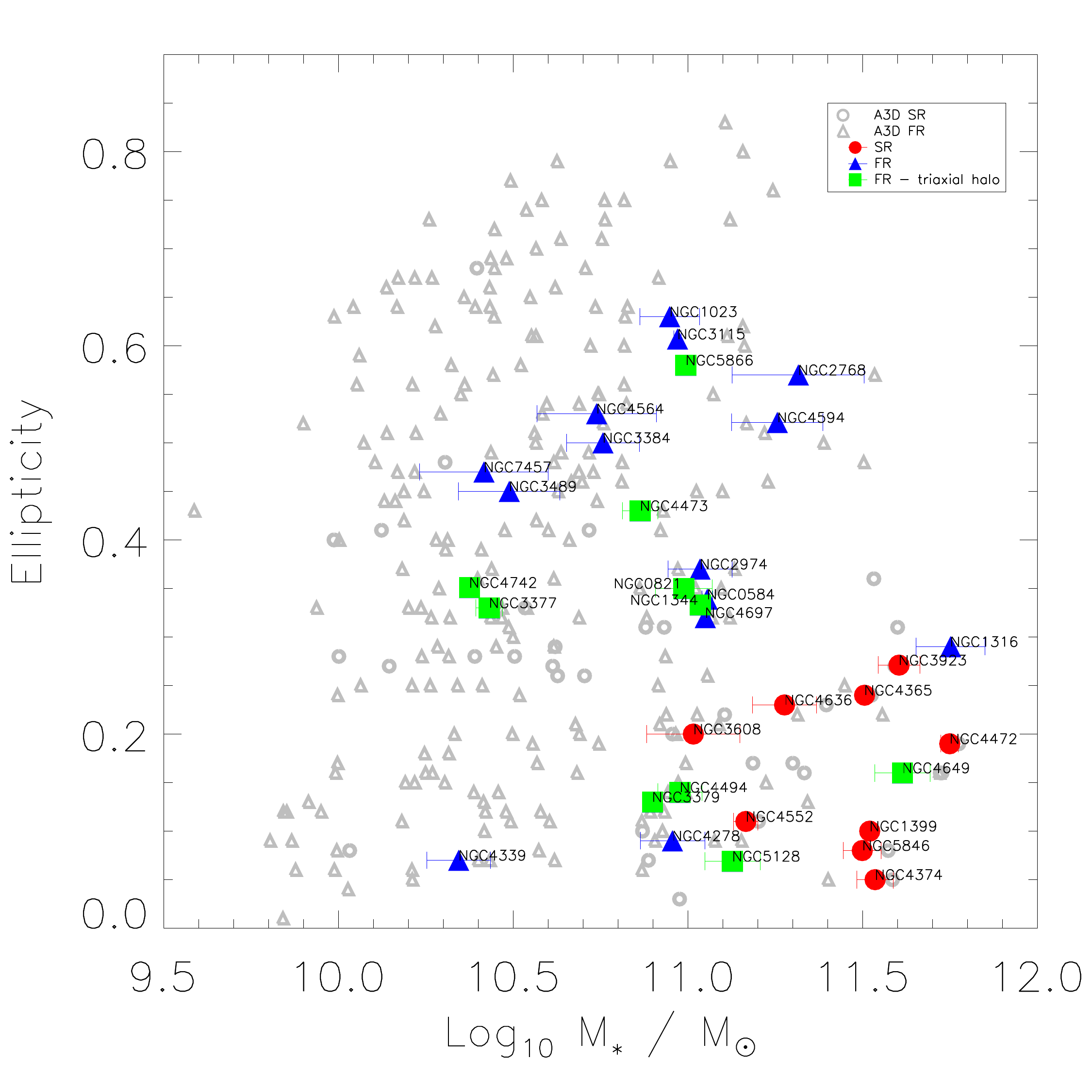} 
   \caption{EPN.S and ATLAS$^{\rm 3D}$ samples compared in the stellar mass versus ellipticity plane. The ATLAS$^{\rm 3D}$ sample of ETGs is
     displayed with gray open symbols (triangles for the FRs, 
     circles for the SRs). For these galaxies we
     show the ellipticity values from \citet{2011MNRAS.414.2923K} and
     $M_{*}=M_{JAM}=\mathrm{(M/L)_{\rm JAM}}\times L_{\rm tot,r}$ from
     \citet{2013MNRAS.432.1709C} as in their equation (28).  The ePN.S
     galaxies are shown with full symbols: red circles for the SRs, blue triangles for the FRs, and green squares
     for the FRs with triaxial halos. For the ePN.S ETGs $M_{*}$ is the
     mean between $\mathrm{M_{\rm JAM}}$ \citep[][when
     available]{2013MNRAS.432.1709C}, and the masses $M_{*1}$ and
     $M_{*2}$ reported in table \ref{tab:smoothing_parameters}. The
     error bars on the mass show the rms scatter between these
     different mass determinations. The values for the ellipticity are
     listed in table \ref{tab:galaxies}.  }
 \label{fig:comparisonsamples}
  
 \end{center}
 \end{figure}

 Figure \ref{fig:comparisonsamples} shows the stellar mass versus
 ellipticity plane for both samples. The ePN.S galaxies are on average
 more massive and less flattened than the A3D ETGs. In particular, the
 ePN.S FRs with triaxial halos are among the most massive A3D FRs. For these galaxies the PNe kinematics generally show fairly
 large twists of the $\PAkin$, as the limited spatial resolution of
 the survey prevents the measurement of small differences in angle. In
 the ePN.S sample 40\% of the FRs show signatures of a
 triaxial halo. The comparison with the photometry in section
 \ref{subsec:DISCUSSION_signatures_of_triaxiality_in_photometry} and
 in appendix \ref{sec:photometric_profiles} suggests that the number
 of FRs with triaxial halos in the ePN.S sample might actually be
 higher.

 Where the data overlap, the PN kinematics agrees well with the IFS
 observations of the central regions, where the rotating disk
 component dominates the kinematics of the FRs.  At larger
 radii both kinematics and photometry suggest that these galaxies show
 a transition to a spheroidal component. If we interpret this
 component with different kinematics and flattening as mainly formed
 by accreted, ex-situ, stars, we find that for the most massive
 galaxies it dominates at smaller fractions of $\re$, in agreement
 with simulations \citep[][see section
 \ref{subsec:DISCUSSION_kinematic_transition_radius}]{2016MNRAS.458.2371R}. This
 dependence on mass, combined with the smaller radial coverage of the
 A3D kinematics, explains the different conclusions drawn by the two
 surveys: low mass FRs are dominated by the disk rotation up
 to large radii; the spheroid is more prominent in the most massive
 FRs, where the accreted envelope reaches smaller fractions
 of $\re$. Even so the radial coverage of the A3D kinematics does not
 reach the transition radius (see section
 \ref{subsec:DISCUSSION_kinematic_transition_radius}), since the
 fraction of $\re$ observed also depends on mass, as discussed above.

 The present work shows that the wider spatial coverage of the
 kinematics by the ePN.S survey delivers new informations about ETGs
 that extend the results of previous surveys. FRs do contain
 inner disk components, but many of them are not simply
 highly flattened oblate objects. The results presented in this paper
 point to a more complex scenario for ETGs than the simple fast/slow
 rotator dichotomy. As predicted by simulations
 \citep{2014MNRAS.444.3357N,2016MNRAS.458.2371R}, FR
 galaxies appear to be more diverse objects than previously thought.

\subsection{Kinematic transition radius} \label{subsec:DISCUSSION_kinematic_transition_radius}

We found that most of the galaxies in the sample show a transition in
their kinematics, marked by a change in the rotation velocity
$V_\mathrm{rot}$ or in the kinematic position angle $\PAkin$. In the
framework of the two-phase formation scenario, we can interpret such
kinematic differences as tracing stellar components with different
origins, such as the in- and the ex-situ components.  From simulations
\citep[][and references therein]{2016MNRAS.458.2371R} it is predicted
that the in-situ stars are concentrated in the central regions of
galaxies, while the accreted stars dominate the halos, and that their
relative contribution is related primarily to the total mass.  Within
this framework we can define a transition radius $R_{T}$ as the
distance at which we observe the described transition in the
kinematics, and compare it with the total stellar mass.

The radial range $R_T\pm\Delta R_T$ is here quantified using\\
(A) in case of a declining $V_\mathrm{rot}$ profile, the interval between the radius at which $V_\mathrm{rot}$ is maximum and the radius at which it decreases by $\sim 50$ \kms \\
(B) in case of an increasing $V_\mathrm{rot}$ profile, the radial range in which $V_\mathrm{rot}$ increases from $\sim 0$ to $\sim 50$ \kms \\
(C) in case of a kinematic twist, the radial range in which $\PAkin$ changes significantly.\\
Table \ref{tab:results} lists the $R_T\pm\Delta R_T$ values measured,
and specifies the criteria (A,B,C) used for deriving it.
$R_T\pm\Delta R_T$ is also plotted in figures
\ref{fig:fitted_velocity_amplitudes} and
\ref{fig:fitted_misalignments} with vertical lines.

For each galaxy we estimate the stellar mass using two different
approaches, designated by $M_{*1}$ and $M_{*2}$, to have also an estimate
of the systematic uncertainties. The values for the masses are listed
in table \ref{tab:smoothing_parameters}, along with the description of
the procedure used for calculating them.

\begin{figure}[h]
\begin{center}

  \includegraphics[width=\columnwidth]{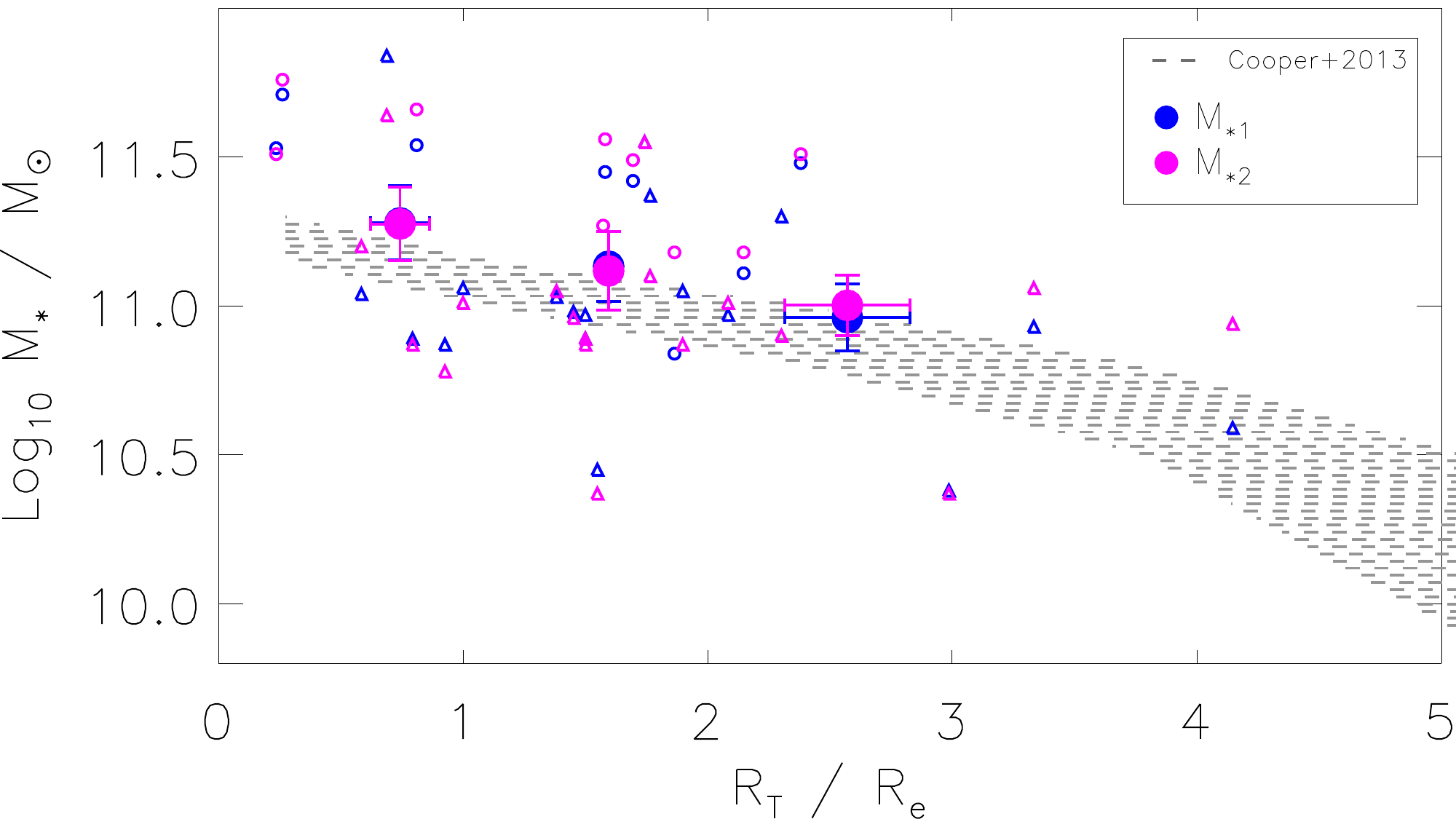}
    
  \caption{Transition radius in units of $\re$ versus total stellar
    mass (open symbols: circles for SRs, triangles for FRs). 
    The full symbols show the same quantities in bins of
    $R_{T}/\re$. Different colors show the results of two different
    procedures for calculating the total stellar mass, see table
    \ref{tab:smoothing_parameters} for details. The shaded region
    shows the corresponding quantities from the simulations of
    \citet{2013MNRAS.434.3348C}, see text.  }
\label{fig:mass_transrad}
 
\end{center}
\end{figure}

Figure \ref{fig:mass_transrad} shows the stellar mass of the galaxies
versus $R_{T}/\re$. The full circles show the same quantities in three
bins of $R_T/\re$; the error bars represent the standard deviation of
the mass and of the $R_T/\re$ ratio in the bin.  A clear correlation
exists between total stellar mass and $R_{T}/\re$, in the sense that
the more massive galaxies tend to have transition radii at smaller
fractions of $\re$.  The shaded region in figure
\ref{fig:mass_transrad} shows the corresponding quantities from the
N-body simulations of \cite{2013MNRAS.434.3348C}. Using their stellar
mass surface density profiles in bins of dark halo virial mass
($M_{200}/M{_\odot}$; their figure 6), we estimated the transition
radius in units of half mass radii as the distance at which the
accreted component overcomes the in-situ component. The total masses
$M_{200}$ are converted to stellar masses via abundance matching,
using the prescription given by \cite{ 2010ApJ...717..379B}.  Both
data and simulation follow the same trend.

\begin{figure}[h]
\begin{center}

  \includegraphics[width=\columnwidth]{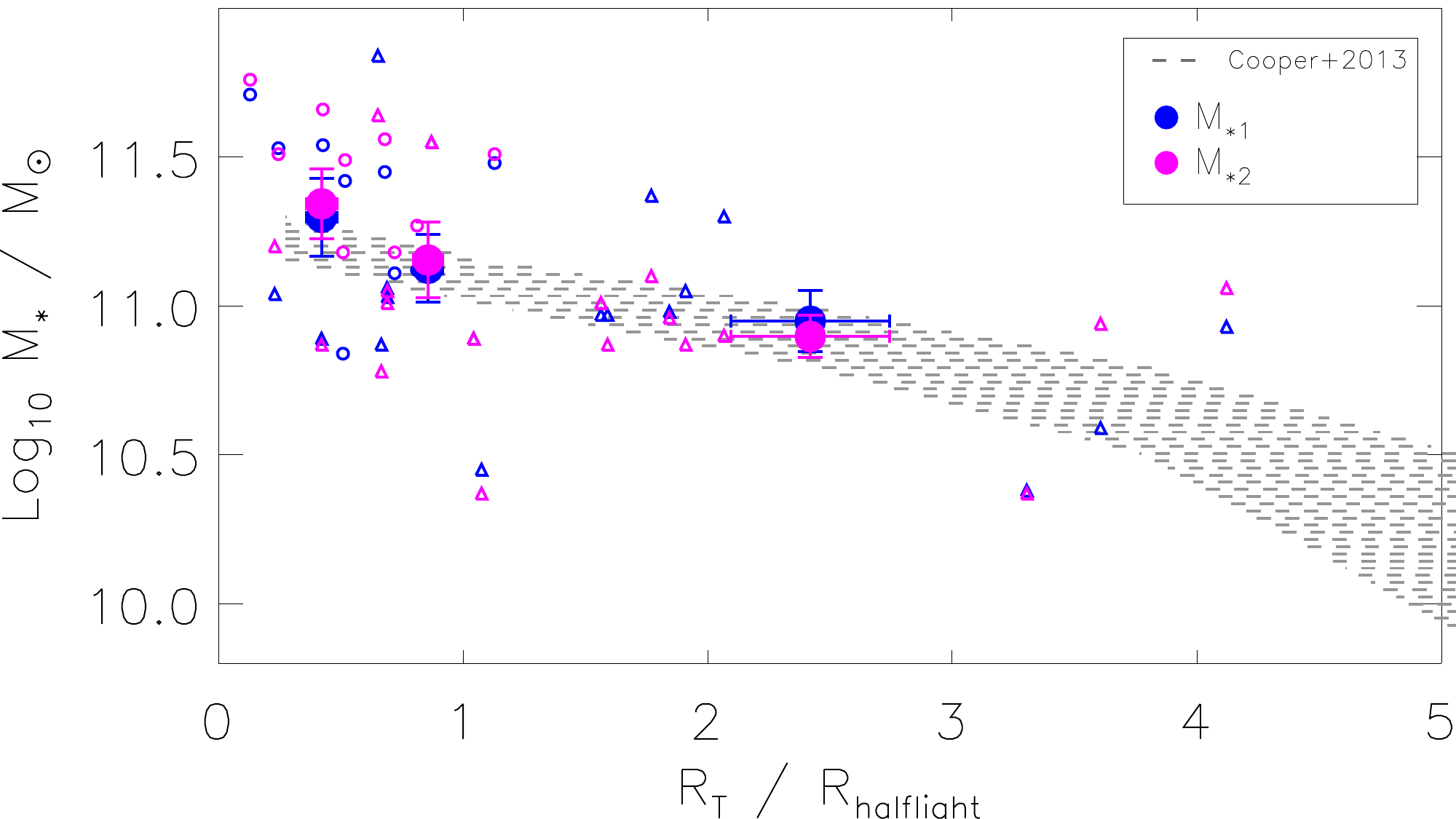}
    
  \caption{Transition radius in units of $R_\mathrm{halflight}$ versus
    total stellar mass (open symbols: circles for SRs,
    triangles for FRs). The full symbols show the same
    quantities in bins of $R_{T}/R_\mathrm{halflight}$. Different
    colors show the results of two different procedures for
    calculating the total stellar mass, see table
    \ref{tab:smoothing_parameters} for details.  }
\label{fig:mass_transrad_RHALF}
 
\end{center}
\end{figure}

We also observe a higher scatter in $R_{T}/\re$ for the lower mass
galaxies, in agreement with the results of \cite{2016MNRAS.458.2371R}
from the Illustris simulations. They defined a three-dimensional
transition radius $r_T$ as the distance at which the accreted stellar
mass fraction overcomes the in-situ, and normalized it by the stellar
half mass radius $r_{\mathrm{half},*}$. Their study shows an
additional dependence of $r_{T}/r_{\mathrm{half},*}$ on morphology,
halo formation time, and recent merger history, and that
$r_{T}/r_{\mathrm{half},*}$ tightly correlates with the ex-situ
stellar mass fraction.

We tested the dependence of the correlation found on the choice of the
value of $\re$. The used values (listed in table \ref{tab:galaxies})
derive mostly from a calibrated average between 2MASS and RC3
determinations \citep{2011MNRAS.413..813C}. Deeper photometric data
would deliver larger values for $\re$ for the brightest galaxies
\citep{2009ApJS..182..216K}. For this reason we carried out a simple
photometric analysis, to be completed in a future study, with the goal
of having a homogeneous determination of $\re$ for all the sample
galaxies which takes into account the shallower surface brightness
profile of the brightest galaxies. Using the surface brightness
profiles available in the literature (references in table
\ref{tab:smoothing_parameters}), we fitted a S\'ersic law in the
outermost regions of the galaxies and extrapolated the profiles up to
very large radii to include all the stellar light. We then evaluated
the half-light radius $R_\mathrm{halflight}$ from the growth curves
(see table \ref{tab:smoothing_parameters}). Figure
\ref{fig:mass_transrad_RHALF} shows that with these half-light radii,
the relation between $R_{T}/R_\mathrm{halflight}$ and total stellar
mass becomes steeper and agrees even better with the predicted trend
from the simulations.

\citet{2017A&A...603A..38S} quantified transition radii for a
subsample of the ePN.S SRs using a parametrization of the
light profiles with a multi-components S\'ersic fit. Their
three-components model describes the superposition of an in-situ
component dominating the central regions, a relaxed, phase-mixed
accreted component, to which is associated the transition radius
$R_{tr1}$, and an outer diffuse component of unrelaxed accreted
material with transition radius $R_{tr2}$. The transition radius is
defined there as the radius at which one component overcomes the
other. The $R_{T}$ derived from the kinematics generally compares with
$R_{tr1}$ for the same galaxies.

\cite{2014MNRAS.443.1433D}, using image stacking of a large sample of
galaxies, defined $R_\mathrm{acc}$ as the radius at which the outer
component of a double S\'ersic model begins to dominate the integrated
stellar light. They found a dependence of $R_\mathrm{acc}$ on the
stellar mass, and, at comparable masses, their photometric values
$R_\mathrm{acc}$ [kpc] are in the same range as our kinematic $R_{T}$
[kpc]. This is interesting considering the different approaches used
to derive these quantities.


\subsection{The diverse halo formation histories}  \label{subsec:DISCUSSION_diverse_halo_formation_histories}

Cosmological simulations \citep[][]{2014MNRAS.444.3357N,
  2017MNRAS.468.3883P} show that FRs, as well as SRs, are the result of different formation pathways,
characterized by factors like the number of mergers, the merger mass
ratio, the timing of mergers, and the gas fraction. We note, however,
that the adopted star formation and AGN feedback models influence the
resulting morphologies and kinematic properties, so that definitive
statements about the origin of FRs and SRs cannot yet be
made \citep[see][for a recent review]{2017ARA&A..55...59N}.  In the
two-phase scenario, if the formation history is dominated by
dissipation, then it is likely that the system becomes a fast
rotator. Dissipation is particularly important in the early stages of
galaxy formation, when the in-situ component generally settles down in
a flattened rotating disk component
\citep[][]{2016MNRAS.458.2371R,2017MNRAS.464.1659Q}, so that before
redshift $\sim1$ the progenitors of the FRs and SRs are
nearly indistinguishable \citep{2017MNRAS.468.3883P}.  Later on, the
evolution of galaxies is determined by the stochasticity of merging
events, which at low redshift are mostly gas-poor
\citep[e.g.][]{2013ApJ...766...38L}.  Major mergers contribute to the
loss of spin of the remnant, while minor mergers are responsible for
the growth in size and mass. The accretion of stars happens
preferentially on radial orbits, and frequent mergers efficiently mix
the in-situ and the accreted stellar populations while changing the
intrinsic shape towards a more spheroidal form
\citep[e.g][]{2005MNRAS.360.1185J,
  2012MNRAS.425.3119H,2013MNRAS.429.2924H, 2014MNRAS.445.1065R}.  Fast
rotating galaxies experience a lower number of mergers when compared
to the more massive slowly rotating systems, so their in-situ
component is preserved in the central regions, in an axisymmetric,
rotationally supported disk \citep[e.g.][]{2014MNRAS.444.3357N,
  2017MNRAS.468.3883P}. Their halos, which host most of the accreted
component, are not necessarily axisymmetric.

\section{Summary and conclusions}
\label{sec:summary_and_conclusions}

In this paper we reconstructed and analyzed the velocity fields of 33
ETGs into their outer halos using planetary nebulae (PNe) as tracers.
For the total ePN.S sample, we combined 25 galaxies from the PN.S ETG
survey with 8 objects observed with counter-dispersed imaging or multi
object spectroscopy from other sources (Arnaboldi et al. in prep.).
In the extended PN.S (ePN.S) sample, 24 galaxies are classified as
FRs \citep[following the definition
by][]{2011MNRAS.414..888E}, and 9 are SRs.  The kinematic
information typically extends to 6 effective radii ($\re$), from a
minimum of 3$\re$ to a maximum of 13$\re$.

We derived kinematic quantities such as the amplitude of rotation, the
kinematic position angle (\PAkin), and the velocity dispersion, using
an adaptive kernel smoothing technique validated with simulated
velocity fields.

In this process, we checked whether the galaxy velocity fields were
point-symmetric as expected for triaxial galaxies in dynamical
equilibrium, and whether the systemic velocity is unchanged in the
outer halo.

We supplemented the PN kinematics with absorption line kinematic data
from the literature (typically limited to the regions inside
$\sim1-2\re$), in order to have a complete picture of the nature of
the objects, and compared the results from different datasets in the
regions of overlap.  Highlights of the results are:

(1) Most FRs (70\%) have decreasing rotation amplitudes with
radius, while a minority shows constant (15\%) or increasing (15\%)
trends.  $60\%$ have $\PAkin$ in their halos aligned with the
photometric major axis at the spatial sampling of the ePN.S survey.  The
SRs display modest but significant rotation in their
outskirts. Among the FRs, 10\% have very weak or no rotation
in their outskirts.

(2) SRs have approximately constant velocity dispersion
profiles with radius, except for NGC 4636 which shows a steep drop in
dispersion beyond 200 arcsec. The FRs display a variety of
velocity dispersion profile shapes: some are approximately constant,
others decrease steeply with radius (figure \ref{fig:sigma_profiles}).

(3) From their velocity fields, all of the SRs and $40\%$ of
the FRs show signatures of a triaxial halo (figure
\ref{fig:fitted_misalignments}). For these galaxies we observe either
a twist in the $\PAkin$ or a constant misalignment with the
photometric major axis.  In addition, two SRs have halos
that are decoupled in their $\PAkin$.  

(4) Five galaxies show deviations from point symmetry. For NGC 2768
and NGC 4594 these effects are small (probably related asymmetries in the light distribution, in the former, and to extinction effects in the
latter) and do not influence our kinematic analysis.  NGC 1316, NGC
4472, and NGC 5128 instead display strong features, that indicate
halos not in dynamical equilibrium.  NGC 5128 is a merger remnant with
clear minor axis rotation, and hence it is a triaxial system.  For NGC
4472 the determination of the $\PAkin$ is influenced by the presence
of an in-falling satellite. While we included this SR among
the triaxial objects, we note that the galaxy would have approximately
major axis rotation once the PNe of the satellite are removed, and we
refer to \citet{2018arXiv180503092H} for a more detailed study. NGC
1316 is a merger remnant with pronounced major axis rotation in the
halo.

(5) In a $V/\sigma(\re)$ versus $V/\sigma(\mathrm{halo})$ diagram
(figure \ref{fig:Vsigma_haloVScenter}), the previous results imply
different locations for the slow and the FRs. The latter
separate further between FRs with slowly rotating outer
spheroid and FRs that rotate rapidly at large radii
also. The high values of $V/\sigma(\mathrm{halo})$ are either due to a
dominating disk component (e.g. NGC 7457) or to a rapidly
rotating outer spheroid (NGC 2768). The FRs with triaxial halos show
approximately equal values of $V/\sigma(\re)$ and
$V/\sigma(\mathrm{halo})$, and span all values of
$V/\sigma(\mathrm{halo})$.

(6) In the 11 FR galaxies with the largest number of PN
tracers, we see a decrease in the rotation velocity profile. In 10
cases this occurs approximately at a transition radius $R_T$ similar
to the scale radius of the known stellar disk
  component\footnote{ Throughout this paper we indicate with "disk
  component" a highly flattened, but definitely three-dimensional,
  oblate, rotating component.}, suggesting that the transition from rapid to
slow rotation is due to the radial fading of the stellar disk in a
slowly rotating outer spheroid. For NGC 4494 there is no photometric
evidence for a disk.  In the other FRs with fewer tracers
that also show a decrease in rotation, the comparison between the slow
rotation of the PN system and the rapid rotation in the absorption
line kinematics, which is related to an inner disk, suggests a similar
transition at smaller radii.

(7) In extended photometry, the ePN.S FRs with triaxial halos show
  small but significant isophote twists, typically of $\sim5-15$
  degrees, lower than for the rounder SRs.  The comparison
  with illustrative, observationally motivated, oblate-triaxial models
  shows that the distribution of observed photometric twists for the
  ePN.S FRs is consistent with the presence of a triaxial
  halo. 
  
(8) For SRs, we estimate a kinematic transition radius $R_T$
from a variation in the rotation or in the $\PAkin$ profile. Combining
with the transition radii of FRs, we find a relation between
the ratio $R_T/\re$ and stellar mass, such that the most massive
galaxies have the lowest $R_T/\re$ (figure
\ref{fig:mass_transrad}). In the framework of the two-phase formation
scenario we can interpret $R_T$ as marking the transition between the
inner regions, dominated by the in-situ stellar component, and the
halo, which is mostly accreted. The comparison with photometric
studies \citep{2014MNRAS.443.1433D} and cosmological simulations
\citep{2013MNRAS.434.3348C, 2016MNRAS.458.2371R} shows good agreement,
in particular the prediction from simulations is quantitatively
confirmed.

(9) In the mass-ellipticity plane (figure \ref{fig:comparisonsamples}),
the ePN.S FRs with triaxial halos are amongst the most massive of the
FRs in the ATLAS$^\mathrm{3D}$ survey. If we interpret the triaxial
envelopes of these galaxies as mainly formed by accreted, ex-situ,
stars, the observed mass dependence is in agreement with simulations
\citep{2016MNRAS.458.2371R} which predict that the accreted component
dominates down to smaller fractions of $\re$ for more massive
galaxies (figure \ref{fig:mass_transrad_RHALF}). The dependence on mass, combined with the fainter survey
limit and smaller radial coverage of the ATLAS$^\mathrm{3D}$ kinematics, explains
the different conclusions drawn by the two surveys: low mass FRs
are dominated by the disk rotation up to large radii;
however, in the more massive FRs the spheroid is more
prominent, where the accreted envelope reaches smaller fractions of
$\re$.  In the massive FRs, the inner disks frequently fade
in slowly rotating outer spheroids which are often triaxial.

In the data presented here, we see 
\begin{itemize}
 \item SRs
 \item FRs without apparent disks (NGC 4494 and NGC 4742)
 \item FRs with only inner disks and slowly rotating
   spheroids (e.g. NGC 3377)
 \item FRs with dominant disks all the way to their
   outermost regions (e.g. NGC 7457)
 \item FRs with inner disks and rapidly rotating spheroids
   (e.g. NGC 2768)
 \item FRs with triaxial halos that are dominated by
   dispersion (e.g. NGC 4649) or that rotate rapidly (NGC 4742 and NGC
   5866)
\end{itemize}

Thus we conclude that ETGs show considerably more diversity in their
halos than is apparent from their central bright regions.  We also see
clear signatures of the two-phase formation scenario: three galaxies
show out of equilibrium kinematics in their halos, and more generally
the inner and outer regions of ETGs often have different kinematic
properties, where the transition radius depends on the stellar mass as
predicted by cosmological simulations.

\begin{acknowledgements}
  \\
  We are grateful to Nigel G. Douglas for his fundamental
  contribution to the foundation of the Planetary Nebula Spectrograph
  instrument.  We thank Luis Ho and Zhao-Yu Li for providing us with
  the Carnagie photometric data for the ePN.S ETG sample.  We thank
  Peter Erwin for sharing his expertise with photometric modeling.
  C.P. thanks I. Soeldner-Rembold and J. Hartke for useful comments
  and discussions. C.P. is also grateful to F. Hofmann for advice and
  suggestions on the manuscript.  AC would like to thank CNPq for the
  fellowship 4150977/2017-4.  A.J.R. was supported by National Science
  Foundation grant AST-1616710, and as a Research Corporation for
  Science Advancement Cottrell Scholar.  C.T. is supported through an
  NWO-VICI grant (project number 639.043.308).  This research has made
  use of NASA's Astrophysics Data System, of the data products from
  the Two Micron All Sky Survey (University of Massachusetts and the
  IPAC/CalTech), of the NASA/ IPAC Infrared Science Archive (Jet
  Propulsion Laboratory, CalTech), of the NASA/IPAC Extragalactic
  Database (NED), and of the VizieR catalogue access tool, CDS,
  Strasbourg, France.  We acknowledge the usage of the HyperLeda
  database (http://leda.univ-lyon1.fr) as well.
\end{acknowledgements}

  \begin{table*}
\caption[]{Measured parameters and typical errors for the smoothed velocity and velocity dispersion fields}
\begin{center}
\begin{tabular}{lllllllllll}
\hline\hline\noalign{\smallskip}
  \multicolumn{1}{l}{Galaxy} &
  \multicolumn{1}{l}{A\tablefootmark{(a)}} &
  \multicolumn{1}{l}{B\tablefootmark{(a)}} &
  \multicolumn{1}{l}{$\delta\vmeas$\tablefootmark{(b)}} &
  \multicolumn{1}{l}{$\langle\Delta{\tilde{V}}\rangle$\tablefootmark{(c)}} &
  \multicolumn{1}{l}{$\langle\Delta{\tilde{\sigma}}\rangle$\tablefootmark{(d)}} &
  \multicolumn{1}{l}{\vsys\tablefootmark{(e)}} &
  \multicolumn{1}{l}{$R_\mathrm{halflight}$\tablefootmark{(f)}} &
  \multicolumn{1}{l}{$\log_{10}\left(\frac{M{*1}}{M_{\odot}}\right)$} &
   \multicolumn{1}{l}{$\log_{10}\left(\frac{M_{*2}}{M_{\odot}}\right)$} &
  \multicolumn{1}{l}{Ref.} \\
  \multicolumn{1}{l}{NGC} &
  \multicolumn{1}{l}{[arcsec$^{-1}$]} &
  \multicolumn{1}{l}{} &
  \multicolumn{1}{l}{[ \kms]} &
  \multicolumn{1}{l}{[ \kms]} &
  \multicolumn{1}{l}{[ \kms]}&
  \multicolumn{1}{l}{[ \kms]}&
  \multicolumn{1}{l}{[arcsec]}&
  \multicolumn{1}{l}{\tablefootmark{(g)}}&
  \multicolumn{1}{l}{\tablefootmark{(h)}}&
  \multicolumn{1}{l}{phot. \tablefootmark{(i)}}\\
\noalign{\smallskip}\hline\noalign{\smallskip}

0584&	  0.37     &	 63.2	  &	21.	&     21.     &     15.     &	  1901.$\pm$27.  &  30     &  11.07   & 11.04  &  (4)	 \\
0821&	  0.50     &	  9.7	  &	21.	&     22.     &     16.     &	  1697.$\pm$16.  &  38     &  10.97   & 10.87  &  (1);(2)  \\
1023&	  0.26     &	 15.3	  &	14.	&     19.     &     13.     &	   618.$\pm$ 9.  &  39     &  10.93   & 11.06  &   (6)   \\
1316&	  0.46     &	 21.3	  &	30.	&     24.     &     21.     &	  1749.$\pm$30.  & 115     &  11.84   & 11.64  &   (4)  \\
1344&	  0.40     &	 10.8	  &	10.	&     21.     &     13.     &	  1190.$\pm$13.  &  44     &  11.06   & 11.01  &   (7)  \\
1399&	  0.84     &	  5.2	  &	37.	&     27.     &     18.     &	  1401.$\pm$19.  & 122     &  11.53   & 11.51  &   (4)  \\
2768&	  0.33     &	 17.5	  &	20.	&     26.     &     22.     &	  1378.$\pm$14.  &  70     &  11.30   & 10.90  &   (15);(19) \\
2974&	  0.57     &	 33.6	  &	21.	&     25.     &     21.     &	  1803.$\pm$42.  &  43     &  10.87   & 11.06  &   (4)  \\
3115&	  0.28     &	 19.6	  &	20.	&     30.     &     21.     &	   624.$\pm$16.  &  73     &  10.98   & 10.96  &   (4)   \\
3377&	  0.32     &	 19.0	  &	20.	&     12.     &      8.     &	   708.$\pm$ 9.  &  51     &  10.45   & 10.37  &  (1);(2);(3);(19)\\
3379&	  0.34     &	 16.2	  &	20.	&     21.     &     14.     &	   934.$\pm$17.  &  58     &  10.89   & 10.89  &   (8);(9) \\
3384&	  0.25     &	 18.9	  &	20.	&     14.     &     10.     &	   722.$\pm$ 9.  &  42     &  10.77   & 10.88  &   (16);(20)  \\
3489&	  0.32     &	 14.7	  &	20.	&     14.     &     11.     &	   707.$\pm$12.  &  22     &  10.56   & 10.61  &   (16)  \\
3608&	  0.37     &	 24.6	  &	21.	&     15.     &     11.     &	  1235.$\pm$16.  & 108     &  10.84   & 11.18  &  (2);(3) \\
3923&	  0.60     &	 24.3	  &	30.	&     36.     &     26.     &	  1677.$\pm$26.  & 164     &  11.54   & 11.66  &   (4)  \\
4278&	  0.88     &	  2.4	  &	18.	&     18.     &     14.     &	   605.$\pm$17.  &  59     &  10.89   & 10.87  &  (12);(19)  \\
4339&	  0.43     &	 20.3	  &	21.	&      8.     &      6.     &	  1300.$\pm$ 7.  &  27     &  10.38   & 10.18  &  (13);(19)  \\
4365&	  0.91     &	  0.2	  &	21.	&     20.     &     14.     &	  1273.$\pm$11.  & 111     &  11.48   & 11.51  &  (14)  \\
4374&	  0.86     &	  1.6	  &	23.	&     27.     &     19.     &	  1050.$\pm$18.  & 122     &  11.45   & 11.56  &  (14)  \\
4472&	  0.74     &	 11.3	  &	20.	&     38.     &     33.     &	   959.$\pm$22.  & 193     &  11.71   & 11.76  &  (14) \\
4473&	  0.91     &	  0.2	  &	21.	&     18.     &     12.     &	  2236.$\pm$13.  &  38     &  10.87   & 10.78  &  (17)  \\
4494&	  0.41     &	 12.8	  &	21.	&     17.     &     12.     &	  1348.$\pm$10.  &  49     &  11.05   & 10.87  &  (10)  \\
4552&	  0.90     &	  0.6	  &	21.	&     23.     &     16.     &	   361.$\pm$15.  & 101     &  11.11   & 11.18  &  (14) \\
4564&	  0.24     &	 16.5	  &	21.	&     19.     &     12.     &	  1169.$\pm$17.  &  24     &  10.59   & 10.94  &   (1)  \\
4594&	  0.54     &	 19.3	  &	30.	&     32.     &     28.     &	  1060.$\pm$16.  & 102     &  11.37   & 11.10  &   (4);(21)  \\
4636&	  0.68     &	  9.5	  &	21.	&     24.     &     16.     &	   903.$\pm$19.  & 172     &  11.12   & 11.27  &   (14) \\
4649&	  0.86     &	  1.5	  &	20.	&     25.     &     18.     &	  1059.$\pm$18.  & 132     &  11.55   & 11.55  &   (14) \\
4697&	  0.45     &	  6.0	  &	35.	&     23.     &     16.     &	  1274.$\pm$11.  & 123     &  11.03   & 11.05  &  (2);(11) \\
4742&	  0.20     &	 16.9	  &	21.	&     13.     &     10.     &	  1305.$\pm$ 9.  &  13     &  10.38   & 10.37  &   (4);(22) \\
5128&	  0.29     &	 25.6	  &	 4.	&     25.     &     21.     &	   536.$\pm$17.  & 414     &  11.04   & 11.20  &   (4)  \\
5846&	  0.90     &	  0.1	  &	21.	&     25.     &     17.     &	  1716.$\pm$19.  & 193     &  11.42   & 11.49  &  (5);(18) \\
5866&	  0.34     &	 18.7	  &	21.	&     15.     &     10.     &	   782.$\pm$10.  &  48     &  10.97   & 11.01  &   (16);(19) \\
7457&	  0.28     &	 10.5	  &	20.	&      7.     &      6.     &	   843.$\pm$ 5.  &  62     &  10.28   & 10.63  &   (16); (19) \\

\noalign{\smallskip}\hline
\end{tabular}
\tablefoot{
 \tablefoottext{a}{Kernel parameters, A and B, used in the smoothing procedure, see section \ref{sec:method}.}\\
 \tablefoottext{b}{Mean error on the measured radial velocities.}\\
 \tablefoottext{c}{Mean error on the smoothed velocity field.}\\
 \tablefoottext{d}{Mean error on the velocity dispersion field.}\\
 \tablefoottext{e}{Subtracted systemic velocity.}\\
 \tablefoottext{f}{Half-light radius, see text.}\\
 \tablefoottext{g}{Logarithm of the stellar mass $M_{*1}$ in solar units.  $M_{*1}$ are derived from the K-band luminosities $M_{K}$ listed in table \ref{tab:galaxies}, corrected for missing flux as in \cite{2013ApJ...768...76S}: $M_\mathrm{Kcorr} = 1.07\times M_{K}+1.53$, and applying a mass-to-light ratio of 1 \citep[appropriate for old stellar population with a Kroupa IMF,][]{2016MNRAS.457.1242F}. }\\
 \tablefoottext{h}{Logarithm of the stellar mass $M_{*2}$ in solar units.  $M_{*2}$ are derived using total luminosities obtained by fitting the surface brightness profiles with a S\'ersic law and integrating till very large radii. The apparent magnitudes obtained are converted to absolute magnitudes using the distances listed in table \ref{tab:galaxies}, corrected for foreground galactic extinction, and homogenized to B magnitude using color indexes from the HyperLeda\footnote{\cite{2014A&A...570A..13M}, http://leda.univ-lyon1.fr/} database and from \cite{1978ApJ...223..707S}. We then applied a stellar mass to light ratio given by the relation from \cite{2003ApJS..149..289B}: $\log_{10}\frac{M}{L_B}=-9.42+1.737\times(B-V)$, where the $(B-V)$ color index is from HyperLeda.}\\
 \tablefoottext{i}{References for the photometry: (1) \cite{1994A&AS..104..179G}, (2) \cite{2005AJ....129.2138L}, (3) \cite{1987MNRAS.226..747J}, (4)\cite{2011ApJS..197...22L}, (5) \cite{2000A&AS..144...53K}, (6) \cite{2008MNRAS.384..943N}, (7) \cite{2007A&A...467.1011S}, (8) \cite{1990AJ.....99.1813C}, (9) \cite{2000AJ....119.1157G}, (10) \cite{2009MNRAS.393..329N}, (11) \cite{2008MNRAS.385.1729D}, (12) \cite{1990AJ....100.1091P}, (13) \cite{1994A&AS..106..199C}, (14) \cite{2009ApJS..182..216K}, (15) \cite{2009ApJS..181..135H}, (16) \cite{2013MNRAS.432.1768K}, (17) \cite{1990A&AS...86..429C}, (18) \cite{2017A&A...603A..38S}, (19) \cite{1993A&AS...98...29M}, (20) \cite{2007AN....328..562M}, (21) \cite{2012MNRAS.423..877G}, (22) \cite{1995AJ....110.2622L}}
 }
 
\end{center}
\label{tab:smoothing_parameters} 
\end{table*}

\vfill\newpage

\begin{table*}
\caption[]{Results of the kinematic analysis}
\begin{center}
\begin{tabular}{lllllllp{0.35\textwidth}}
\hline\hline\noalign{\smallskip}
  \multicolumn{1}{l}{Galaxy} &
  \multicolumn{1}{l}{class  \tablefootmark{(a)}} &
  \multicolumn{1}{l}{point} &
  \multicolumn{1}{l}{features in} &
  \multicolumn{1}{l}{multiple}&
  \multicolumn{1}{l}{triaxial \tablefootmark{(e)}} &
  \multicolumn{1}{l}{$R_T$ \tablefootmark{(f)}}&  
  \multicolumn{1}{p{0.4\textwidth}}{NOTES \tablefootmark{(f)}} \\
  \multicolumn{1}{l}{NGC} &
  \multicolumn{1}{l}{} &
  \multicolumn{1}{l}{symm.\tablefootmark{(b)}} &
  \multicolumn{1}{l}{\PAkin(R)\tablefootmark{(c)}} &
  \multicolumn{1}{l}{kin. comp.\tablefootmark{(d)}} &
  \multicolumn{1}{l}{} &
  \multicolumn{1}{l}{arcsec} &
  \multicolumn{1}{p{0.4\textwidth}}{} \\
\noalign{\smallskip}\hline\noalign{\smallskip}
   0584	& F    &	y	&     n       &       n       &       n       &  {---}{---}{---} \par	 &	   small number of tracers    		\\
   0821 & F    &	y	&     y       &       n       &       y       &  $60 \pm 30$     &	  (A,C) 				\\
   1023 & F    &	y	&     n       &       n       &       n       &  $160 \pm 20$	 &	  (A)				       \\
   1316 & F    &	n	&     y       &       y       &       n       &  $75 \pm 25$	 &	  (A) - merger remnant  		\\
   1344 & F    &	y	&     y       &       n       &       y       &  $30 \pm 15$	 &	  (C) - prolate rotation	       \\
   1399 & S    &	y	&     y       &       y       &       y       &  $30 \pm 15$	 &	  (B) - kinematically decoupled halo   \\
   2768 & F    &	n	&     n       &       n       &       n       &  $145 \pm 45$	 &	  (A) - fast rotating halo	       \\     
   2974 & F    &	y	&     n       &       n       &       n       &  {---}{---}{---} \par  	 &	  small number of tracers  	       \\	 
   3115 & F    &	y	&     y       &       n       &       n       &  $135 \pm 45$	 &	  (A) - photometry reveals a perturbed disk at the interface with the spheroid component \\
   3377 & F    &	y	&     y       &       y       &       y       &  $55 \pm 15$	 &	  (A)				       \\
   3379 & F    &	y	&     y       &       y       &       y       &  $60 \pm 20$	 &	  (A,C) - hints for growing rotation in the outskirts\\
   3384 & F    &	y	&     n       &       n       &       n       &   {---}{---}{---} \par	 &		  extended disk 	       \\
   3489 & F    &	y	&     n       &       n       &       n       &   {---}{---}{---} \par	 &		  small number of tracers         \\
   3608 & S    &	y	&     y       &       n       &       y       &  $55 \pm 35$	 &	  (B)				       \\
   3923 & S    &	y	&     y       &       n       &       y       &  $70 \pm 30$	 &	  (B)				       \\
   4278 & F    &	y	&     n       &       n       &       n       &  $25 \pm 15$	 &	  (A)				       \\
   4339 & F    &	y	&     n       &       n       &       n       &  {---}{---}{---} \par	 &	  small number of tracers  	       \\
   4365 & S    &	y	&     y       &       y       &       y       &  $125 \pm 25$    &	 (C) - kinematically decoupled halo   \\
   4374 & S    &	y	&     y       &       n       &       y       &  $83 \pm 33$     &	 (B)				      \\
   4472 & S    &	n	&     y       &       y       &       y       &  $25 \pm 10$     &	 (B) - recent merger; kinematics dominated by in-falling satellite\\	
   4473 & F    &	y	&     y       &       n       &       y       &  $25 \pm 13$	 &	 (A) - minor axis rotation observed by \citet{2013MNRAS.435.3587F}\\
   4494  & F    &	y	&     y       &       y       &       y       &  $93 \pm 43$	 &	   (A) \\
   4552 & S    &	y	&     y       &       y       &       y       &  $73 \pm 58$	 &	   (B,C)				\\
   4564 & F    &	y	&     n       &       n       &       n       &  $85 \pm 45$	 &	  (A) - fast rotating halo\\
   4594 & F    &	y	&     n       &       n       &       n       &  $180 \pm 90$	 &	  (A) - no detections in the plane of the disk (dust absorption) \\
   4636 & S    &	y	&     y       &       y       &       y       &  $140 \pm 40$	 &	   (C)  				\\
   4649 & F    &	y	&     y       &       y       &       y       &  $115 \pm 15$	 &	   (A) - no detections on the west side of the galaxy\\
   4697 & F    &	y	&     y       &       n       &       n       &  $85 \pm 65$	 &	   (A) - hints for growing rotation in the outskirts; note that this PN sample contains a secondary PN population out of dynamical equilibrium \citep {2006AJ....131..837S}\\
   4742 & F    &	y	&     y       &       n       &       y       &  $43 \pm 32$	 &	  (C)     \\
   5128 & F    &	n	&     y       &       y       &       y       &  $95 \pm  55$	 &	   (C) - recent merger  		\\
   5846 & S    &	y	&     y       &       n       &       y       &  $100 \pm 50$	 &	   (B)  				\\
   5866 & F    &	y	&     y       &       n       &       y       &  $75 \pm 25$	 &	   (A) - fast rotating halo		\\
   7457 & F    &	y	&     n       &       n       &       n       &  {---}{---}{---} \par &	  extended disk 		       \\
   
\noalign{\smallskip}\hline
\end{tabular}
\tablefoot{
\tablefoottext{a}{The sample is divided into slow (S) and FRs (F), according to the definition of \citet{2011MNRAS.414..888E}.}\\
\tablefoottext{b}{Point symmetry properties of the PN smoothed velocity fields: "y" marks point symmetric galaxies while "n" marks galaxies showing asymmetries, see section \ref{sec:Point_symmetry_analysis}}\\
\tablefoottext{c}{Significant kinematic twist or misalignment of $\PAkin$ with respect to $\PAphot$: "y" marks galaxies showing at least one of these, see section \ref{subsec:triaxiality}}\\
\tablefoottext{d}{For the galaxies marked with "y", the PN smoothed velocity fields reveal rotation along two perpendicular axes or a halo which is kinematically decoupled with respect to the innermost regions, see sections \ref{sec:results_per_family} and \ref{subsubsec:triax_decoupledhalo}. }\\
\tablefoottext{e}{Galaxies marked with "y" have been classified as triaxial in section \ref{subsec:triaxiality}.}\\
\tablefoottext{f}{$R_T\pm\Delta R_T$ is the radial range of the kinematic transition.}\\
\tablefoottext{g}{Criteria (A, B, C) used to define the radial range $R_T\pm\Delta R_T$ as described in the text (section \ref{sec:discussion}), and notes on individual galaxies (see also the dedicated appendix \ref{sec:notes_on_single_galaxies}).}
 }
\end{center}
\label{tab:results}
\end{table*}

\vfill\newpage

\bibliographystyle{aa}
\bibliography{auto,general}

  \begin{appendix}

\section{Testing the adaptive kernel smoothing procedure on simulated data}
\label{subsec:discussion_about_the_method}

Our procedure for the measurement of the mean velocity and velocity dispersion fields from the observed LOS velocities of the PNe is based on performing a weighted local average, where the weights depend on the local density of tracers. The derived quantities depend on how well the detected PNe statistically sample the parent galaxy in the phase space. In particular the spatial distribution of the PNe plays an important role, as the smoothing averages over objects that are close together on the sky. Appendix \ref{subsubsec:completeness_function} describes what determines the observed PN spatial distribution and how this, combined with a smoothing procedure, may affect the results. Appendices \ref{subsubsec:tests_on_simulations} and \ref{subsubsec:statistical_tests} study and quantify these effects on a simulated galaxy (with variable number of tracers and $\mathrm{V}/\sigma$ ratio) and on simulated velocity fields, with the same spatial distribution as the observed galaxies. We found that the smoothing procedure does not create artifacts in the velocity fields above the $1\sigma$ level; any features above this threshold are probably real.

\subsection{PN spatial distribution and completeness}
\label{subsubsec:completeness_function}
The observed PN spatial distribution depends on several factors. The most evident one is the incompleteness in the central high surface brightness regions because of the difficulty in detecting faint point sources against a bright background \citep[Arnaboldi et al. in prep., ][]{2009MNRAS.394.1249C}. This is a strong function of the radius and it is quantified by the \emph{completeness function}, the fraction of detected objects in different radial ranges.

In addition, the PN number density is proportional to the local surface brightness of the galaxy. This means that the number of tracers decreases with radius following the steep decrease of the surface brightness, and features in the surface brightness distribution may generate under- or over-densities of PNe. Of course at our typical number of tracers ($\sim200$) these features in density are barely distinguishable from statistical effects.

The adopted smoothing procedure takes into account the natural non-uniformity of the PNe distribution through the use of an adaptive kernel technique, which optimizes the size of the kernel to the local density of tracers.
Nevertheless a local average of the velocities over a non-uniform spatial distribution, combined with the statistics in the LOSVD sampling, may in principle generate features in the kinematic maps or biases in the estimate of the kinematic parameters.
On the other hand, averaging over voids or over sparsely distributed tracers unavoidably leads to over-smoothed velocity amplitudes and to under-resolved velocity gradients. The residual velocities from the unresolved gradients may artificially boost the dispersion , creating local maxima in the velocity dispersion maps. In appendices \ref{subsubsec:tests_on_simulations} and \ref{subsubsec:statistical_tests} we assess the impact of such biases using simulated data.

\begin{figure*}[t]
  \centering
  \begin{minipage}[b]{0.3\linewidth}
    \includegraphics[width=\linewidth]{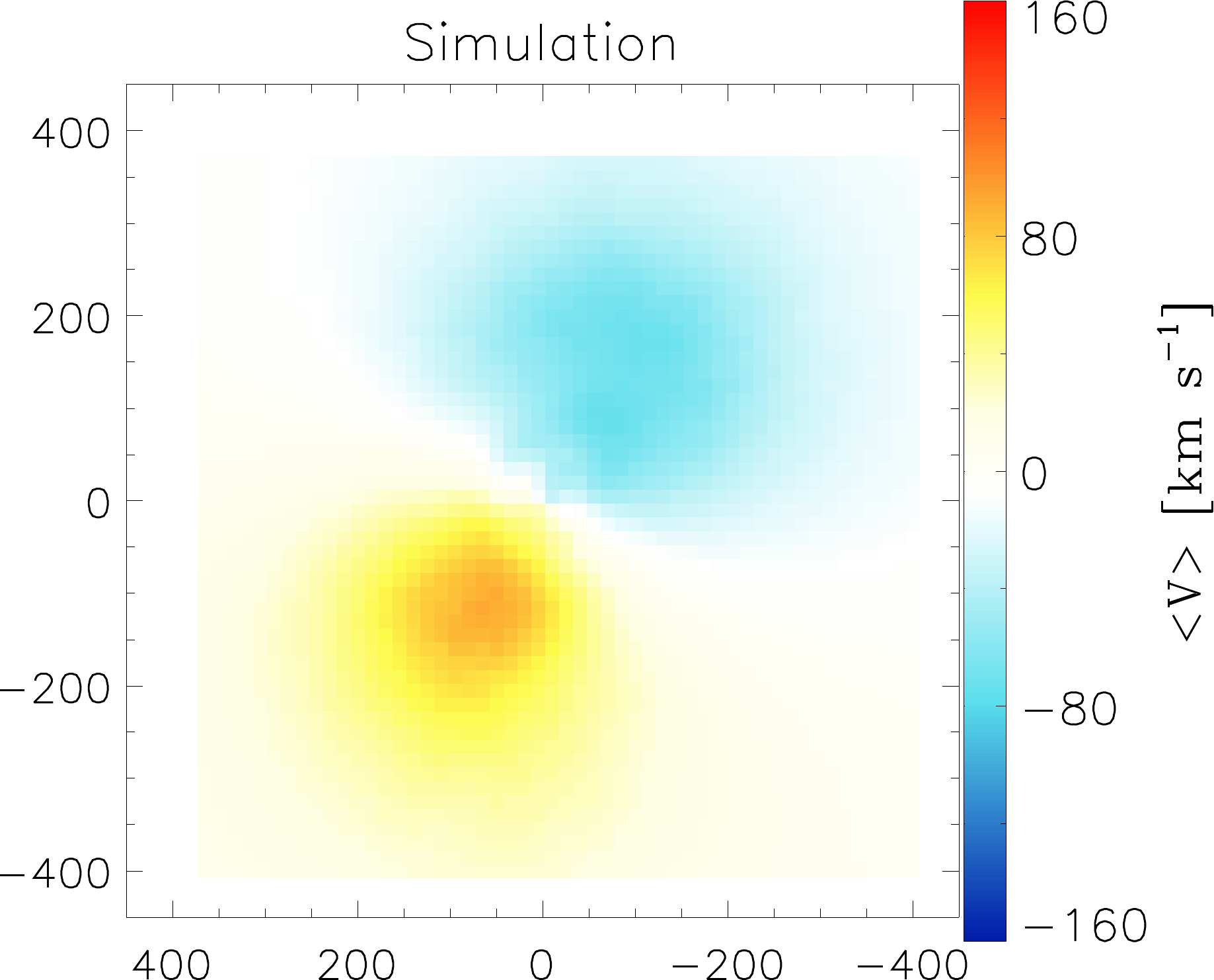}
 
     \includegraphics[width=\linewidth]{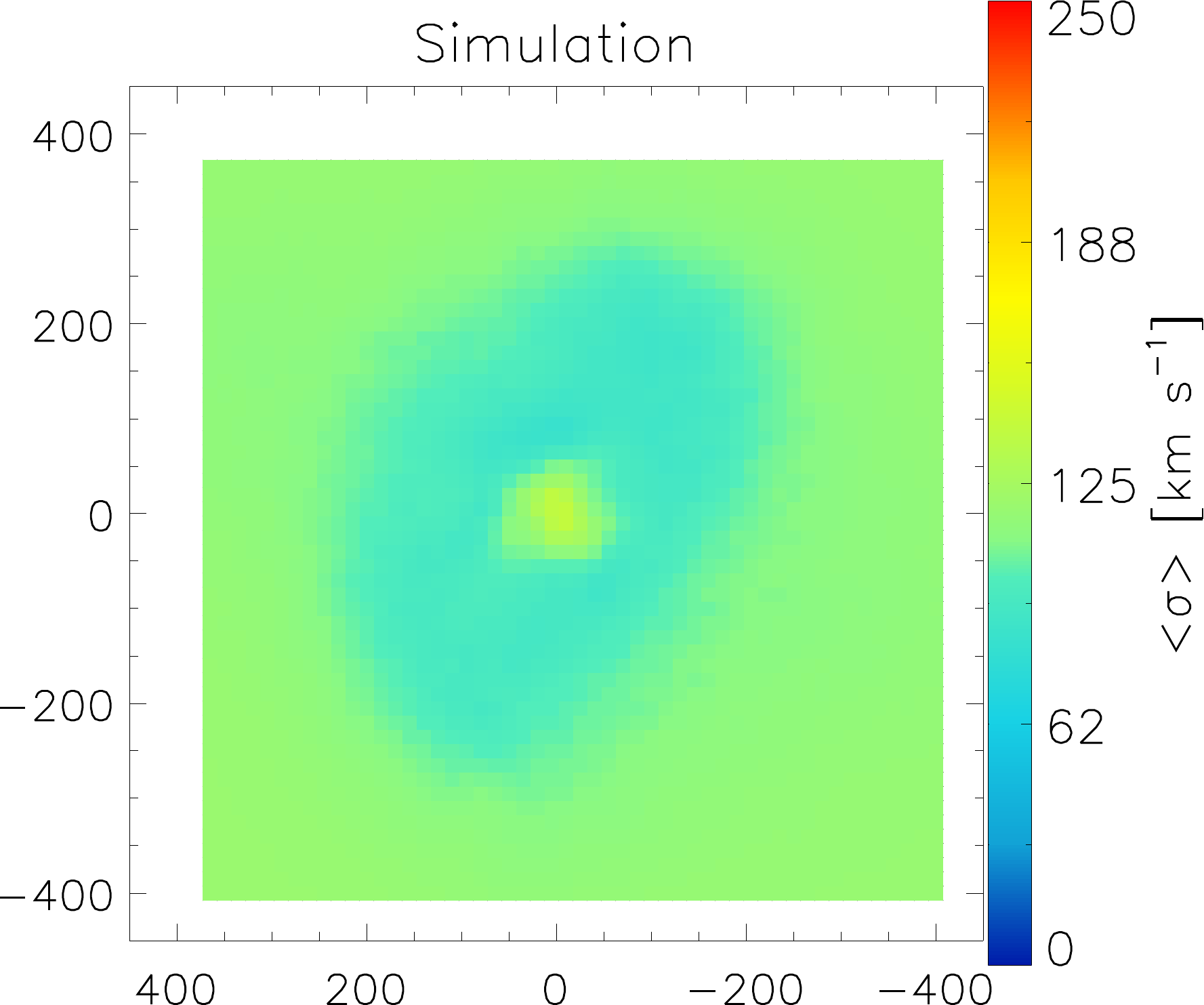}
  \end{minipage}
  \begin{minipage}[b]{0.3\linewidth}
    \includegraphics[width=\linewidth]{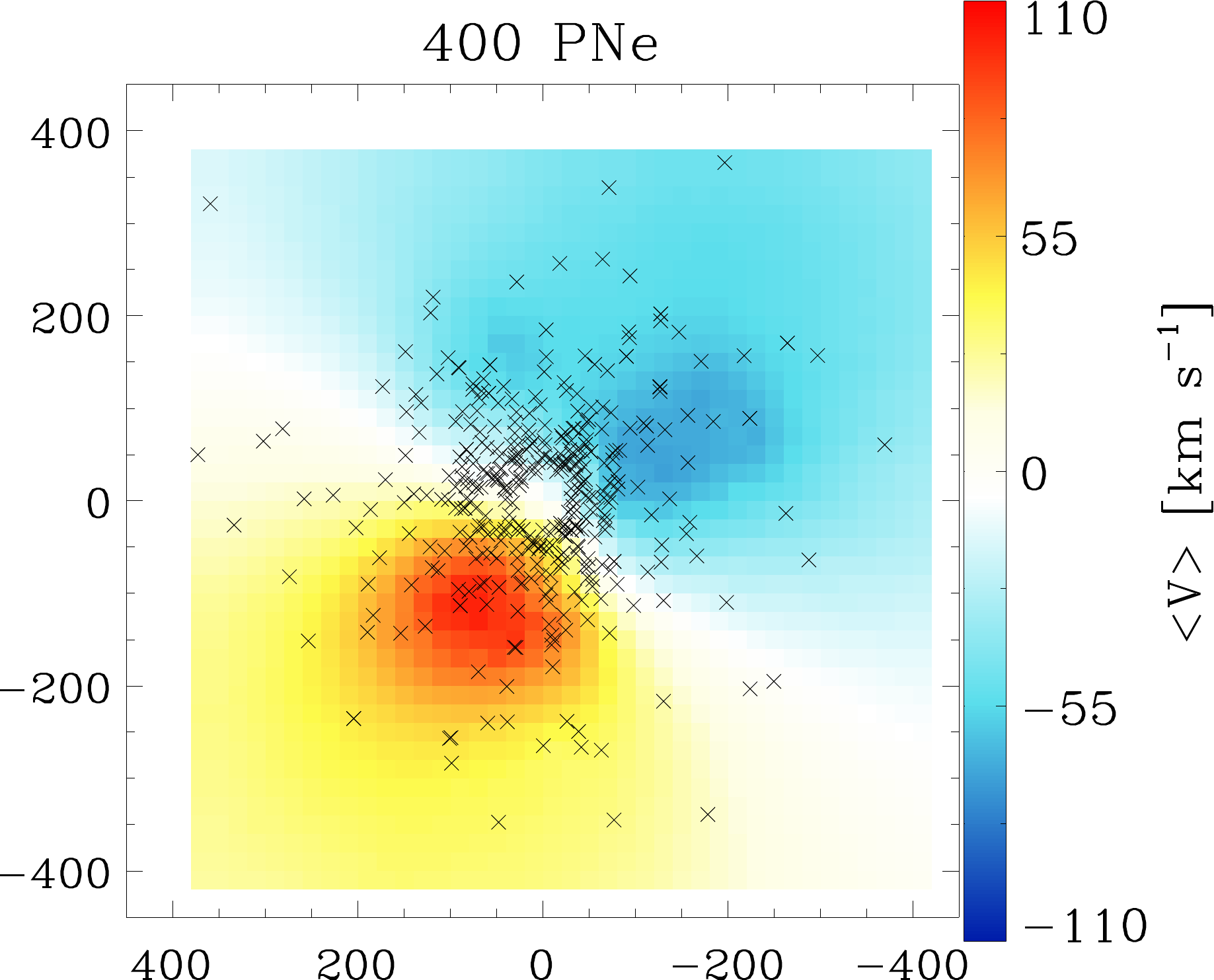}
 
     \includegraphics[width=\linewidth]{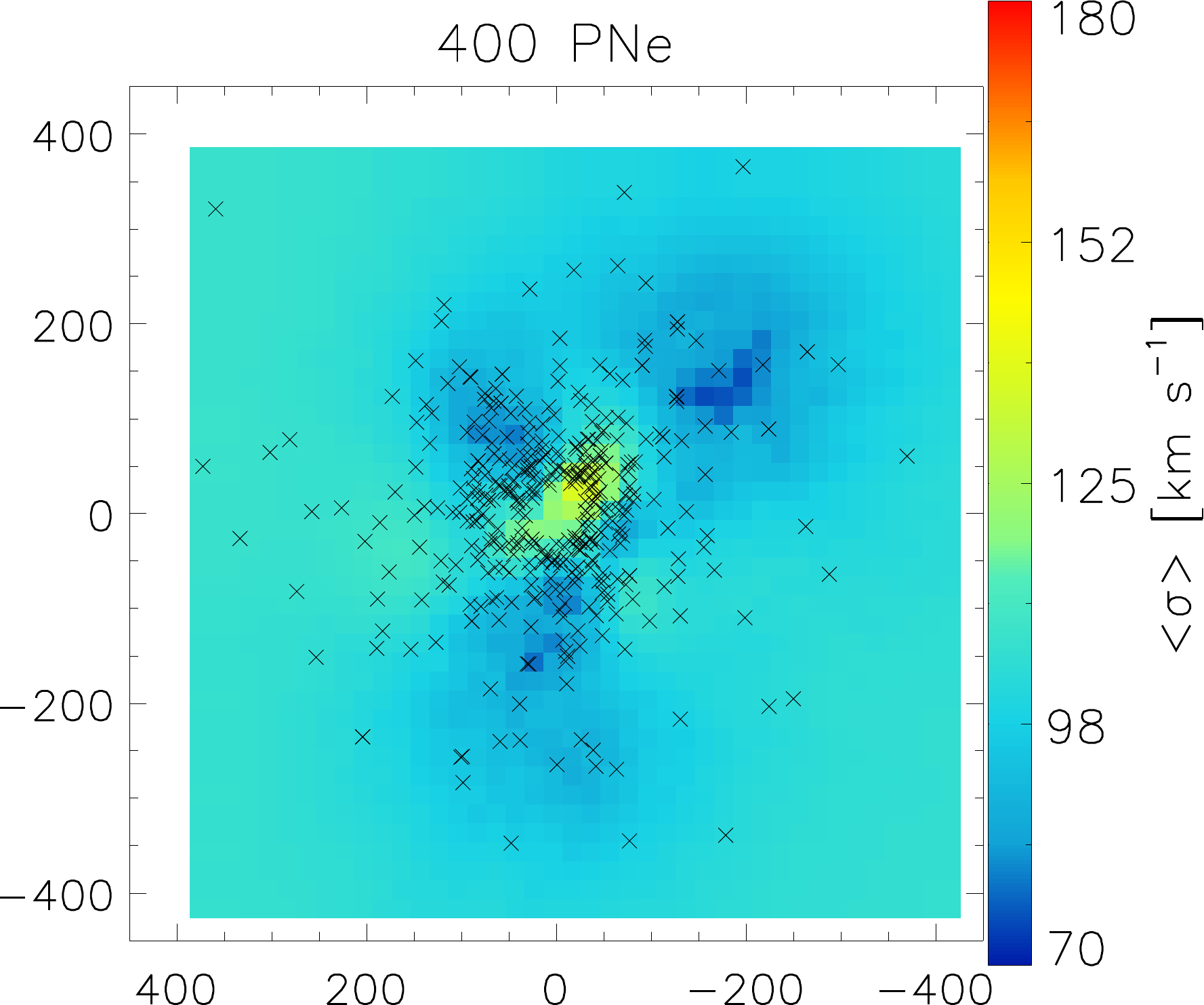}
  \end{minipage}
   \begin{minipage}[b]{0.3\linewidth}
    \includegraphics[width=\linewidth]{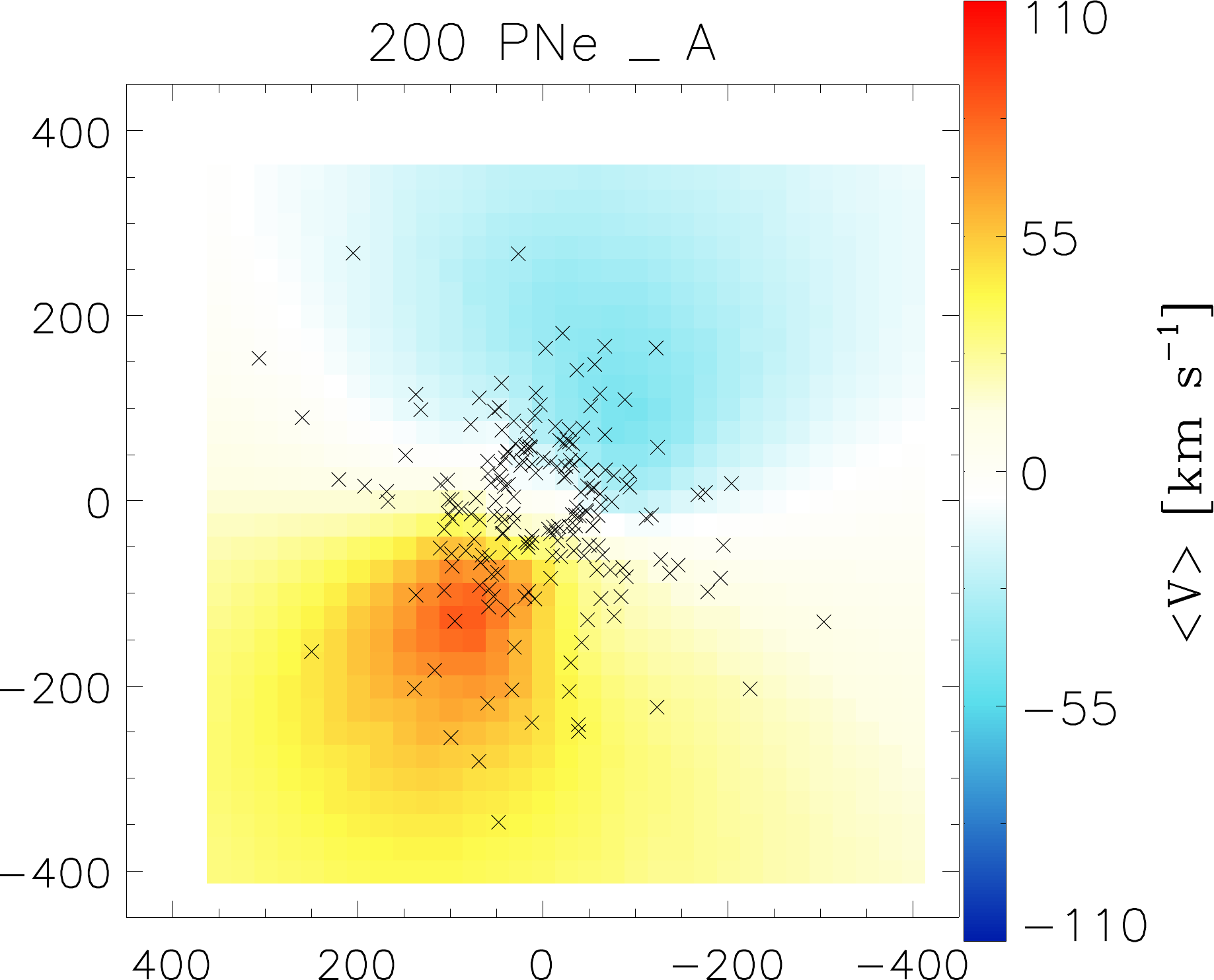}
 
     \includegraphics[width=\linewidth]{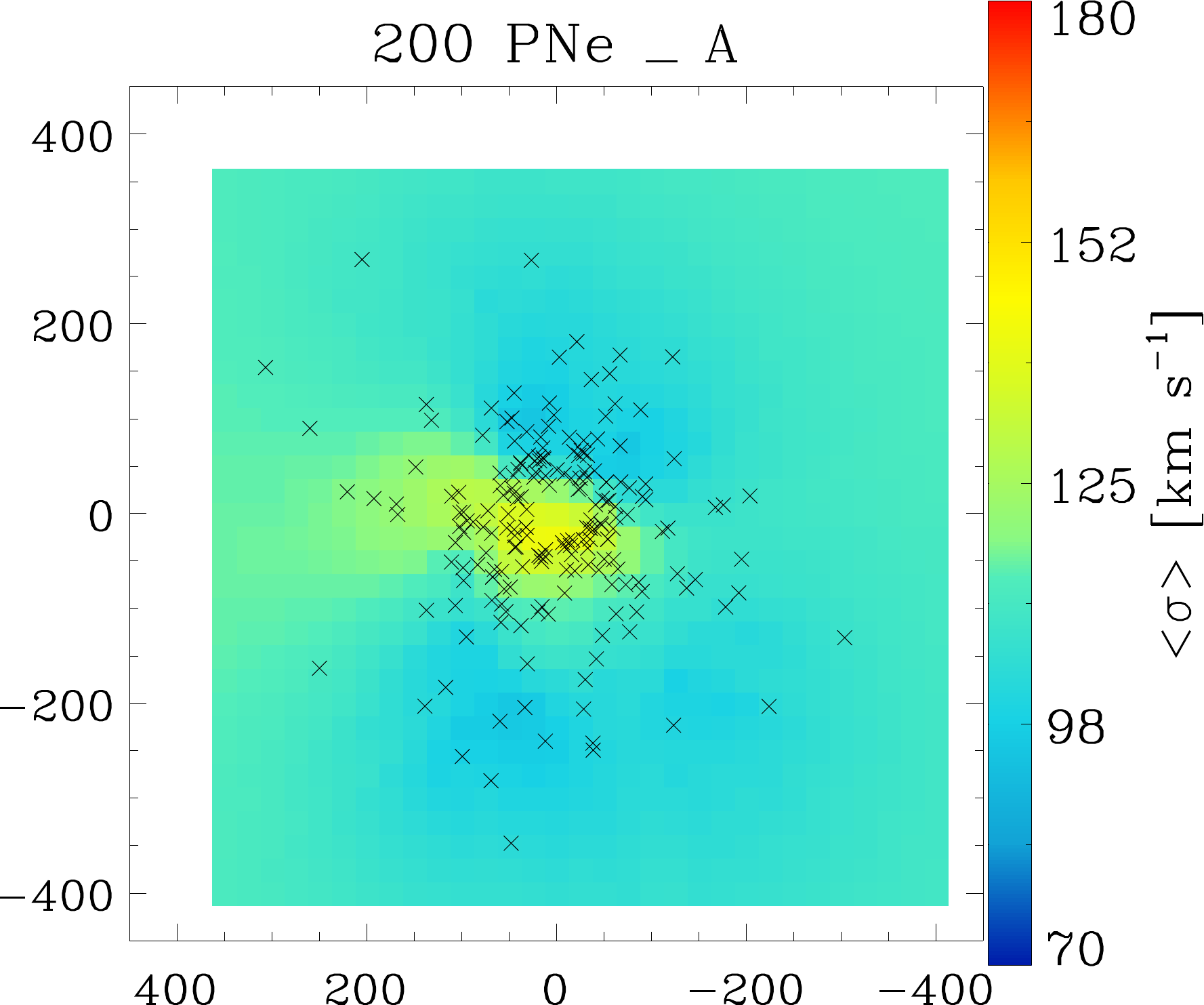}
  \end{minipage}
   \begin{minipage}[b]{0.3\linewidth}
    \includegraphics[width=\linewidth]{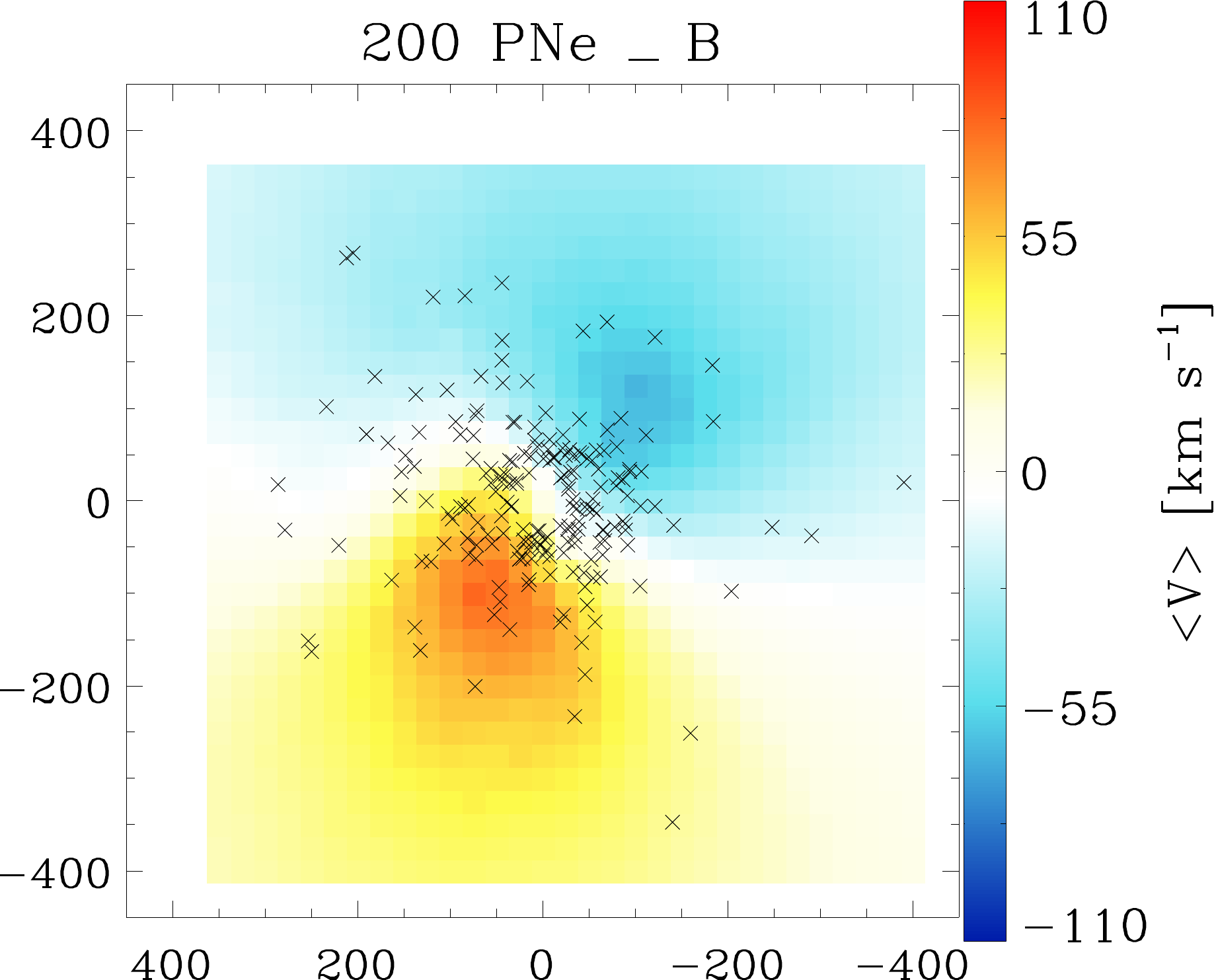}
 
     \includegraphics[width=\linewidth]{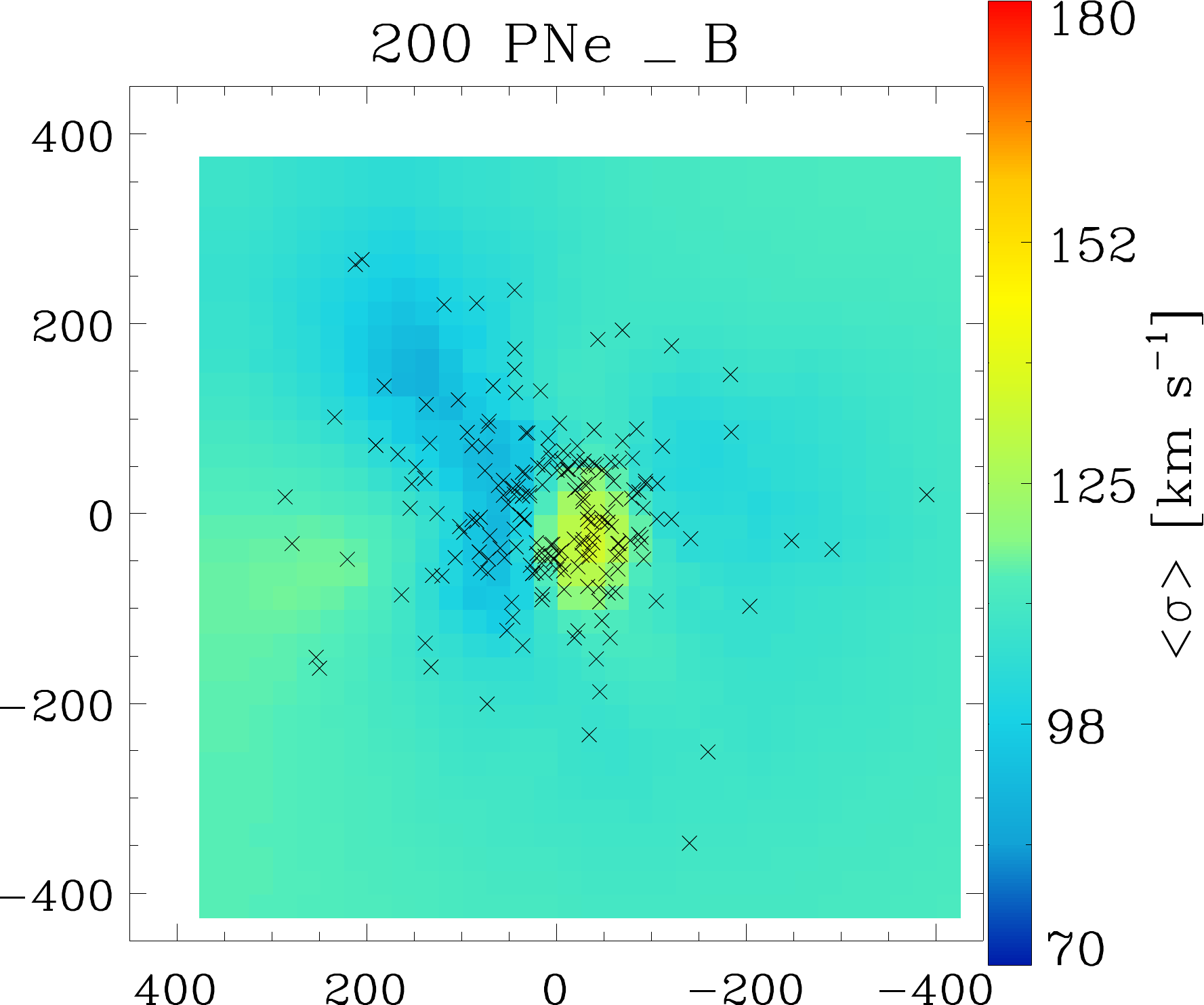}
  \end{minipage}
  \includegraphics[width=0.6\linewidth]{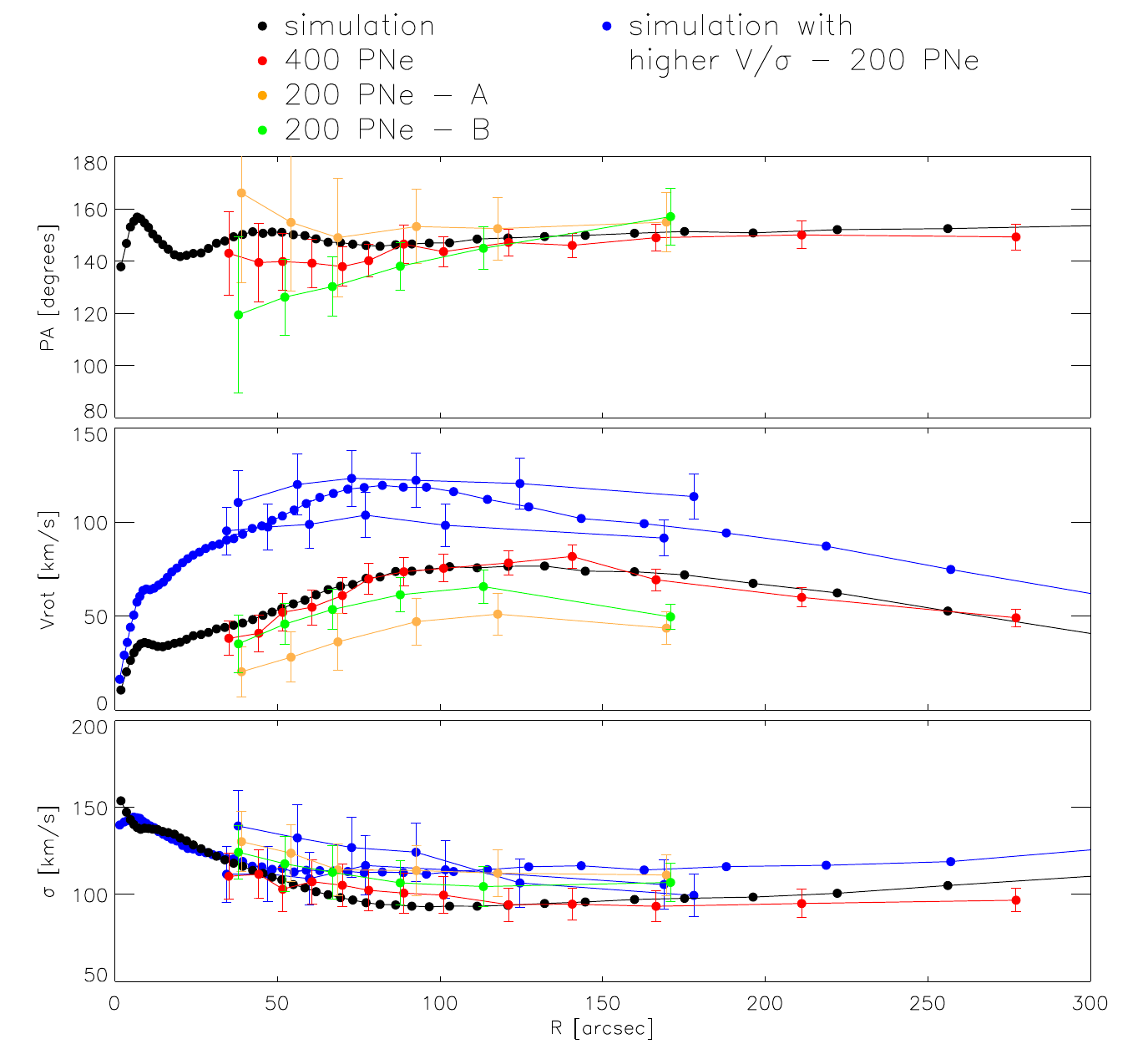}
  \caption{Smoothed velocity and velocity dispersion fields in the ePN.S field of view for a simulated ETG, traced by 10000 particles, 400, and 200 stars after the convolution with a completeness function. For the 200 particle case, we show two typical statistical realizations (A and B) of the extracted sample, to illustrate the effects of the limited statistics on the smoothed quantities.   
  The global features of the simulation are preserved in the different fields but, as expected, the spatial and velocity resolution of the local gradients is downgraded when the number of tracers decreases. The lower right panels display the fitted kinematic position angle $PA$, the rotation velocity $V_\mathrm{rot}$, and the azimuthally averaged velocity dispersion $\sigma$ in circular radial bins as function of radius. It is evident that the radial gradients are less resolved with fewer tracers, and that the amplitude of rotation for the case with 200 stars is oversmoothed. This happens because the kernel width is larger than the real spatial scale of the velocity gradients due to the sparse distribution of the tracers.  The blue full circles show a simulation with higher $V/\sigma$ ratio and two of its typical statistical realizations sampled by 200 objects to illustrate how of the accuracy of the reconstructed velocity amplitudes depends on the statistical noise. Errors for the kinematic profiles are determined from Monte Carlo simulations using the smoothed velocity and velocity dispersion fields as described in section \ref{subsubsec:Errors_on_the_fitted_parameters}.
  }
   \label{fig:models}
  \end{figure*}
  
\subsection{Tests with data from a simulated galaxy} 
\label{subsubsec:tests_on_simulations}

We used a simulated major merger remnant, sampled by 10000 particles. We reproduced a realistic observed dataset by locating the simulated galaxy at a distance of 20 Mpc, and applying to it a completeness function similar to those observed (see Arnaboldi et al. in prep.). We observed the galaxy in a field of view similar to that of the ePN.S, $800\times800$ $\mathrm{arcsec}^2$ wide centered on to the galaxy. At this point, the simulated galaxy can be used as a test case for the procedure: we can explore how the results change by varying the number of tracers at our disposal, by selecting randomly typical numbers of observed PNe; different random extractions of tracers give us different statistical realizations of the system, and different projections of the galaxy on the sky plane give us different $V/\sigma$ ratios.

Figure \ref{fig:models} shows an example of such an experiment. The smoothed velocity and velocity dispersion fields are plotted for the 10000 particle simulation, together with the maps for 400 stars and 200 stars. For the 200 particles case we show two typical statistical realizations (A and B) in order to illustrate the effects of limited statistics on the kinematic quantities.
The panels at the bottom right show the fitted  kinematic position angles, the amplitudes of the rotational velocity, and the azimuthally averaged velocity dispersion profiles in circular bins. The kinematic quantities are in general well recovered within the errors. The simulation is point symmetric in the region selected, and so are the limited statistical realizations. The statistical fluctuations that appear in the smoothed fields usually do not cause a deviation from point symmetry nor a variation of the kinematic position angle larger than the error bars. The velocity dispersion may be marginally boosted at the center of the galaxy, associated with the lack of detections, (see the case of 200 stars in figure \ref{fig:models}), but the dispersion profile is recovered within the errors. So we conclude that the procedure to calibrate the smoothing parameters through Monte Carlo simulations (section \ref{subsec:derive_A_B}) works very well in reconstructing the mean properties of the velocity and velocity dispersion fields. Even though the smoothing procedure may create artificial peaks or deeps in the velocity maps, all these features are generally within the $1\sigma$ level. Any radial trends can therefore be considered significant only if it is outside the statistical uncertainties. 

However, we can notice that as the number of tracers drops or the ratio $V/\sigma$ decreases, the amplitude of rotation is systematically underestimated, while the velocity dispersion is systematically higher \citep[see also][]{2001A&A...377..784N}.
This happens because lower numbers of stars or higher statistical noise require larger smoothing scales, which eventually become larger than the actual spatial scales of the velocity gradients. 
Oversmoothed velocity amplitudes, i.e. lower $\langle \vmeas\rangle$, imply higher $\tilde{\sigma}$ in equation \eqref{eq:sigma}.
There is no way to estimate this offsets unless the mean velocity and velocity dispersion fields of the galaxy are known from independent data, since the effects of the smoothing jointly depend on the local density of tracers, on the $V/\sigma$ ratio, and on the particular statistical realization (e.g. A and B in figure \ref{fig:models}), hence they vary from case to case. It is however safe to say that for the galaxies with limited statistics the recovered kinematics provides a lower limit to the amplitude of rotation and an upper limit for the velocity dispersion in the halos.

\subsection{Statistical tests with simulated velocity fields on real galaxies} 
\label{subsubsec:statistical_tests}

Simulated velocity fields were used to verify the results obtained for the ePN.S sample of galaxies, such as assessing the significance of deviations from point symmetry or the radial trends of the kinematic parameters. These simulated velocity fields are featureless and trend-less (fixed kinematic position angle and constant velocity dispersion) and are built for each galaxy using the positions of the observed PNe. This means that the incompleteness function is already built into the simulated catalogs and is identical to that for the real galaxy under study.  
In this way we can check how the smoothing procedure correlates spatially close PNe, as the degree of spatial correlation in the models will be identical to that of the real datasets.

The simulated velocity field at the positions $(x_i,y_i)$ of the ith PNe, from the real datasets, is interpolated with the simple function 
\begin{equation}
 V(\phi_i, R_i)=V_\mathrm{rot}\cos(\phi_i-\PAphot) \arctanh(R_i/h)
\end{equation} 

\noindent
where $\phi_i$ is the position angle of the ith PN with coordinates $(x_i,y_i)$, $R_i$ its distance from the center, and $h$ is a scale-length that defines the steepness of the central velocity radial gradient. $V_\mathrm{rot}$ is chosen to be equal to the maximum fitted rotational amplitude for the galaxy in consideration, while $h$ has the dimension of the central void of the detections ($\sim 50$ arcsec). We choose to align the rotation of the simulated velocity field with the real photometric position angle of the galaxy, in order to simulate the velocity field of a regular, point-symmetric disk.
The model velocity field $V(x_i,y_i)$ is sampled with a constant dispersion $\sigma$, which is equal to the mean velocity dispersion of the galaxy under study. This is done by extracting random values from a Gaussian distribution centered at 0 and with dispersion $\sigma$, and adding them to $V(x_i,y_i)$. 
The assumption of constant dispersion is certainly unrealistic for the central regions of the galaxies, but it is reasonable at large radii.
We produced 100 statistical realizations for each galaxy, and each of those is treated as a real dataset for obtaining a total of 100 smoothed velocity fields. 

The smoothed velocity fields obtained in this way were used to assess whether an observed feature in the galaxy velocity or velocity dispersion field is real, by studying whether it can be produced by statistical effects in a featureless velocity field, and how typically this happens (see sections \ref{subsec:test_point-symmetry} and \ref{subsec:velocity_dispersion_profiles}). Also, the 100 models can provide a statistic for the fitted kinematic parameters (equation \ref{eq:fourier3}) to check the probability that statistical noise combined with the smoothing procedure artificially produce effects like twists of the kinematic position angle or misalignments with the photometric axes (section \ref{subsec:triaxiality}).  
 
\section{Photometric profiles and models}
\label{sec:appendix_photometry}
 In this appendix we report extended photometric data from the
literature for three quarters of the ePN.S galaxies to compare with
the PN kinematic analysis of this paper, as discussed in section
\ref{subsec:DISCUSSION_signatures_of_triaxiality_in_photometry}. Furthermore
we present illustrative photometric models which reproduce the main
trends seen in the photometric profiles of the galaxies. 

\subsection{Photometric profiles from the literature}
\label{sec:photometric_profiles}

  Figure \ref{fig:photometry} shows ellipticity and $\PAphot$
  profiles from the literature for the subsample of the ePN.S galaxies
  (references in table \ref{tab:smoothing_parameters}) with
  photometric data reaching at least 4 $\re$, excluding the merger
  remnants NGC 1316, NGC 4472, and NGC 5128. For three galaxies,
  NGC 0584, NGC 2768, and NGC 5866, we marked the outermost photometric measurements
  that may be affected by perturbations from ongoing interactions (see appendix \ref{sec:notes_on_single_galaxies}). For NGC 1023 and NGC 3384 the presence of multiple components like e.g. the bar affect the $\PAphot$ and ellipticity profiles. Therefore for these galaxies we flagged the corresponding regions, as described in appendix \ref{sec:notes_on_single_galaxies}. These regions are highlighted with open symbols and black vertical lines in figure \ref{fig:photometry}.
  
  The $\PAphot$ profiles in figure \ref{fig:photometry} show more or less pronounced variations with
  radius for most of the galaxies.  All SRs have isophote twists, typically $\sim 10-40$
  degrees with a very large twist in NGC 4374. For the fast rotating galaxies classified as triaxial
  in section \ref{subsec:triaxiality} photometric twists range from $\sim 3-20$ degrees with the
  largest twist in NGC 4494 ($\sim22$ degrees).  For the other FRs the photometric twists range from $\sim 0-28$
  degrees, with two galaxies NGC 4278 and NGC 0584 having twists of $\sim28$ degrees. The latter objects are consistent with axisymmetry at the
  resolution of the kinematic survey, but the photometry suggests that some of them might be
  triaxial as well.

  The ellipticity profiles show that while most of the SRs
  become flatter in the outskirts, the opposite is true for the FRs.
  Most of the FRs shown in figure
  \ref{fig:photometry} have decreasing ellipticity profiles at large
  radii, approximately at the distance where $V_\mathrm{rot}$ is
  observed to drop. The vertical lines indicate the radial range of
  the kinematic transition discussed in section
  \ref{subsec:DISCUSSION_kinematic_transition_radius}. This is also evident from the distribution of the maximum ellipticity and of the outermost measured values in figure \ref{fig:maxtwist_maxell}, and it is
  consistent with a picture where the FRs are dominated by a
  disk component in the central regions, embedded in a rounder, dispersion
  dominated outer component which is triaxial for a fraction of the
  sample. The exceptions to the decreasing trend in ellipticity at
  large radii are NGC 1344, a prolate rotator in the halo, NGC 3379
  and NGC 4494 which are both very round at all radii, and NGC 2974, NGC 3384 and NGC 4339 for which the small number of tracers does not allow us to resolve
  any kinematic transition.

\subsection{Photometric models of a FR with triaxial halo}
\label{sec:photometric_models}

  Here we describe triaxial photometric models for FR
  galaxies which explain the magnitude of the photometric twists seen
  in figure \ref{fig:maxtwist_maxell}, and which are used in the 
  discussion in section
  \ref{subsec:DISCUSSION_signatures_of_triaxiality_in_photometry}.
  These simple models are not meant to provide an exact description of
  the intrinsic shape of a galaxy. Their purpose is to obtain a
  reasonable representation of the structure of FRs with triaxial halos, to derive what is the maximum isophotal twist expected, and
  compare it with the observations. 

  To build these models we use the results of previous studies on the
  intrinsic shapes of FRs that constrain their central
  regions. \citet{2014MNRAS.444.3340W} and \citet{2017MNRAS.472..966F}
  found that FRs are close to oblate ($p\sim1$), and rather
  flat ($q\sim0.3$). Guided by the PN kinematic results, we assume
  that the intrinsic shape of the FR model changes from flat
  and oblate in the innermost regions, to rounder and mildly triaxial
  beyond the transition radius in the outer regions, as discussed in section \ref{subsec:DISCUSSION_signatures_of_triaxiality_in_photometry}. \citet{2014MNRAS.444.3340W} and
  \citet{2017MNRAS.472..966F} found that SRs have
  $q\sim0.6$, but the p value is not well constrained (their analysis
  failed to converge). The main reason for this is the rather small
  number of SRs in the $\mathrm{ATLAS^{3D}}$ and in the SAMI
  samples. In addition, as argued by \citet{2017MNRAS.472..966F}, the
  family of SRs comprises both truly spheroidal pressure
  dominated systems and flattened non-regular rotators, which might
  prevent the analysis to converge.  
 
  We present here four different photometric models, that have
  similar axis ratios in the central regions but
  different halos intrinsic shapes, as well as different realizations in how the transition between central-oblate regions and triaxial halos occurs:
 
  - the \emph{S\'ersic model} (Model 1) is a one component model built
  using the three-dimensional S\'ersic deprojection proposed by
  \cite{1999MNRAS.309..481L} with S\'ersic index $n=4$ and variable
  flattening.  The effective radius of the two-dimensional profile is
  set equal to the average of the observed $R_e = 50$ arcsec.  The
  transition radius is also the average of the measured values for the
  FRs: $R_T = 1.8 R_e$. We used $\mathrm{(q = 0.3, p = 1)}$
  for the inner axis ratios, as measured by \cite{2014MNRAS.444.3340W}
  and \citet{2017MNRAS.472..966F}. For the outer regions we choose
  $\mathrm{(q=0.6, p=0.9)}$, close to values found by
  \cite{2014MNRAS.444.3340W} for the SRs.  The variation
  between inner and outer values of flattening is modeled by an
  arctangent function; the width of the radial transition is chosen to
  be 50 arcsec (mean $\Delta R_T$, section
  \ref{subsec:DISCUSSION_kinematic_transition_radius}) .
 
  - the \emph{cored S\'ersic halo plus disk model} (Model 2) is a two
  component model built from the sum of deprojections of an $n=4$
  spheroid and $n=1$ spheroidal-exponential disk as in
  \cite{1999MNRAS.309..481L}, but we added a core with radius
  $R_\mathrm{core} = R_T = 1.8 \times 50$ arcsec to the deprojected $n=4$ profile. 
  Both the spheroid and the disk have constant flattening:
  $\mathrm{(q=0.6, p=0.9)}$ for the former, $\mathrm{(q=0.3, p=1)}$
  for the latter. The horizontal scale-height of the disk is such that 
  $1.67 h = R_T$, i.e. the transition radius is equal to the
  half-light radius of the disk, as qualitatively observed in section
  \ref{subsubsec:embedded_disks}. The effective radius of the
  two-dimensional profile of the spheroid is $R_e=150$ arcsec.
  The luminosity density of the disk is scaled so that it equals the
  luminosity density of the spheroid at $r=R_T$, hence in this model
  the disk component dominates the luminosity density from the center
  out to $R_T$. This choice makes the total integrated light of the
  disk inside $r=R_T$ circa $3.5$ times the total luminosity of the
  spheroidal component inside $R_T$. The intrinsic axis ratios of the
  spheroid are $\mathrm{(q=0.6,p=0.9)}$, those of the disk are
  $\mathrm{(q=0.3, p=1)}$. 
  
  - the \emph{maximally triaxial cored S\'ersic halo plus disk model} (Model 3) is built as Model 2 but with axis ratios $\mathrm{(q=0.6,p=0.8)}$ for the halo.
  
  - Model 4 has a \emph{strongly flattened triaxial halo} plus disk. The model is built as Model 2 with a cored S\'ersic halo, but the axis ratios are $\mathrm{(q=0.4,p=0.8)}$.
  
  Each model is observed at 100 random viewing angles. We chose a
  coordinate system with axes $\mathrm{(x,y,z)}$ aligned,
  respectively, with the major, the intermediate, and the minor axis
  of model ellipsoids, and called $\phi$ and $\theta$ the azimuthal
  and polar angles that define the direction of the line-of-sight
  (LOS). We choose the LOS by random samplings of the solid
  angle centered on the center of the ellipsoids
  $d\Omega = \sin\theta d\theta d\phi$. This makes the probability of
  observing the galaxy face-on (i.e. $\theta=0$) lower with respect to
  edge-on ($\theta=90$).  The models, projected on a plane orthogonal
  to the LOS, are fitted with ellipses using the IRAF task ELLIPSE
  \citep{ 1993ASPC...52..173T}, to derive surface brightness,
  ellipticity and position angle profiles. 
 
\begin{figure*}[!h]
 \begin{center}

   \includegraphics[width=0.27\linewidth]{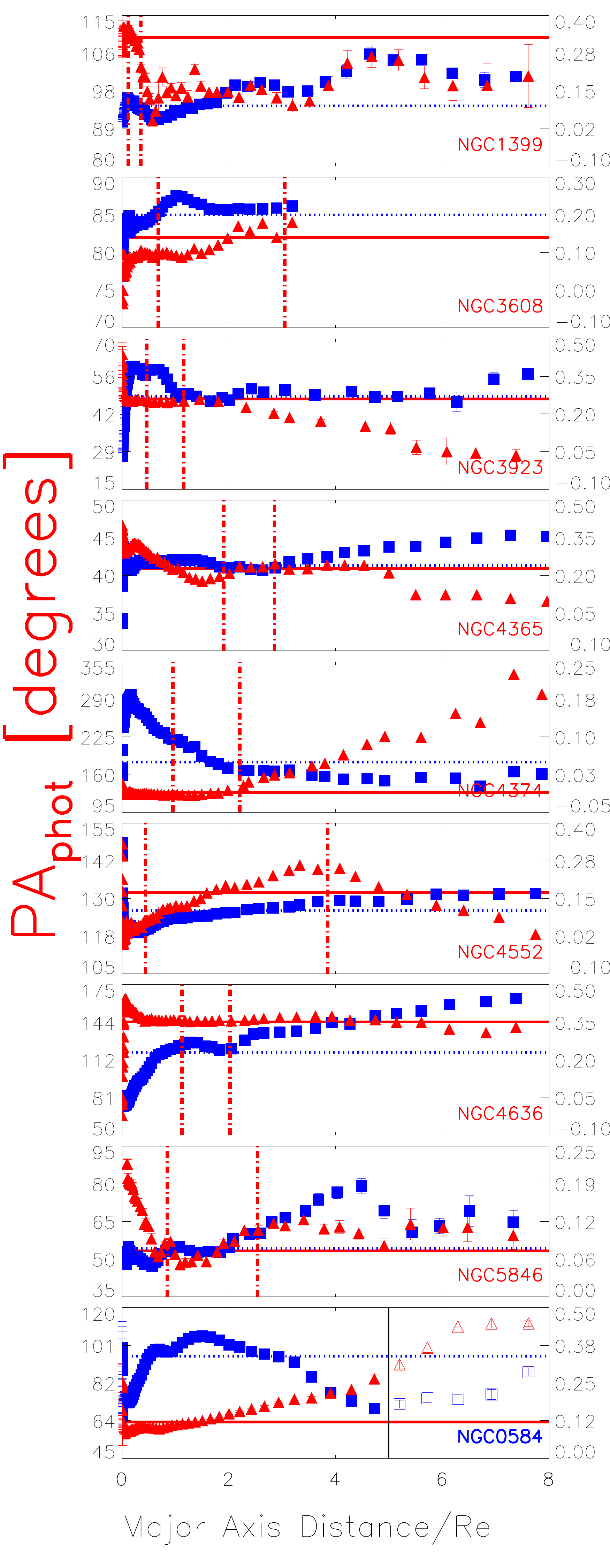}
   \includegraphics[width=0.27\linewidth]{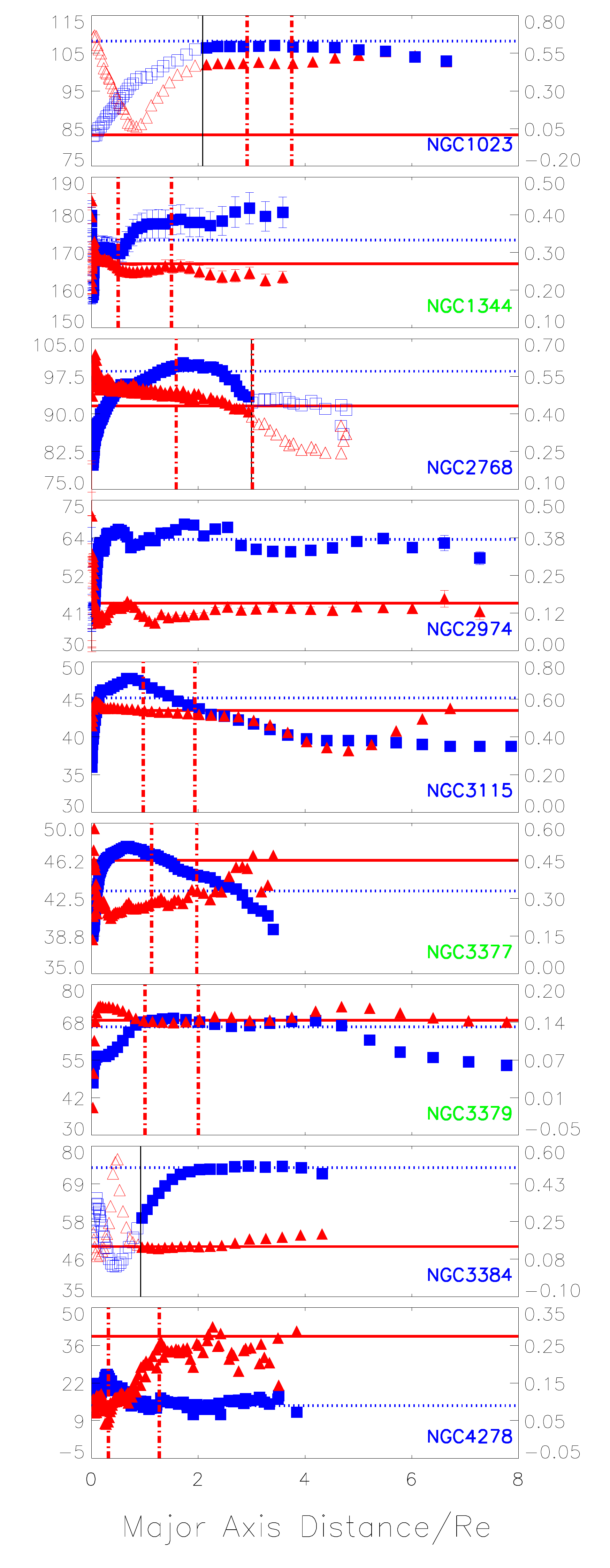}
   \includegraphics[width=0.27\linewidth]{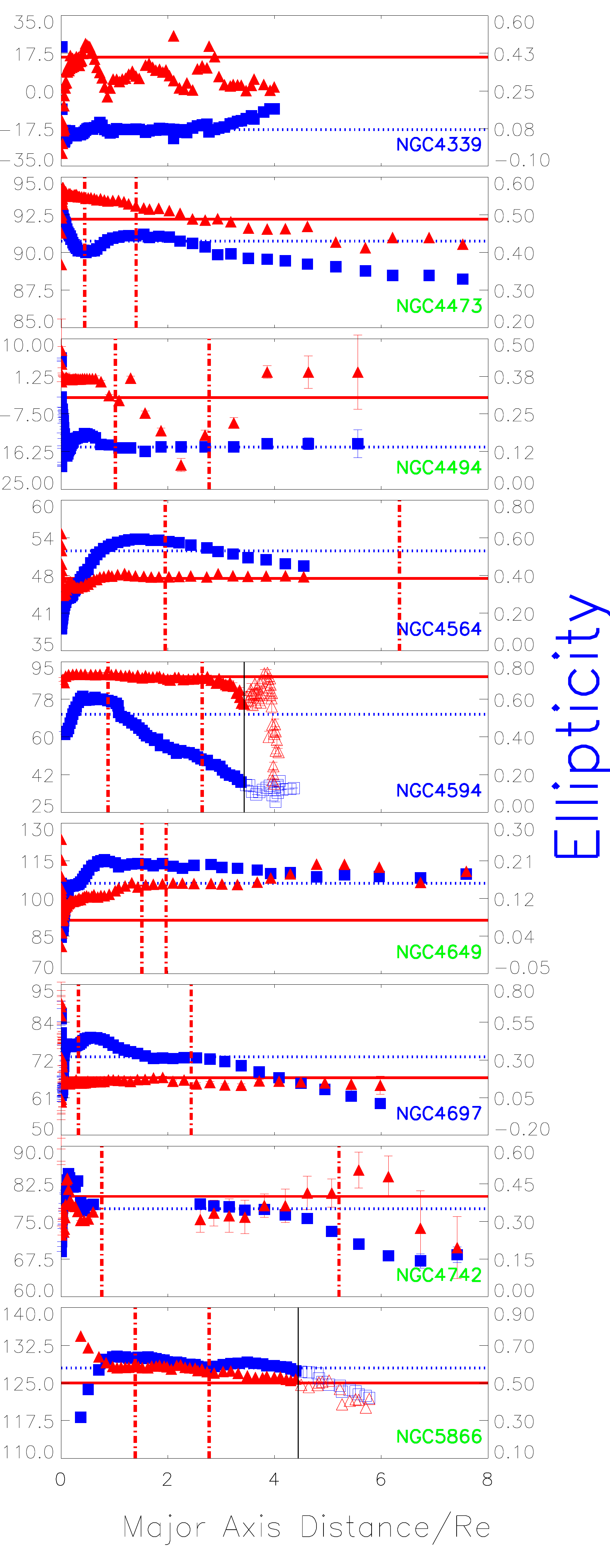}	 
  
 \end{center}
   \caption{ Position angle (red triangles, y axis on the left) and ellipticity profiles (dark blue
     squares, y axis on the right) for the subsample of the ePN.S galaxies with literature
     photometry reaching at least $4R_e$ (references in table \ref{tab:smoothing_parameters}). The
     NGC numbers of the galaxies are shown in red for the SRs, in blue for the FRs, and in green for the FRs with triaxial halos. The open symbols and the black vertical line for NGC 0584, 2768,
       5866 mark regions that are probably affected by ongoing interactions, and for NGC 1023 and NGC 3384 the regions affected by the presence of other photometric components, e.g. the bar (see appendix \ref{sec:notes_on_single_galaxies}). For NGC 4594 we excluded the data beyond 350" where the small ellipticity corresponds to large uncertainties in the $\PAphot$ measurements. The
     horizontal lines report the $\PAphot$ (solid line) and $\epsilon$ (dotted line) with values
     listed in table \ref{tab:galaxies}. The red dot-dashed vertical lines show the kinematic
     transition range $R_T\pm\Delta R_T$ from section
     \ref{subsec:DISCUSSION_kinematic_transition_radius}. }
 \label{fig:photometry}
   \end{figure*}

\begin{figure*}[b]
   \begin{center}  
     \includegraphics[width=0.29\linewidth]{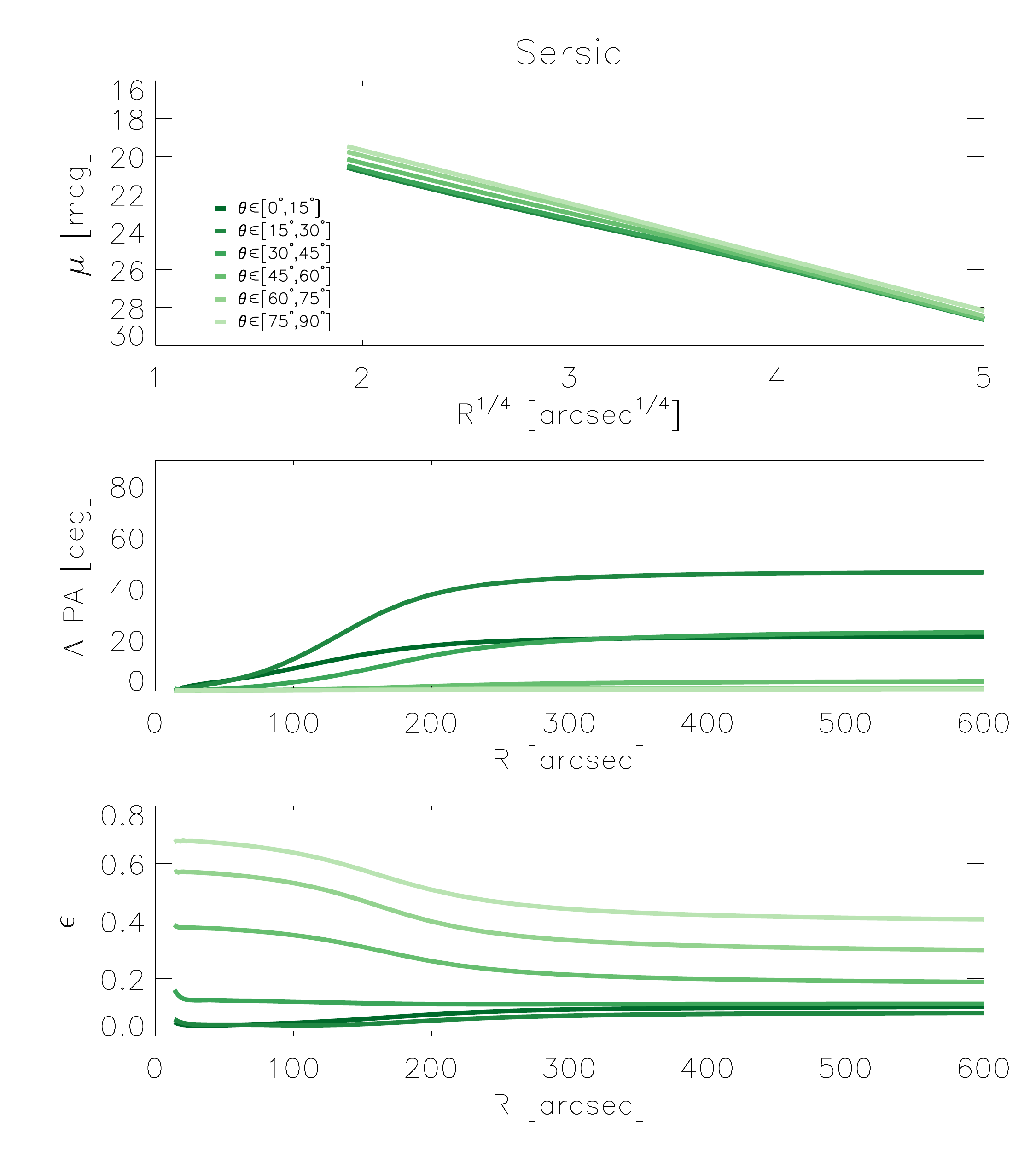}
     \includegraphics[width=0.29\linewidth]{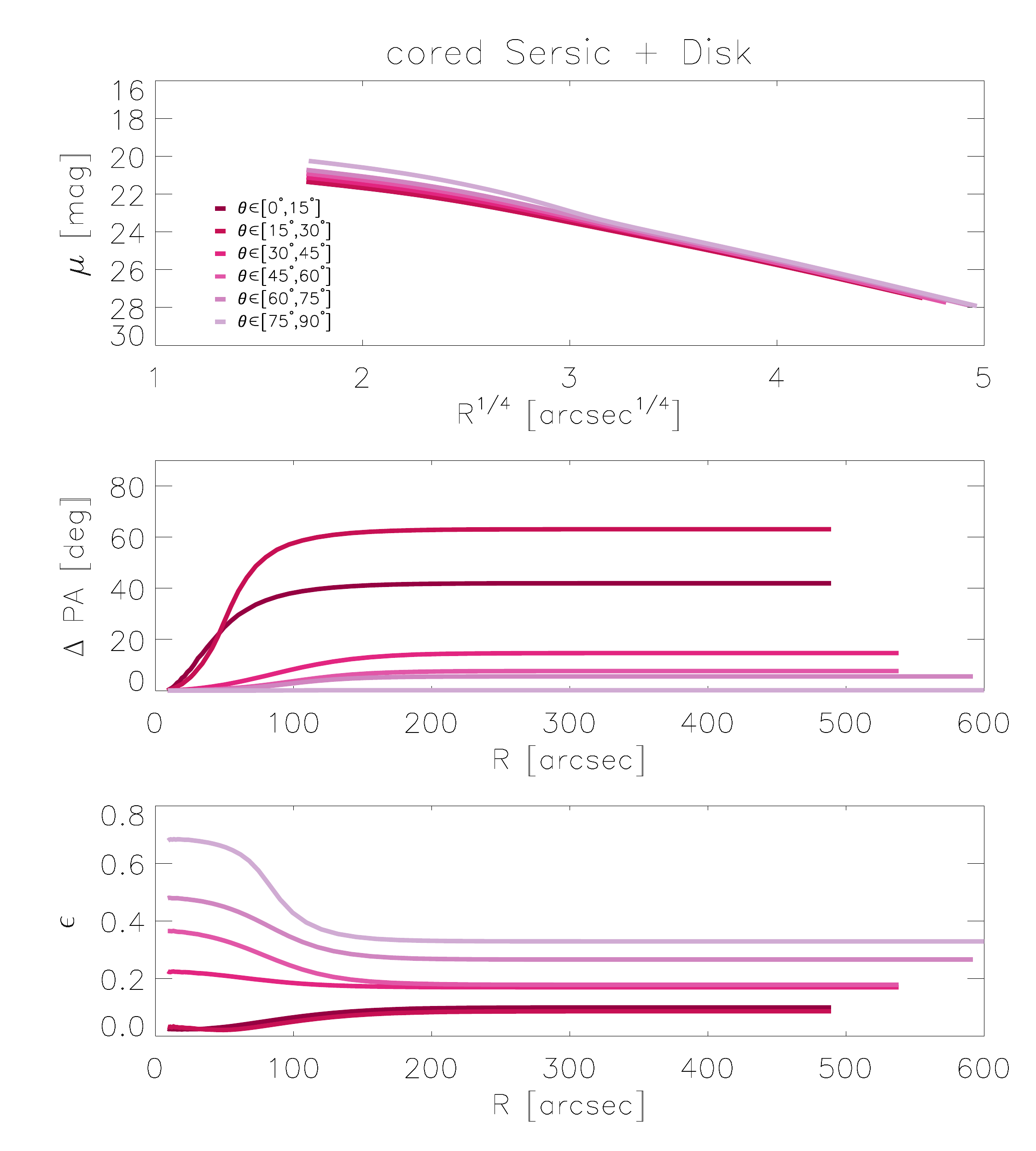}
     \includegraphics[width=0.29\linewidth]{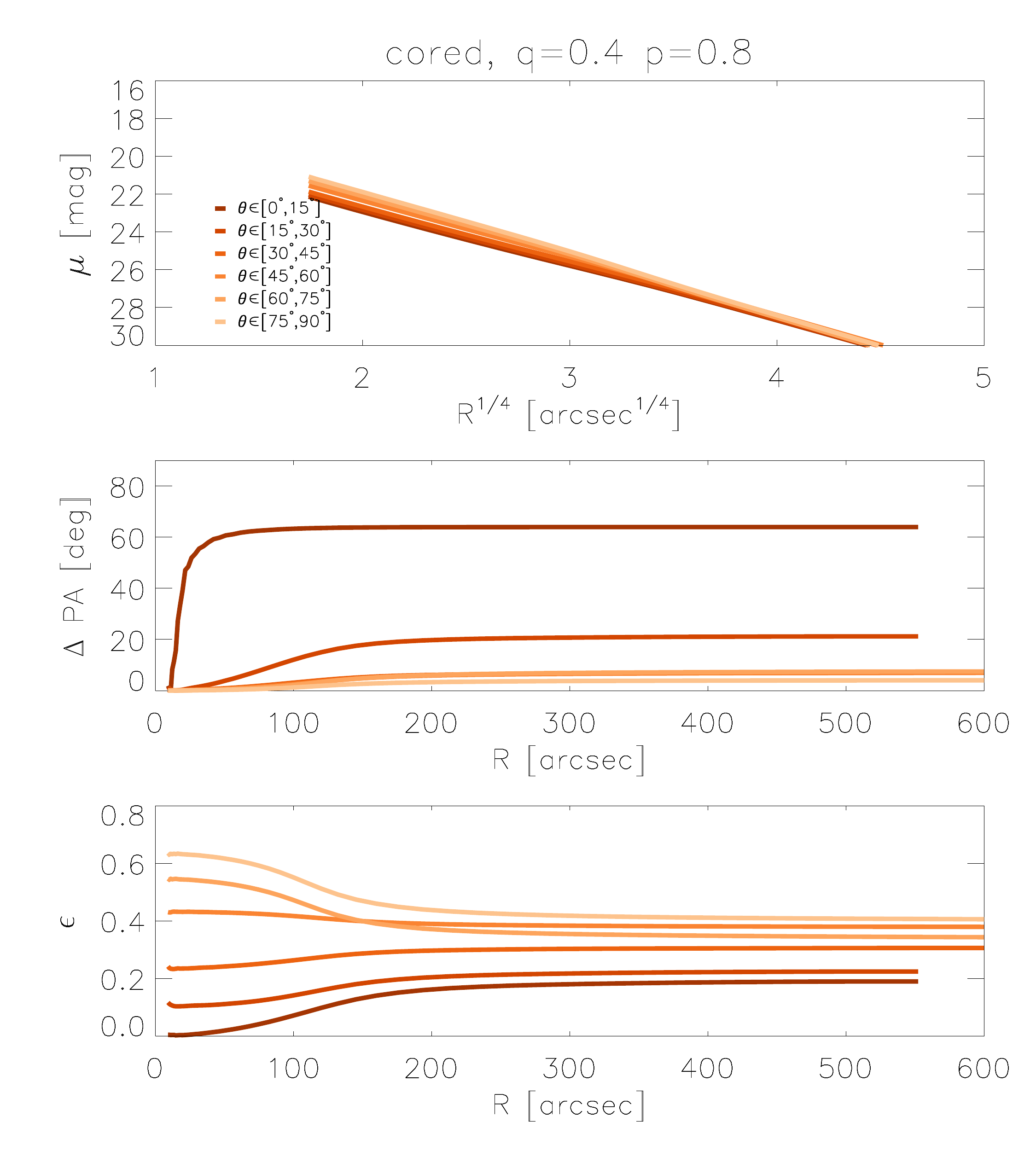} 
   \end{center}
   \caption{ Surface brightness, position angle, and ellipticity
     profiles for three photometric models described in this
     appendix \ref{sec:photometric_models}, at different inclinations
     $\theta$ ($\theta=0$ for face-on, $\theta=90$ edge-on).  For each
     photometric model, we show one random projection in each of six
     intervals $\Delta \theta = 15$ degrees, starting from near
     face-on ($\Delta\theta=0-15$ degrees, darkest color profile), through
     $\Delta\theta=15-30$ degrees, etc., to near edge-on
     ($\Delta\theta=75-90$ degrees, lightest profile). These $\Delta\theta$
     intervals represent 3\%, 10\%, 20\%, 19\%, 21\%, 27\% of 100
     random directions, from face-on (darkest) to edge-on (lightest
     profiles).}
 \label{fig:photometric_models}
 \end{figure*}

  Figure \ref{fig:photometric_models} shows the result of this
  analysis on three models for 6 different values of $\theta$ in
  bins of 15 degrees. The shapes of the model ellipticity and position
  angle profiles are qualitatively similar to the observed profiles in
  figure \ref{fig:photometry}. Large photometric twists ($>10$
  degrees) are observed for viewing angles $\theta \lesssim 50$ degrees for Models 1 and 2, and for $\theta \lesssim 65$ degrees for Models 3 and 4. The rate of occurrence of large twists is of order $\sim 30\%$ for our models (45\% for Model 3), as shown also in
  figure \ref{fig:maxtwist_maxell} in section
  \ref{subsec:DISCUSSION_signatures_of_triaxiality_in_photometry}
  where we report the maximum twist versus the mean ellipticity. The distribution of the models in this diagram represent well the location of the ePN.S FRs, for which we observe a fraction $37\%$ having photometric twists larger than $10$ degrees. This analysis based on simple triaxial photometric
  models shows that the presence of small photometric twists, as
  typically observed for most of the FRs, is consistent with
  the presence of a triaxial halo, i.e. it is likely that 
  detailed individual models of these galaxies can be constructed that are consistent
  with the measured kinematic and photometric twist angles.

\section{Notes on single galaxies and comparison with the literature}
\label{sec:notes_on_single_galaxies}

\subsection{Fast rotators} 
\label{subsec:notes_FR}
\begin{itemize}

\item {\bf NGC 0584}. This is the brightest member of a small cluster
  containing mostly E galaxies \citep[][]{1961AJ.....66..562V}. It is
  classified as a E4 in the NED catalog, but seems to contain a
  disk structure \citep[][]{1994A&AS..105..481M}.  Our PN kinematic data
  show that the galaxy halo rotates with at least $\sim 50 $ \kms along
  the photometric major axis, and has a velocity dispersion of
  $\sim 100 $ \kms. The galaxy has central ellipticity $\epsilon\sim0.4$ which
  falls to $\sim0.1$ at $150''\sim 5R_e$ from whereon $\epsilon$  increases
  again. The photometry shows pronounced isophote twist starting
  at $\sim 1R_e$, which steepens beyond $5R_e$. Both the outer increase
  in ellipticity and the rapid outer isophote twist may be related to 
  the nearby companion galaxy, NGC 0586; therefore we consider only the
  region where the ellipticity decreases as intrinsic to the galaxy.
 
\item {\bf NGC 0821}. This is a field E6 galaxy. The $\mathrm{ATLAS^{3D}}$ velocity field
  \citep[][]{2004MNRAS.352..721E} maps the inner regions of the galaxy where it behaves like a
  regular rotator along the photometric major axis. The PNe sample, instead, probes the kinematics
  from $1$ to $5 \re$ and show a sustained rotation along the photometric minor axis with a maximum
  amplitude of $51 \pm 16$ \kms, suggesting the triaxiality of the object. No rotation is detected
  along the major axis.  The long-slit data from \citet{2009MNRAS.398...91P} and the integral field
  map of \citet{2014ApJ...791...80A} agree in detecting rotation along the photometric major axis,
  with decreasing amplitude ($80$ \kms at $0.3\re$ to very slow rotation at $R<2\re$). They do not
  detect any rotation along the minor axis, but their data do not extend far beyond $1.5\re$ along that
  axis. The kinemetry of \cite{2016MNRAS.457..147F}, based on the same data used by
    \citet{2014ApJ...791...80A} but with a more careful treatment of outliers, shows a radial twist
    of the $\PAkin$ between 40 to 60 arcsec towards the same position angle measured from the PNe.
  Long slit data from \citet{2009MNRAS.398..561W} and \citet{2010ApJ...716..370F} also show some
  minor axis rotation at large radii. PN data are consistent with the kinematics of the blue GCs
  \citep[see][]{2013MNRAS.428..389P}, which are found to rotate at $\sim 85$ \kms along a direction
  consistent with the photometric minor axis. The red GCs instead appear to faintly counter-rotate
  with respect to the host galaxy stars.  The outer gradient of the velocity dispersion radial
  profile has been debated in the literature, with implications on the inferred dark matter content
  of the galaxy. The PN data combined with the long slit stellar kinematics show a decrease of the
  velocity dispersion as a function of radius, consistently as both red and blue GCs show, while
  \citet{2009MNRAS.398...91P} and \citet{2010ApJ...716..370F} found a flat profile within 100
  arcsec.
 
\item {\bf NGC 1023}. This is a barred lenticular galaxy (SB0),
  brightest member of a group of 13 galaxies
  \citep[][]{2001ApJ...546..681T}. It has a small companion NGC 1023A, which is probably interacting with the main galaxy \citep{1984MNRAS.210..497S}, but it is likely too small to cause significant disruptions in the dynamical state of the main galaxy \citep{2008MNRAS.384..943N}. The $\mathrm{ATLAS^{3D}}$ velocity
  map \citep[][]{2004MNRAS.352..721E} covers the innermost regions of
  the galaxy, up to half effective radii, and displays rotation with maximum amplitude of $\sim120$ \kms and strong twist of the zero-velocity line. The
  stellar kinematics along the major axis of
  \citet{2002MNRAS.332...65D} and \citet{2017MNRAS.471.4005B} show
  that the rotation continues to increase mildly in the outer regions
  exceeding $200$ \kms beyond $2\re$. Both the PN and SLUGGS
  \citep[][]{ 2014ApJ...791...80A} velocity maps show decreasing
  rotation beyond $3\re$, \citep[see also][]{2008MNRAS.384..943N,
    2009MNRAS.394.1249C,2013MNRAS.432.1010C}. The galaxy does not show
  any twist of the kinematic position angle or any misalignment with
  the photometric axis. The non zero third order harmonics amplitudes
  describe a rather cylindrical velocity field. The velocity
  dispersion map shows minimum values ($\sigma\sim40$ \kms) at the
  location of the disk, while the outermost (bulge) PNe are hotter
  ($\sigma\sim120$ \kms). The photometric position angle profile is influenced by the presence of the bar at $R< 100"$ \citep{1975A&A....42..103B}, therefore we excluded these regions from our discussion of the photometric twists.

 \begin{figure*}[ht]
 \begin{center}
  \includegraphics[width=8cm]{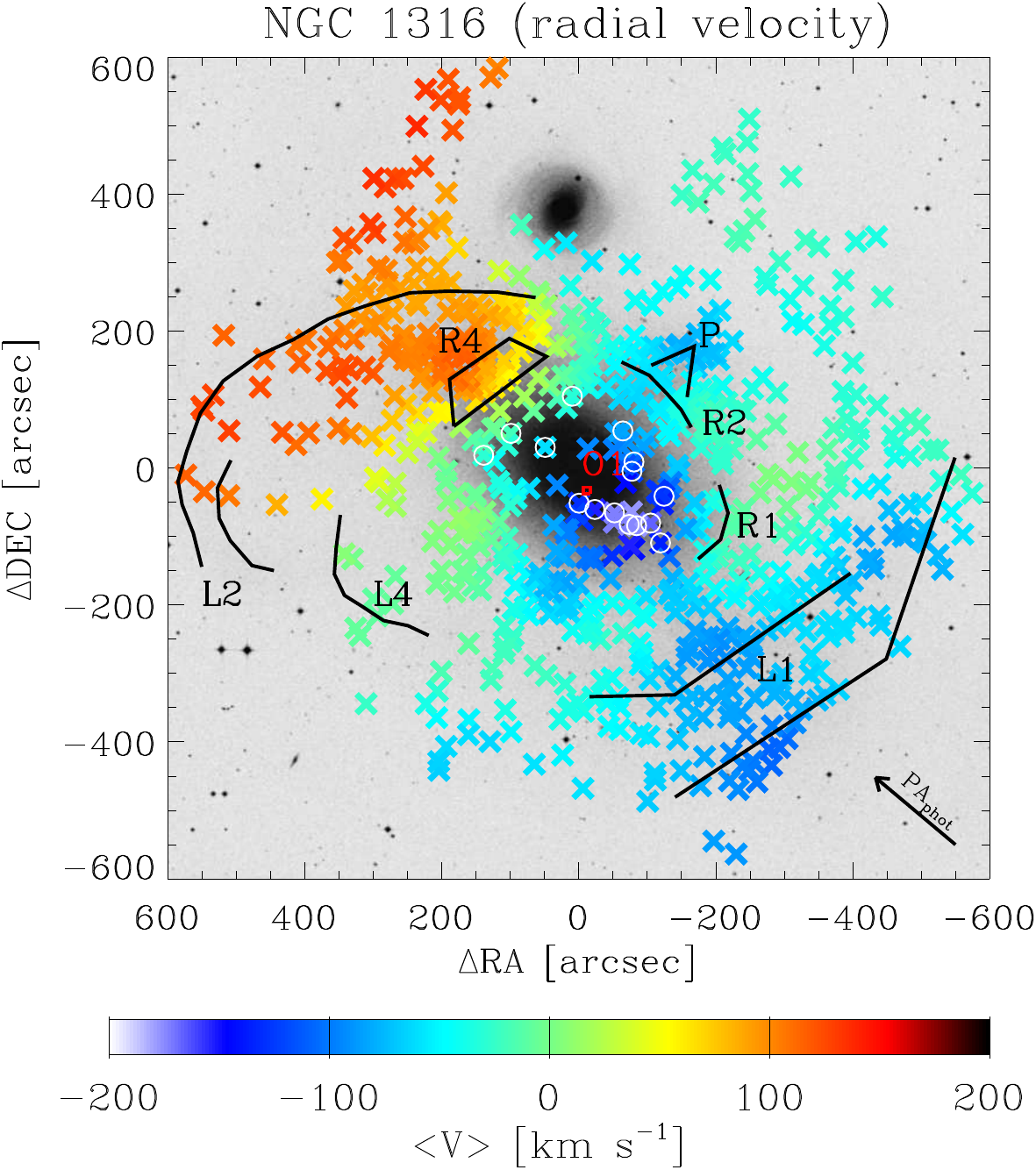}
  \hspace{0.1cm}
  \includegraphics[width=8cm]{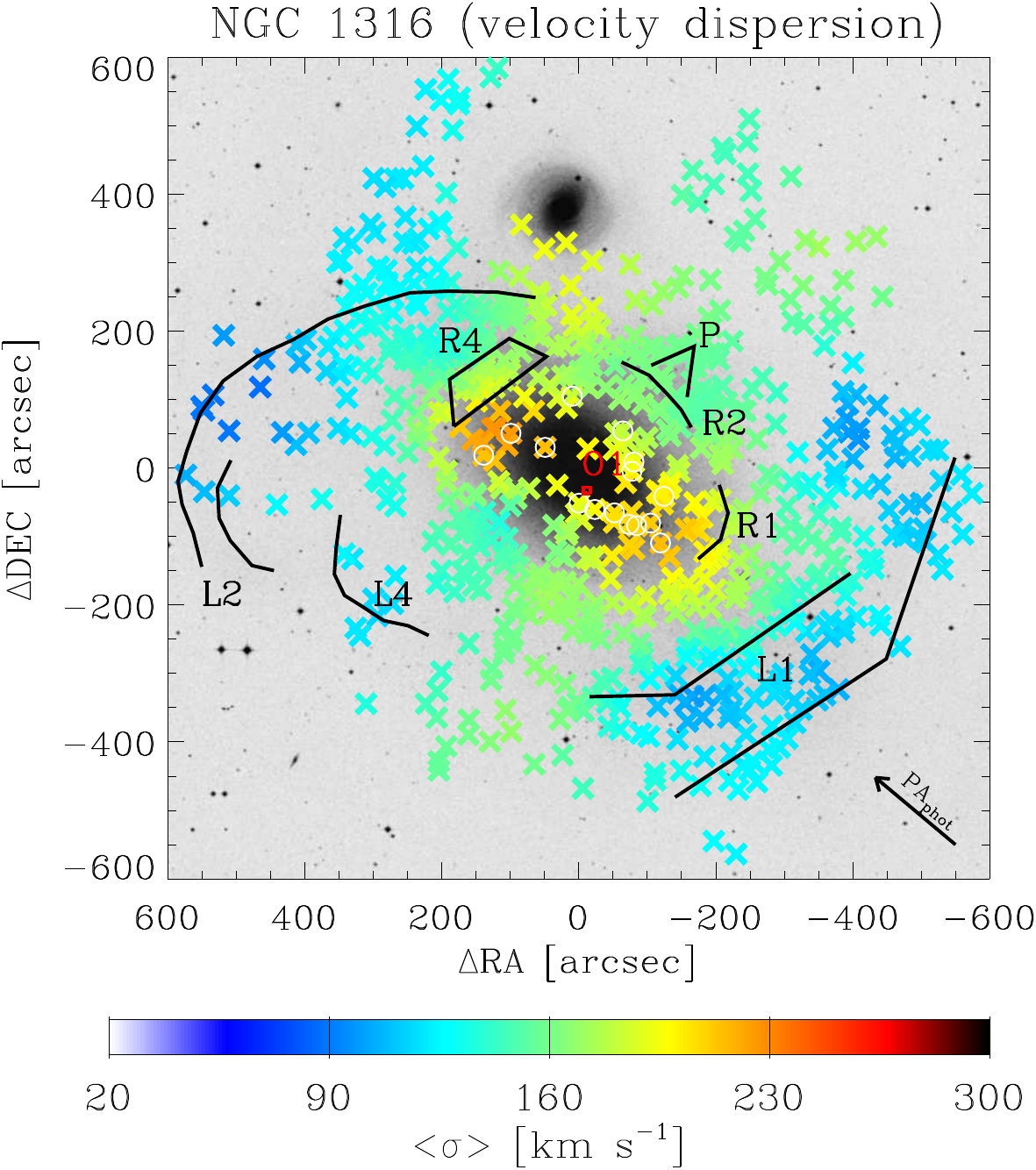}
\caption{Smoothed velocity field and velocity dispersion field of NGC 1316 shown on the DSS image; north is up, east is left. The circles highlight the PNe belonging to a structure in the phase space, discussed in the text. The black contours trace features in the surface brightness distribution \citep[designations from][]{1980ApJ...237..303S}. In red is marked the position of the feature O1 observed by \cite{2014A&A...569A..41R}. The arrow at the bottom right shows the direction of the $\PAphot$.}
\end{center}
\label{fig:N1316}
\end{figure*}

 \item {\bf NGC 1316}. This is a giant elliptical (also classified as a peculiar S0), member of the Fornax cluster (Fornax A). 
 NGC 1316 is a known merger remnant. Its photometric substructures were mapped for the first time by \cite{1980ApJ...237..303S}, and were recently studied by \cite{2017ApJ...839...21I}. The PN velocity field is highly asymmetric, showing that the galaxy is in a non-equilibrium phase. The modeling with a point-symmetric rotation model gives an approximate description of the properties of the galaxy. The fit of the $a_0 (R)$ parameter of the rotation model (see equation \ref{eq:fourier_definitivo}) in radial bins shows the presence of peculiar bulk motions in groups of PNe.
 There is no kinematics from integral field spectroscopy in literature, and the comparison with the long-slit observations from \citet{2006MNRAS.371.1912B} shows general good agreement with our data in the regions of overlap. 
 The PNe in the halo of NGC 1316 (figure \ref{fig:N1316}) reveal that the galaxy rotates along the major axis with an amplitude of $\sim65$  \kms, as already observed by  \citet{1998ApJ...507..759A,2012A&A...539A..11M}. The structures in the velocity field can be spatially associated with corresponding surface brightness features in the same location. There is an overdensity of PNe on the north-eastern part, co-located spatially with the ripple R4 \citep[the designations of the photometric features are from ][and their positions are indicated in black in figure \ref{fig:N1316}]{1980ApJ...237..303S}, with velocities that do not have a point symmetric counterpart. Other groups of PNe that result in non-point-symmetric velocities are those possibly related with the loop L1, on the south-west side, and to the plume P in the north-west. 
 Figure \ref{fig:point_sym} shows that the PNe of NGC 1316 in the innermost radial bin ($\sim 100$ arcsec) have a velocity offset of $\gtrsim150$ \kms with respect to the systemic velocity. This is produced by a group of relatively slower PNe with mean velocity $\sim1325$ \kms and dispersion $\sim60$ \kms \citep[highlighted in figure \ref{fig:N1316} with white circles, along with the photometric feature O1 found by][in red]{2014A&A...569A..41R}. These PNe are not associated with any substructures in the light, and the fact that they are very localized in the phase space challenges (but does not rule out) the possibility that they could be contaminants. More likely this kinematic feature is the result of a combination of spatial incompleteness and differential absorption from localized dust in the galaxy, producing local lacks of detections in the phase space and so more negative smoothed velocities $\tilde{\mathrm{V}}$. A more detailed study is beyond the scope of this paper.

\item {\bf NGC 1344}. This galaxy, also known as NGC 1340, is an E5
  elliptical belonging to the Fornax cluster. It is characterized by
  the presence of internal and external concentric shells \citep[][]{1980Natur.285..643M}, as consequences of a recent merger
  activity.  The smoothed velocity field does not deviate from point
  symmetry, however it shows that the galaxy in the outskirts rolls
  around the photometric major axis.  The velocity dispersion field
  shows a gently decreasing profile.  Unfortunately there is no
  integral field stellar kinematics available for a comparison with
  the PNe kinematics.  The long slit stellar kinematics
  \citep[][]{2005ApJ...635..290T} show sustained rotation along the
  photometric major axis that reaches $\sim 100$ \kms at $1.3 \re$ and
  a decreasing velocity dispersion.  The $\sigma$ values from the
  spectroscopy are in agreement with those found with the PN data. We
  do not detect such an strong rotation, probably because of the
  different radial coverage of the PN data.

\item {\bf NGC 2768}. This S0 is a field galaxy \citep[][]{2004ApJ...607..810M} or member of a poor
  group \citep[][]{1992A&AS...93..211F}.  NGC 2768 has a cylindrical velocity field
  \citep[][]{2004MNRAS.352..721E} whose amplitude remains constant at around $\sim 130$ \kms from
  the innermost regions \citep[][]{2004MNRAS.352..721E} to the outskirts \citep[see][]{
    2009MNRAS.398...91P, 2013MNRAS.428..389P}. Our velocity map is in good agreement with that
  presented by \citet{2014ApJ...791...80A} and extends it beyond 6$\re$.  There is some discrepancy
  between the PN $V_\mathrm{rot}$ and the kinemetry fit from \cite{2016MNRAS.457..147F} in the
  innermost radial bins where the disk strongly dominates. This is due to the fact that the smoothed
  PN velocity field in these bins is an average over the fast rotating PNe in the disk and the more
  slowly rotating PNe from the spheroid. A kinematic study of the different components of this
  galaxy was published by \cite{2013MNRAS.432.1010C}. At larger radii ($r\gtrsim3\re$) the
  contribution of the disk weakens and the PN kinematics trace the spheroid. In this region we find
  a constant sustained ($V\sim130$ \kms) rotation along the photometric major axis. The velocity
  dispersion profile is flat at $\sim140$ \kms.  The PN smoothed velocity field of NGC 2768 shows
  localized, small scale deviations from point symmetry. These asymmetries do not influence the
  kinematic analysis, but we used the unfolded catalog to build the velocity fields. The light
    distribution shows asymmetries beyond $\sim 20$ kpc radius \citep[][]{2015MNRAS.446..120D} which
    cause an increased isophote twist beyond $3R_e=186''\sim 20$ kpc. The PN velocity field is
    aligned with the major axis except in the last bin which could be related to these
    asymmetries. Therefore the galaxy is classified as consistent with axisymmetry. The PN
  kinematics agrees with that of GCs. \citet{2013MNRAS.428..389P} found significant rotation only
  for the red GCs along the photometric major axis and with constant amplitude. Their velocity
  dispersion decreases with radius.

\item {\bf NGC 2974}. This is an elliptical (E4) field galaxy. Its
  kinematics was mapped by the SAURON survey
  \citep[][]{2004MNRAS.352..721E}, showing that the galaxy is rotating
  at $\sim 120$ \kms along the photometric major axis, while the
  velocity dispersion has a steep decreasing profile. PN kinematic
  data show that the galaxy halo rotates at least at $\sim70 $ \kms
  along the photometric major axis, and has a velocity dispersion of
  $150 $ \kms.

\item {\bf NGC 3115}. This is an S0 associated with a loose group of galaxies
  \citep[][]{1992A&AS...93..211F}.  It has a nearly edge-on disk ($i=86$ degrees) which flares
  outwards \citep[][]{1987AJ.....94.1519C}. The long slit spectroscopy from
  \citet{2006MNRAS.367..815N} and the IFS observations of \cite{2016A&A...591A.143G} show that
    the rotation velocity increases steeply from the innermost regions to $\sim 230$ \kms around
    $\sim1\re=93''$. The major axis profile from \cite{2016A&A...591A.143G} offset by 20 arcsec
    along the minor axis show that also the spheroidal component has pronounced rotation. The
    velocity dispersion field is colder ($\sigma\sim120$ \kms) in the region dominated by the disk,
    and hotter along the minor axis ($\sigma\sim220$\kms). The kinemetry of
    \citet{2016MNRAS.457..147F} on SLUGGS data show that the amplitude of rotation decreases from
    $\sim 1$ Re outwards.  The PN data show a cylindrical velocity field with gently decreasing
  rotation amplitude reaching $\sim120$ \kms at $\sim3 \re$, where the spheroid dominates.  The
    $\PAkin$ has a constant misalignment of $\sim 10\pm5$ degrees with respect to $\PAphot$. By
    comparison the kinemetry from \citet{2016MNRAS.457..147F} in figure
    \ref{fig:fitted_misalignments} has $\sim 5\pm10$ degrees misalignment at 200'' major axis
    radius, slightly offset from the $\PAphot$ in the same direction as the PNe, showing that both
    datasets agree within the errors. This misalignment is probably related to perturbations at the
    interface between the disk and the spheroid, visible as deviations from axisymmetry in the
    photometry of the disk component. \citet[][]{1987AJ.....94.1519C} found that the spheroidal
    component of this galaxy shows a photometric twist of $\sim 4$ degrees, from $\sim 45$ degrees
    at $10$ arcsec to $\sim 41$ degrees at 350 arcsec. The disk instead flares \citep[figure 23
      from][]{1987AJ.....94.1519C} and has a warp of few degrees (their figure 24) in the same
    direction of the PN velocity field. The same features are visible in the photometry of
    \citet{2016A&A...591A.143G} (their figure 11) and in new observations with the VST telescope
    (VEGAS survey, Iodice private communication). This shows that NGC 3115 is indeed a complicated
    case, and we therefore do not include it in the sample of galaxies with triaxial halo.

\item {\bf NGC 3377}. It is an E6 galaxy in the Leo I group.  The absorption line kinematics
  \citep[][]{2009MNRAS.394.1249C,2004MNRAS.352..721E,2016MNRAS.457..147F} shows that the galaxy
  displays a disk-like rotation in the inner regions, which starts to decrease at $\sim 1.5\re$. Our
  analysis of the PN kinematics shows that the decreasing trend continues out to large radii, where
  it becomes $16\pm8$ \kms at $\sim 4\re$ and $11\pm7$ \kms beyond $\sim7 \re$.  The kinematic major
  axis position angle twists with radius in the outskirts in agreement with
    \citet{2016MNRAS.457..147F}.  The velocity dispersion at large radii decreases very gently with
  radius, from $\sim80$ \kms at 1 $\re$ and to $\lesssim60$ \kms at 8 $\re$.  This is consistent
  with the GC study of \citet{2013MNRAS.428..389P}, whose kinematics reaches $\sim 8\re$. While the
  red GCs show a constant rotation of $\sim50$ \kms, the blue ones do not rotate significantly in
  agreement with the PNe. The velocity dispersion of the whole GC sample is consistent with that
  from PNe.
 
\item {\bf NGC 3379}. This is the largest galaxy in the Leo I group,
  classified as an E1. The PN system of this galaxy is has been widely
  studied in literature \citep[][]{1993ApJ...414..454C,
    2003Sci...301.1696R, 2006AJ....131.2089S}.  The velocity maps from
  \citet{2004MNRAS.352..721E} reveal that the galaxy rotates regularly
  along its photometric major axis ($V_\mathrm{rot}^{MAX}\sim63$
  \kms), while the velocity dispersion has a central peak of $\sim220$
  \kms and it decreases to $180$ \kms around $0.25 \re$. Then, the
  long slit data of \citet{1999AJ....117..839S} and
  \citet{2009MNRAS.398..561W} show that the profile gently declines to
  $\sim 120$ \kms at $\sim2\re$.  The PN smoothed velocity field is in
  good agreement with the inner kinematics and extends it to $6
  \re$.
  We find that the galaxy has a rotation along an axis that twists
  from the photometric major axis position angle ($\sim70^{\circ}$ at
  $\sim15$ arcsec) towards higher values at large radii. We observe
  the onset of a rotation along the minor axis at $\sim 2.5 \re$. This
  is reflected in the trend of the fitted kinematic position angle and
  in the amplitudes of the third order harmonics. The amplitude of
  this additional rotation is $\sim 30$ \kms. At the moment there are
  no integral field kinematic data covering these large distances from
  the center of the galaxy.  The velocity dispersion map shows that
  the decline continues up to $6 \re$ where it reaches values around
  70 \kms.

 \item {\bf NGC 3384}. This lenticular galaxy is a member of the Leo I group \citep[][]{1992A&AS...93..211F}. We find a rotation velocity of $160\pm7$  \kms at $\sim7\re$, consistent with the rotation amplitude published by \citet{2011MNRAS.414.2923K}. We do not observe any twist of the kinematic position angle or any misalignment with the photometric one, and the motion is that of a regular rotator. The velocity dispersion map in \citet{2004MNRAS.352..721E} shows that the $\sigma$ has a double peak at the center ($\sigma\sim155$  \kms), and a declining profile that reaches $\sim 80$  \kms at $ \sim 0.5 \re$. The PN data show a velocity dispersion of $\sim 80$ \kms. \cite{2013MNRAS.432.1010C} performed a kinematic analysis of the disk and spheroid component separately. The photometric study of \cite{2007AN....328..562M} showed the presence of an "inner component" and and "elongated" component, visible in the $\PAphot$ profiles as a local minimum at $R=3"$ and a local maximum at $R=15"$. Therefore we considered only the regions at $R>30"$ in our discussion in section \ref{subsec:DISCUSSION_signatures_of_triaxiality_in_photometry}.
 
 \item {\bf NGC 3489}. This S0 galaxy belongs to the Leo I group. The $\mathrm{ATLAS^{3D}}$ kinematic map \citep[][]{2011MNRAS.414.2923K} shows a fast rotation along the photometric major axis, whose amplitude reaches $\sim115$  \kms at $\sim 1\re$. PN data show that the halo of this galaxy is also rotating at $>40 $ \kms. The dispersion is around 80 \kms, consistent with \citet{2011MNRAS.413..813C}. We refer to \cite{2013MNRAS.432.1010C} for a kinematic analysis of the the disk and spheroid component separately.
  
 \item {\bf NGC 4278}. This a rather round elliptical galaxy (E1-2), member of the Coma I group.
 The $\mathrm{ATLAS^{3D}}$ kinematic map \citet{2011MNRAS.414.2923K} shows that stars are regularly rotating only within $0.5 \re$ along a direction slightly offset from the photometric major axis, while there is very weak or no rotation farther out. The kinemetry of \cite{2016MNRAS.457..147F} shows that the galaxy does not significantly rotate beyond $1 \re$. The PN velocity field does not show any significant rotation, hence also the kinematic position angle is not very well defined. \citet{2013MNRAS.428..389P} do not find any evidence of rotation for the metal rich GCs, while the blue metal poor GCs display a weak rotation in the outer regions, along a direction between the major and the minor axes. 
 The velocity dispersion falls rapidly with radius from a central value of $\sim 270$  \kms to $\sim 120$  \kms at $\sim 2\re$ \citep[][]{1994MNRAS.269..785B, 2004MNRAS.352..721E,2016MNRAS.457..147F}. Our map shows values of $\sigma$ almost constant at $130$  \kms, in good agreement with the literature.

 \item {\bf NGC 4339}. This is a member of the Virgo cluster, classified as intermediate type between E0 and S0. It is regular FR, showing no misalignment between the photometric and kinematic major axis \citep[][]{2011MNRAS.413..813C}. The PN motion agrees with the stellar kinematics showing a rotation of $>20$ \kms and a dispersion of $30 $ \kms.

 \item {\bf NGC 4473}. This Virgo galaxy is an E5 disky elliptical. It is well known for its multicomponent central kinematics \citep[][]{2004MNRAS.352..721E}: it is a fast but non-regular rotator ($V_\mathrm{rot}^{MAX}=63$  \kms, along the photometric major axis) and the velocity dispersion map displays a double peak, which have been interpreted as the presence of a counter-rotating disk \citep[][]{2007MNRAS.379..418C}, maybe formed in binary disk major mergers \citep[][]{2011MNRAS.416.1654B}.
 We detect a low amplitude rotation of $\sim 30$  \kms  along a direction tilted with respect to the photometric major axis. The kinematics from integrated light of \citet{2013MNRAS.435.3587F} already proved this object to be triaxial. Our velocity dispersion map does show the double peaked structure along the major axis, where $\sigma\sim160$  \kms. The velocity dispersion decreases to $\sim 100$  \kms at $10 \re$. 
 
 \item {\bf NGC 4494}. This is an E1 galaxy in the NGC 4565 group. The velocity map from \citet{2011MNRAS.414.2923K} shows a kinematically distinct core. The galaxy rotates almost constantly at $\sim 60$ \kms up to $2 \re$ \citep[][]{2009MNRAS.393..329N,2016MNRAS.457..147F}. In the halo we find that this amplitude decreases with radius. There is evidence for an additional component rotating along the photometric minor axis at $\sim 25$  \kms, as also shown by the twist of the $\PAkin$. The velocity dispersion field shows a peak at the center ($\sigma\sim150$  \kms), and then decreases to $\sim 80$  \kms at $R\gtrsim 5 \re$, in very good agreement with the stellar kinematics. The four-fold structure that appears in the $\sigma$ map is a result from smoothing procedure, and reflects some unresolved gradient in the velocity field. The trend of the rotation amplitude from PN kinematics seems to be confirmed by the kinematic map from \citet{2014ApJ...791...80A}, but their data do not extend out in radii enough to show any rotation along the minor axis. The kinematics of GCs is studied by \citet{2011MNRAS.415.3393F} out to $\sim3.5 \re$. They found that the metal poor GCs rotate with similar amplitude as the galaxy stars, while the metal-rich GCs show marginal rotation. The velocity dispersion profile of the global GC sample is consistent with that of the stars.

 \item {\bf NGC 4564}. NGC 4564 is an E4 galaxy in the Virgo cluster. It shows a prominent stellar disk along the apparent major axis, so that it has also been referred as an S0 \citep[][]{1994A&AS..105..481M}. We agree with the stellar kinematic from \citet{2016MNRAS.457..147F} and we extend it beyond $6\re$, where the galaxy is still clearly rotating along its photometric major axis ($V_\mathrm{rot}\sim95$  \kms). The velocity dispersion profile is decreasing, from the central peak of 180  \kms \citep{2004MNRAS.352..721E} to $\sim 70$  \kms at $6\re$.

  \item {\bf NGC 4594}. NGC 4594, also known as "the Sombrero" (M 104), is an SAa galaxy in the
    Virgo cluster. It has an extremely bright bulge, while its disk contributes only 10\% of the
    light \citep[][]{1986AJ.....91..777B} and has pronounced dust lanes. Kinematics from
    line-profile measurements show that the galaxy rotates along its major axis reaching $\sim250$
    \kms at $\sim1\re$, where the rotation curve flattens to a constant amplitude
    \citep[][]{1982ApJ...256..460K,1994MNRAS.268..521V}. The velocity dispersion declines from the
    central value of $\sim 280$ \kms to values close to $100$ \kms where the disk component
    dominates. The bulge, instead, has a dispersion remarkably constant at $210$ \kms.  Our PN data
    extend beyond $4\re$, but there is a void in the detections along the major axis because of the
    dust lane. For this reason the PNe do not trace the fast rotating disk, and their kinematics is
    in very good agreement with the integrated light kinematics of \cite{1989ApJ...338..752K} from a
    slit parallel to the major axis, offset by 30 arcsec.  Extinction effects are probably also
    responsible for the slight asymmetry of the smoothed velocity field, so the kinematic analysis
    is performed on the unfolded catalog. These asymmetries are localized and do not significantly
    influence the kinematic results.  The only studies of the kinematics of this galaxy at large
    radii is from \citet{2007ApJ...658..980B} and \citet{2014AJ....147..150D} using GCs. The latter
    sample distances out to 24' with 360 tracers, finding little or no evidence of rotation in the
    GC system as a whole, and within the red and blue subpopulations. The photometry shows
      ellipticity less than 0.1 and a correspondingly uncertain large isophote twist beyond 350''
      which we do not include in the photometric analysis.

  \item {\bf NGC 4649}. Also known as M 60, it is an E2 galaxy in the Virgo cluster. It forms a close optical pair with NGC 4647, but the lack of evidence of strong tidal interaction features \citep[e.g.][]{2006ApJ...650..166Y} suggests that the galaxies are at different distances. The $\mathrm{ATLAS^{3D}}$ velocity map shows a regular, cylindrical rotation of $\sim95$  \kms \citep[][]{2011MNRAS.414.2923K}, while the dispersion has a central peak of $\sim370$  \kms and it decreases to $\sim260$  \kms at 0.5$\re$. The kinematic maps from \citet{2014ApJ...791...80A} shows that the cylindrical rotation extends up to $2\re$, while there the dispersion is $\sim120$  \kms.
  The PN data extend out to  5 $\re$. The smoothed PN velocity field shows that the amplitude of the rotation along the photometric major axis decreases with radius, and no rotation along the major axis is detected beyond $2\re$. In the outer regions, instead, the system rotates along a nearly orthogonal direction, suggesting a triaxial halo. The velocity dispersion is roughly constant at $\sim 200$ \kms from 1 to 5 \re. We see a slight bump in the dispersion profile at 3 $\re$, but it may be associated to an unresolved velocity gradient.
  There are several studies of  the GC kinematics in NGC 4649. The most recent, with the largest sample is from \citet{2015MNRAS.450.1962P}. The red GCs display a nearly constant rotation with radius along the photometric major axis. Interestingly they observe a dip in rotation velocity between 100 and 200 arcsec, where we observe the inner rotation decreasing also. The blue GCs also show hints of rotation at all radii, with lower amplitude, and generally along the major axis of the galaxy. Blue GCs also appear to have a minor axis rotating component at 300 arcsec, coinciding with the one found with the PNe. The velocity dispersion of GCs is constant at $\sigma=240\pm30$ \kms.

 \item {\bf NGC 4697}. This is an E6 galaxy in the Virgo southern extension, belonging to a group of 18 galaxies \citep[][]{ 1993A&AS..100...47G}. This galaxy has 535 PN detections by \citet{2001ApJ...563..135M}. In this dataset \citet{2006AJ....131..837S} found evidence for two separate PN populations, with different luminosity functions, spatial distributions, and radial velocities. In particular the secondary population of PNe is found to be azimuthally unmixed and not in dynamical equilibrium. The velocity field shows a rotation along the photometric major axis with amplitude decreasing with radius from the maximum $\sim90$  \kms at $1\re$ to $\sim 15$  \kms at $2.5$. There is a hint for an increase in the amplitude of rotation at larger distances ($\sim 40$  \kms at 4.5$\re$). The kinematic position angle is constant, aligned with the photometric major axis. There is a variation of the $\PAkin$ profile at $\sim 3 \re$, but we did not interpreted it as signature of a triaxial halo, as the present dataset still includes the secondary not-in-equilibrium population of PNe.  The velocity dispersion steady declines with radius, from $\sim180$  \kms at 0.5$\re$ to $\sim100$  \kms at 4.5$\re$. The kinematic maps from \citet{2011MNRAS.414.2923K} and \citet{2015MNRAS.452...99S} show a regular disk-like rotation with a maximum amplitude of 111 \kms, confirming the trend for the rotation amplitudes that we find. \citet{2014ApJ...791...80A} also observes a clear decline of the rotation and of the velocity dispersion, in agreement with PN data.  
 
 \item {\bf NGC 4742}. This is a field E4 galaxy. It is a faint object, so the number of PN
   detections is relatively low. We detect a rotation of at least $\sim74$ \kms, in agreement with
   the stellar kinematics of \citet{1983ApJ...266...41D}. The direction of rotation is along a
   direction misaligned with respect to the photometric position angle. The dispersion decreases
   from 100 \kms at 1.5 arcsec \citep[][]{1983ApJ...266...41D} to below 50 \kms in the
   outskirts. The photometric PA profile from \citet{2011ApJS..197...22L} shows a jump inside $\sim 40''$ which is in disagreement with \citet{1995AJ....110.2622L} based on HST data. Because the PA from \citet{1995AJ....110.2622L} agrees with the outer profile from \citet{2011ApJS..197...22L} we therefore use combined ellipticity and PA
     profiles from both sources.

\begin{figure*}[ht]
\begin{center}
  \includegraphics[width=8cm]{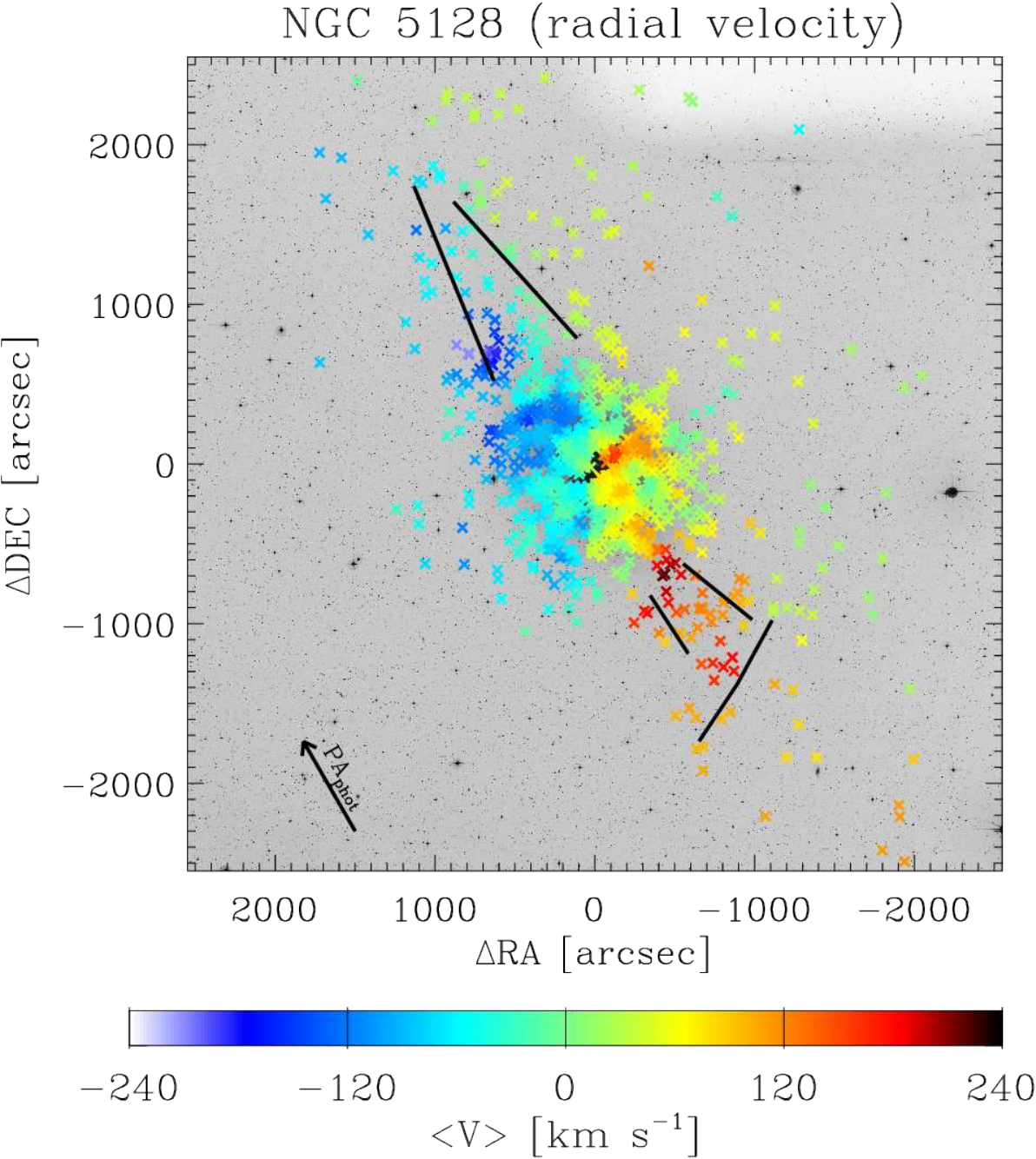}
  \hspace{0.1cm}
  \includegraphics[width=8cm]{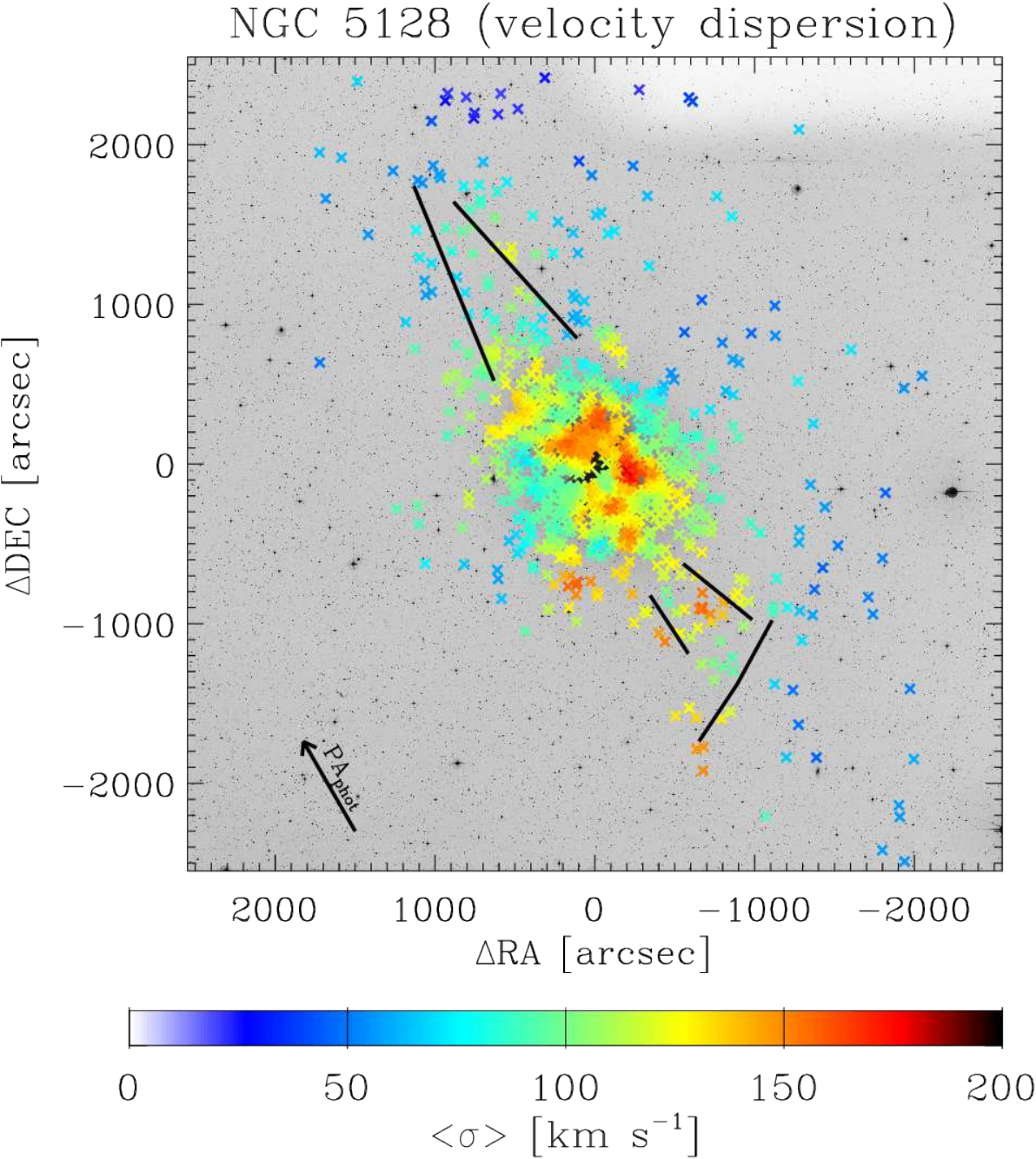}
\caption{Smoothed velocity field and velocity dispersion field of NGC 5128 shown on the DSS image; north is up, east is left. Features in the surface brightness distribution spatially consistent with features in the PN kinematics are contoured in black. The arrow at the bottom left shows the direction of the $\PAphot$.}
\label{fig:N5128}
\end{center}
\end{figure*} 
  
 \item {\bf NGC 5128}. Also known as Centaurus A, it is a giant elliptical (often classified as a peculiar lenticular) in the Centaurus group. It shows clear signs of its accretion history in tidal streams and shells \citep[][]{1983ApJ...272L...5M,2002AJ....124.3144P,2016ApJ...823...19C}, evidences of major nuclear activity \citep[see e.g.][]{2010PASA...27..449N}, and jet-induced star forming regions \citep[e.g.][]{2002ApJ...564..688R}. Its kinematics probed by PNe reflects the presence of regions which are not in equilibrium, showing a highly asymmetric and rich in subcomponents velocity field.
 The fit of the $a_0 (R)$ parameter of the rotation model \eqref{eq:fourier_definitivo} in radial bins shows the presence of peculiar bulk motions in groups of PNe.  
 The galaxy shows a rotation along the photometric major axis (NE direction) with increasing amplitude, from $\sim50$ \kms at $\sim200$ arcsec ($\sim1.5\re$) till at least $\sim 150$ \kms at 800 arcsec ($\sim5\re$). The system also displays strong rotation ($\sim100$  \kms) along the minor axis, revealing its triaxiality \citep[see also][]{2004ApJ...602..685P}.
 The deviation from point-symmetry of the kinematics in almost all the radial bins is accompanied by several features in the light. In particular the outermost regions, from 700 arcsec outwards, show the most important kinematic features \citep[][]{2013MNRAS.436.1322C}, in correspondence to the elongated structure (sketched with black contours in figure \ref{fig:N5128}) in light extending roughly along the radio jets \citep{1983PASAu...5..241H}.

 \item {\bf NGC 5866}. NGC 5866 is an SA0 galaxy belonging to a loose group \citep[][]{
   2006MNRAS.369..360S}. It has a narrow, clear-cut dark lane, running along the disk at an angle
   with respect to the photometric major axis. It is a regular, disk-like rotator
   \citep[][]{2011MNRAS.414.2923K} with a maximum amplitude of $\sim150-180$ \kms
   \citet{2016MNRAS.457..147F} at $\sim2\re$, beyond which the rotation decreases to $110$ \kms at
   $\sim3$ \re. Our velocity map shows that the galaxy keeps rotating at approximately constant
   speed ($\sim85$ \kms) at large radii (up to 9 \re), with a progressive twist of the kinematic
   position angle. The PN velocity dispersion map and velocity dispersion profile have values around
   90 \kms that are constant within the errors from 2 up to 9 \re.  We refer to Cortesi et al. (in
   prep.) for a kinematic study of the disk and spheroid components. The photometric position angle
   is constant from 30''-80'', then decreasing by $\sim2$ degrees until 160'' = 4.5 $\re$ \citep{1993A&AS...98...29M}.
   Beyond 160'' the profile becomes more irregular which coincides with a flare or X-like structure
   on the deep image by \citet[][]{2015MNRAS.446..120D} that could be due to tidal debris along
   a radial orbit nearly aligned with the disk.

 \item {\bf NGC 7457} This is a field S0. The velocity maps from \citet{2004MNRAS.352..721E} and \cite{2014ApJ...791...80A} show that this galaxy is a regular rotator along the major axis. Our PN velocity map shows that the rotation velocity is growing with radius, reaching $100$  \kms at $\sim3.5 \re$. No twisting of the kinematic major axis is observed. The velocity dispersion is relatively low, around $\sim 60$  \kms at $1 \re$, $\sim40$  \kms at $\sim3\re$.
 There is good agreement between the PN kinematics and the kinemetry from \citet{2016MNRAS.457..147F}.
 \end{itemize}

\subsection{Slow rotators} 
\label{subsec:notes_SR}
\begin{itemize}

 \item {\bf NGC 1399}. It is a CD galaxy in the Fornax I cluster. 
 We observe a very low amplitude ($30\pm$15  \kms) rotation inside $1 \re$ along the photometric minor axis, and a slow rotation also in the halo ($R\sim4\re$), which almost counter-rotates with respect to the inner regions. The rotation velocity is also very low in the innermost regions as reported by integral light kinematics studies \citep[e.g.][]{1998A&AS..133..325G,2000AJ....119..153S,2008MNRAS.391.1009L,2014MNRAS.441..274S}, which are in good agreement with the PN data. In particular the integral field map from \citet{2014MNRAS.441..274S} shows that NGC 1399 does not display important ordered motions in the central regions, while at a radius of $\sim30$\arcsec it rotates along the minor axis ($\sim45$ \kms). 
 The velocity dispersion, instead, is relatively high: it rises steeply in the inner $10$ arcsec, reaching $\sim 370$  \kms at the center \citep[e.g.][]{2000AJ....119..153S}, while at large radii (from $\sim 1\re$) the PN $\sigma$ flattens at $\lesssim 200$  \kms. 
 \citet{2010A&A...518A..44M} showed that kinematics of the red GCs of NGC 1399 is in excellent agreement with the PNe.

 \item {\bf NGC 3608}. This is an E2 galaxy in the Leo II group. It is known to have a counter-rotating core \citep[][]{1988ApJ...330L..87J} in the central $0.27 \re$, aligned with the photometric major axis and rotating at $\sim20$  \kms \citep[][]{2001MNRAS.326..473H,2004MNRAS.352..721E}. From $\sim 1\re$ galaxy starts to rotate at approximately along the photometric major axis \citep{2016MNRAS.457..147F}. The PN kinematics shows that the growing trend continues at larger radii, where the amplitude of rotation reaches $\sim65$ \kms. The twist of the $\PAkin$ reveals that the system is triaxial.
 The velocity dispersion decreases from the central value of $\sim220$  \kms to 100  \kms at $\sim 1\re$ \citep[][]{2001MNRAS.326..473H}. We find a constant velocity dispersion profile in the halo ($\sim90$  \kms).
  
 \begin{figure*}[ht]
\begin{center}
  \includegraphics[width=8cm]{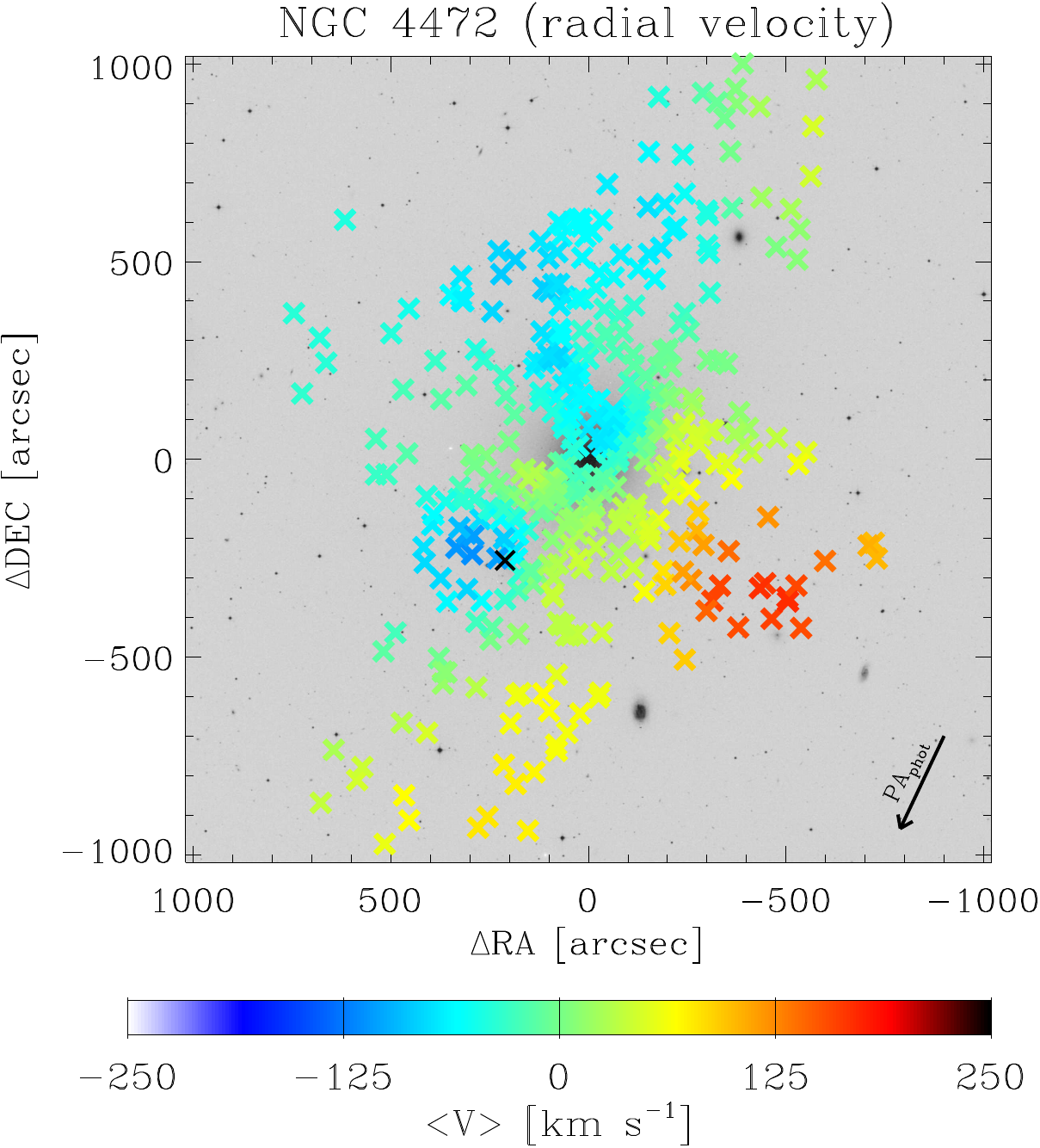}
  \hspace{0.1cm}
  \includegraphics[width=8cm]{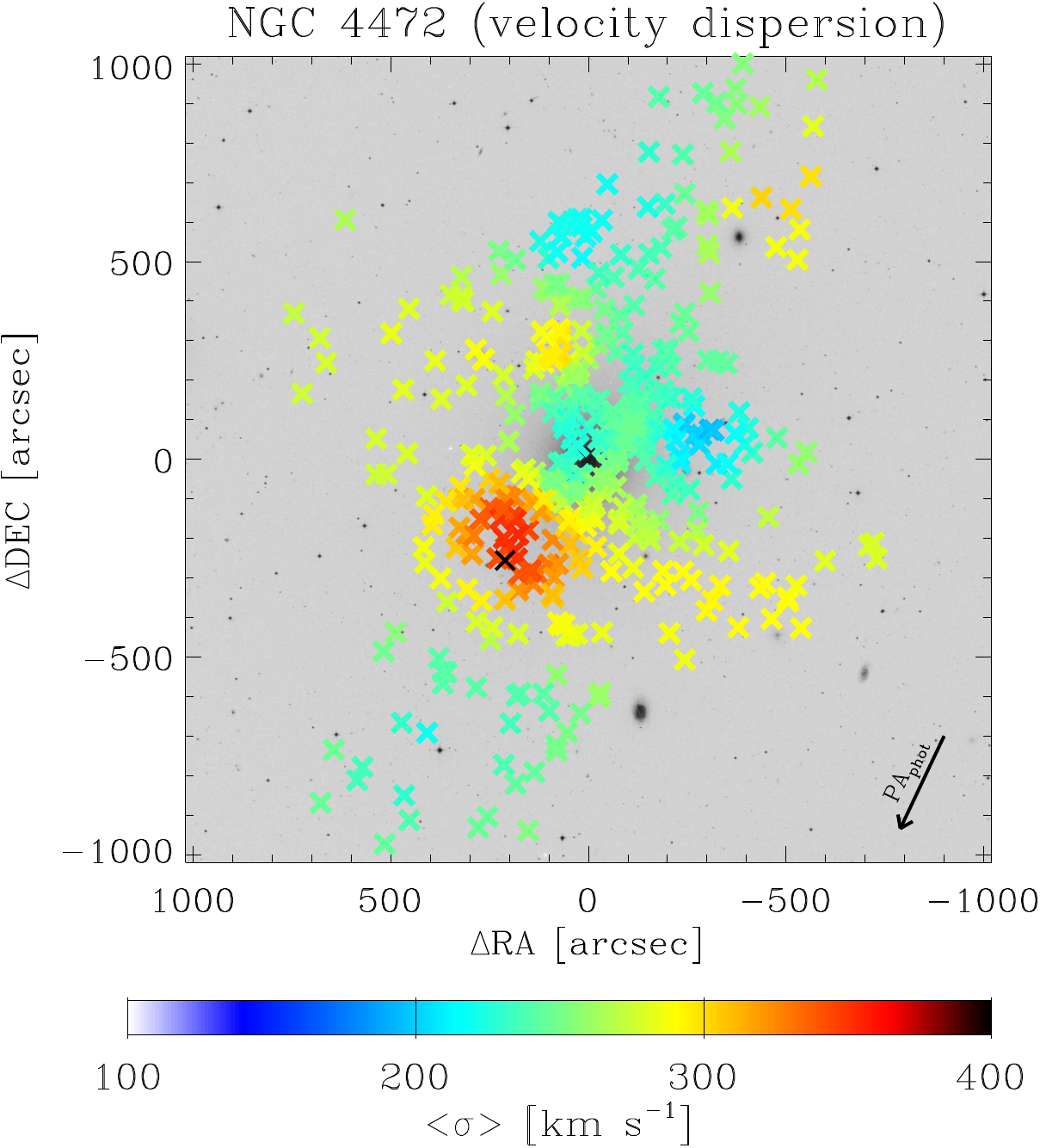}
\caption{Smoothed velocity field and velocity dispersion field of NGC 4472 shown on the DSS image; north is up, east is left. The black cross marks the position of the in-falling dwarf UGC 7636. The arrow at the bottom right shows the direction of the $\PAphot$.}
\label{fig:N4472}
\end{center}
\end{figure*}

 \item {\bf NGC 3923}. This an E4-5 galaxy in a group of 8 galaxies \citep[][]{2006MNRAS.370.1223B}. We detect a cylindrical velocity field along an intermediate direction between the minor and the major axes, which rotate at $\sim75$  \kms. 
 The velocity dispersion is constant around $235$ \kms. Long-slit kinematics from \citet{1998MNRAS.294..182C} inside $\sim1 \re$ shows that NGC 3923 does not rotate significantly along the photometric major axis for $R\lesssim0.5 \re$, but it displays a weak rotation ($\sim20$ \kms) along the photometric minor axis. 
 They also found a falling dispersion profile, from the central value of $\sim285$  \kms to $\sim240$ \kms at $0.7\re$. 
 \citet{2012MNRAS.421.1485N} studied the kinematics of 79 GC/UCD extended to more than $6\re$. They do not show appreciable rotation, and their velocity dispersion seem constant with radius.

 \item {\bf NGC 4365}. It is an E3 galaxy, one of the brightest member of the Virgo cluster. It has a complex kinematic structure with a counter-rotating core \citep[][]{1988A&A...202L...5B}, aligned with the photometric major axis, and a prolate rotation of $\sim61$  \kms \citep[][]{2001ApJ...548L..33D}, which indicates a triaxial potential.  
 Our PN velocity field extends to $6\re$. The inner kinematics ($R\lesssim2\re$) is compatible with no rotation. We do not detect the rolling about the minor axis that \citet{2014ApJ...791...80A} report, probably because of the smoothing over the inner velocity gradients. 
 We do measure a significant outer ($R\gtrsim3 \re$) rotation of $\sim 50$  \kms along the major axis, counter-rotating with respect to the kinematically decoupled core.
 The velocity dispersion map has almost constant values, around $\sim150$  \kms, with slightly lower values in the region along the major axis ($\sim140$  \kms).
 The velocity dispersion measured by \citet{2001ApJ...548L..33D} shows a inner value of $>250$  \kms which decreases to $\sim185$  \kms at $0.5\re$. \citet{2014ApJ...791...80A} measure a $\sigma\sim150$  \kms at $2\re$, consistent with the values found with PNe. 
 \citet{2012MNRAS.426.1959B}, studying the GC kinematics, found that the intermediate metallicity subpopulation (green) rotates along the photometric minor axis, as the stellar population at intermediate radii, while the red GCs rotate only at large radii along the photometric major axis in the opposite direction of the core of the galaxy, in agreement with the PN kinematics at large radii.

 \item {\bf NGC 4374}. Also known as M84, it is a bright E1 galaxy in the Virgo cluster. The galaxy does not show significant rotation in the innermost $\sim0.4\re$, as reported by \citet{2004MNRAS.352..721E} and \cite{2016MNRAS.457..147F}, while the velocity dispersion steeply decreases from 310 \kms at the center to $\sim200$ \kms at $\sim 1\re$. The PN velocity field shows evidence for rotation only at large radii ($R\gtrsim2\re$), with constant amplitude $\sim 60$ \kms and along a direction compatible with the minor axis.
 The velocity dispersion is rather constant around $\sim220$  \kms up to 6\re. Our kinematics is in agreement with that from the SLUGGS velocity maps \citep{2014ApJ...791...80A,2016MNRAS.457..147F}, but their data extend only to $\sim2\re$. The long slit kinematics of \citet{2009MNRAS.394.1249C} reaches almost $3\re$ and agrees in showing no signs of rotation along the major axis, the onset of rotation along the minor, and constant velocity dispersion in the outer regions around $\sim230$  \kms.

 \item {\bf NGC 4472}. NGC 4472 (M 49) is an E2 galaxy, the brightest object in the Virgo cluster. It is  surrounded by a complex  system  of  diffuse shells and other features  in light which are evidences of its recent and still ongoing accretion activity \citep[see][]{2010ApJ...715..972J,2012A&A...543A.112A,2018arXiv180503092H}. It is classified as a non regular rotator by \citet{2011MNRAS.414.2923K}, who also found evidence for a counter-rotating core. From $R\sim0.2 \re$ the galaxy rotates at $\sim60$  \kms \citep{2011MNRAS.413..813C,2017MNRAS.464..356V}.  Beyond $\sim3 \re$ the PN data show a complicated and out of equilibrium kinematics.
 The halo rotates along a direction compatible with the galaxy minor axis. This motion is neither point- nor axi-symmetric, and is dominated by the kinematics of the in-falling satellite, the dwarf UGC 7636. A visual inspection of the smoothed velocity field reveals that the main body of the galaxy would show major axis rotation once the PNe belonging to the satellite are excluded \citep[see][for a detailed study]{2018arXiv180503092H}. While the velocity dispersion profile is flat over $4\re$ at $\sim250$ \kms, we notice a high dispersion feature with $\sigma\sim350$  \kms in a position corresponding to the coordinates of UGC 7636. 
  
 \item {\bf NGC 4552}. NGC 4552 (M 89) is an E0-1 galaxy in Virgo.
 It known to possess a kinematically distinct core, with the innermost region rotating at $\sim30$  \kms, and the region outside 0.3$\re$ having very weak rotation \citep[][]{1997A&AS..126...15S,2004MNRAS.352..721E}. Our PN smoothed velocity field shows that the galaxy start rotating beyond $3\re$. The smoothed velocity field reveal that the rotation is about two perpendicular directions. This is reflected in the twist of the kinematic position angle with radius.
 The velocity dispersion map from the data is consistent with the one from the integrated light \citep[][]{2004MNRAS.352..721E}, and it shows constant values with radius, around $180$ \kms.

 \item {\bf NGC 4636}. This is an E0-1 galaxy in the Virgo cluster. The velocity map from \citet{2011MNRAS.414.2923K} does not show significant rotation in the innermost region ($R<0.25 \re$). The velocity dispersion, instead, has a central value of $\sim240$  \kms and it decreases to 190 \kms at 0.25$\re$. The long slit kinematics from \citet{2011RAA....11..909P} show that the galaxy starts rotating along the minor axis at $\sim40$ arcsec (0.2 $\re$), and along the major axis at $\sim85$ arcsec ($\sim1 \re$). The $\sigma$ profile is gently decreasing along the minor axis, reaching $\sim100$ \kms at 0.5 $\re$, while it is rather flat along the major, around 190  \kms, then it decreases to $\sim 150$  \kms where the rotation starts.
 We detect a weak rotation ( $40\pm15$ \kms) along a direction compatible with the minor axis. In addition, in the smoothed velocity map there is evidence for a rotation of comparable magnitude at larger radii ($R\sim 1\re$) around the major axis, in very good agreement with  \citet{2011RAA....11..909P}. The PN velocity dispersion profile decreases abruptly from $\sim 180$ \kms at $R<2 \re$ to $\sim110 $ \kms at $3\re$.
 \citet{2012A&A...544A.115S} find a strong rotation  signature for  the blue GCs ($V\sim88$ \kms along the major axis, but in opposite direction with respect to the outermost PNe). The red GCs have a weaker rotation ($V\sim30$  \kms) along an axis that changes with being aligned with the minor axis to the photometric major axis, in agreement with the PN kinematics. For radii beyond $1.5 \re$ the axis of rotation remains constant along the direction of the major axis. The dispersion profile of the blue GC declines with radius, while the red GCs show values of $\sigma\sim135$ \kms.

 \item {\bf NGC 5846}. This is an E0 galaxy, the brightest member in a group of ten. \citet{2004MNRAS.352..721E} detect a low rotation of $\lesssim30$ \kms and a high central dispersion of 255 \kms. 
 PN smoothed velocity field is consistent with low amplitude rotation. We find constant values for the velocity dispersion around 180 \kms.
 The SLUGGS kinematic maps \citet{2014ApJ...791...80A,2016MNRAS.457..147F} are consistent with PN data. \citet{2013MNRAS.428..389P} showed that neither the GC subpopulations show significant rotation. The velocity dispersion of the red GCs has a flat profile, around values comparable with those found for the PNe. The blue GCs have systematically higher $\sigma\sim260$ \kms.
 
\end{itemize}

\section{Velocity and Velocity Dispersion Fields} \label{sec:figures_vel_fields}

The following figures show the smoothed velocity and the velocity dispersion fields for all the galaxies of the
sample. The maps are interpolated on a regular grid of coordinates, as described in section
\ref{subsec:velocity_fields}. The fields are rotated such that the photometric major axis is horizontal ($\PAphot$
values are listed in table \ref{tab:galaxies}); the orientation is specified by the direction of the arrows indicating
North and East. The X and Y axes units are arc seconds. The elliptical contours displayed on the velocity fields
  are ellipses of constant ellipticity $\epsilon$ as given in table \ref{tab:galaxies}, and of major axes radii equal to
  even multiples of $\re$ (table \ref{tab:galaxies}).

\begin{figure*}[t]
  \centering
  \begin{minipage}[b]{1.60\columnwidth}
    \includegraphics[width=\columnwidth]{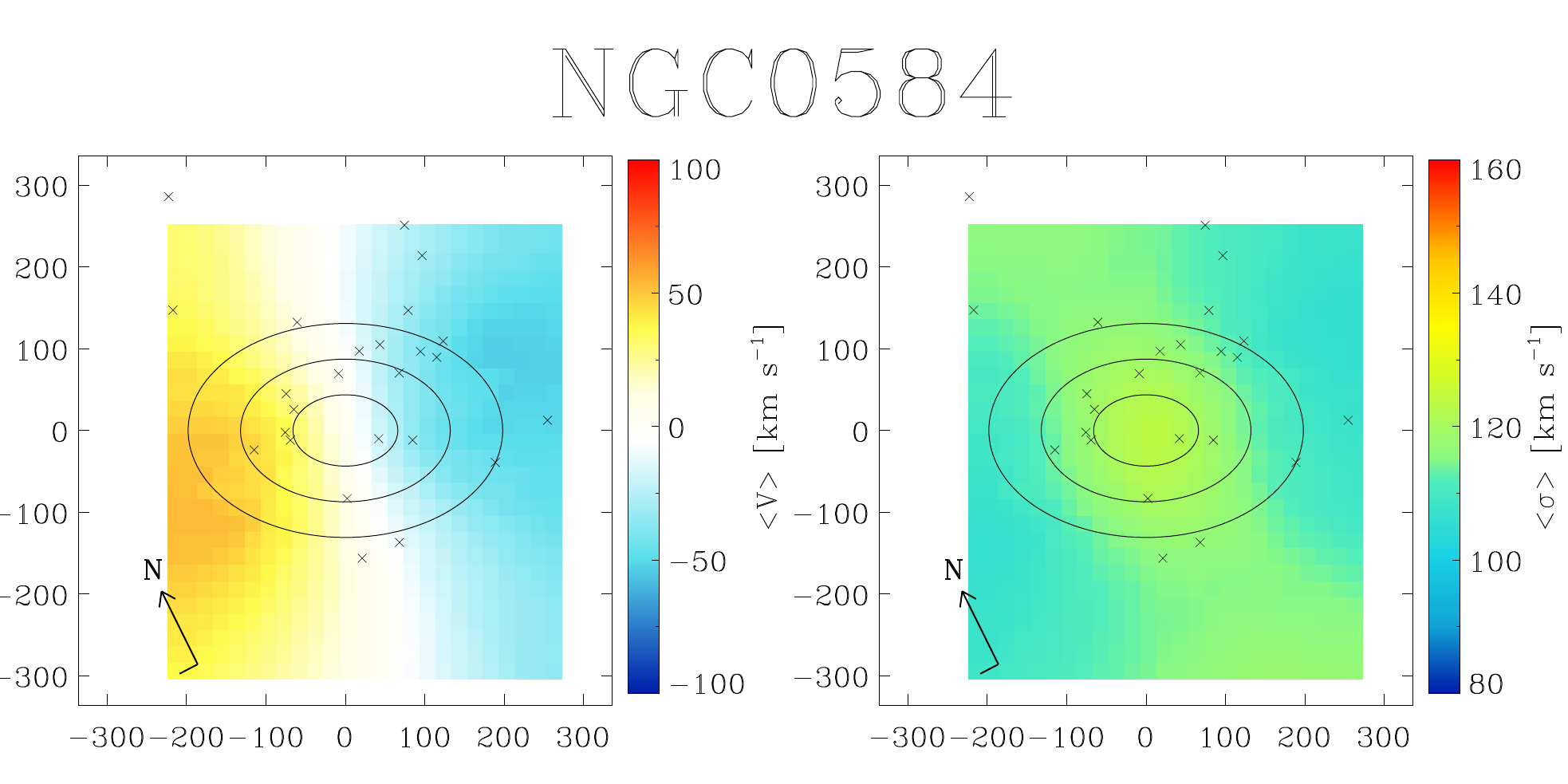}
    \vspace{0.2cm}
     \includegraphics[width=\columnwidth]{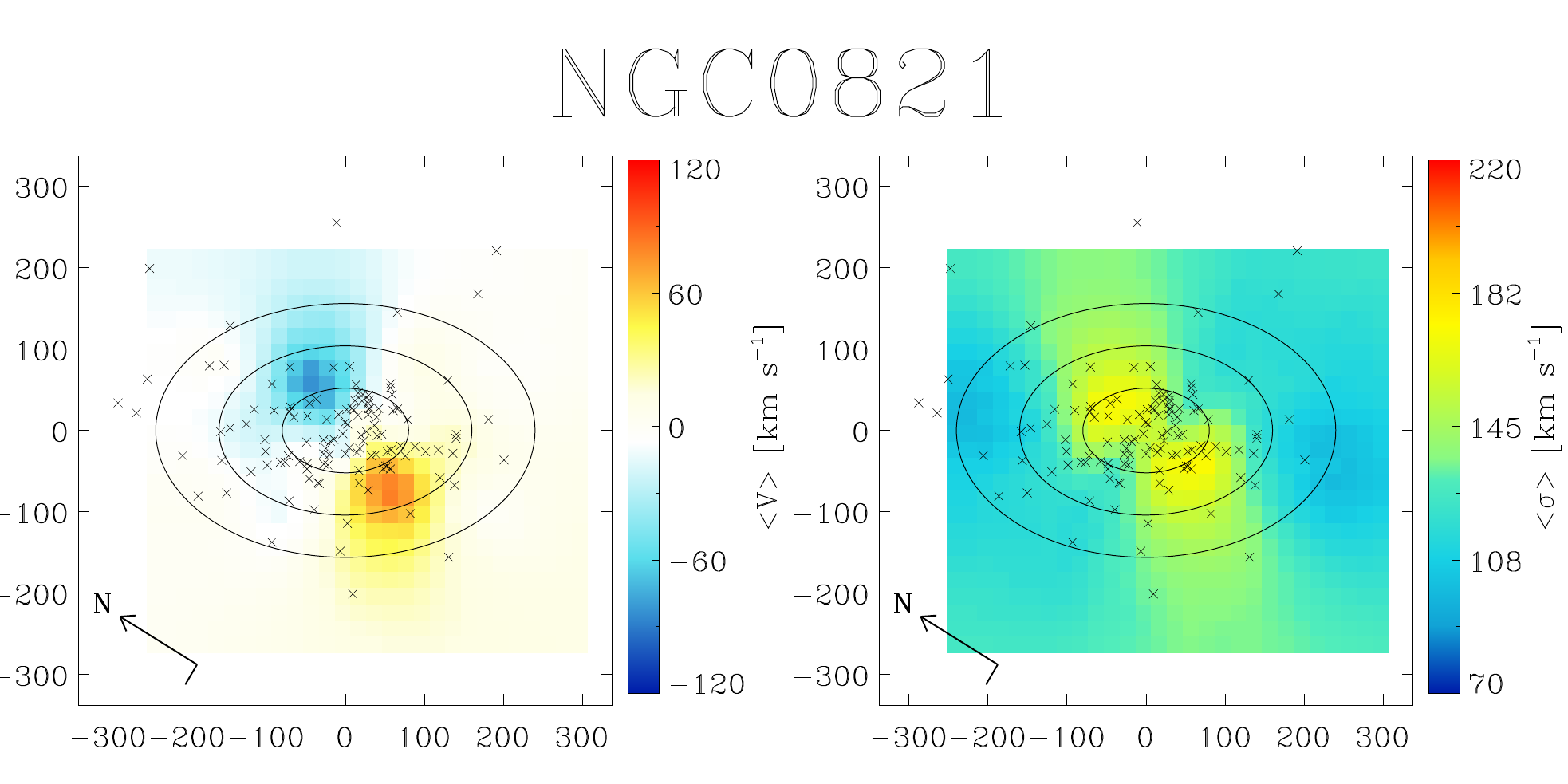}
    \vspace{0.2cm}
     \includegraphics[width=\columnwidth]{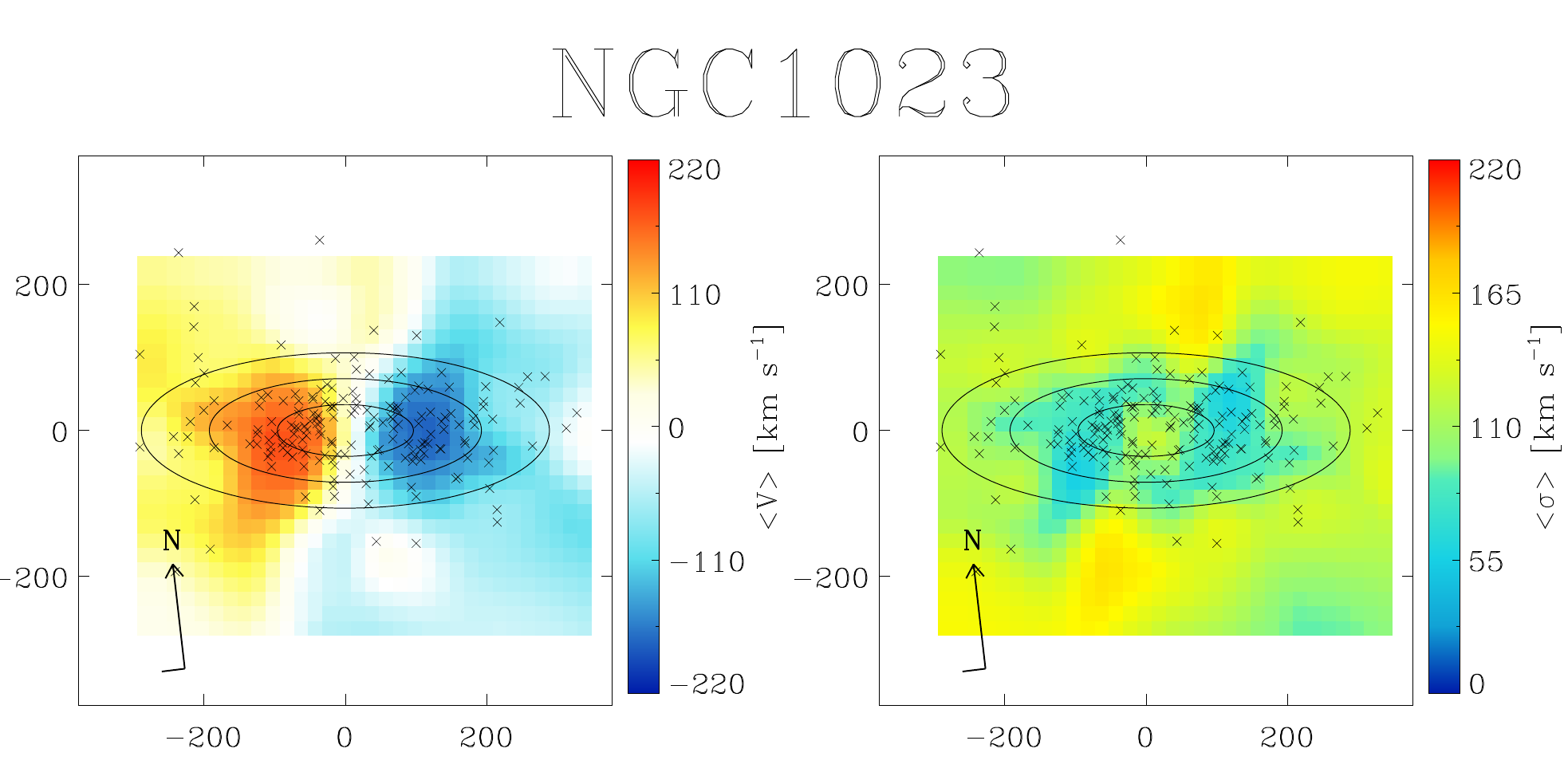}
  \end{minipage}
  
 \end{figure*}

\addtocounter{figure}{-1} 
\begin{figure*}[t]
  \centering
  \begin{minipage}[b]{1.7\columnwidth}
    \includegraphics[width=\columnwidth]{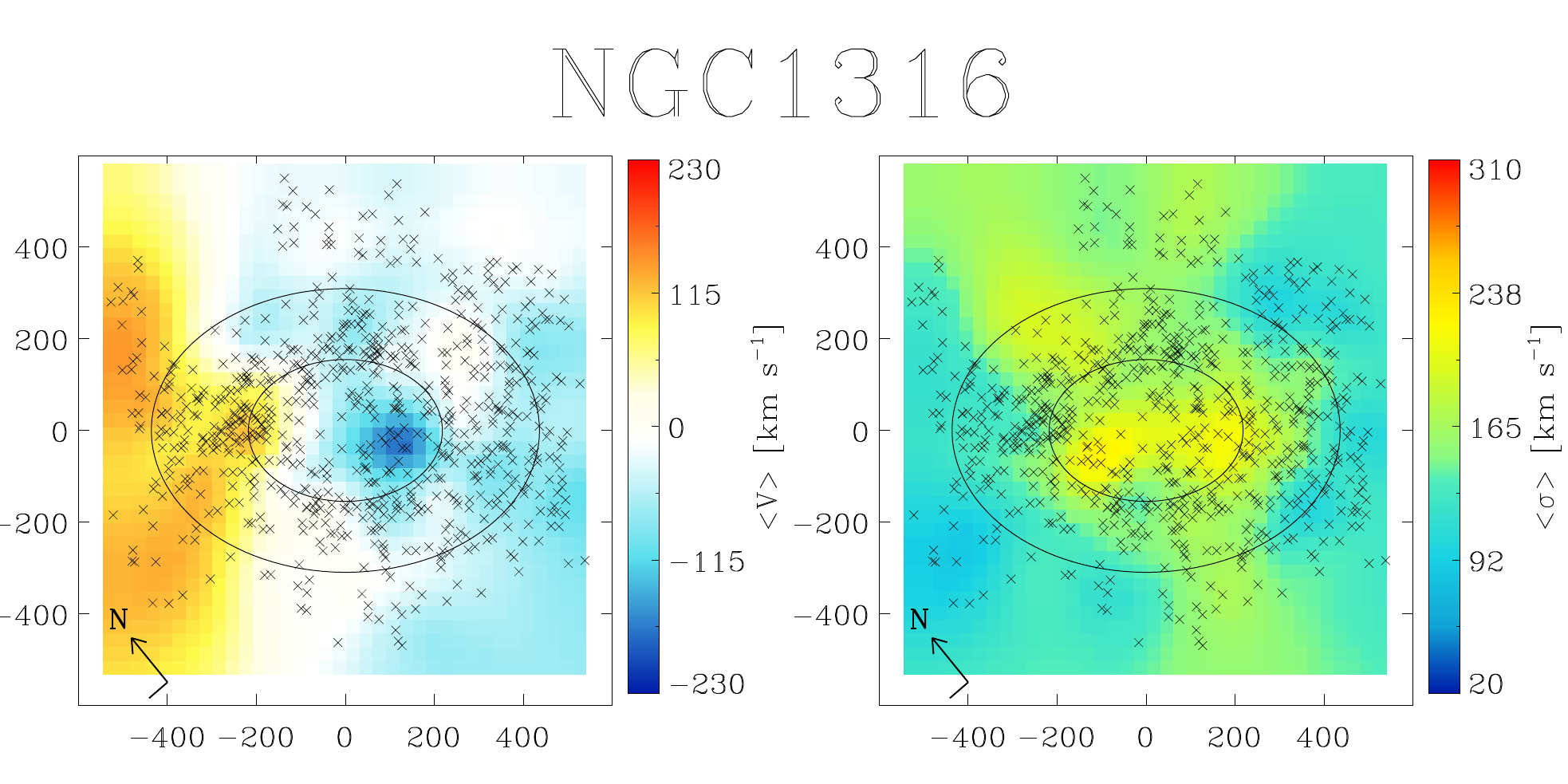}
    \vspace{0.2cm}
     \includegraphics[width=\columnwidth]{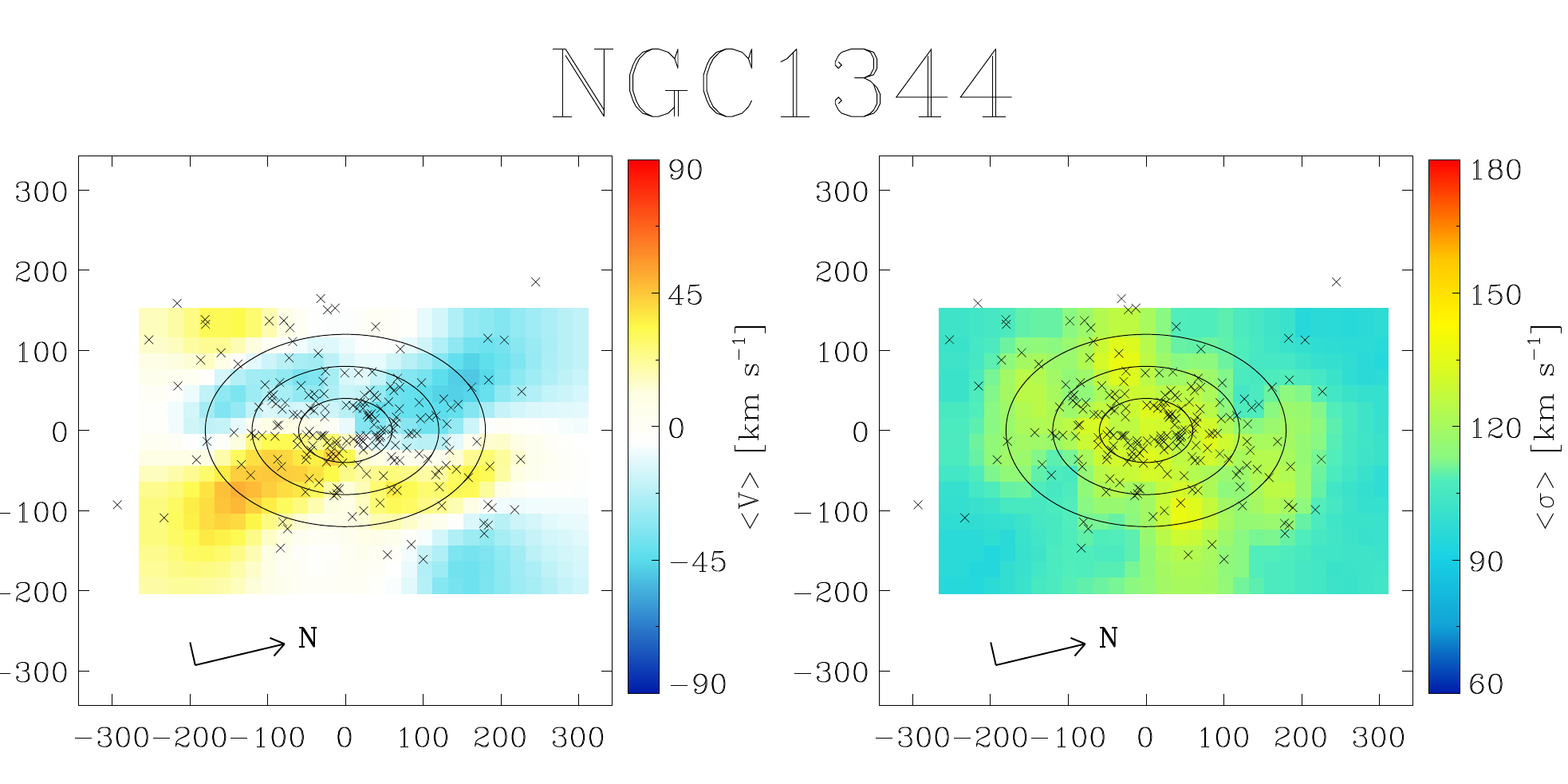}
    \vspace{0.2cm}
     \includegraphics[width=\columnwidth]{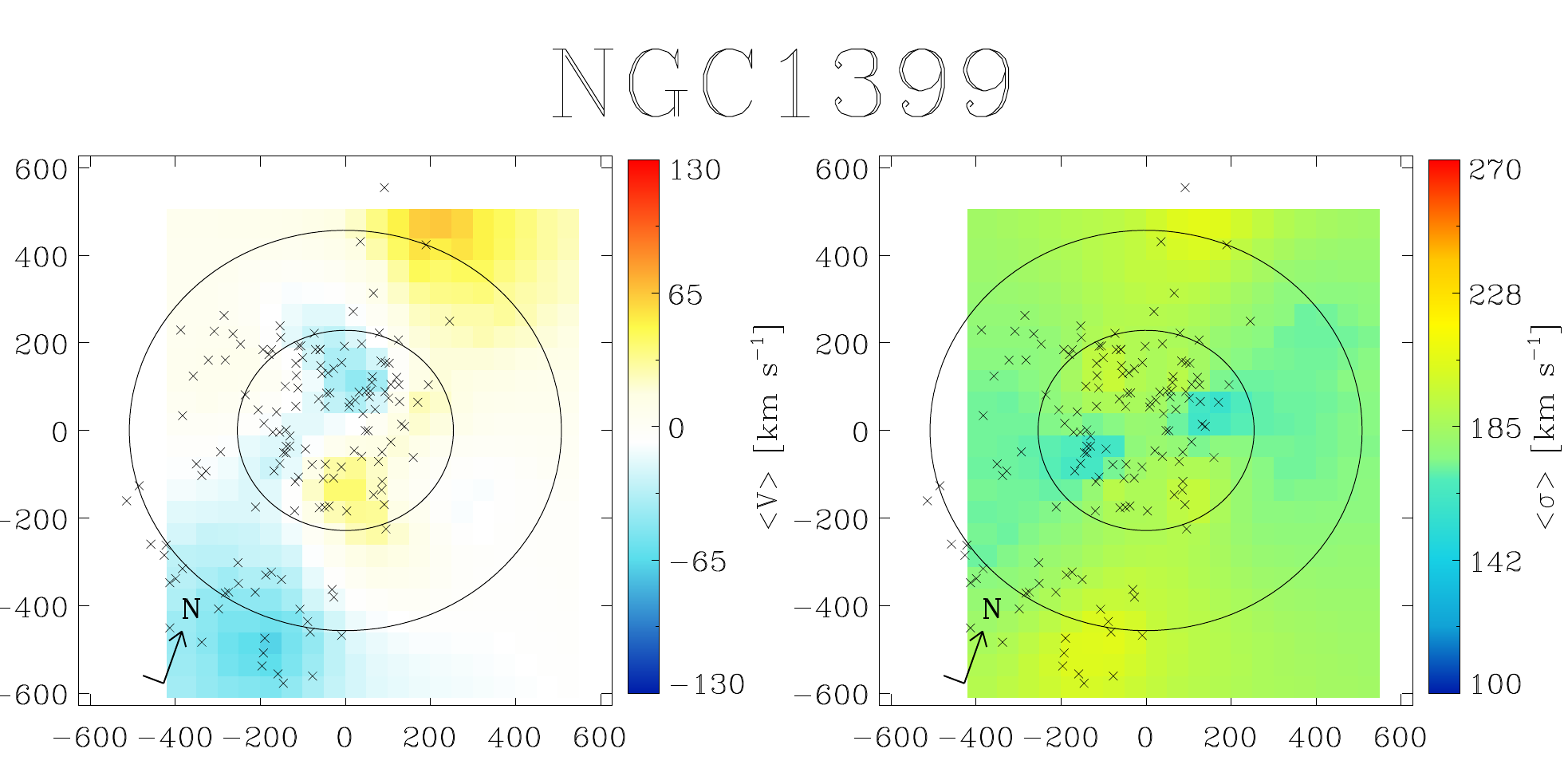}
  \end{minipage}

 \end{figure*}
 
\addtocounter{figure}{-1} 
\begin{figure*}[t]
  \centering
  \begin{minipage}[b]{1.7\columnwidth}
    \includegraphics[width=\columnwidth]{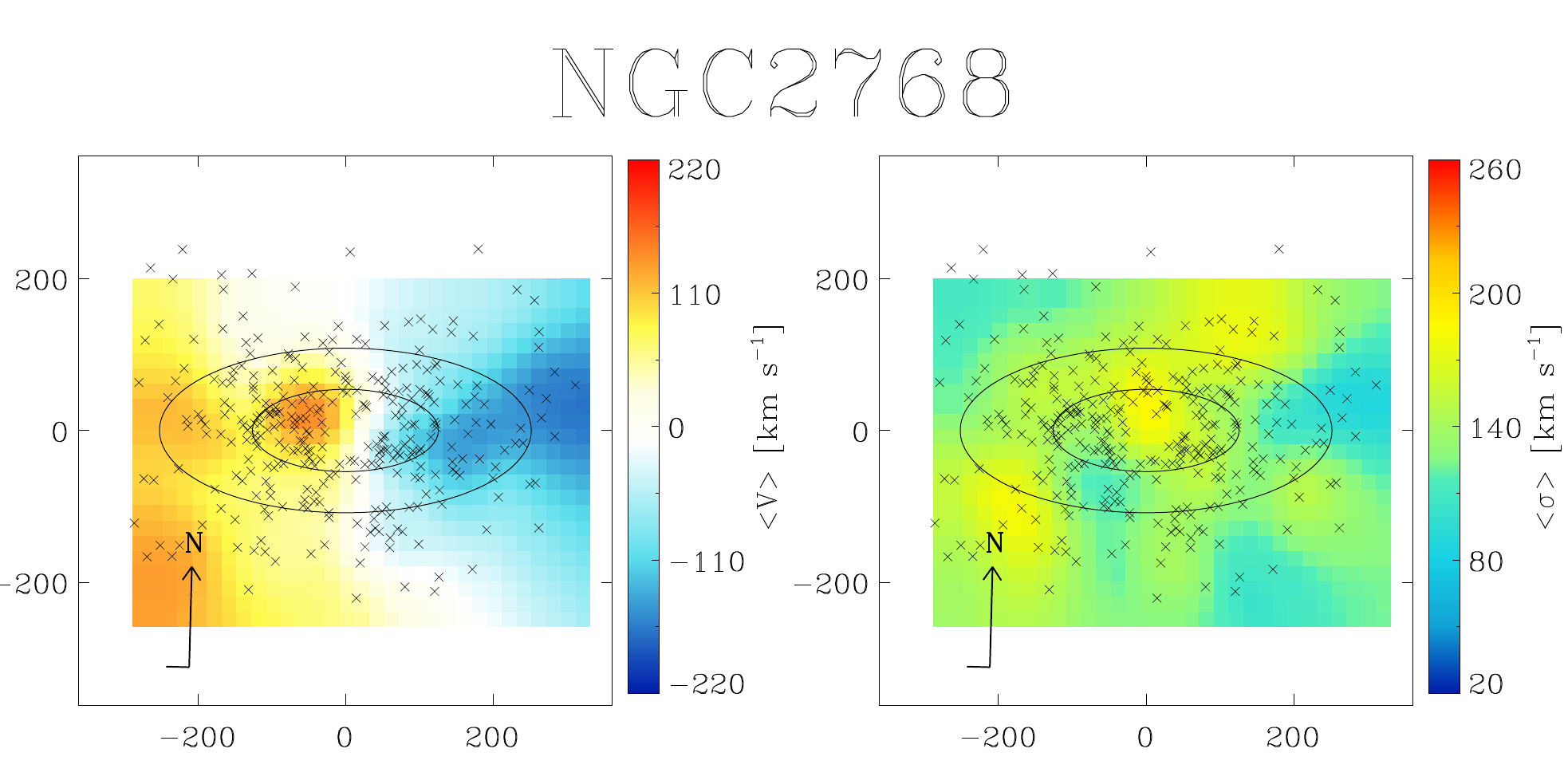}
    \vspace{0.2cm}
     \includegraphics[width=\columnwidth]{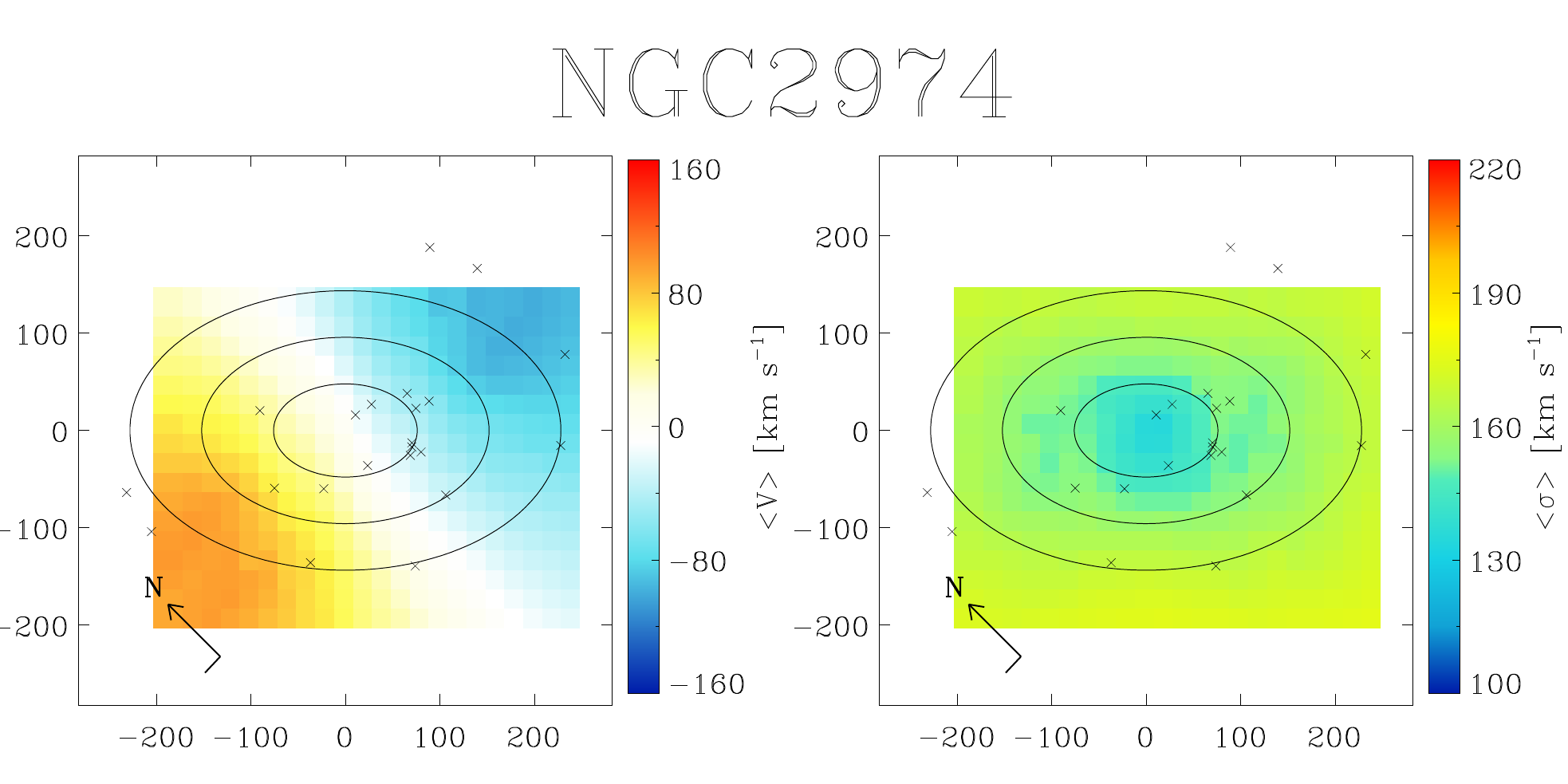}
    \vspace{0.2cm}
     \includegraphics[width=\columnwidth]{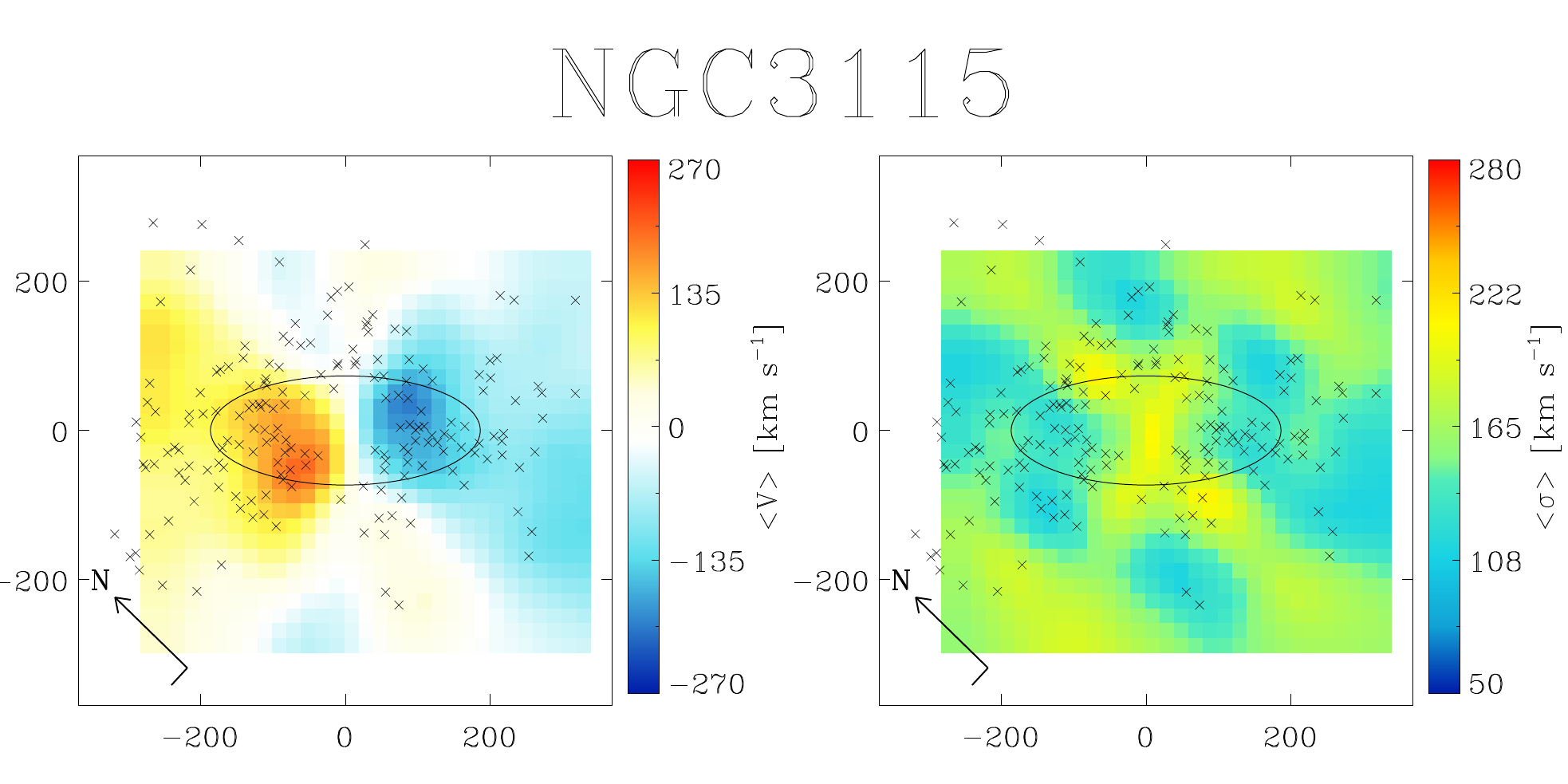}
  \end{minipage}

 \end{figure*}

\addtocounter{figure}{-1} 
\begin{figure*}[t]
  \centering
  \begin{minipage}[b]{1.7\columnwidth}
    \includegraphics[width=\columnwidth]{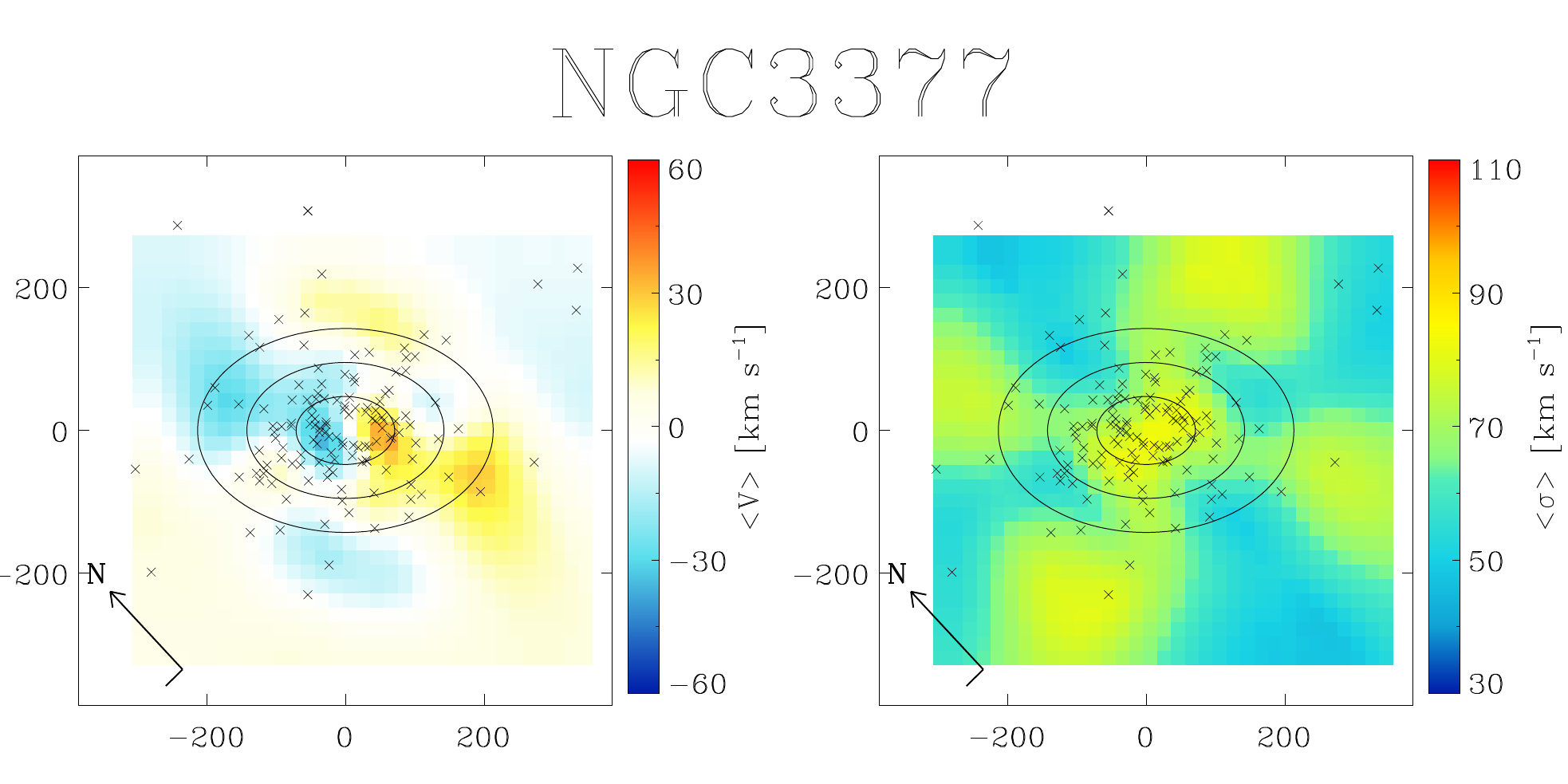}
    \vspace{0.2cm}
     \includegraphics[width=\columnwidth]{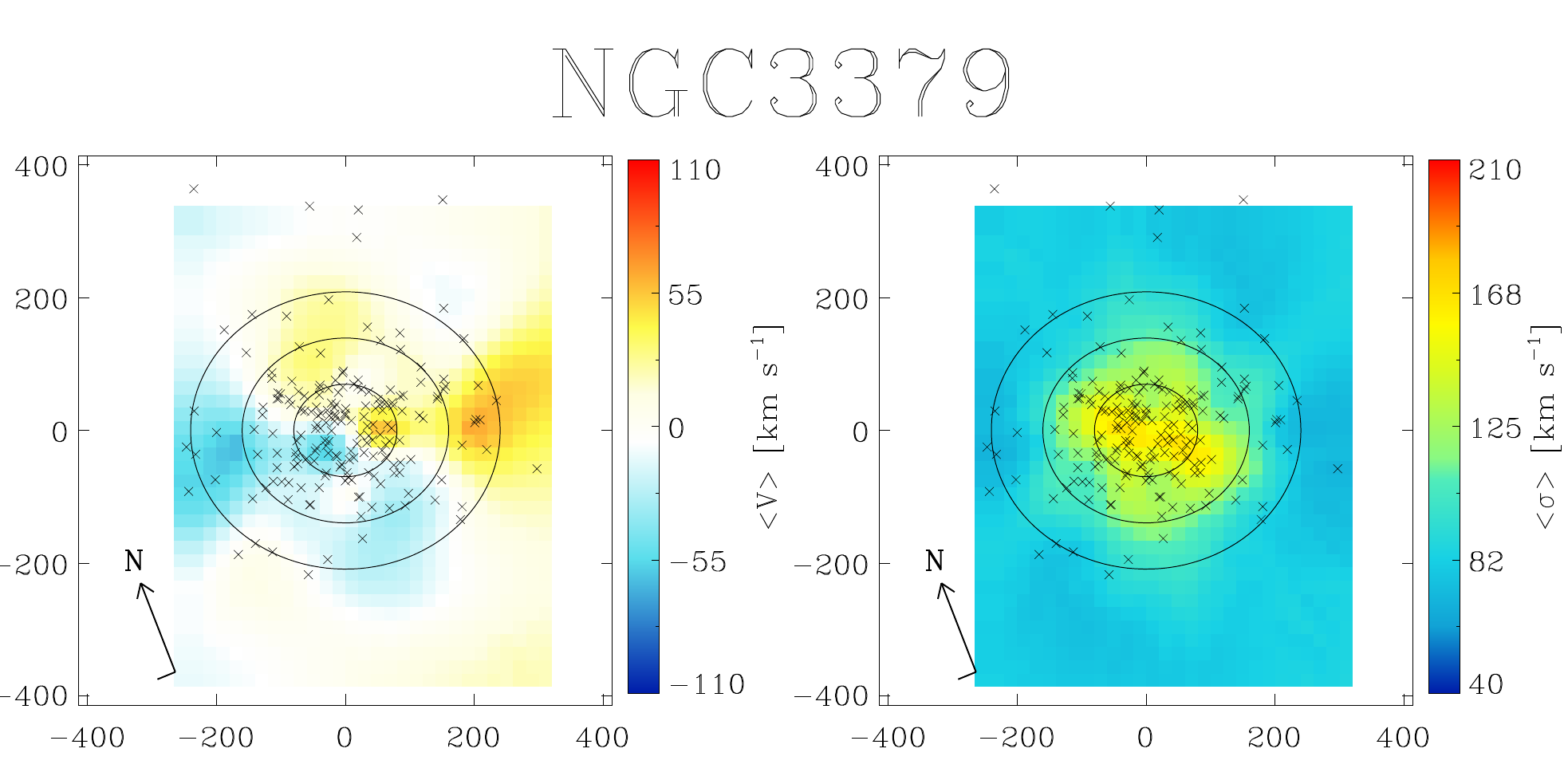}
    \vspace{0.2cm}
     \includegraphics[width=\columnwidth]{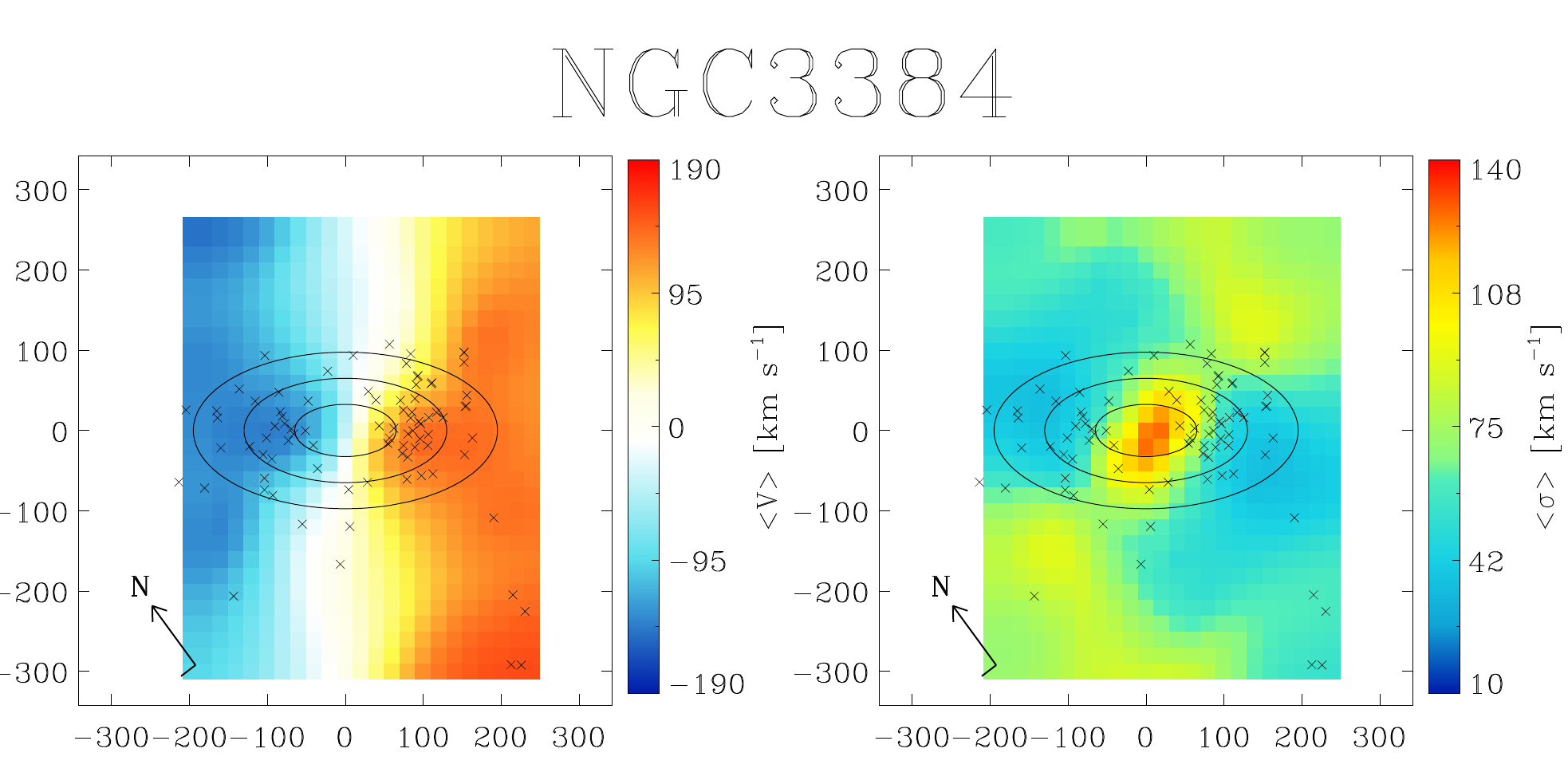}
  \end{minipage}

 \end{figure*}

\addtocounter{figure}{-1} 
\begin{figure*}[t]
  \centering
  \begin{minipage}[b]{1.7\columnwidth}
    \includegraphics[width=\columnwidth]{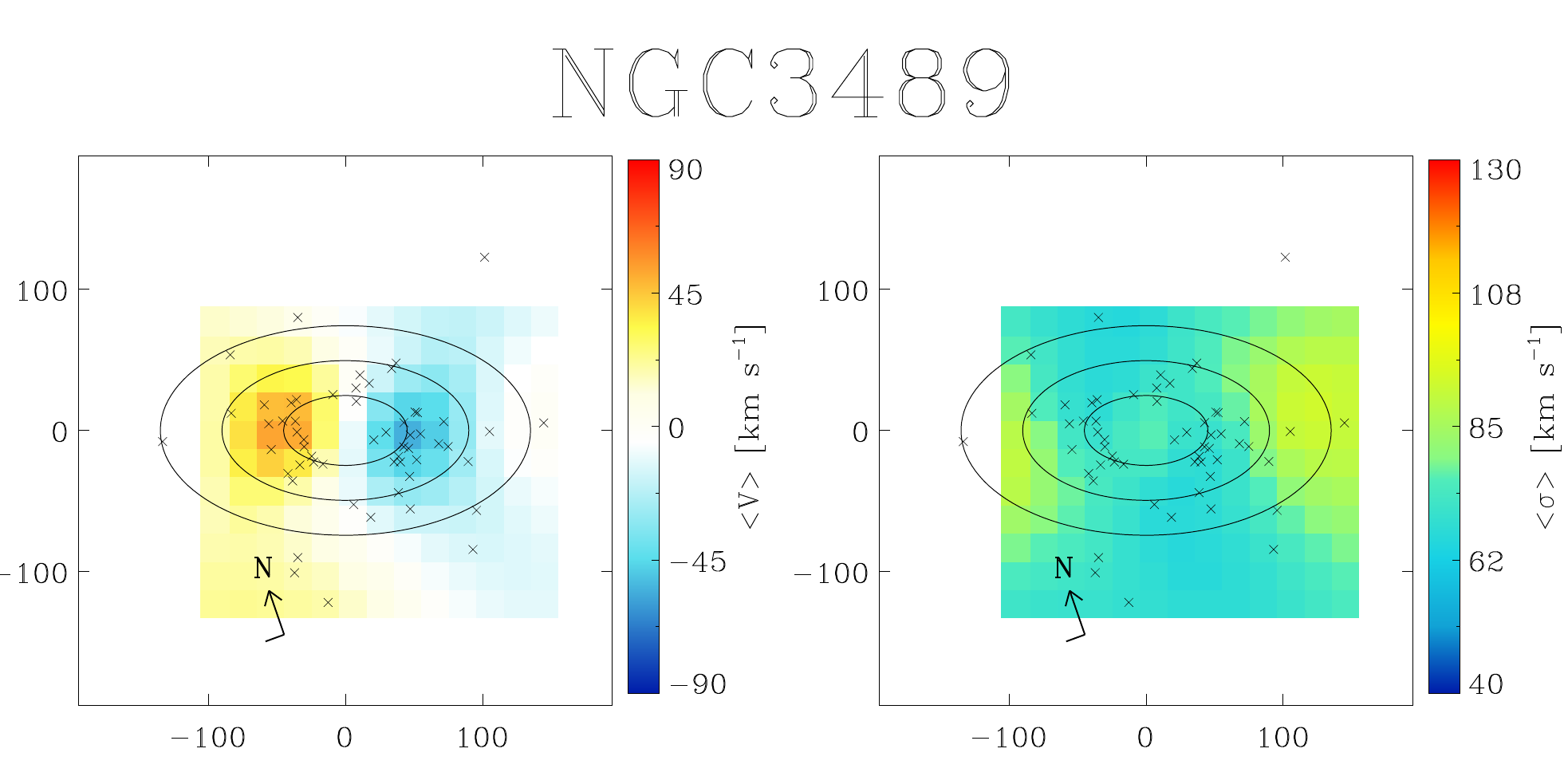}
    \vspace{0.2cm}
     \includegraphics[width=\columnwidth]{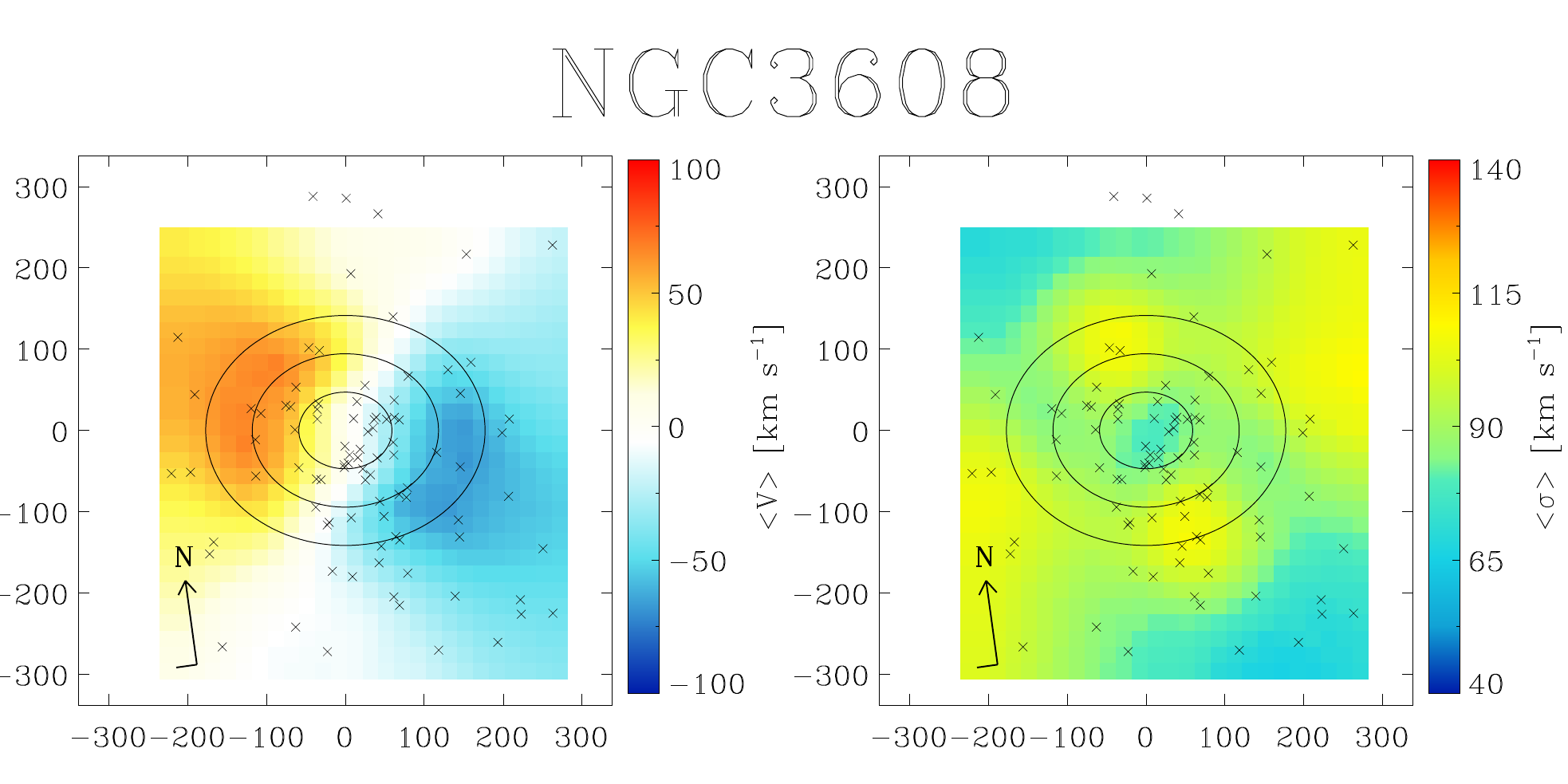}
    \vspace{0.2cm}
     \includegraphics[width=\columnwidth]{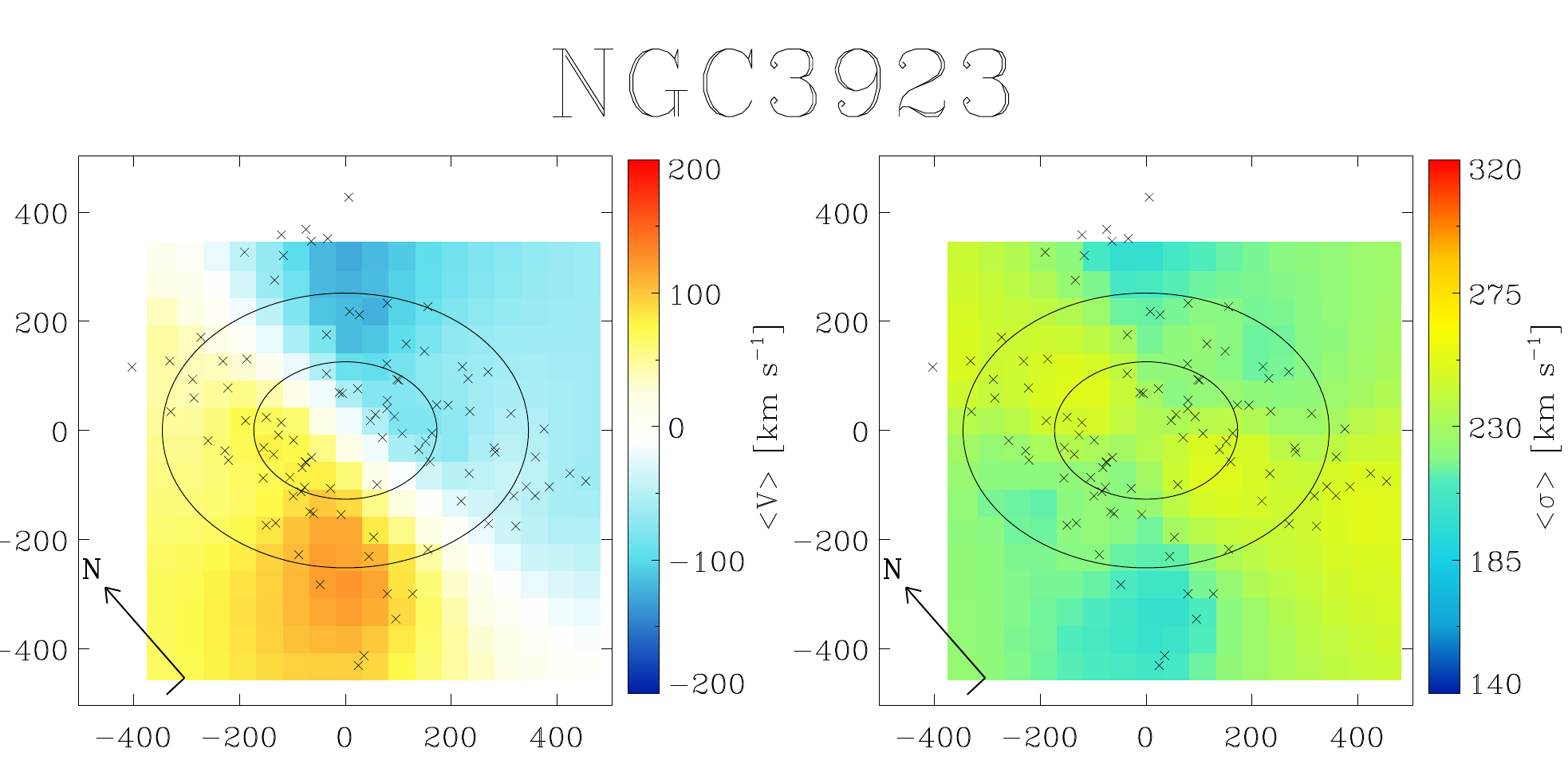}
  \end{minipage}

 \end{figure*}

\addtocounter{figure}{-1} 
\begin{figure*}[t]
  \centering
  \begin{minipage}[b]{1.7\columnwidth}
    \includegraphics[width=\columnwidth]{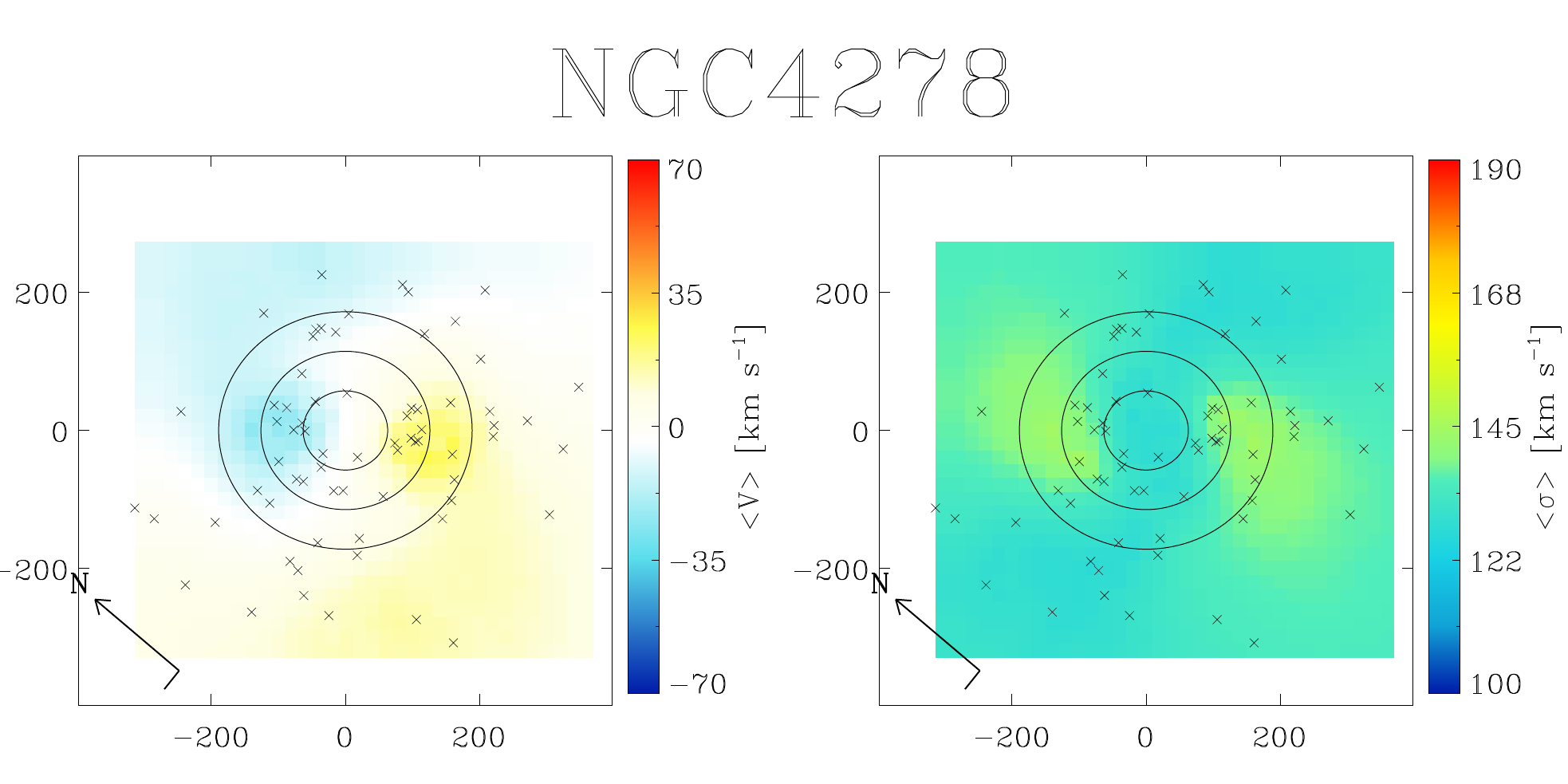}
    \vspace{0.2cm}
     \includegraphics[width=\columnwidth]{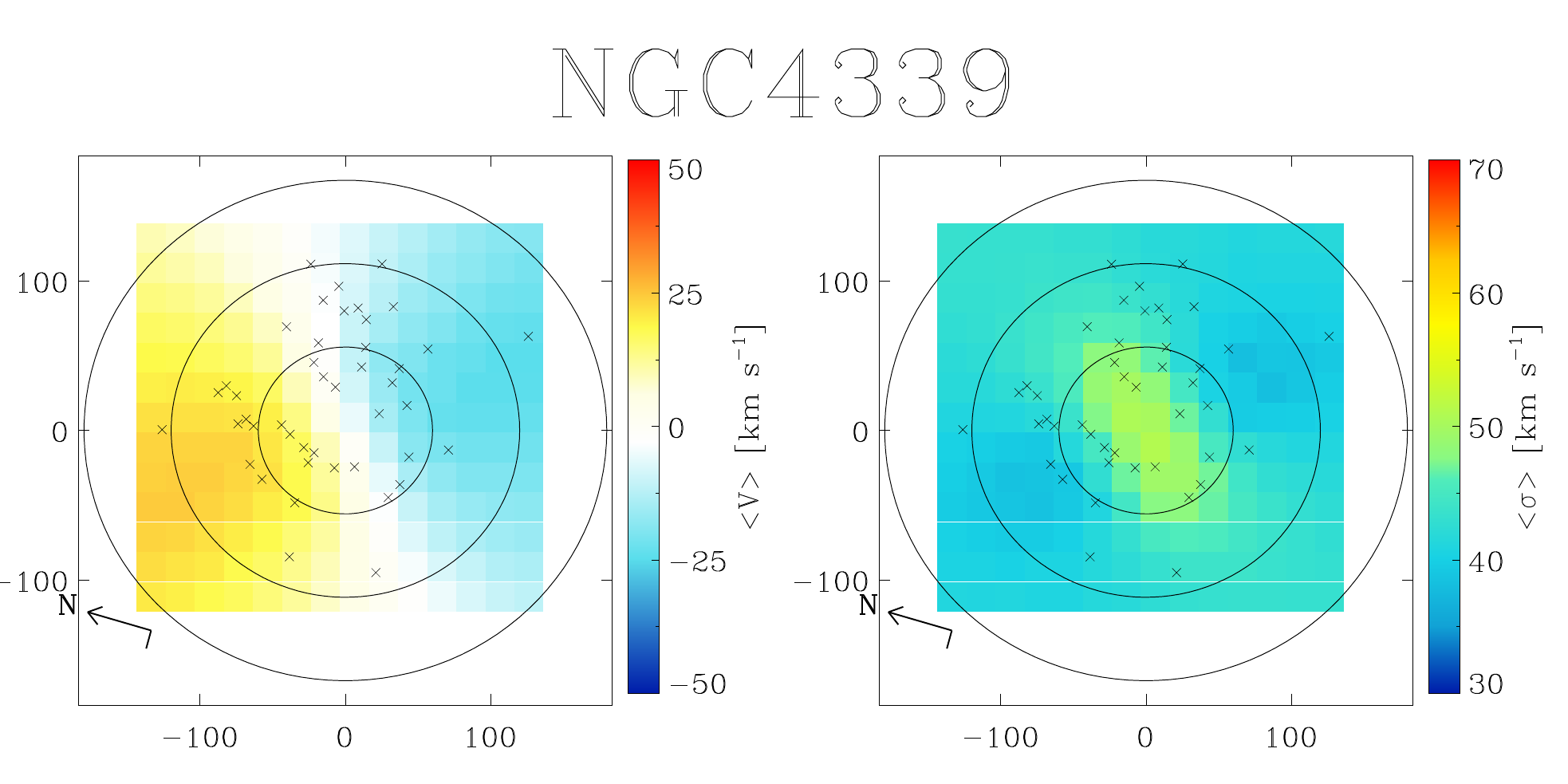}
    \vspace{0.2cm}
     \includegraphics[width=\columnwidth]{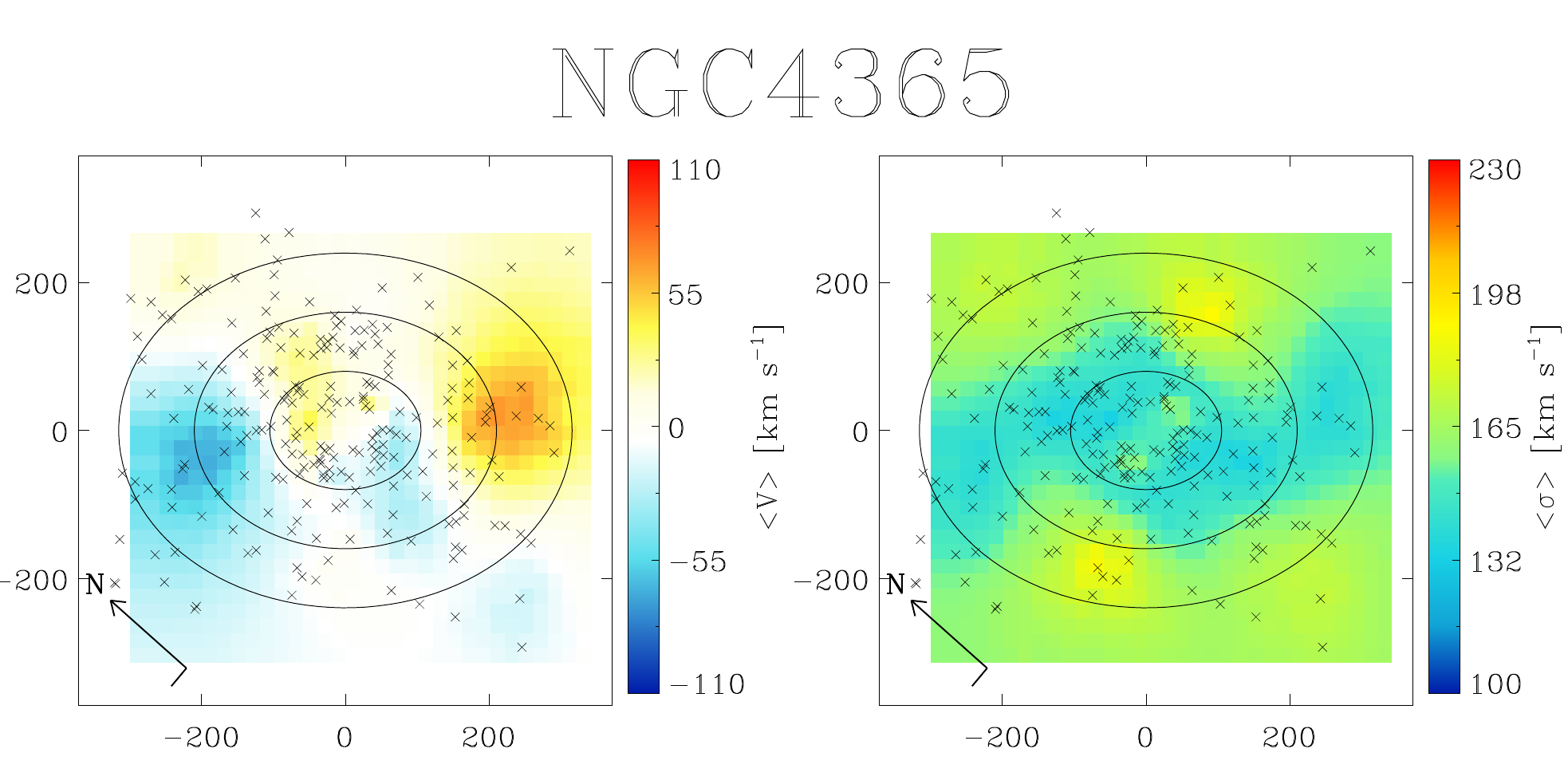}
  \end{minipage}

 \end{figure*}

\addtocounter{figure}{-1} 
\begin{figure*}[t]
  \centering
  \begin{minipage}[b]{1.7\columnwidth}
    \includegraphics[width=\columnwidth]{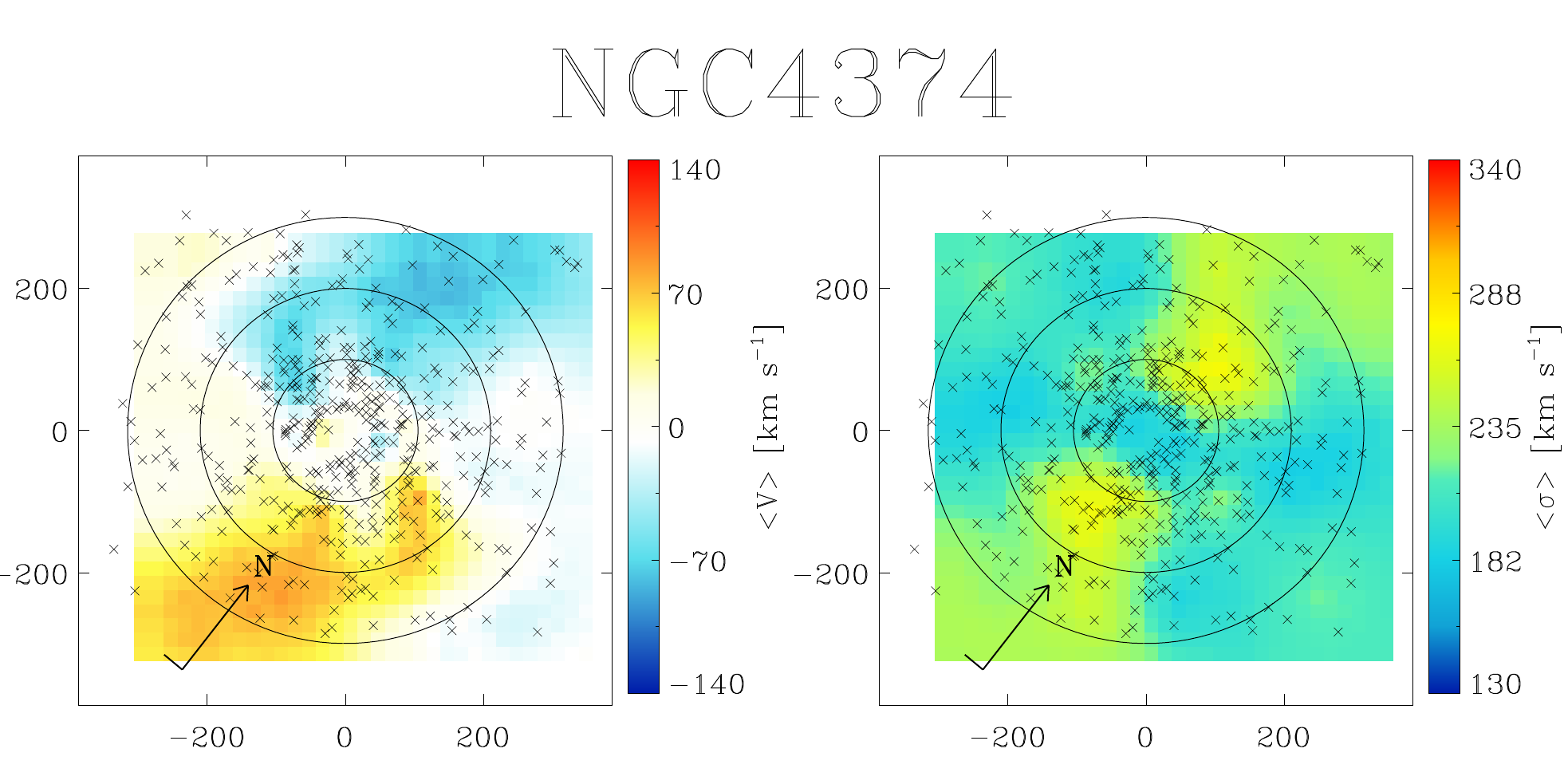}
    \vspace{0.2cm}
     \includegraphics[width=\columnwidth]{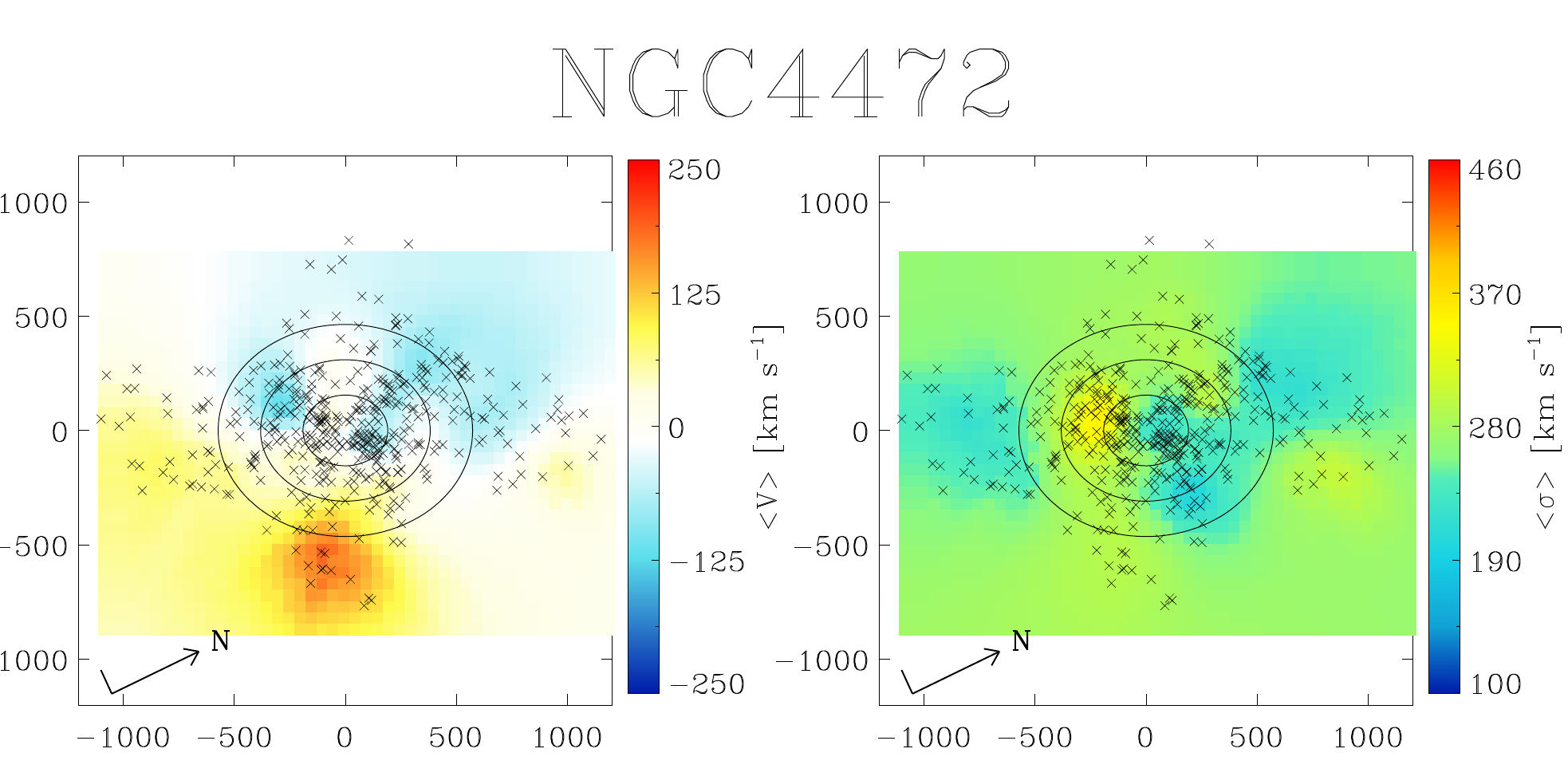}
    \vspace{0.2cm}
     \includegraphics[width=\columnwidth]{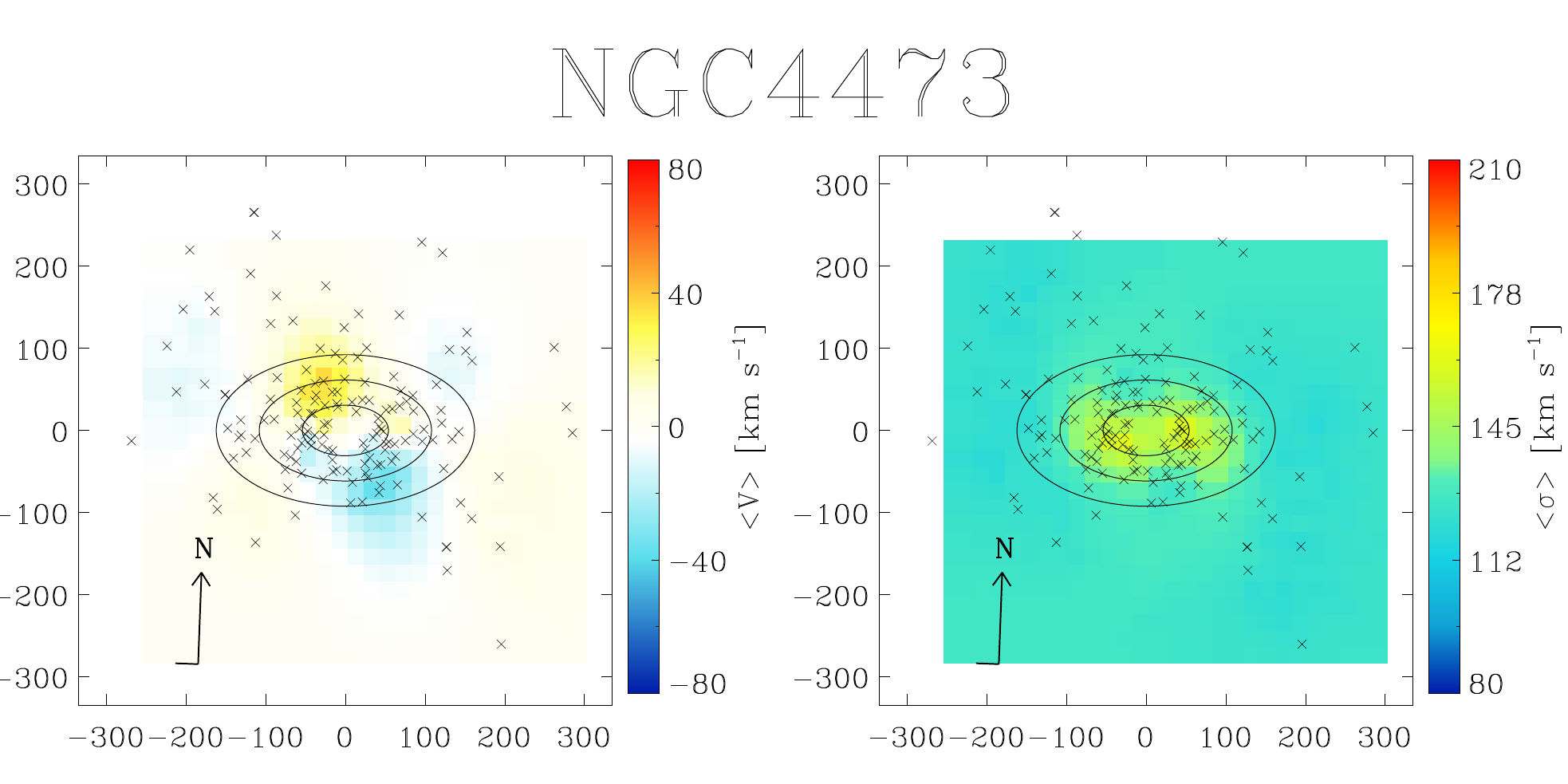}
  \end{minipage}

 \end{figure*}

\addtocounter{figure}{-1} 
\begin{figure*}[t]
  \centering
  \begin{minipage}[b]{1.7\columnwidth}
    \includegraphics[width=\columnwidth]{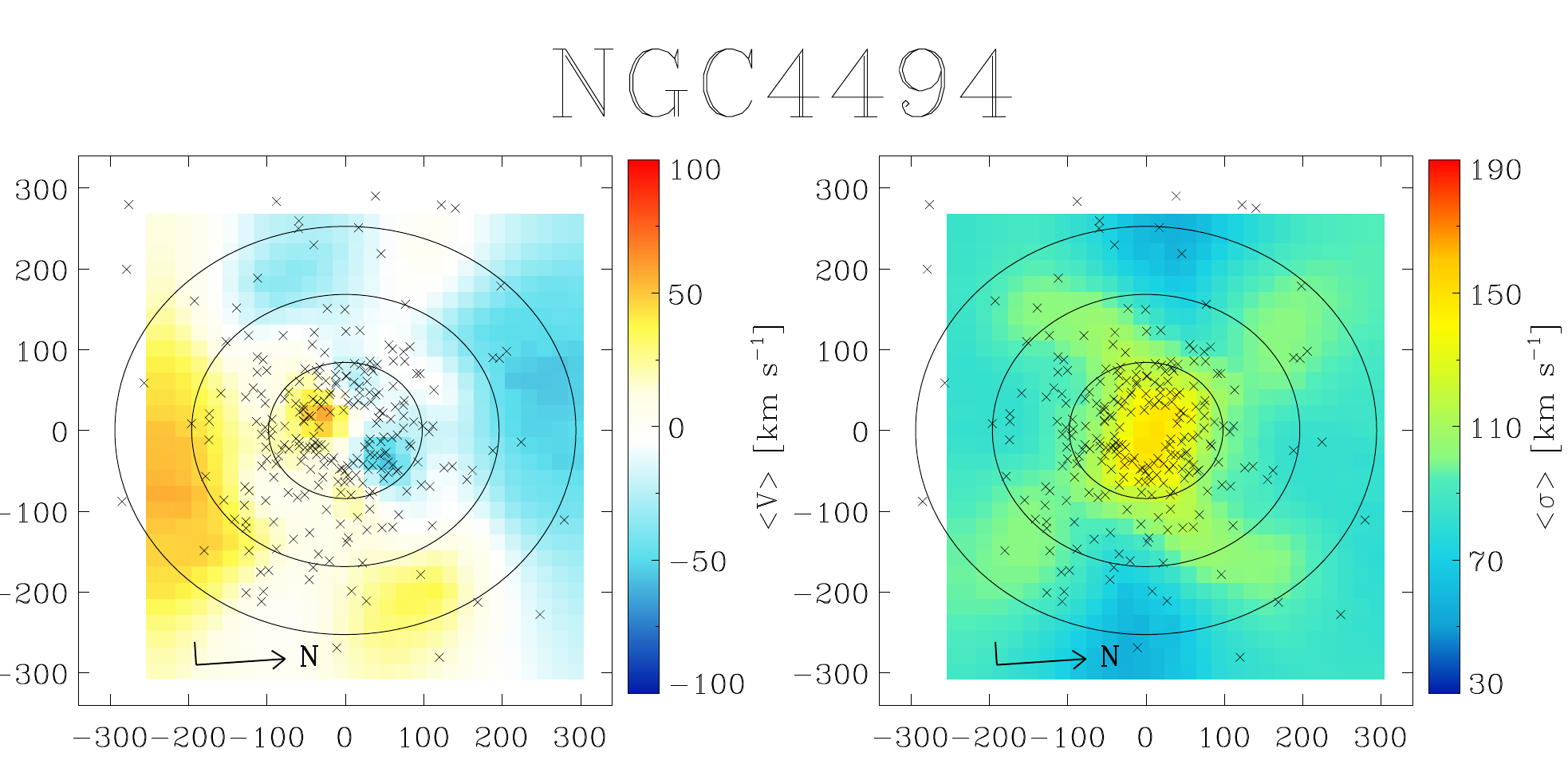}
    \vspace{0.2cm}
     \includegraphics[width=\columnwidth]{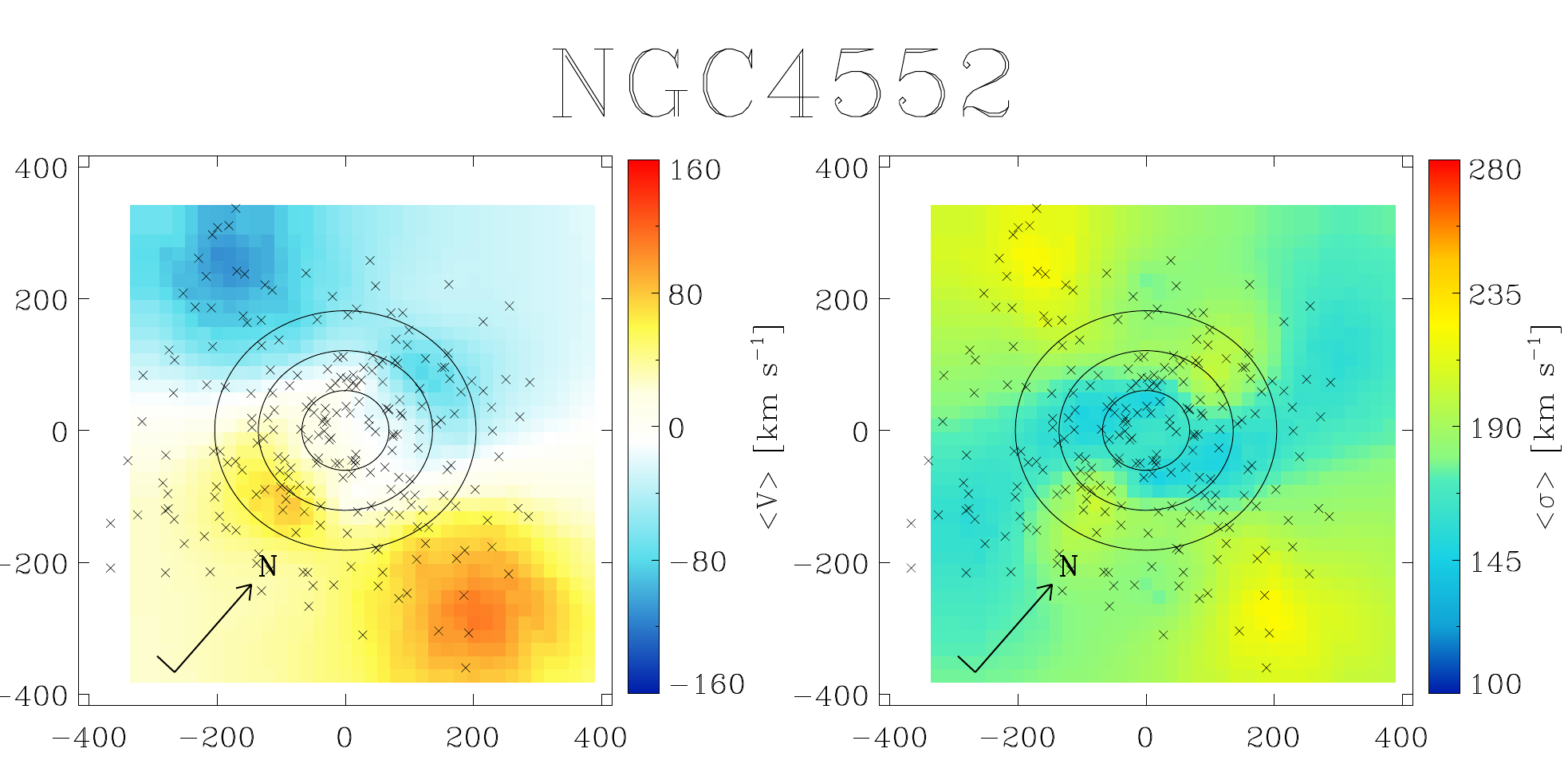}
    \vspace{0.2cm}
     \includegraphics[width=\columnwidth]{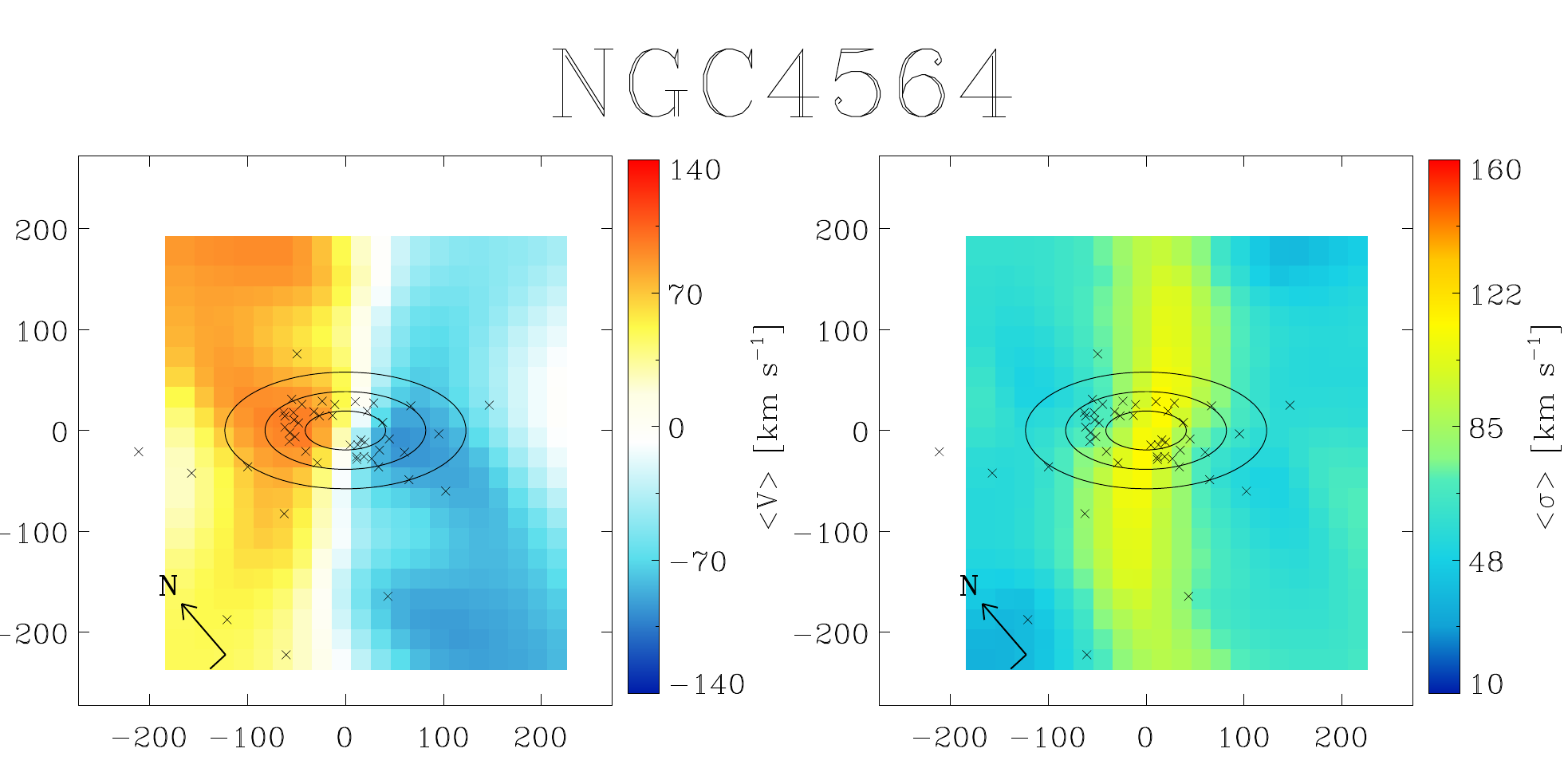}
  \end{minipage}

 \end{figure*}

\begin{figure*}[t]
  \centering
  \begin{minipage}[b]{1.7\columnwidth}
    \includegraphics[width=\columnwidth]{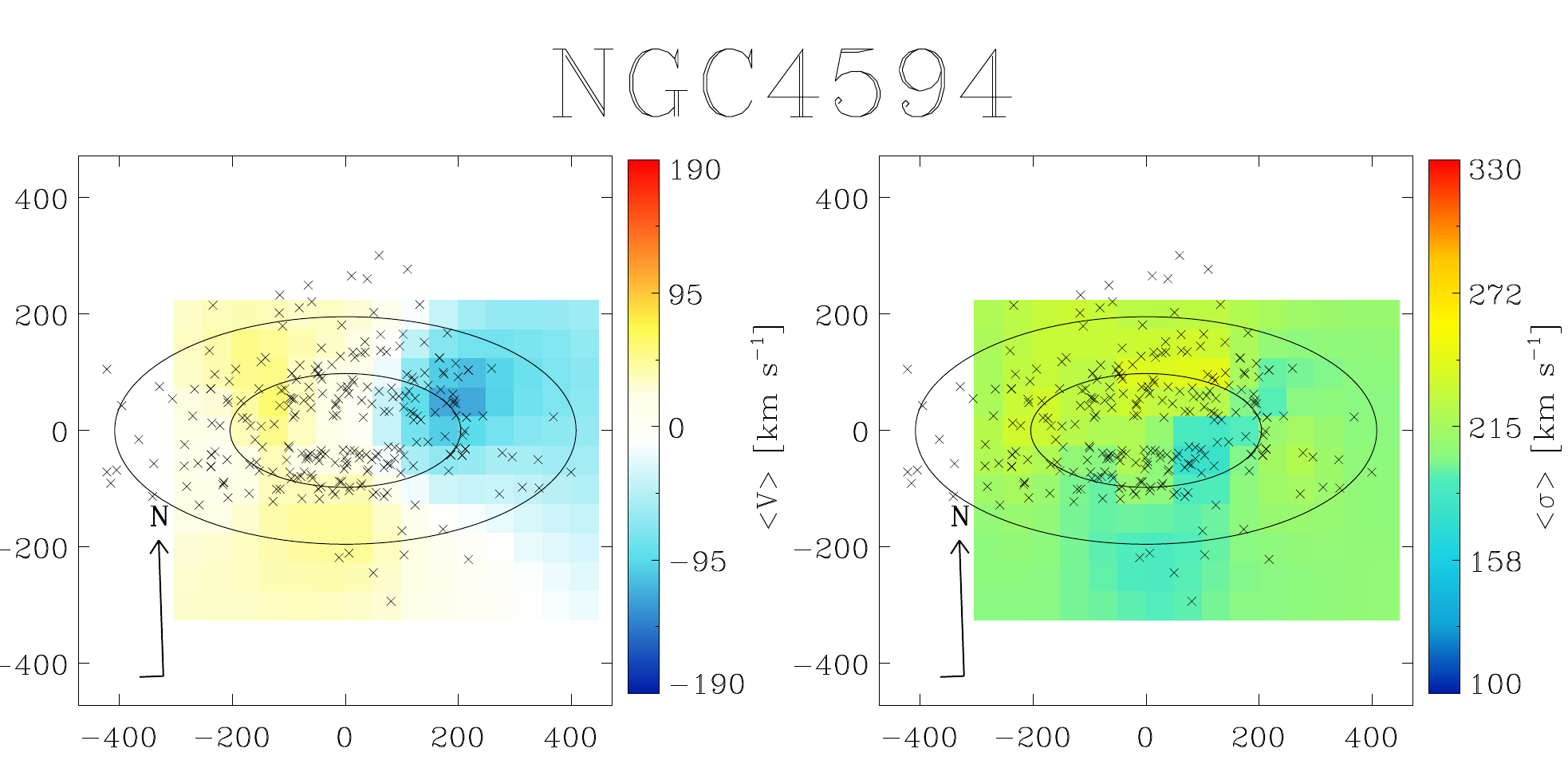}
    \vspace{0.2cm}
     \includegraphics[width=\columnwidth]{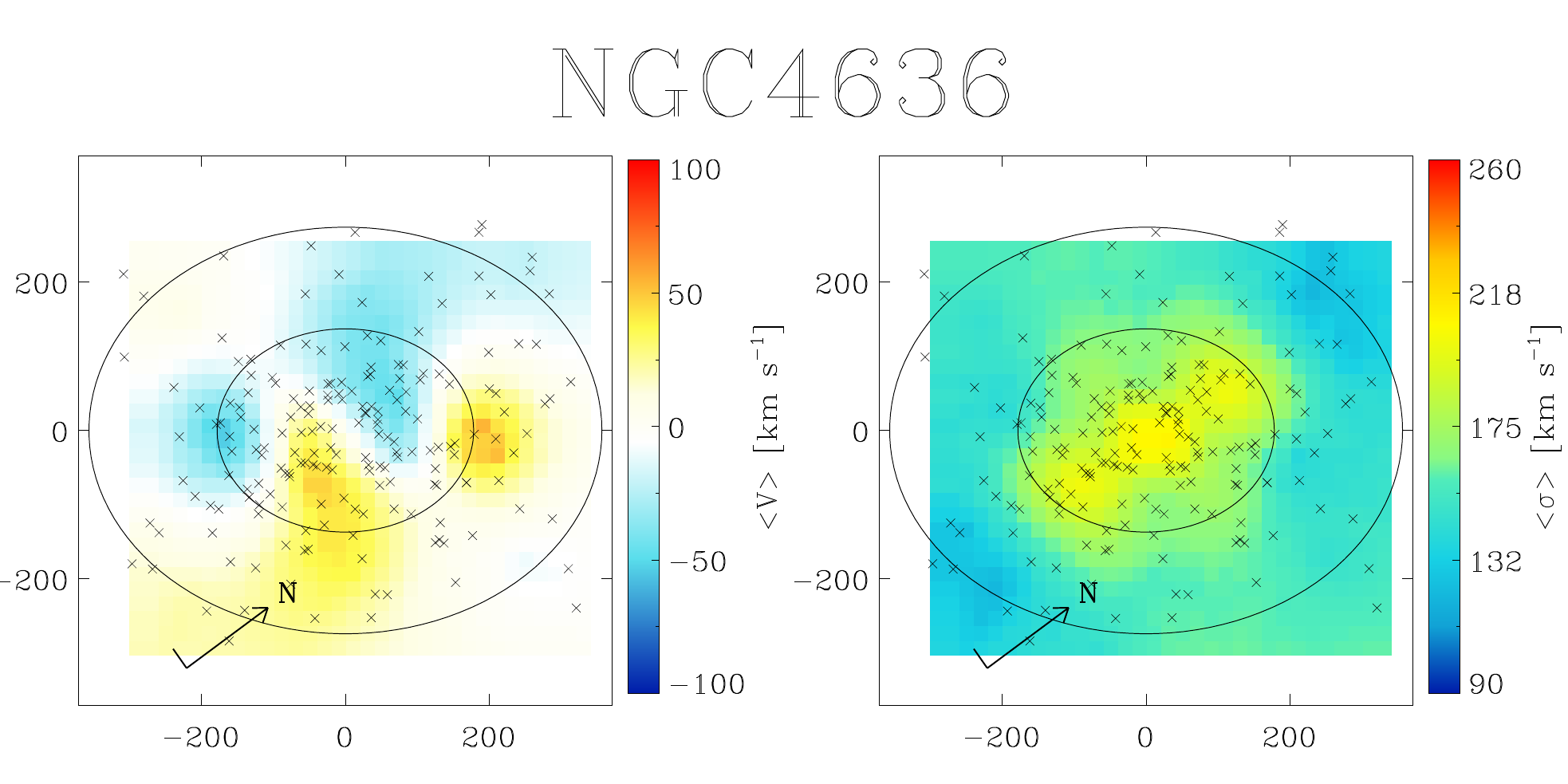}
    \vspace{0.2cm}
     \includegraphics[width=\columnwidth]{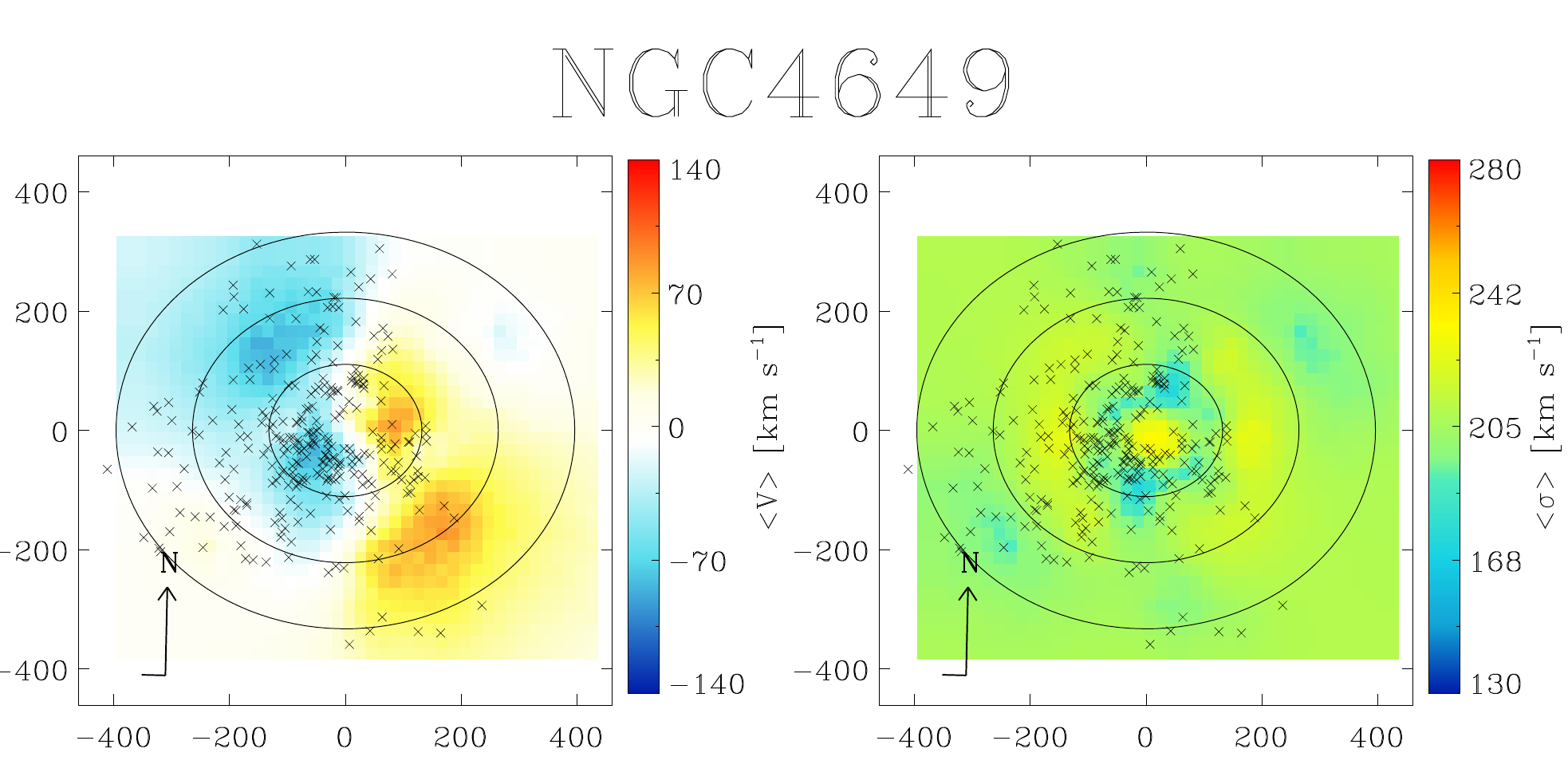}
  \end{minipage}

 \end{figure*}

\addtocounter{figure}{-1} 
\begin{figure*}[t]
  \centering
  \begin{minipage}[b]{1.7\columnwidth}
    \includegraphics[width=\columnwidth]{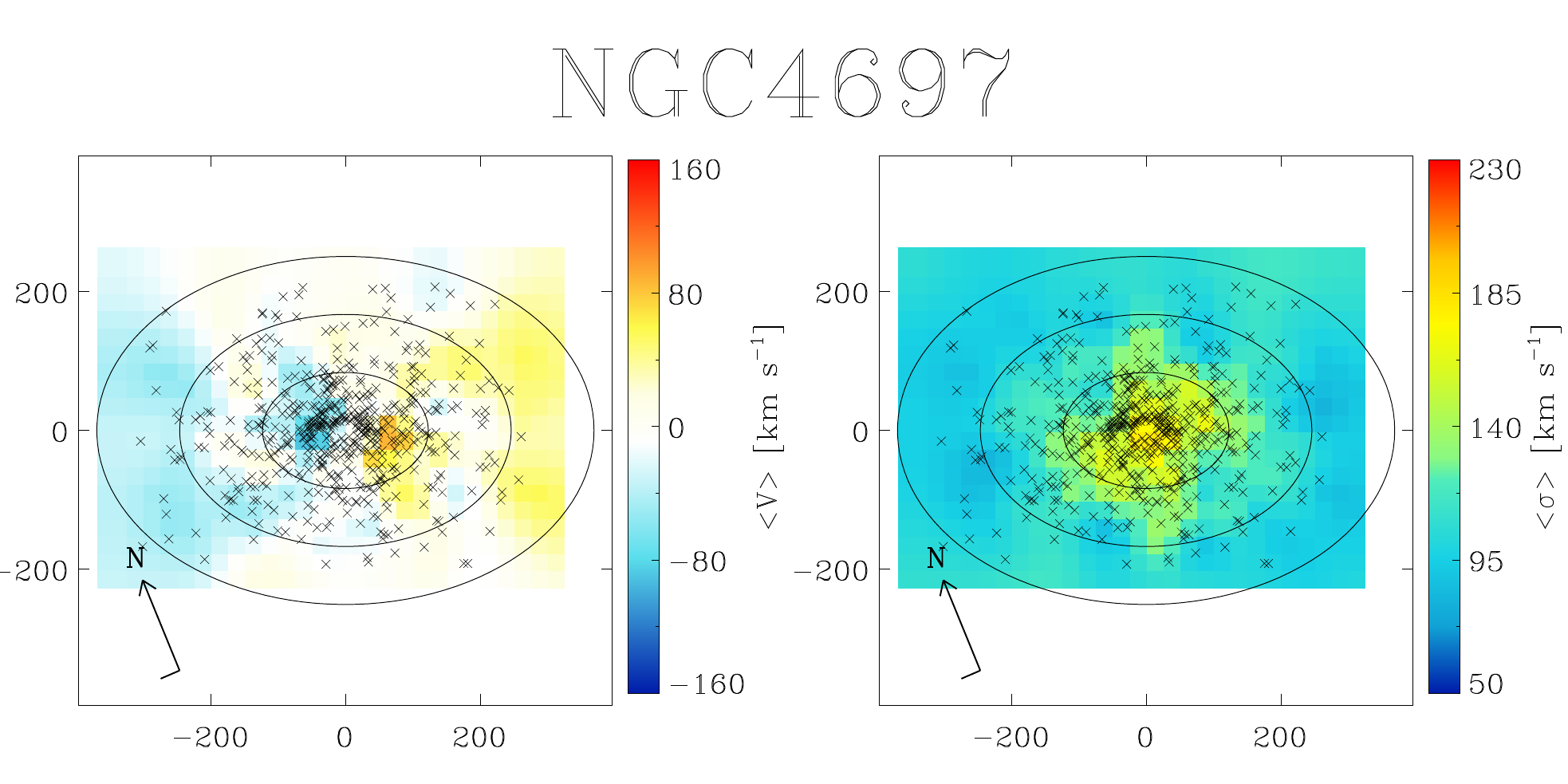}
    \vspace{0.2cm}
     \includegraphics[width=\columnwidth]{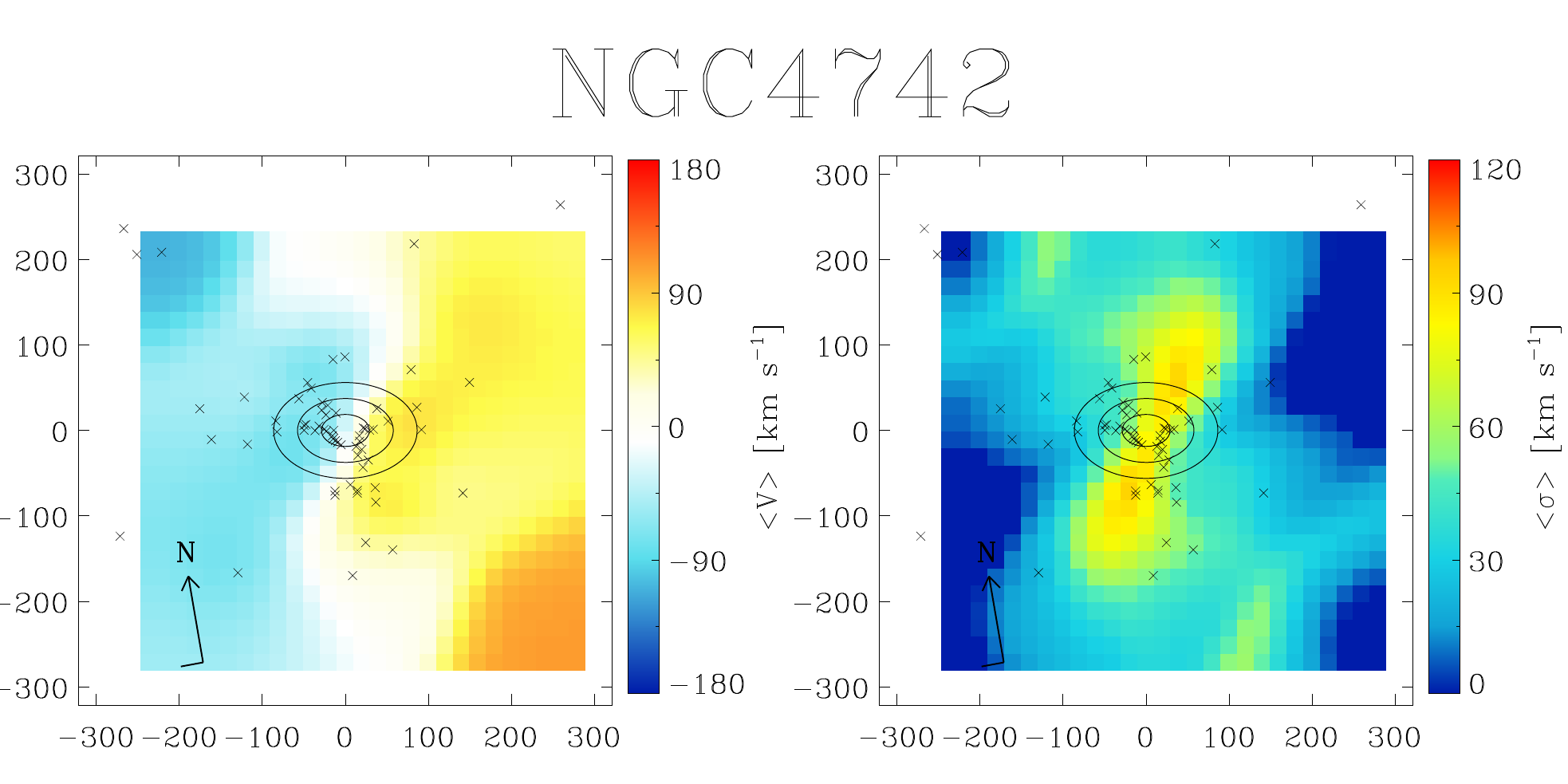}
    \vspace{0.2cm}
     \includegraphics[width=\columnwidth]{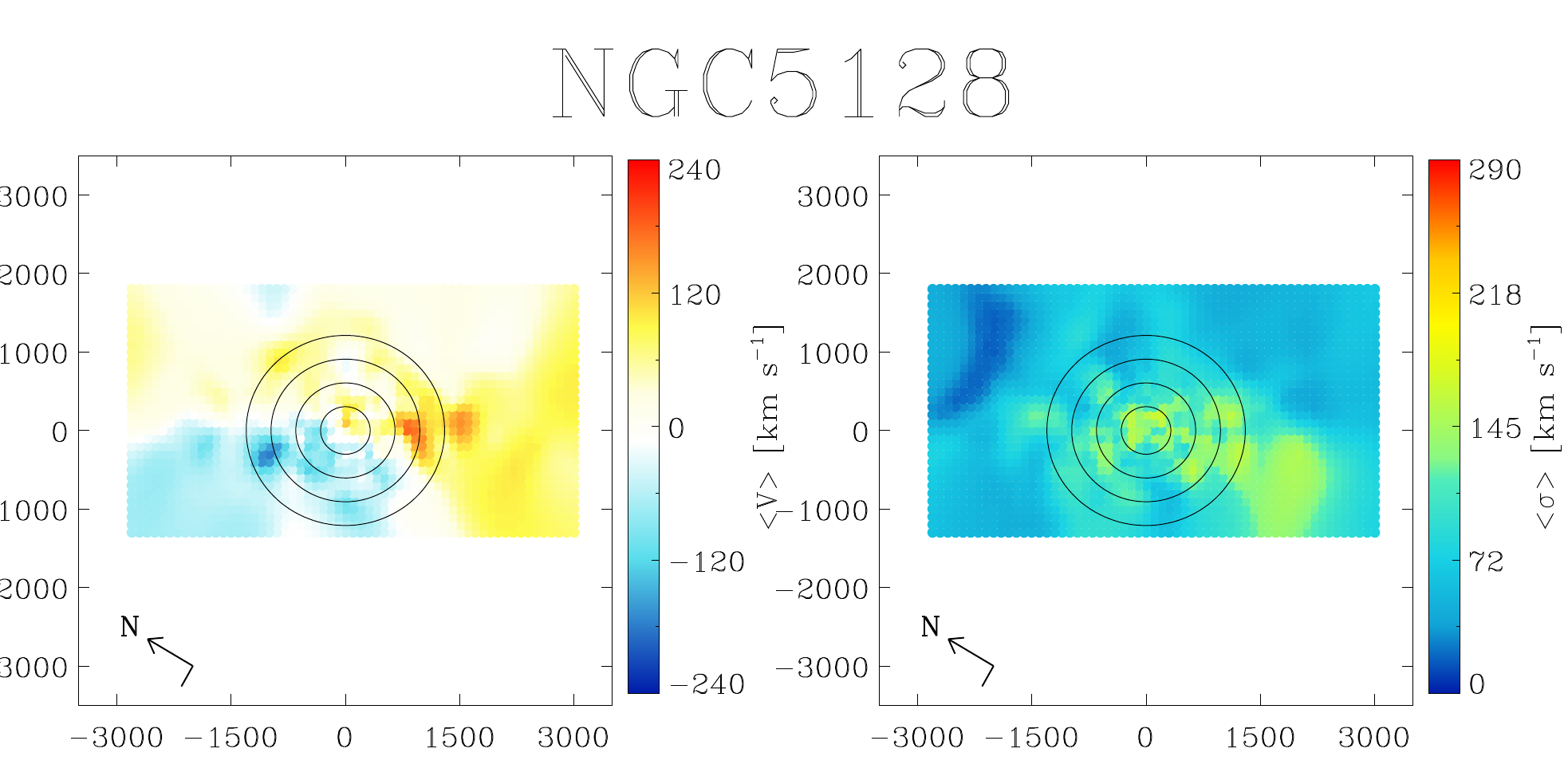}
  \end{minipage}

 \end{figure*}

\addtocounter{figure}{-1} 
\begin{figure*}[t]
  \centering
  \begin{minipage}[b]{1.7\columnwidth}
    \includegraphics[width=\columnwidth]{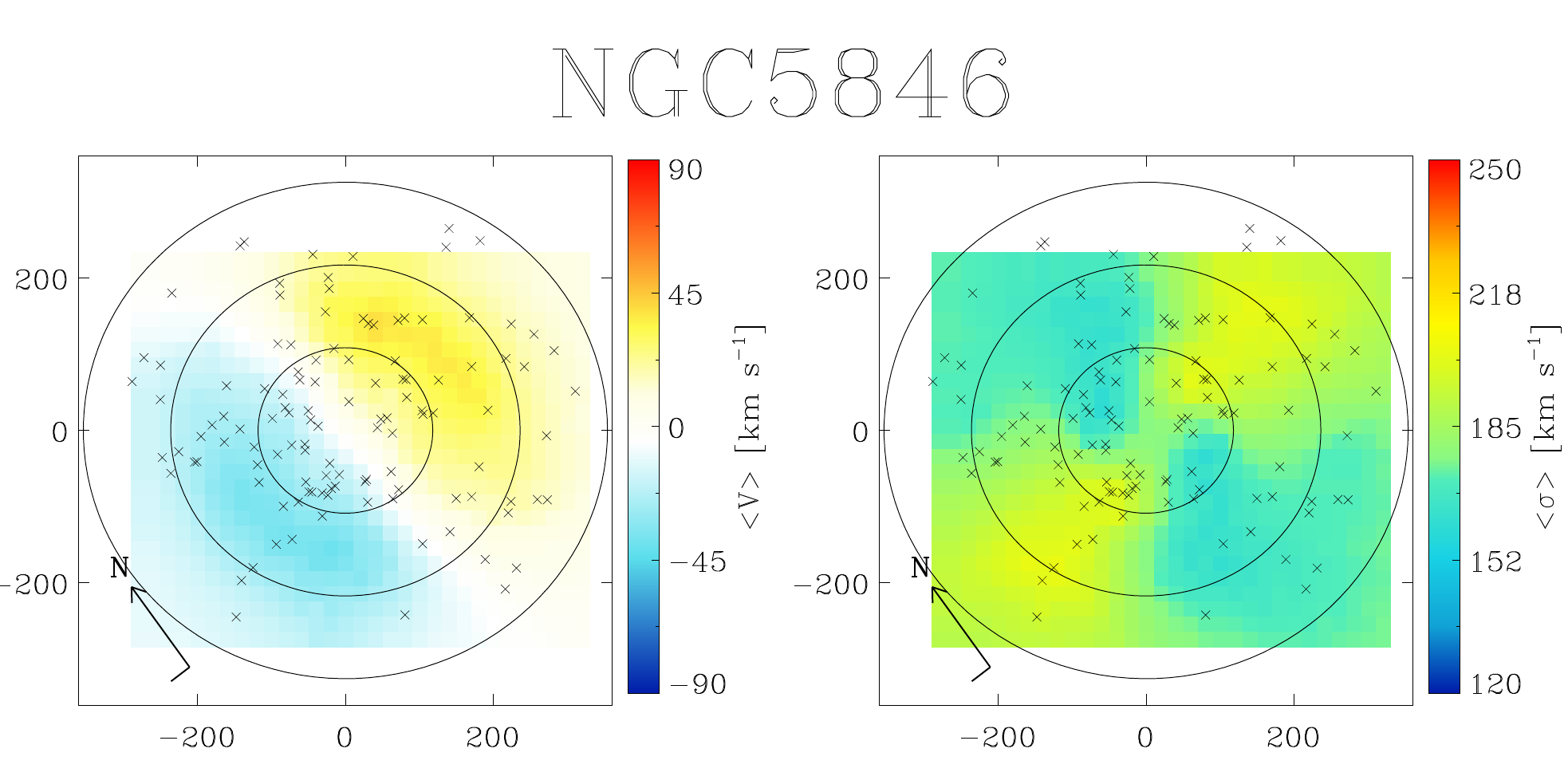}
    \vspace{0.2cm}
     \includegraphics[width=\columnwidth]{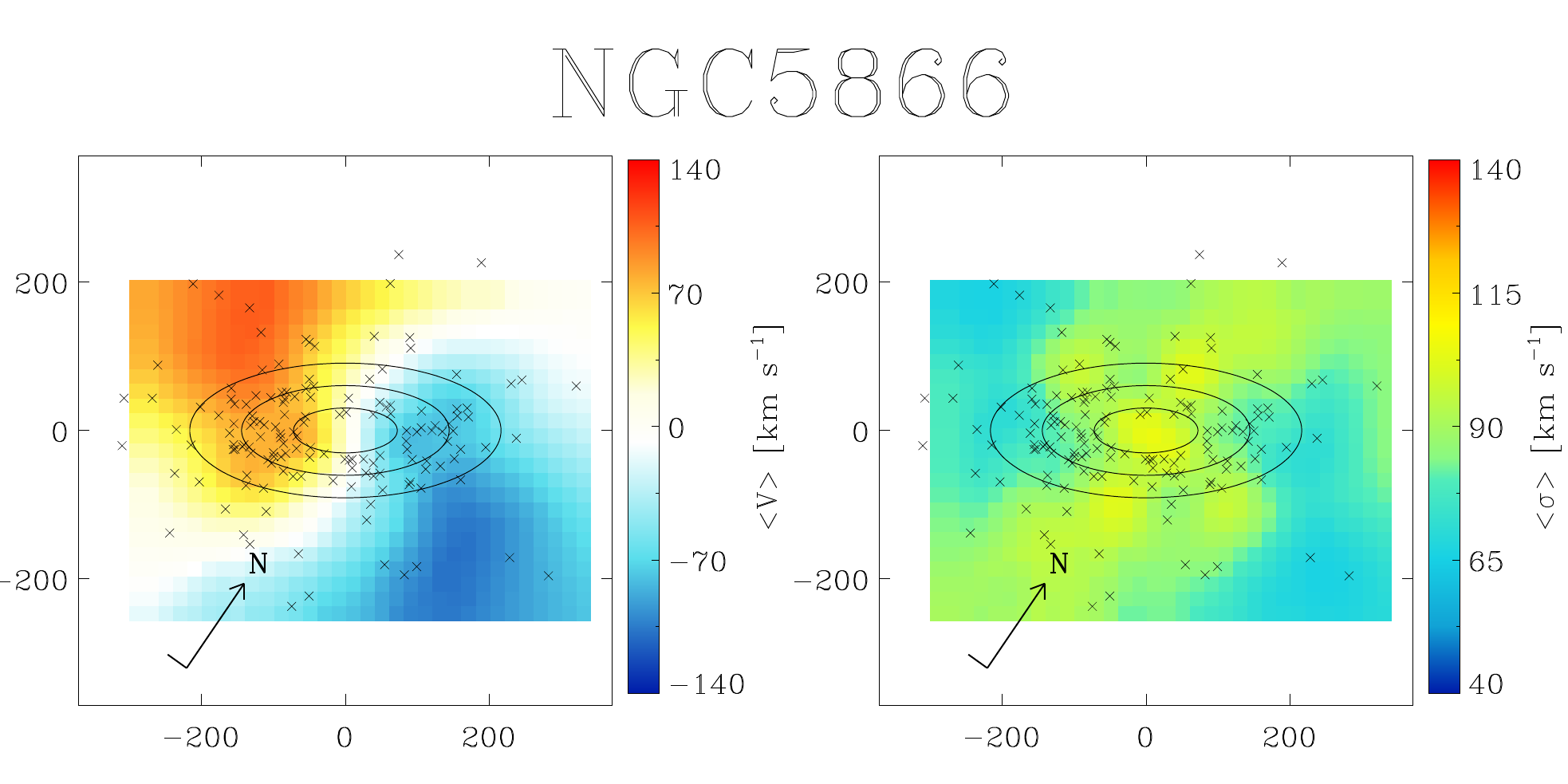}
    \vspace{0.2cm}
     \includegraphics[width=\columnwidth]{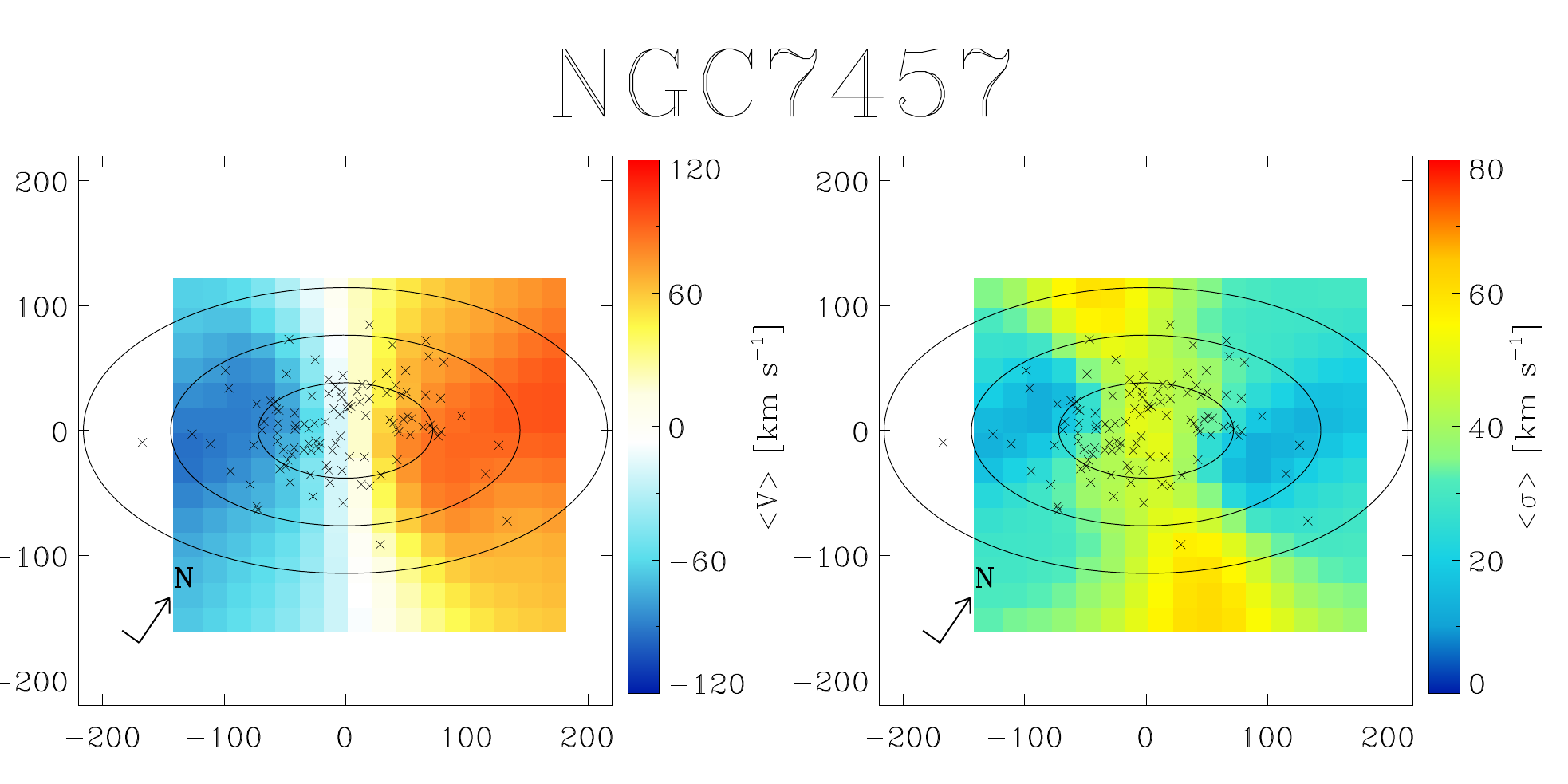}
  \end{minipage}

 \end{figure*}
 
 \end{appendix}

\end{document}